\documentclass[structabstract,english]{aa}

\usepackage{mypackages}

\usepackage[dvipsnames]{xcolor}
\usepackage{amsmath}
\usepackage{graphicx}
\usepackage{booktabs}
\usepackage{placeins}
\usepackage{txfonts}
\usepackage{natbib}
\bibpunct{(}{)}{;}{a}{}{,}
 
\usepackage{ulem}

\begin{document}

\title{CHANG-ES XVI: An in-depth view of the cosmic-ray transport in the edge-on spiral galaxies NGC\,891 and NGC\,4565\thanks{Based on observations with the 100-m telescope of the Max-Planck-Institut f\"ur Radioastronomie (MPIfR) at Effelsberg and the Karl G. Jansky Very Large Array (VLA) operated by the NRAO. The NRAO is a facility of the National Science Foundation operated under agreement by Associated Universities, Inc.}}

\authorrunning{P. Schmidt}
\titlerunning{An in-depth view of the cosmic-ray transport in NGC\,891 and NGC\,4565}
\subtitle{}
\author{Philip Schmidt\inst{1}
    \and
       Marita Krause\inst{1}
    \and
       Volker Heesen\inst{2}
    \and
       Aritra Basu\inst{3}
    \and
       Rainer Beck\inst{1}
    \and
       Theresa Wiegert\inst{4}
    \and
       Judith A. Irwin\inst{4}
    \and
       George Heald\inst{5}
    \and
       Richard J. Rand\inst{6}
    \and
       Jiang-Tao Li\inst{7}
    \and
       Eric J. Murphy\inst{8}
\institute{Max-Planck-Institut f\"ur Radioastronomie, Auf dem H\"ugel 69, 53121 Bonn, Germany \\
\email{pschmidt@mpifr-bonn.mpg.de,mkrause@mpifr-bonn.mpg.de}
\and Hamburger Sternwarte, Universit\"{a}t Hamburg,  Gojenbergsweg 112, 21029 Hamburg, Germany
\and Fakult\"{a}t f\"ur Physik, Universit\"{a}t Bielefeld, Universit\"{a}tsstr. 25, 33615 Bielefeld
\and Department of Physics, Engineering Physics, \& Astronomy, Queen's University, Kingston, Ontario K7L 3N6, Canada
\and CSIRO Astronomy and Space Science, PO Box 1130, Bentley WA 6012, Australia 
\and Department of Physics and Astronomy, University of New Mexico, 800 Yale Boulevard, NE, Albuquerque, NM 87131, USA 
\and Department of Astronomy, University of Michigan, 311 West Hall, 1085 S. University Ave, Ann Arbor, MI 48109-1107, USA 
\and National Radio Astronomy Observatory, 520 Egmont Road, Charlottesville, VA 22903, USA 
}
}
\date{Received 31 December 2018 / Accepted 01 June 2019}

\abstract{Cosmic-ray electrons (CREs) originating from the star-forming discs of spiral galaxies frequently form extended radio  haloes 
that are best observable in edge-on galaxies, where their 
properties can be directly investigated as a function of vertical height above the disc.
}
{
For the present study, we selected two nearby edge-on galaxies from the Continuum Halos in Nearby Galaxies -- an EVLA Survey (CHANG-ES), NGC\,891 and 4565, which differ 
largely in their detectable halo extent and their star-formation rates (SFRs). Our aim is to figure out how such differences are related to the (advective and/or 
diffusive) CRE transport in the disc and in the halo.
}
{
We use wide-band 1.5 and 6\,GHz Very Large Array (VLA) observations obtained in the B, C, and D configurations, and combine the 6\,GHz images with Effelsberg observations to correct 
for missing short spacings. After subtraction of the thermal emission, we investigate the spatially resolved synchrotron spectral index distribution in terms of CRE 
spectral ageing. We further compute total magnetic field strengths assuming equipartition between the cosmic-ray (CR) energy density and the magnetic field, and measure synchrotron 
scale heights at both frequencies. Based on the fitted vertical profiles of the synchrotron intensity and on the spectral index profile between 1.5 and 6\,GHz, 
we create purely advective and purely diffusive CRE transport models by numerically solving the 1D diffusion--loss equation. In particular, we investigate for the first 
time the radial dependence of synchrotron and magnetic field scale heights, advection speeds, and diffusion coefficients, whereas previous studies of these two galaxies 
only determined global values of these quantities.
}
{
We find that the overall spectral index distribution of NGC\,891 is mostly consistent with continuous CRE injection. In NGC\,4565, many of the local synchrotron spectra 
(even in the disc) feature a break between 1.5 and 6\,GHz and are thus more in line with discrete-epoch CRE injection (Jaffe-Perola (JP) or Kardashev-Pacholczyk (KP) models). This implies that CRE 
injection time-scales are lower than the synchrotron cooling time-scales. The synchrotron scale height of NGC\,891 increases with radius, indicating that synchrotron losses are significant. NGC\,891 is probably dominated by advective CRE transport at a 
velocity of $\gtrsim150\,\mathrm{km\,s^{-1}}$.  
In contrast, NGC\,4565 is diffusion-dominated up to $z=1$\,kpc or higher, with a diffusion coefficient of $\geq2\times10^{28}\,\mathrm{cm^2\,s^{-1}}$. 
}


\maketitle

\defcitealias{dumke97}{D97}

\captionsetup[table]{aboveskip=0pt}

\section{Introduction}

Over the past four decades, evidence has accumulated that the majority of star-forming spiral galaxies show  haloes emitting radio continuum, which often extend up to 
several kpc from the disc in vertical direction. Recently, a study of 35 edge-on galaxies in the framework of Continuum Halos in Nearby Galaxies -- an EVLA Survey 
(CHANG-ES, \citealt{irwin12a}) showed that spiral galaxies without significant extraplanar radio emission are, if anything, a rare exception \citep{wiegert15}. 
Still, the brightness and vertical extent of radio haloes vary considerably between individual galaxies.

Radio continuum emission in  haloes is primarily synchrotron radiation of cosmic-ray electrons (and hence a tracer of extraplanar magnetic fields), which have been 
accelerated to relativistic energies in shock fronts of supernova (SN) remnants in the disc. 
That galactic  haloes are closely related to star-forming activity in the underlying disc was first indicated by the discovery of an 
extraplanar diffuse ionised gas component in several star-forming galaxies \citep{rand90,dettmar92a,dahlem94,dahlem95,rossa03a,rossa03b}. 
Since then, disc--halo interactions have largely been investigated by considering models describing the blowout of SN-generated superbubbles in the interstellar medium (ISM) \citep{maclow99}, 
galactic chimneys \citep{norman89}, and galactic fountains, in which the blown-out material eventually condenses and falls back to the disc \citep{shapiro76}.
For radio-continuum emission in particular, the connection to star formation is evident from the far-infrared (FIR)--radio correlation \citep{murphy06,murphy08,li16}, as SN progenitors (i.e. massive stars) are also prominent sources of dust heating.

In spite of substantial evidence for a disc--halo connection via star formation, the vertical scale height of a radio continuum halo is not directly linked 
to star-forming activity \citep{dumke98,krause09,krause11,krause18}. On the other hand, \citet{krause18} found for a CHANG-ES sub-sample of 13 galaxies that the halo scale height increases linearly with the 
diameter of the radio disc. 
The average scale heights they obtained  
are $1.4\pm0.7$\,kpc at 1.5\,GHz and $1.1\pm0.3$\,kpc at 6\,GHz.
The total magnetic field strength is usually highest in the central regions of galaxies, implying higher synchrotron loss rates of cosmic-ray electrons (CREs) at small galactocentric 
radii. A common consequence of this are dumbbell-shaped radio haloes, a prime example of which is that of NGC\,253 \citep{heesen09a}. 

Since the cosmic rays (CRs) are bound to the magnetic field lines and are coupled via the so-called streaming instability to the ionised gas \citep{kulsrud69}, in 
disc--halo outflows all three components are to a certain degree transported together by advection. In addition, diffusion of CRs is possible along or across magnetic 
field lines \citep[e.g.][]{buffie13}.  
The brightness and shape of radio continuum  haloes is influenced considerably by the relative amounts of advective and diffusive vertical cosmic-ray electron (CRE) transport. In 
advection-dominated galaxies, outflow speeds are typically of the order of several hundred  $\mathrm{km\,s^{-1}}$ \citep{heesen18}. 
If the advection speed exceeds the escape velocity of the galaxy, the outflowing gas leaves the gravitational potential of the galactic disc as a galactic wind (along with the CRs), rather than circulating in a galactic fountain flow.

The idea that CRs could in fact be the driving force in galactic winds was initially brought up by \citet{ipavich75}. \citet{hanasz+13} demonstrated that an energy deposit of $10 \% $ from type II supernovae into the ISM as cosmic rays can trigger the local formation of a strong low-density galactic wind driving and maintaining vertically open magnetic field lines at least in star-forming high-redshift galaxies. Analytical models \citep[e.g.][]{breitschwerdt91, zira+96, socrates+08, dorfi12, mao+18} and numerous (magneto-) hydrodynamic simulations \citep[e.g.][]{jubelgas+08, uhlig+12, hanasz+13, girichidis+16, pakmor+16, pfrommer+17, mao+18} of galactic outflows indicate that the formation of a CR-driven wind is possible in star-forming galaxies under a variety of physical conditions in the disc. A CR-driven wind seems to be more effective than a thermal wind, at least at larger distances from the disc \citep{girichidis+18, mao+18}. Owing to the coupling between the CREs 
and the ionised gas, such a wind is expected to remove considerable amounts of mass, energy, and angular momentum \citep{ptuskin97, mao+18} and can thus significantly influence galactic evolution, for example in terms of altered metallicity gradients and star-formation rates (SFRs). Galactic winds are furthermore being debated as a possible mechanism of magnetisation of the intergalactic medium \citep{kronberg99, hanasz+13, farber+18}. 
Evidence of how far magnetic fields in galaxy  haloes extend into the intergalactic medium is, however, observationally limited by the energy losses of CREs.

A detailed investigation of CRE transport in galaxy  haloes requires multi-frequency observations of edge-on galaxies with high spatial resolution. The nearby galaxies NGC\,891 and NGC\,4565 are ideal targets for this purpose.
In this paper, we discuss CHANG-ES observations of NGC\,891 and 4565 at 1.5 and 6\,GHz.

Our paper is structured as follows: 
The galaxies are introduced in Sect.~\ref{galaxies}. In Sect.~\ref{datareduction} we describe the observational setup and the basic data reduction procedure. 
We present and discuss the morphologies in the total intensity images in Sect.~\ref{totalpower}, while in Sect.~\ref{thermalsep} we provide an estimate for the spatial distribution of thermal free--free emission.
Sect.~\ref{spectralindex} deals with 
the spatially resolved total and non-thermal spectral index distribution. 
In Sect.~\ref{scaleheights} we determine the vertical radio scale heights at 1.5 and 6\,GHz and their radial distributions,
and model the vertical emission distribution by solving the diffusion--loss equation 
in Sect.~\ref{CRtransport}. We discuss the implications of our results in Sect.~\ref{discussion}, followed by a summary and outlook 
in Sect.~\ref{sumcon}.

\begin{table*}[h]
\caption{Basic properties of NGC\,891 and 4565}
{\small
\begin{center}
\begin{tabular}{lcccc}
\toprule
\toprule[0.3pt]
 & NGC\,891 & Ref. & NGC\,4565 & Ref. \\
\hline
RA (J2000)  &  02$^{\mathrm{h}}$ 22$^{\mathrm{m}}$ 33$^{\mathrm{s}}$.41 & (1)  &  12$^{\mathrm{h}}$ 36$^{\mathrm{m}}$ 20$^{\mathrm{s}}$.78  & (1)   \\       
DEC (J2000) &  +42\degr 20$^{\mathrm{\prime}}$ 56$^{\mathrm{\prime\prime}}$.9 & (1)  &  +25\degr 59$^{\mathrm{\prime}}$ 15$^{\mathrm{\prime\prime}}$.6  & (1)   \\           
$D$ [Mpc]                                                 &  9.1    &  (2)     &  11.9    &   (2)        \\
Morph. type                                               &  SBb    &  (3)     &  SB(r)b &  (4)         \\    
Inclination [$\degr$]                                     &  89.8   &  (5)     &  86.3    &  (6)         \\
Position angle [$\degr$]                                  &  23.25  &  (5)     &  44.5    &  (6)         \\
$d_{25}\,^{a}$ [$^{\mathrm{\prime}}$]                     &  12.2   &  (1)     &  16.2    &  (1)        \\                                           
SFR [$\mathrm{M_{\odot}\,yr^{-1}}$]                       &  3.3 / 1.55   &  (7)/(8)     &  1.3 / 0.74   &   (7)/(8)        \\      
$\Sigma_{\mathrm{SFR}}$ [$10^{-3}\mathrm{M_{\odot}yr^{-1}kpc^{-2}}$] &  3.13    &  (8)     &   0.73   &   (8)        \\ 
$V_{\mathrm{rot}}$ [$\mathrm{km\,s^{-1}}$]            &  225    &  (9)     &   245   &   (10)        \\      
Total atomic gas mass$^{b}$ [$10^{9}\mathrm{M_{\odot}}$]  &  5.6    &  (11)     &  9.9    &  (12)         \\      
\hline                                    
\end{tabular}
\label{tab:basicparms}
\end{center}
}
{\footnotesize 
$^{a}$ Observed blue diameter at the 25th mag arcsec$^{-2}$ isophote. \\
$^{b}$ Includes multiplication by a factor of 1.36 to account for neutral He. \\
References: (1) \citet{irwin12a}, (2) \citet{radburn-smith11}, (3) \citet{garcia-burillo95}, (4) \citet{kormendy10}, (5) \citet{kregel05}, (6) this work, (7) \citet{krause11}, (8) \citet{wiegert15}, (9) \citet{swaters97}, (10) \citet{heald11}, (11) \citet{oosterloo07}, (12) \citet{zschaechner12}
}
\end{table*}

\section{The galaxies}
\label{galaxies}

\subsection{NGC\,891}

NGC\,891 has been repeatedly dubbed as a twin of the Milky Way, due to its similar optical luminosity \citep{vanderkruit81} and rotation velocity \citep[$225\,\mathrm{km\,s^{-1}}$,][]{rupen91,swaters97}. Depending on methodology, the SFR is found to lie within values of $1.55\,\mathrm{M_{\odot}\,yr^{-1}}$ \citep{wiegert15} and $3.3\,\mathrm{M_{\odot}\,yr^{-1}}$ \citep{krause11}. NGC\,891 contains approximately twice the amount of CO gas as in the Milky Way \citep{scoville93}, therefore it is very likely to have a higher SFR than our Galaxy. Owing to its similarities to our Galaxy, its almost perfectly edge-on inclination, and its easily detectable extraplanar emission in various wavelength regimes, NGC\,891 is a preferred target for investigations of halo properties and the disc--halo connection, and arguably the most extensively studied edge-on galaxy to date. Basic physical parameters of NGC\,891 are listed in Table~\ref{tab:basicparms}.

Early observations of NGC\,891 at 610\,MHz, 1.4\,GHz, and 5\,GHz with the Westerbork Synthesis Radio Telescope (WSRT) by \citet{allen78} allowed for the first time to separate the radio continuum emission of a galaxy into a thin disc (or just `disc') and a thick disc (or halo) component. On the other hand, the vertical spectral index profiles turned out too steep to be explained by basic CRE transport models \citep{strong78}. Observations with the Effelsberg 100-m telescope at 8.7\,GHz \citep{beck79} and at 10.7\,GHz \citep{klein84a} showed that a lack of large-scale structure in the 5\,GHz map of \citet{allen78} due to missing short-spacings was responsible for this inconsistency. Later, \citet{hummel91a} observed NGC\,891 at 327 and 610\,MHz with the WSRT and at 1.49\,GHz with the Karl G. Jansky Very Large Array (VLA). The observed vertical extent of the halo at each frequency and the resulting spectral index profiles showed first indications of consistency with the theoretical models for disc--halo advection flows by \citet{lerche82}. 
\citet{dumke98} for the first time combined interferometric (VLA) and single-dish (Effelsberg) data of NGC\,891 (at 4.85 GHz) and measured average exponential scale heights of 0.27\,kpc for the disc and 1.82\,kpc for the halo. 

Recently, NGC\,891 was observed with the Low Frequency Array (LOFAR) at 146\,MHz \citep{mulcahy18} and is hence the first edge-on galaxy investigated with LOFAR. The scale height of the non-thermal halo emission at 146\,MHz varies between 1.6 and 3.6\,kpc, increasing with decreasing magnetic field strengths, which is a signature of dominating synchrotron losses of CREs. Comparison with the CHANG-ES data at 1.5\,GHz gave discrepant results whether CRE propagation in the halo is diffusive or advective.

The geometry of the magnetic field is known to be plane-parallel in the disc \citep{sukumar91,dumke95}, while in the halo it is X-shaped \citep{krause09,krause11}.
Using a depolarisation model, \citet{hummel91b} estimated an average magnetic field scale height of 3.6\,kpc for the halo.
Substantial extraplanar emission has further been detected from neutral hydrogen (H\,{\sc i}) \citep{oosterloo07}, diffuse ionised gas 
\citep[e.g.][]{dettmar92b,rand98}, dust 
\citep[e.g.][]{hughes14},
polycyclic aromatic hydrocarbons (PAHs) \citep{whaley09}, and X-ray emitting hot ionised gas 
\citep[e.g.][]{li13b}.

\begin{table*}[h]
\caption{VLA observations of NGC\,891: observation, calibration, and imaging parameters}
{\small
\begin{center}
\begin{tabular}{lccccccc}
\toprule
\toprule[0.3pt]
Frequency$^{a}$  & \multicolumn{4}{c}{1.5\,GHz (L-band)} & \multicolumn{3}{c}{6\,GHz (C-band)} \\
\cmidrule(lr){2-5}  \cmidrule(lr){6-8}
Configuration & B & C & D & B+C+D & C & D & C+D \\ 
\toprule[0.3pt]
Obs. date                                 & 2012/06/24 & 2012/02/11 & 2011/12/16 & N/A  & 2012/02/06--07 & 2011/12/09 & N/A  \\
                                          &            & 2012/04/01 & 2013/03/17 &  &               & 2011/12/18    \\
Obs. time$^{b}$                           & 121 min    & 47 min     & 30 min     & N/A & 368 min       & 79 min     & N/A \\
Primary calibrator                           & 3C 48 & 3C 48 & 3C 48 & N/A  & 3C 48 & 3C 48 & N/A  \\
Secondary calibrator                          & J0230+4032 & J0314+4314 & J0314+4314 & N/A  & J0251+4315 & J0230+4032 & N/A  \\
$b_{\mathrm{maj}}$ [$\arcsec$]$^{c}$         & 3.16 & 10.64 / 12.00 & 36.42 & 4.09 / 12.00 & 2.77 & 9.00 / 12.00 & 3.88   \\
$b_{\mathrm{min}}$ [$\arcsec$]$^{c}$         & 2.90 &  9.66 / 12.00 & 32.48 & 3.73 / 12.00 & 2.61 & 8.81 / 12.00 & 3.50   \\
Beam position angle [$\degr$]                 & 54.22 & 88.06 & $-74.29$ & 50.65  & 85.93 & $-79.46$ & 80.49  \\
Pixel size [$\arcsec$]                       & 0.5 & 2.5 & 5 & 0.5 & 0.5 & 2.5 & 0.5   \\
Scales for MS-\texttt{CLEAN} [$\arcsec$]           & 0, 6, 15 & 0, 10, 20, 40 & 0, 20, 100, 200 & 0, 6, 15,  & 0, 6, 14 & 0, 10, 20, 40 & 0, 6, 15  \\
                                          &          &               &                 & 50, 100, 200 &  &  & \\
Self-calibration                          & 1x phase & no & no & N/A$^{d}$  & 1x amp+phase & 1x amp+phase & N/A$^{d}$  \\
Outer $uv$-taper [k$\lambda$]                         & $-$ & $-$ & $-$ & 30 & $-$ & $-$ & 34    \\
Noise rms [$\mu$Jy\,beam$^{-1}$]                  & 16 & 28 & 60 & 14.5 / 24$^{e}$  & 3.1 & 6.5 / 5.9$^{e}$ & 2.7  \\
\bottomrule[0.3pt]                                
\end{tabular}
\label{tab:imgparsN891L}
\end{center} 
}
{\footnotesize 
$^{a}$ The central frequency of the respective band is given. The total bandwidth is 512\,MHz for L-band and 2.048\,GHz for C-band. \\
$^{b}$ On-source time (before flagging) \\
$^{c}$ Major- and minor-axis FWHM of the synthesised beam (the intermediate-resolution maps were smoothed to $12\arcsec$ FWHM after deconvolution). \\
$^{d}$ N/A in this case means that the specified self-calibrations were already applied to the single-array data sets before producing the combined-array images. \\
$^{e}$ Noise rms before and after smoothing the map to $12\arcsec$ FWHM
}
\end{table*}

\begin{table*}[h]
\caption{VLA observations of NGC\,4565: observation, calibration, and imaging parameters}
{\small
\begin{center}
\begin{tabular}{lcccccccc}
\toprule
\toprule[0.3pt]
Frequency$^{a}$  & \multicolumn{4}{c}{1.5\,GHz (L-band)} & \multicolumn{3}{c}{6\,GHz (C-band)} \\
\cmidrule(lr){2-5}  \cmidrule(lr){6-8}
Configuration & B & C & D & B+C+D & C & D & C+D \\ 
\toprule[0.3pt]
Obs. date                                 & 2012/06/02--03 & 2012/04/02 & 2011/12/30 & N/A  & 2012/04/16 & 2011/12/29 & N/A  \\
Obs. time$^{b}$                           & 125 min    & 43 min     & 19 min     & N/A & 356 min       & 76 min     & N/A\\
Primary calibrator                           & 3C 286 & 3C 286 & 3C 286 & N/A  & 3C 286 & 3C 286 & N/A \\
Secondary calibrator                          & J1221+2813 & J1221+2813 & J1221+2813 & N/A  & J1221+2813 & J1221+2813 & N/A  \\
$b_{\mathrm{maj}}$ [$\arcsec$]$^{c}$               & 3.31 & 10.49 / 12.00 & 34.50 & 10.16 / 12.00 & 2.63 & 9.02 / 12.00 & 3.57   \\
$b_{\mathrm{min}}$ [$\arcsec$]$^{c}$               & 3.01 & 10.01 / 12.00 & 32.32 & 9.40 / 12.00  & 2.59 & 8.82 / 12.00 & 3.52  \\
Beam position angle [$\degr$]                 & 45.50 & 64.64 & $-89.06$ & 47.75  & $-61.72$ & 85.48 & $-83.98$  \\
Pixel size [$\arcsec$]                       & 0.5 & 2.5 & 5 & 0.5 & 0.5 & 2.5 & 0.5   \\
Scales for MS-\texttt{CLEAN} [$\arcsec$]           & 0, 6, 15 & 0, 10, 20 & 0, 35, 80 & 0, 6, 15, & 0, 5, 13 & 0, 10, 40 & 0, 5, 13    \\
                                          &          &           &           & 35, 70, 105 &  &  &  \\
Self-calibration                          & no & 1x amp+phase & 1x phase & N/A$^{d}$  & 1x phase & no & N/A$^{d}$  \\
Outer $uv$-taper [k$\lambda$]             & $-$ & $-$ & $-$ & 50 & $-$ & $-$ & 34   \\
Noise rms [$\mu$Jy\,beam$^{-1}$]                  & 14 & 20.5 & 34 & 20  & 3.2 & 7.4 / 7.0$^{e}$ & 2.5  \\
\bottomrule[0.3pt]                                
\end{tabular}
\label{tab:imgparsN4565L}
\end{center}
}
{\footnotesize 
$^{a}$ The central frequency of the respective band is given. The total bandwidth is 512\,MHz for L-band and 2.048\,GHz for C-band. \\
$^{b}$ On-source time (before flagging) \\
$^{c}$ Major- and minor-axis FWHM of the synthesised beam (the intermediate-resolution maps were smoothed to $12\arcsec$ FWHM after deconvolution). \\
$^{d}$ N/A in this case means that the specified self-calibrations were already applied to the single-array data sets before producing the combined-array images. \\
$^{e}$ Noise rms before and after smoothing the map to $12\arcsec$ FWHM
}
\end{table*}

\subsection{NGC\,4565}

Compared with NGC\,891, NGC\,4565 has a distinctly lower SFR ($1.3\,\mathrm{M_{\odot}\,yr^{-1}}$, \citealt{krause11}; or even as low as $0.74\,\mathrm{M_{\odot}\,yr^{-1}}$, \citealt{wiegert15}) and, at $86\fdg 3$, is somewhat less inclined. Further basic properties of the galaxy are given in Table~\ref{tab:basicparms}.

The first radio continuum detection of NGC\,4565 (WSRT at 610\,MHz and 1.4\,GHz, \citealt{hummel84}), indicated the presence of a thick radio disc, with follow-up observations \citep{broeils85} allowing for a tentative detection of a thin disc. 
From combined VLA and Effelsberg data at 4.85\,GHz, \citet{dumke98} determined average exponential scale heights of 0.28\,kpc for the thin disc and 1.68\,kpc for the thick disc or halo, which is similar to what they measured for NGC\,891, in spite of the considerably smaller minor to major axis ratio (not only at radio wavelengths) of NGC\,4565. 
  
NGC\,4565 does not have a particularly massive or extended H\,{\sc i} halo \citep{zschaechner12}, suggesting that there is no significantly strong disc--halo interaction. 
H$\alpha$ imaging by \citet{rand92} showed that the H\,{\sc ii} content of the galaxy is basically limited to the compact star-forming regions in the disc plane and hence does not include any notable extraplanar diffuse ionised gas component. This appears to be consistent with its low SFR and its low extraplanar H\,{\sc i} content.

\section{Observations and data reduction}
\label{datareduction}

\subsection{VLA}
\label{vlareduction}

The CHANG-ES observations were carried out with the VLA, using the L-band (1.5\,GHz) and C-band (6\,GHz) receivers, with bandwidths of 512\,MHz 
and 2.048\,GHz, 
respectively. In order to probe a variety of spatial scales, the B-, C-, and D-array configurations were used at 1.5\,GHz, and the C- and D-array configurations at 6\,GHz. 
The FWHM of the VLA primary beam $\theta_{\mathrm{PB}}$ is $30^{\prime}$ at 1.5\,GHz and $7\farcm 5$ at 6\,GHz. Therefore, the C-band observations of all CHANG-ES galaxies with $d_{25}>1.3\times\theta_{\mathrm{PB}}=9\farcm 75$ (which includes both NGC\,891 and 4565) were performed with two pointings, which were placed on the major axis on either side of the galaxy centre, separated by $\approx\theta_{\mathrm{PB}}/2$. 
Observational parameters specific to each array configuration and frequency band are given in Tables~\ref{tab:imgparsN891L} and \ref{tab:imgparsN4565L}.

We carried out most of the basic data reduction as described comprehensively in \citet{irwin13} and \citet{wiegert15}, using the Common Astronomy Software Applications (\texttt{CASA}) package \citep{mcmullin07}.
After applying standard procedures for RFI excision and calibration, we imaged the data in total intensity (Stokes $I$) with the multi-scale multi-frequency synthesis algorithm \citep[MS--MFS \texttt{CLEAN},][]{rau11} in wide-field mode \citep{cornwell08b}. For MFS we used two Taylor terms, to take account of the spectral behaviour and produce maps of the spectral index $\alpha$ (we will refer to these as in-band spectral index maps throughout this paper).  
Briggs robust-0 weighting \citep{briggs95}, which results in a good compromise between angular resolution and sensitivity, was applied to the visibility data before imaging. Self-calibration was done in those cases where it improved the image quality (see Tables~\ref{tab:imgparsN891L} and \ref{tab:imgparsN4565L}). 

All maps were corrected for primary-beam (PB) attenuation, including a correction of the $\alpha$ maps for the frequency dependence of the PB. As we discuss in Sect.~\ref{revpbcor}, the NRAO-supplied PB model at 6.0\,GHz turned out to be unsuitable for the correction of our C-band D-array maps. Here, we could achieve more accurate results by using PB models at 6.25--6.6\,GHz instead. 

In addition to imaging each array configuration separately, images were also produced from a combination of the (B-), C- and D-array data for each frequency band. 
Tables~\ref{tab:imgparsN891L} and \ref{tab:imgparsN4565L} contain a summary of the imaging parameters for each data set. To obtain images at a common angular resolution at both frequencies, we produced C-band D-array, L-band C-array, and L-band B+C+D-array total intensity and spectral index maps with a beam full-width at half-maximum (FWHM) of $12\arcsec$. 

\begin{table*}[h]
\caption{Integrated flux densities of  NGC\,891 and 4565 contained in the various total intensity images. All values are given in mJy.}
{\small
\begin{center}
\begin{tabular}{lccc}
\toprule
\toprule[0.3pt]
 & NGC\,891 & NGC\,4565$^{a}$ \\ 
\hline
L-band B-array & $322\pm 16$ & $31.8\pm 1.6$ \\ 
L-band C-array & $560\pm 28$ & $73.7\pm 3.7$ \\ 
L-band B+C+D-array & $695\pm 35$ & $135\pm 7$ \\ 
L-band B+C+D-array merged with D-array & $737\pm 37$ & $139\pm 7$ \\ 
L-band D-array & $737\pm 37$ & $146\pm 7$ \\ 
\hline
Thermal (1.5\,GHz) & $53\pm 23$ & $15\pm 7$ \\ 
Non-thermal$^{b}$ (1.5\,GHz) & $684\pm 44$ & $124\pm 10$ \\ 
\hline
\hline
C-band C-array & $125\pm 6$ & $14.9\pm 0.7$ \\ 
C-band C+D-array & $157\pm 8$ & $22.2\pm 1.1$ \\ 
C-band D-array & $207\pm 10$ & $36.6\pm 1.8$ \\ 
C-band D-array merged with 6\,GHz Effelsberg & $252\pm 27$ & $39.8\pm 5.9$ \\ 
C-band C+D-array merged with 6\,GHz Effelsberg & $252\pm 27$ & $39.5\pm 5.9$ \\ 
\hline
6\,GHz Effelsberg & $255\pm 14$ & $48.2\pm 3.9$ \\ 
Thermal (6\,GHz) & $46\pm 20$ & $13\pm 6$ \\ 
Non-thermal$^{b}$ (6\,GHz) & $206\pm 34$ & $27\pm 9$ \\ 
\hline                                    
\end{tabular}
\label{tab:fluxdensities}
\end{center}
}
{\footnotesize 
$^{a}$ Flux densities of NGC\,4565 were determined after subtracting the central source of the galaxy from the VLA images. \\
$^{b}$ For each galaxy, to obtain the non-thermal map at 1.5\,GHz, the thermal map was subtracted from the combination of the L-band D-array and B+C+D-array maps. At 6\,GHz, the thermal map was subtracted from the combination of the C-band D-array and Effelsberg maps.  \\
}
\end{table*}

\subsection{Effelsberg}

\begin{figure*}[h]
 \centering
 \topinset{\it a)}{\includegraphics[scale=0.163,clip=true,trim=186pt 145pt 0pt 0pt]{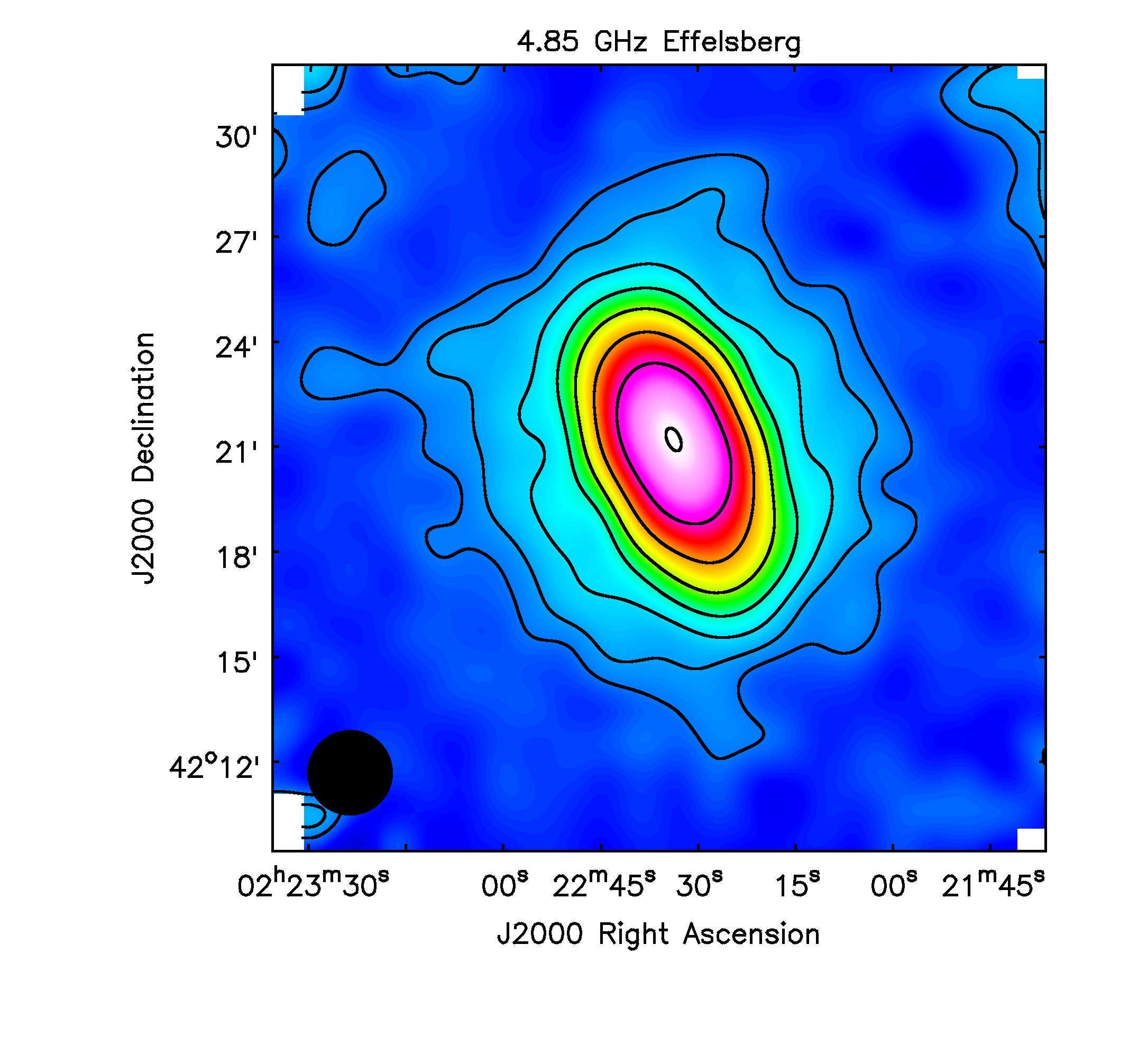}}{5pt}{0pt}\topinset{\it b)}{\includegraphics[scale=0.168,clip=true,trim=226pt 146pt 0pt 32pt]{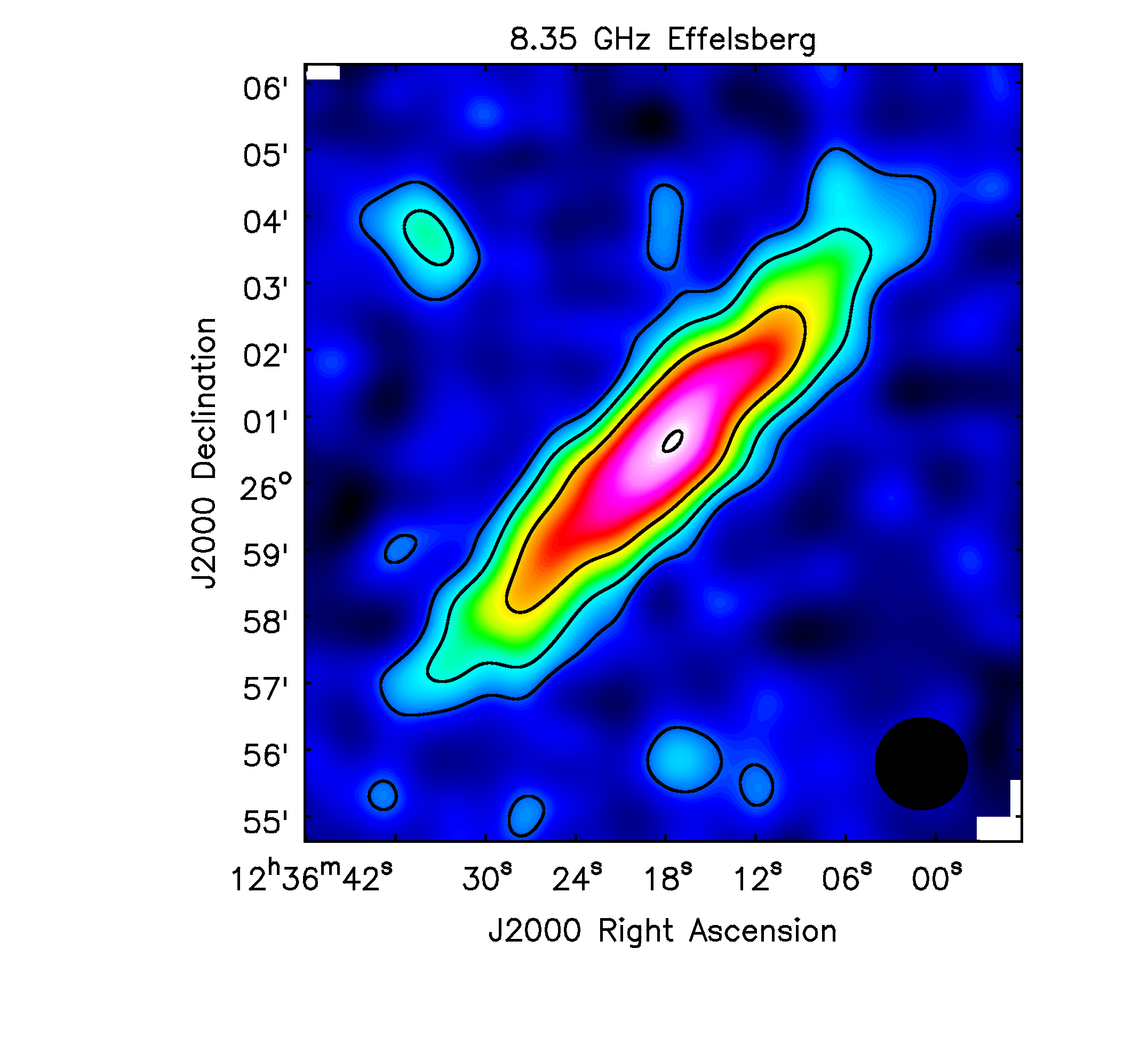}}{0pt}{0pt}
 \caption{
 \it a\normalfont: Effelsberg map of the total intensity distribution of NGC\,891 at 4.85\,GHz (FWHM of the synthesised beam: $147^{\prime\prime}$, rms noise: $\sigma=325\,\mu\mathrm{Jy\,beam^{-1}}$).
 \it b\normalfont: Effelsberg map of the total intensity distribution of NGC\,4565 at 8.35\,GHz (FWHM of the synthesised beam: $83\farcs 58$, rms noise: $\sigma=260\,\mu\mathrm{Jy\,beam^{-1}}$).
 Contour levels in both maps are $\sigma \times (3, 6, 12, 24, \rm{etc})$. Filled black circles indicate the size of the synthesised beam.
 }
 \label{fig:tp0}
\end{figure*}

To fill in missing short spacings\footnote{The largest angular size scale detectable by the VLA (D-array) is $16\farcm 2$ at 1.5\,GHz and $4\arcmin$ at 6\,GHz.} at 6\,GHz, we used observations with the Effelsberg 100-m telescope at 4.85 and 8.35\,GHz. The 4.85\,GHz map of NGC\,4565 was already published by \citet{dumke97}, the 8.35\,GHz map of NGC\,891 by \citet{krause09}. The observations of NGC\,4565 at 8.35\,GHz and of NGC\,891 at 4.85\,GHz 
are published here for the first time, and are shown in Fig.~\ref{fig:tp0}.

NGC\,891 was observed in 2014 at 4.85\,GHz using the dual-beam receiver. We took 21 coverages, with a final map size of $35^{\prime}\times25^{\prime}$. 
The 8.35\,GHz observations of NGC\,4565 were performed in 2003 with 31 coverages and a total scanned area of $20\arcmin \times 15\arcmin$, orientated along the major axis of the galaxy. 
In both cases, the data reduction was performed with the \texttt{NOD3} software package \citep{mueller17}. Flux calibration was done based on 3C\,286 and the flux density scale of \citet{baars77}.

\subsection{Short-spacing corrections}
\label{shortspacings}

Using the Effelsberg observations, a map of the spectral index between 
4.85 and 8.35\,GHz was computed  
for each galaxy (after smoothing the 8.35\,GHz map to the resolution at 4.85\,GHz, i.e. $147\arcsec$). 
Based on this spectral index map, we rescaled the 4.85\,GHz map to 6\,GHz, and combined the rescaled map with the VLA C-band D-array map.

No single-dish observations at (or near) 1.5\,GHz with sufficient sensitivity are presently available for either galaxy. On the other hand, we expect that the D-array data are not too severely affected by missing short spacings, since the largest angular scale (LAS) we can recover is $16\arcmin$, which is not exceeded by our galaxies. 
However, at higher resolutions (even when combining B- and/or C-array with D-array data), it turned out impossible to achieve the same integrated flux density as in the D-array maps, even if large MS-\texttt{CLEAN} scales are used (see Table~\ref{tab:fluxdensities}). At 1.5\,GHz therefore, we used the D-array maps to at least partly correct for missing short-spacings in the combined B+C+D-array maps. We keep in mind that the missing spacings problem is possibly still present to some degree in these 
maps, and will discuss its possible effects on our science results where appropriate. 

The short-spacing corrections at both 1.5 and 6\,GHz were performed in the image domain, using the \texttt{ImMerge} task in \texttt{NOD3}. 
This program convolves the interferometric image with the beam of the single-dish observation and subtracts this convolved map from the single-dish map. The resulting difference map, which is essentially a map of the large-scale emission missing from the interferometric observation, is normalised to the interferometric resolution 
by multiplication with the ratio of the solid angles of the two observing beams $\theta_{\mathrm{interf.}}^{2}/\theta_{\mathrm{SD}}^{2}$. The normalised difference map is then added to the interferometric map to yield the short-spacing corrected image.

\begin{figure*}[h]
 \centering
 \topinset{\it a)}{\includegraphics[scale=0.163,clip=true,trim=264pt 145pt 0pt 0pt]{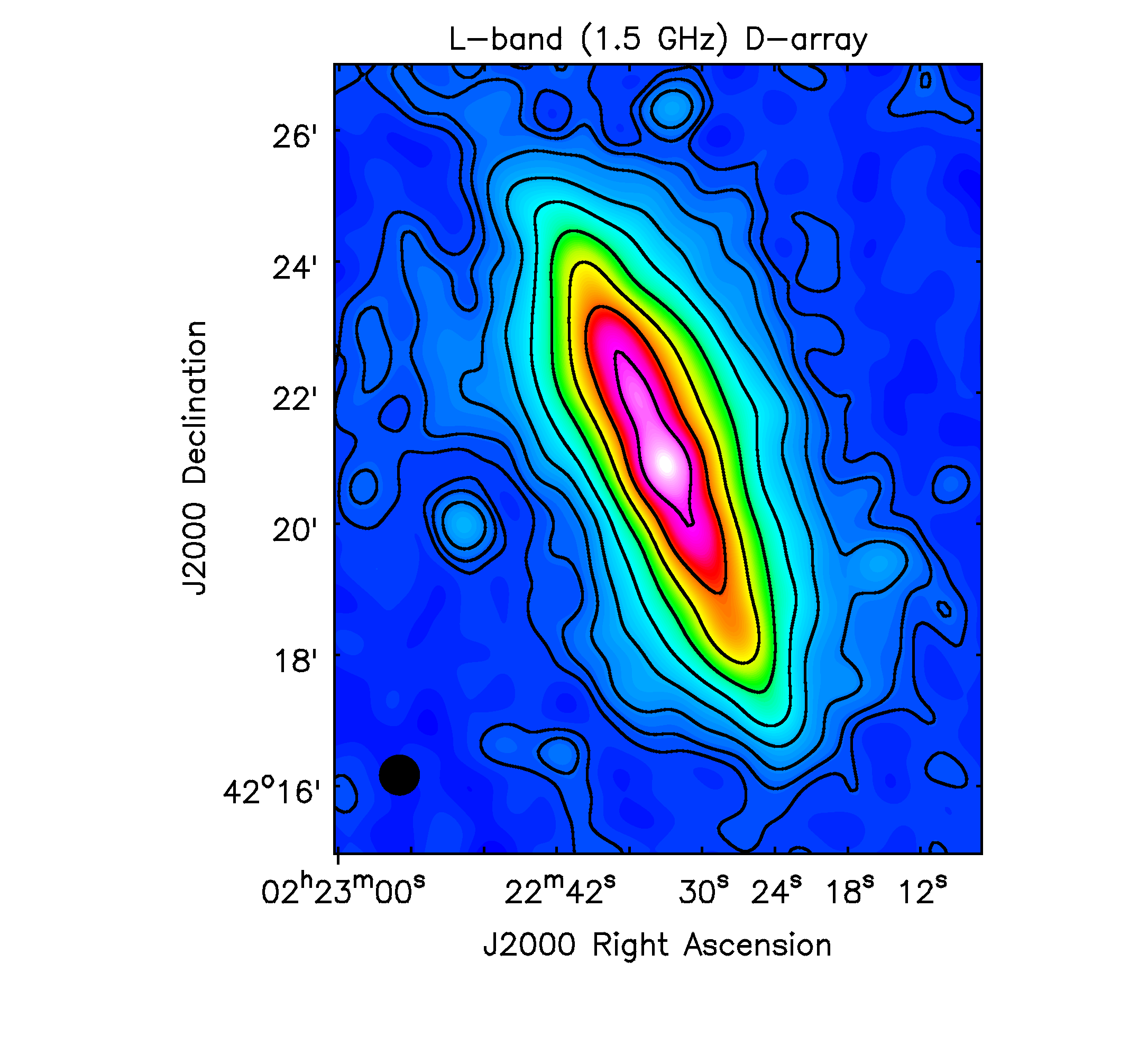}}{5pt}{0pt}\topinset{\it b)}{\includegraphics[scale=0.163,clip=true,trim=226pt 146pt 0pt 32pt]{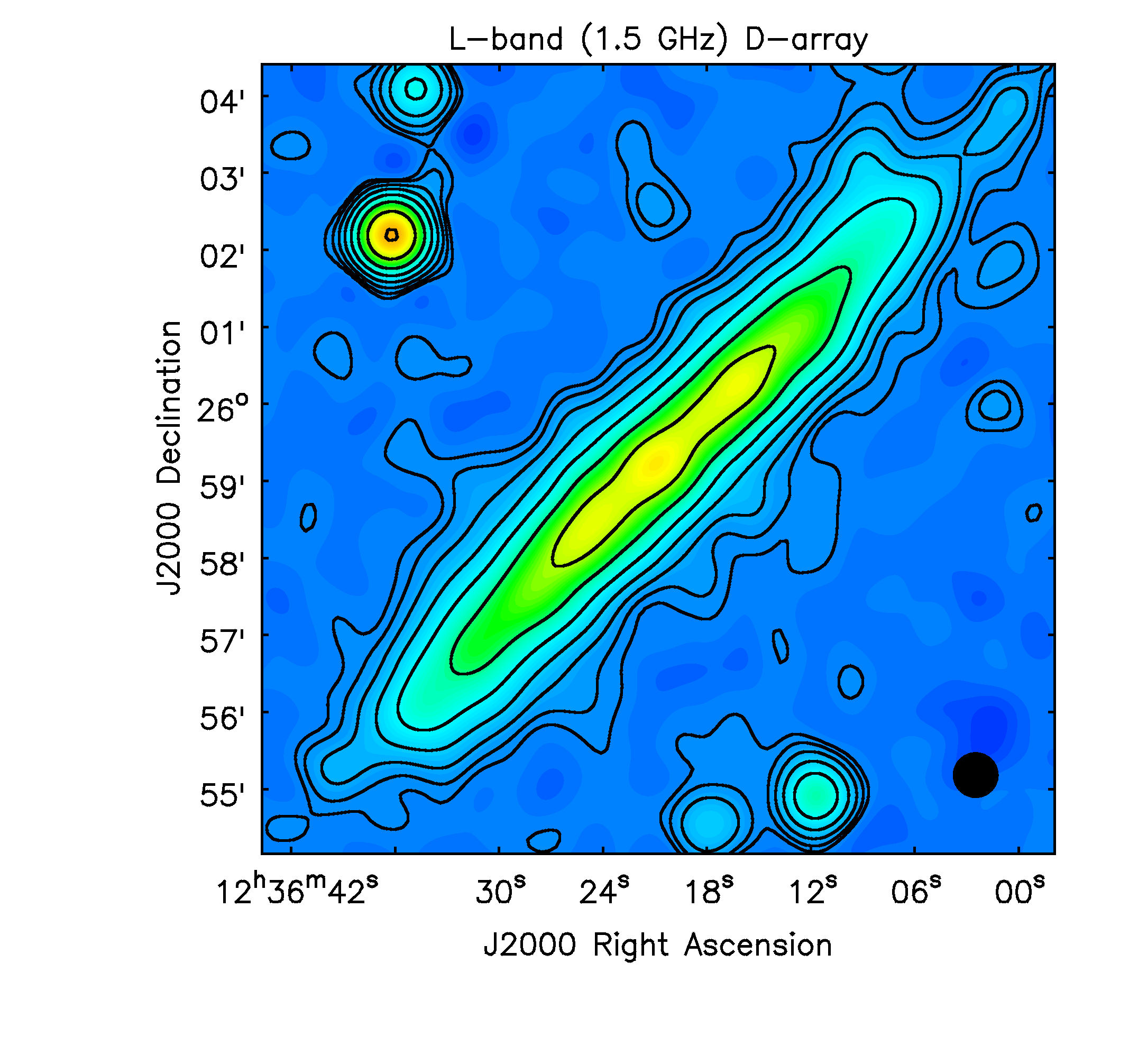}}{0pt}{0pt}
 \caption{L-band (1.5\,GHz) D-array total intensity distribution.
 \it a\normalfont: NGC\,891 (FWHM of the synthesised beam: $37\farcs 2$, rms noise: $\sigma=60\,\mu\mathrm{Jy\,beam^{-1}}$),
 \it b\normalfont: NGC\,4565 (FWHM of the synthesised beam: $35\farcs 4$, rms noise: $\sigma=34\,\mu\mathrm{Jy\,beam^{-1}}$).
 Both maps have been PB-corrected. Contour levels are $\sigma \times (3, 6, 12, 24, \rm{etc})$. Filled black circles indicate the size of the synthesised beam. 
 }
 \label{fig:tp1}
\end{figure*}

\subsection{Ancillary data}
\label{ancillary}

To estimate the contribution of thermal radio continuum emission (Sect.~\ref{thermalsep}), we use H$\alpha$ images 
taken at the Kitt Peak National Observatory  
\citep[KPNO, see][]{heald11}.
We prepared the maps by first removing foreground stars, interpolating the surrounding intensities across the removed areas. The maps were  
smoothed to a resolution of $12^{\prime\prime}$ FWHM
and clipped below the $3\sigma$ level, with $\sigma$ defined as the average noise rms in emission-free regions of the map.

To allow correction of the H$\alpha$ maps for internal extinction, we make use of \emph{Spitzer} Multiband Imaging Photometer (MIPS) images at 24\,$\mu\mathrm{m}$. We smoothed these maps, which have a $6\farcs 5$ FWHM point-spread function (PSF), to a Gaussian beam of $12^{\prime\prime}$ FWHM, using the appropriate convolution kernel provided by \citet{aniano11}. According to these authors, the application of this kernel lies between a `moderate' and a `very safe' convolution, meaning that not too much energy is moved from the wings of the original PSF into the Gaussian core, which would possibly amplify any potential image artefacts. 
The smoothed images were clipped at $10\sigma$ in case of NGC\,891 and at $3\sigma$ in case of NGC\,4565.

\section{Morphology of the total radio continuum emission}
\label{totalpower}

\subsection{NGC\,891}
\label{TPN891}

\begin{figure*}
 \centering
 \topinset{\it a)}{\includegraphics[scale=0.166,clip=true,trim=264pt 126pt 36pt 0pt]{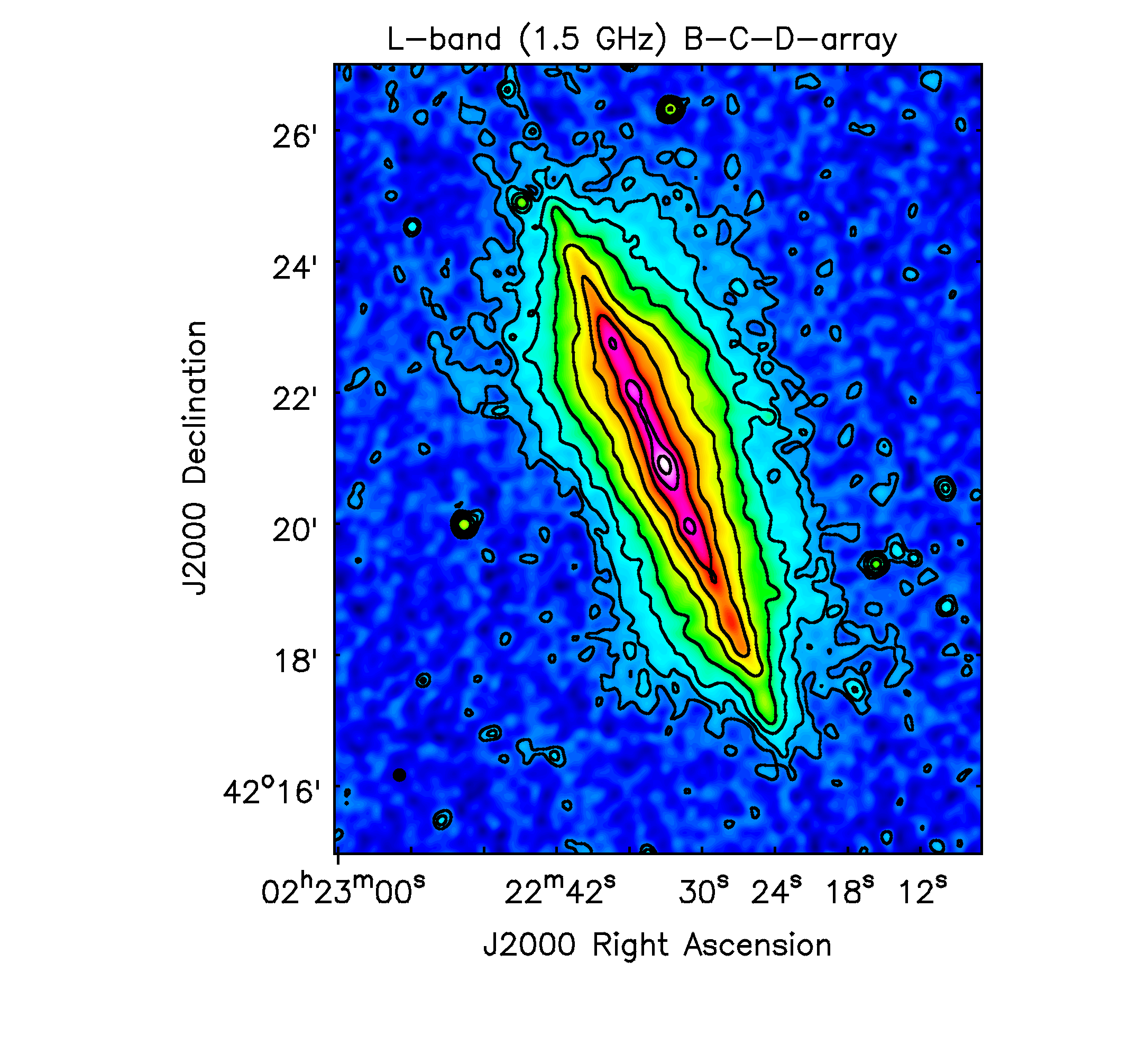}}{5pt}{0pt}\topinset{\it b)}{\includegraphics[scale=0.166,clip=true,trim=264pt 126pt 36pt 0pt]{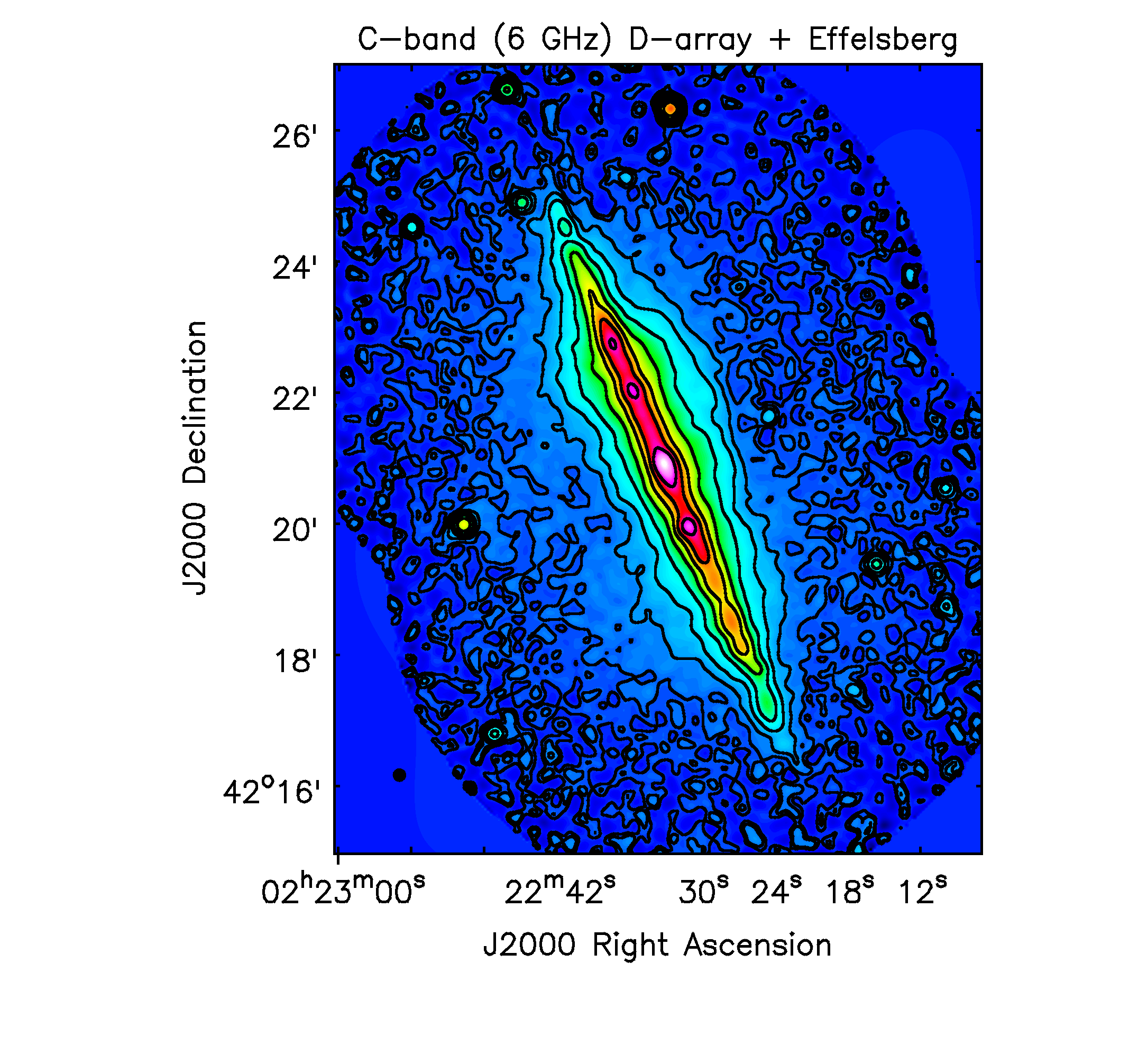}}{5pt}{0pt}
 \caption{Short-spacing corrected total intensity images of NGC\,891 at a resolution of $12^{\prime\prime}$ FWHM. \it a\normalfont: 1.5\,GHz (rms noise: $\sigma=24\,\mu\mathrm{Jy\,beam^{-1}}$), \it b\normalfont: 6\,GHz (rms noise: $\sigma=5.9\,\mu\mathrm{Jy\,beam^{-1}}$). 
 All individual VLA maps have been PB-corrected before short-spacing corrections were applied. Contour levels are $\sigma \times (3, 6, 12, 24, \rm{etc})$. The $3\sigma$ contours near the edges of the C-band map are due to the radially increasing noise levels after PB correction. Filled black circles indicate the size of the synthesised beam.
 }
 \label{fig:tp2} 
\end{figure*}

Figure~\ref{fig:tp1} \it a \normalfont shows the total intensity distribution of NGC\,891 at our lowest resolution VLA map (1.5\,GHz, D-configuration), while short-spacing corrected 1.5 and 6\,GHz maps at intermediate and high angular resolution are displayed in Fig.~\ref{fig:tp2}. 
The large extent of the radio halo emission of NGC\,891 is best visible in the Effelsberg single-dish observations at 4.85~GHz in Figure~\ref{fig:tp0} \it a \normalfont and in L-band D-array. While the inner radio emission in all images roughly follows the asymmetric distribution observed in H$\alpha$ \citep[cf.][]{dahlem94}, it becomes more peanut- or dumbbell-shaped towards larger $z$ heights in the L-band D-array map. A more quantitative description of the shape of the halo will be provided in Sect.~\ref{scaleheights}.

The integrated flux density for  each map is given in Table~\ref{tab:fluxdensities}. 
In L-band D-array, the flux density  
is in close agreement with the value 
found by \citet{hummel91a}. 
As single-dish-based flux density measurements of NGC\,891 of sufficient sensitivity at this frequency range do not exist in the literature (e.g. \citet{white92} measured only 658\,mJy at 1.4\,GHz), the true total flux density of the galaxy in this frequency range remains uncertain.

\subsection{NGC\,4565}
\label{TPN4565}

\begin{figure*}[h]
 \centering
 \topinset{\it a)}{\includegraphics[scale=0.16,clip=true,trim=226pt 126pt 0pt 0pt]{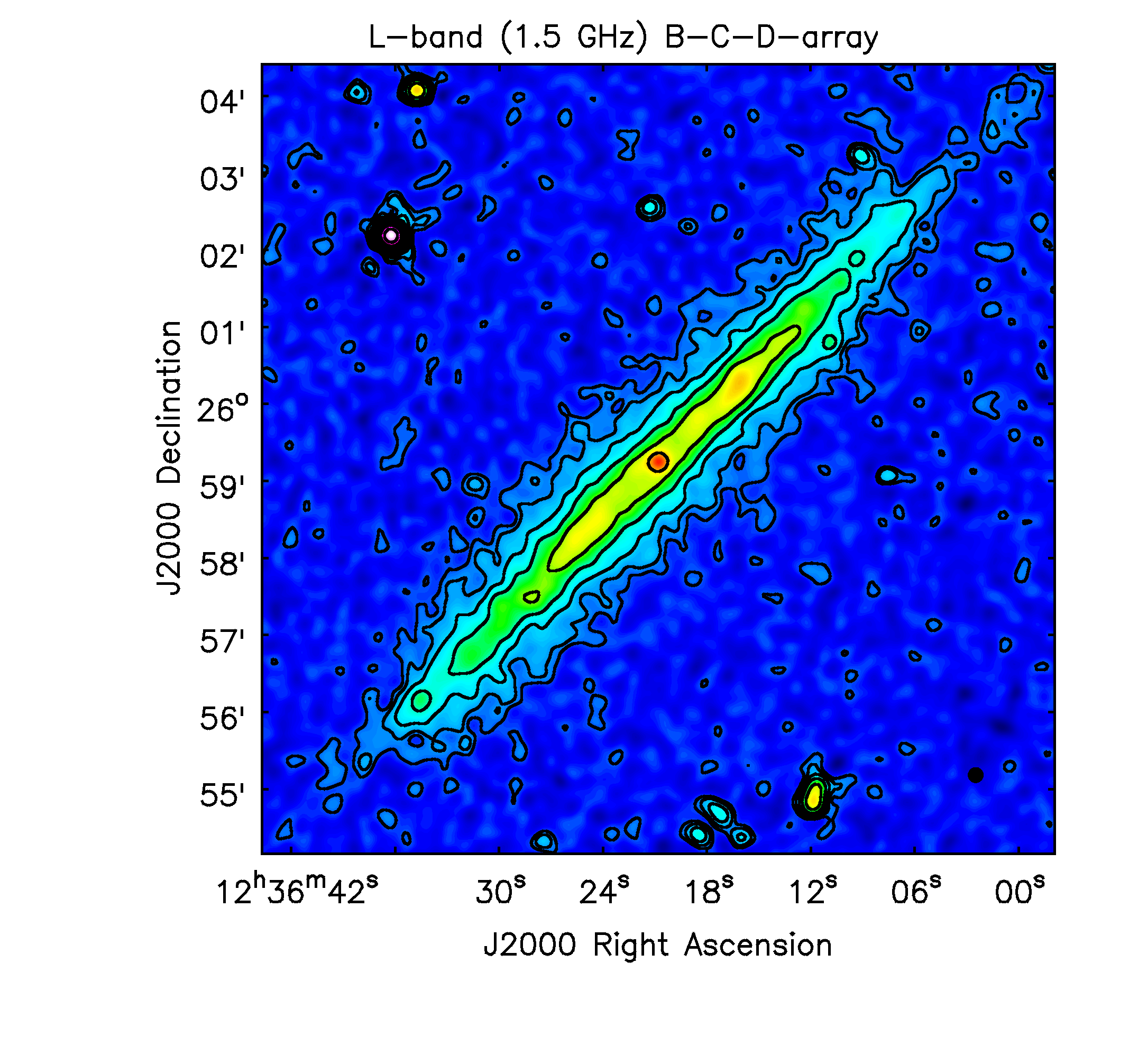}}{0pt}{0pt}\topinset{\it b)}{\includegraphics[scale=0.16,clip=true,trim=224pt 126pt 0pt 0pt]{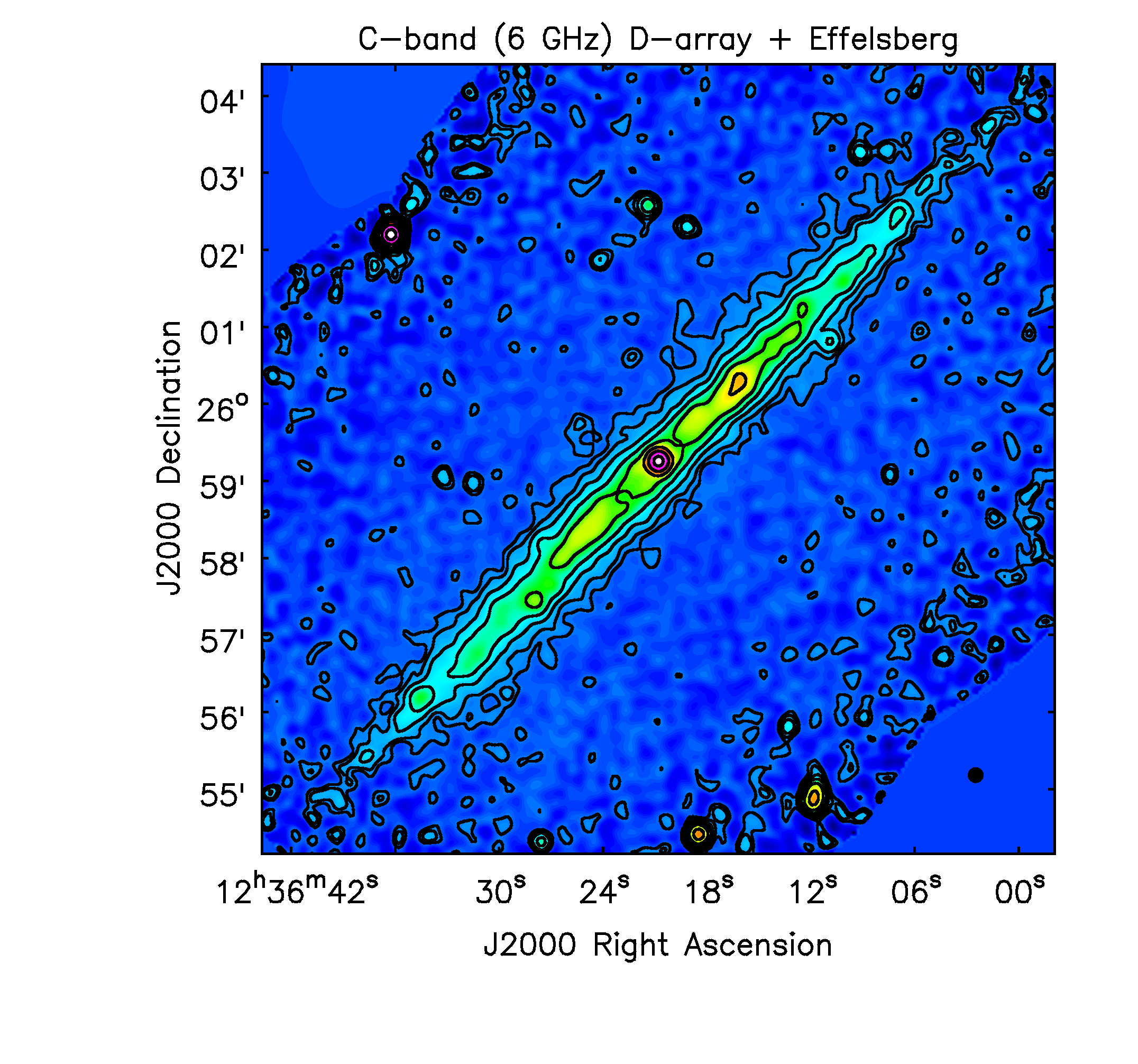}}{0pt}{0pt}
 \topinset{\it c)}{\includegraphics[scale=0.16,clip=true,trim=226pt 146pt 0pt 0pt]{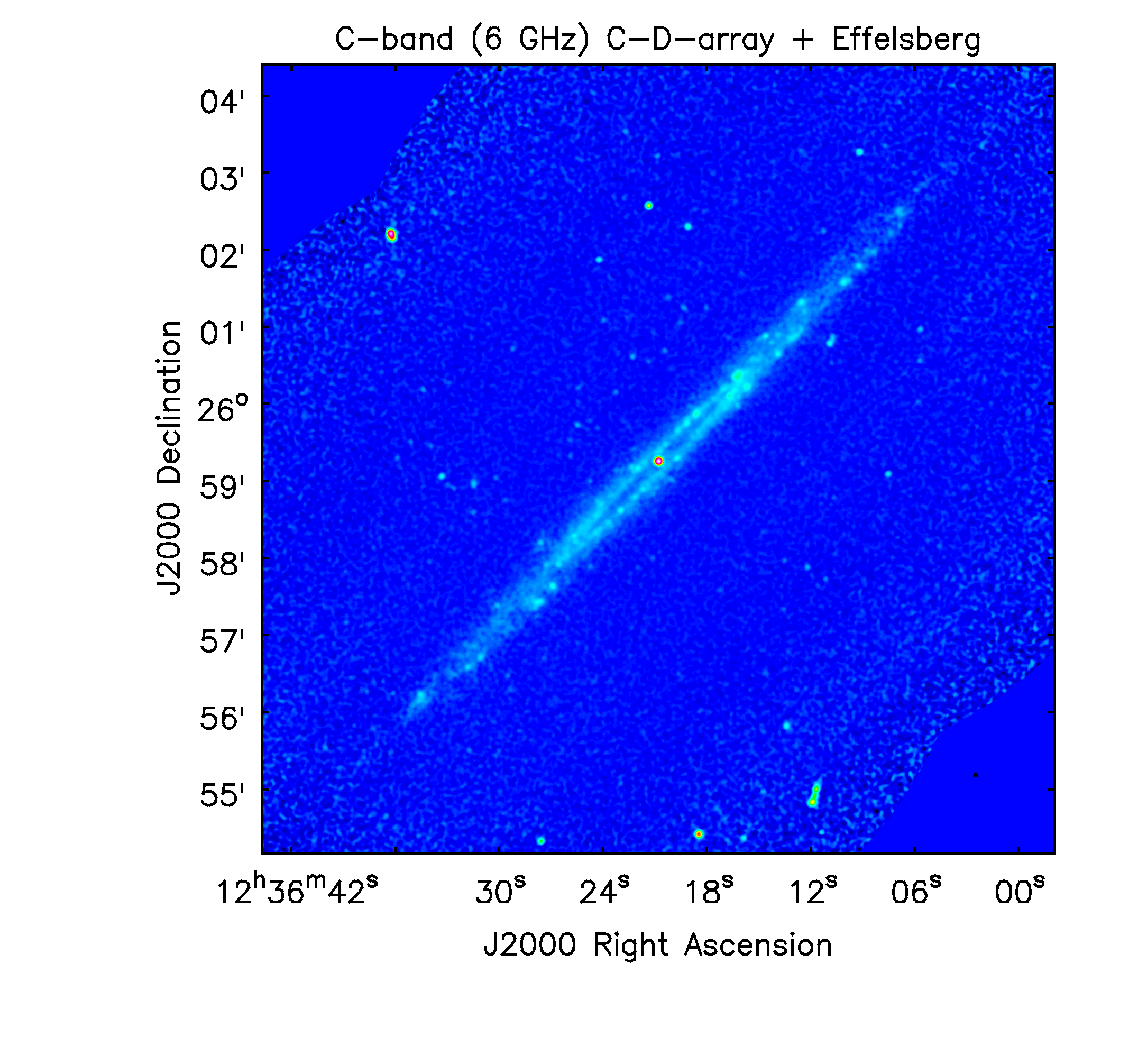}}{0pt}{0pt}\topinset{\it d)}{\includegraphics[scale=0.162,clip=true,trim=224pt 146pt 0pt 0pt]{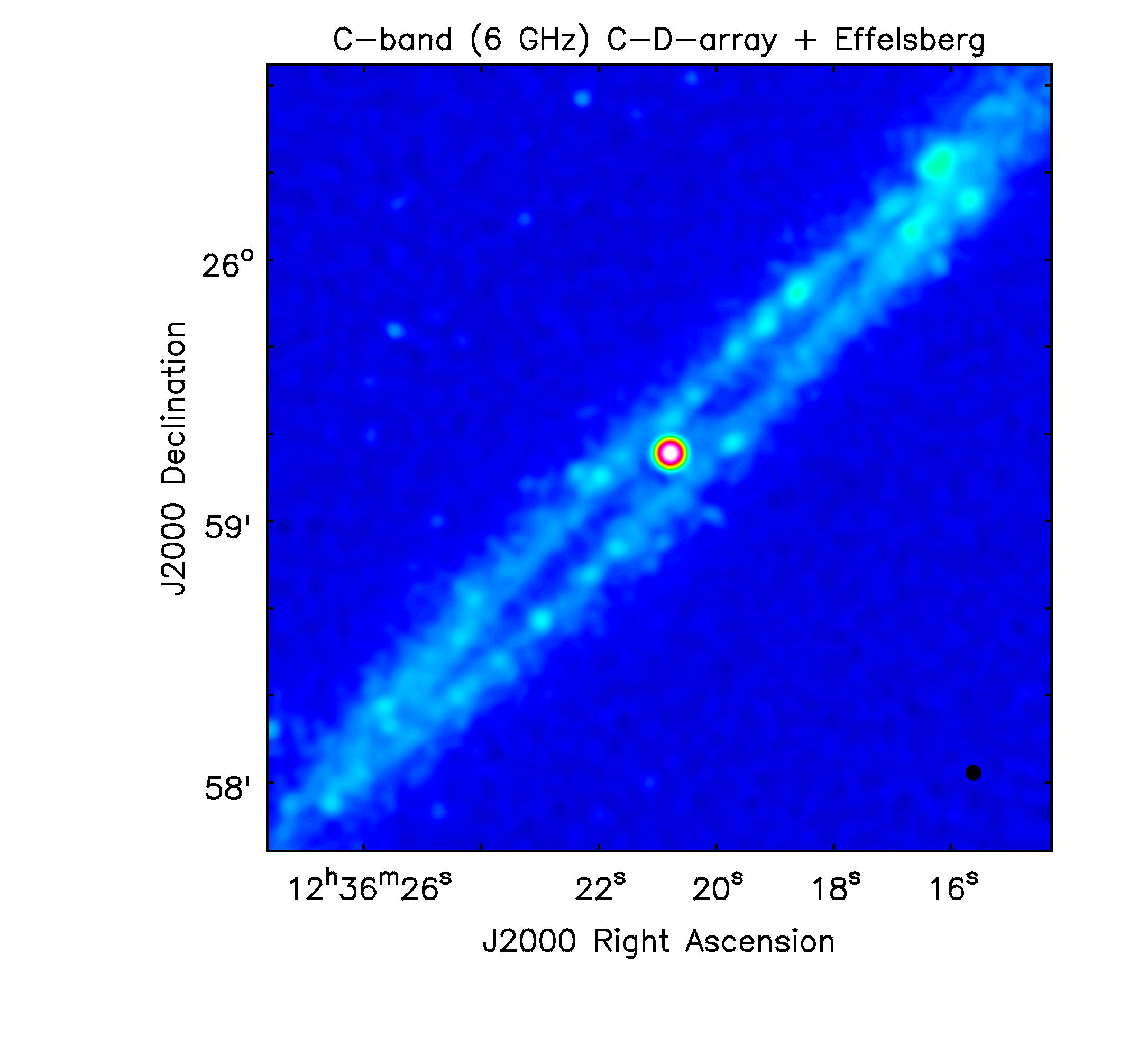}}{0pt}{0pt}
 \caption{Short-spacing corrected total intensity images of NGC\,4565 at 1.5 and 6\,GHz. The synthesised beam FWHM is $12^{\prime\prime}$ in panels \it a \normalfont and \it b\normalfont, and $3\farcs 6\times3\farcs 5$ in panels \it c \normalfont and \it d\normalfont. The rms noise $\sigma$ is $20\,\mu\mathrm{Jy\,beam^{-1}}$ in panel \it a\normalfont, $7.0\,\mu\mathrm{Jy\,beam^{-1}}$ in panel \it b\normalfont, and $2.5\,\mu\mathrm{Jy\,beam^{-1}}$ in panels \it c \normalfont and \it d\normalfont.
 All individual VLA maps have been PB corrected before short-spacing corrections were applied. Contour levels are $\sigma \times (3, 6, 12, 24, \rm{etc})$. The $3\sigma$ contours near the edges of the C-band maps are due to the radially increasing noise levels after PB correction. Panel \it d \normalfont shows the same map as panel \it c\normalfont, but zoomed in on the ring-shaped central part of the radio disc. Filled black circles indicate the size of the synthesised beam. }
 \label{fig:tp5}
\end{figure*}

The L-band D-array total intensity map of NGC\,4565 is displayed in Fig.~\ref{fig:tp1} \it b\normalfont, and short-spacing corrected images are shown in Fig.~\ref{fig:tp5}.
In each image, the ratio of vertical to radial extent of the radio emission is much lower than for NGC\,891, and has not significantly increased compared with earlier observations. Moreover, even after applying the short-spacing correction at 6\,GHz (Figs.~\ref{fig:tp5} \it b\normalfont--\it d\normalfont), practically all of the added large-scale emission still falls below the $3\sigma$ level. Hence, at least in the high-resolution maps, only emission from the disc of the galaxy is detected.

The combination of an earlier VLA observation with the Effelsberg map at 4.85\,GHz \citep{dumke97}, which had been obtained using the Astronomical Image Processing System\footnote{\texttt{http://www.aips.nrao.edu/index.shtml}} (\texttt{AIPS}) task \texttt{IMERG}, shows dumbbell-like vertical extensions of the otherwise narrow halo component, most prominently on the south-eastern side of the galaxy. No such feature is visible in any of the images presented here. Errors in any of the merging procedures cannot be ruled out. 
However, after testing different merging routines on both NGC\,891 and 4565 \citep{schmidt16}, we found that \texttt{IMERG} is more likely to produce filamentary, often dumbbell-shaped, extraplanar features than the corresponding routines in \texttt{CASA} or \texttt{NOD3}. 

\par In the lower-resolution maps (Figs.~\ref{fig:tp0} \textit{b} 
and \ref{fig:tp1} \textit{b}
) we detect an extended source $\approx$4$^\prime$ south of the centre of NGC\,4565. This source is bright in X-ray emission (see Fig.~4~Cv in \citealt{li13a}) and hence is probably the radio relic of a background cluster.

\begin{figure*}
 \centering
 \topinset{\it a)}{\includegraphics[scale=0.17,clip=true,trim=131pt 126pt 60pt 0pt]{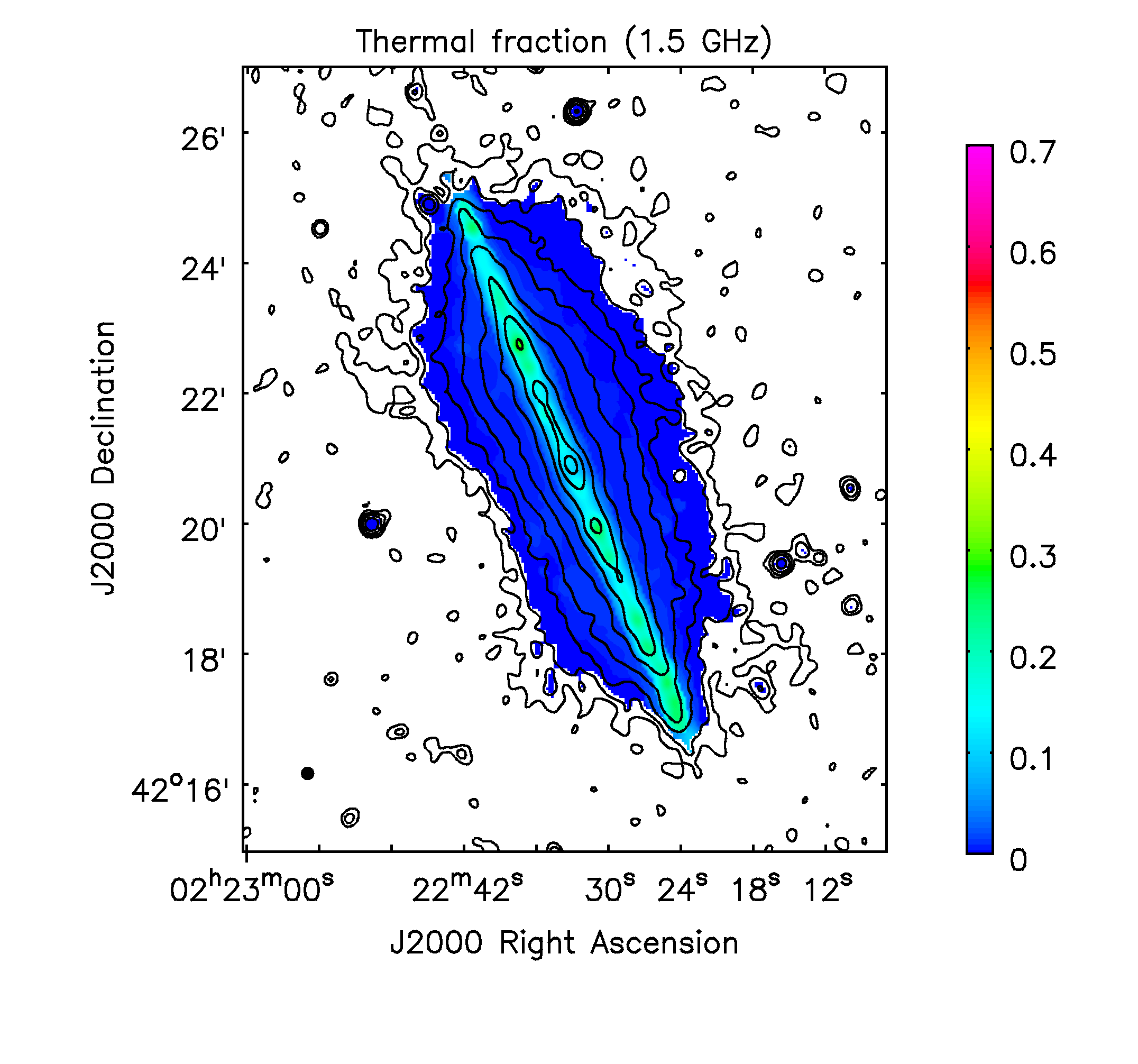}}{5pt}{0pt}\topinset{\it b)}{\includegraphics[scale=0.17,clip=true,trim=190pt 126pt 58pt 0pt]{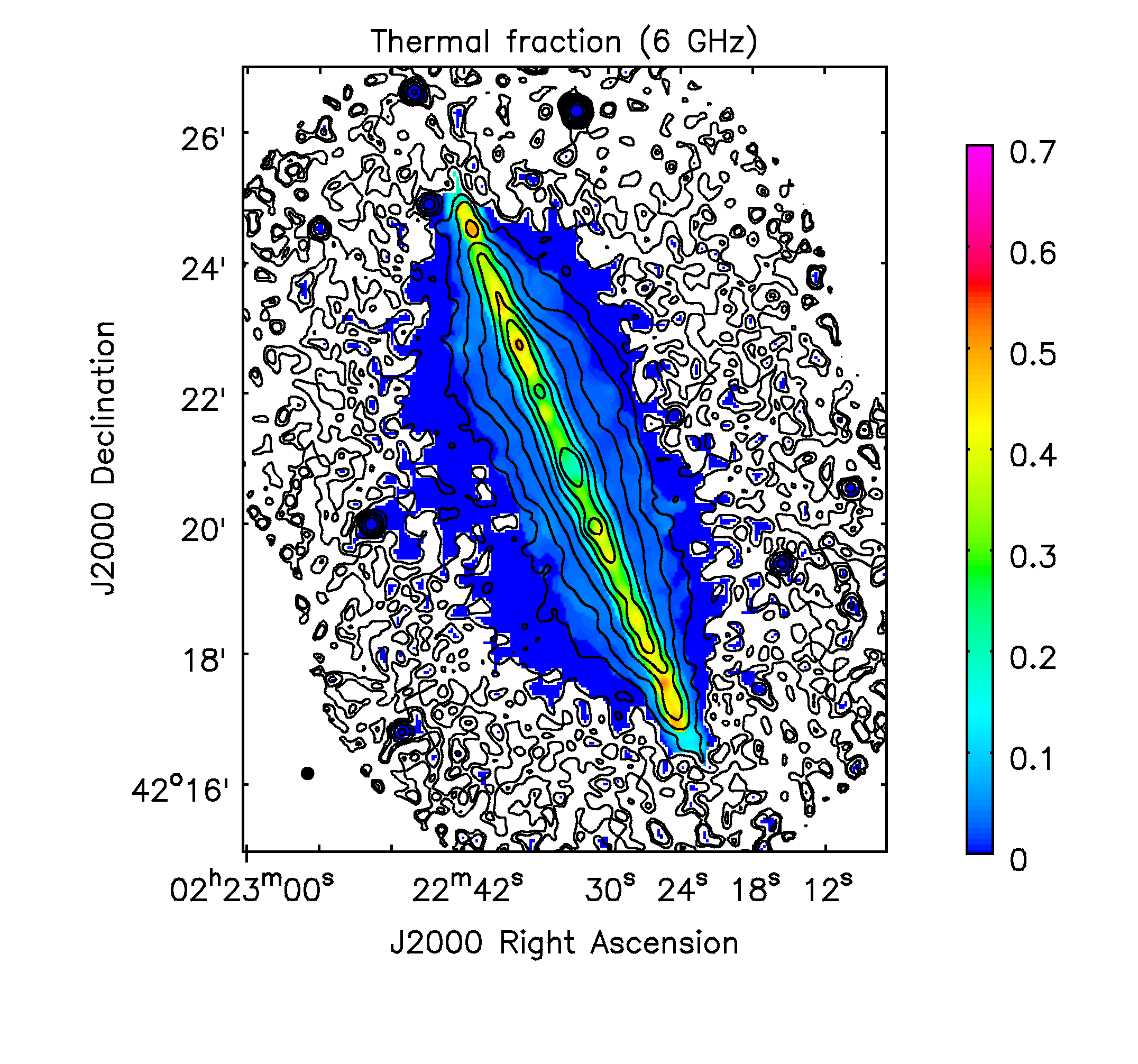}}{5pt}{4pt}
 \topinset{\it c)}{\includegraphics[scale=0.17,clip=true,trim=131pt 252pt 0pt 86pt]{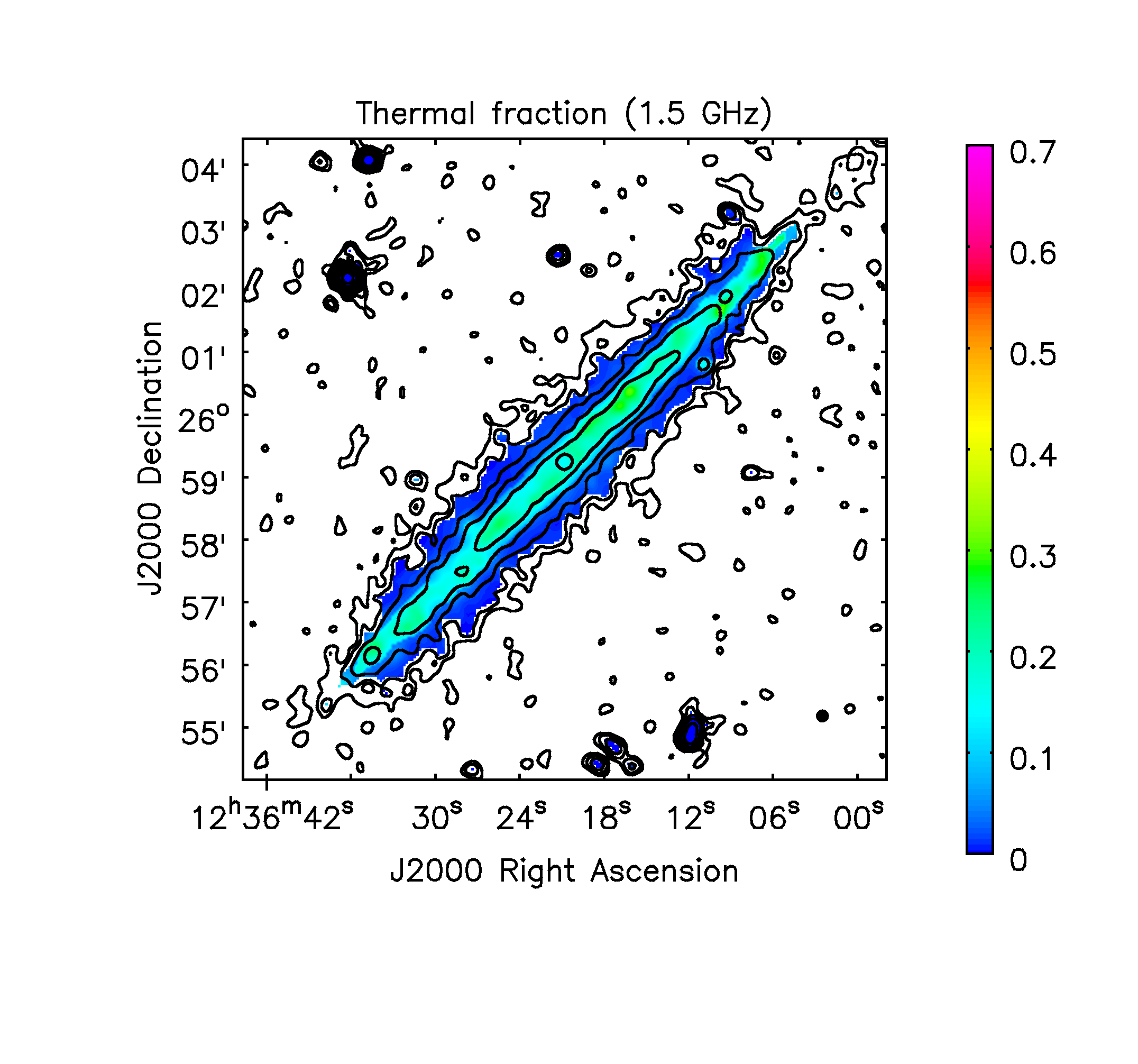}}{7pt}{0pt}\topinset{\ it d)}{\includegraphics[scale=0.17,clip=true,trim=245pt 252pt 50pt 86pt]{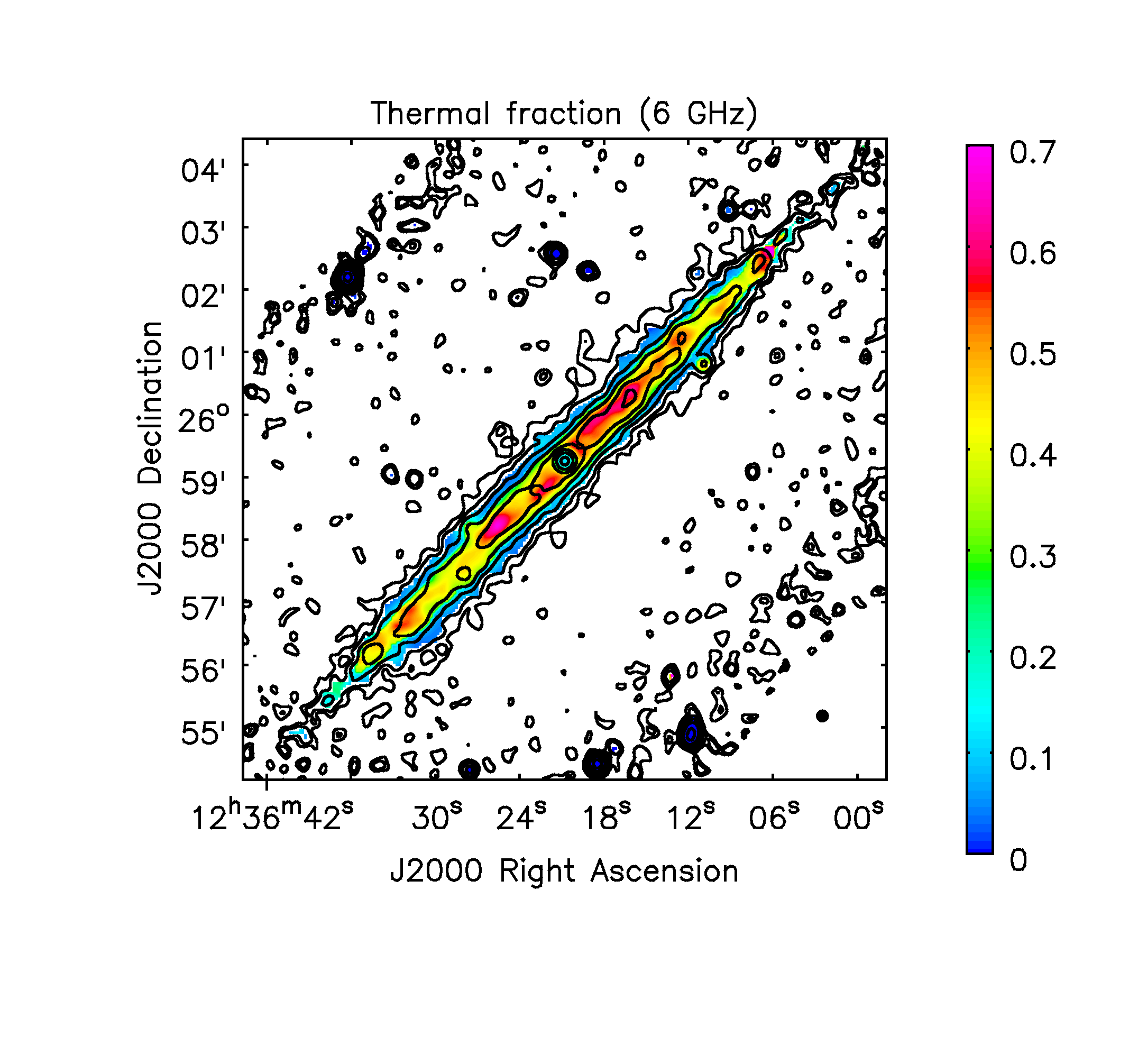}}{7pt}{0pt}
\caption{Fraction of thermal radio continuum emission in NGC 891 and 4565 at 1.5 and 6\,GHz. The angular resolution in all panels is $12^{\prime\prime}$. Each map is overlaid with contours of the total radio emission at the respective frequency, identical to those shown in shown in Figs.~\ref{fig:tp2} \it a \normalfont and \it b \normalfont and Figs.~\ref{fig:tp5} \it a \normalfont and \it b \normalfont, respectively. Contour levels are $\sigma \times (3, 6, 12, 24, \rm{etc})$ of the corresponding total emission maps (see Tables~\ref{tab:imgparsN891L} and \ref{tab:imgparsN4565L}). Filled black circles indicate the size of the synthesised beam.
 }
 \label{fig:tp7}
\end{figure*}

\subsection{The radio ring of NGC\,4565}
\label{4565spiral}

NGC\,4565 is known to have a prominent ring of dust emission with a radius of $\approx$5.5\,kpc \citep{laine10,kormendy10}, coinciding with a ring of molecular gas \citep{sofue94,neininger96,yim14}. Our 6\,GHz C-array observations (as included in Fig.~\ref{fig:tp5} \it c \normalfont and \it d\normalfont) are the first to resolve this ring at radio frequencies.  
The observed axis ratio of the ring in radio continuum implies an inclination of $i=86.3^{\circ}\pm0.4^{\circ}$ for this galaxy.

We estimate the width and thickness of the ring based on the $10\sigma$ level of the (short-spacing corrected) combined-array C-band image.  
If $x$ is the total extent of the ring along the major axis, its observed minor axis extent is $y=x\cos{i}+h\sin{i}$, where $(h\sin{i})$ corresponds to its projected vertical thickness. With $x\approx240^{\prime\prime}$, $y\approx20^{\prime\prime}$, $i=86.3^{\circ}$, and a distance of $D=11.9$\,Mpc, one obtains a vertical thickness of $h=261\mathrm{\,pc}$. From the projected width $y_{1}$ of the near or far side of the ring along its minor axis, one can determine its width in the galactic plane: $w=(y_{1}-h\sin{i})/\cos{i}$. The typical width of the $10\sigma$ contour (excluding the nuclear radio source) near the minor axis of the ring is found to be 
$y_{1}\approx7^{\prime\prime}$, which results in $w\approx2.2\,\mathrm{kpc}$. 
However, we note that the 
results are highly sensitive to the assumed inclination. For our lower error limit in inclination, that is $i=85\fdg 9$, we obtain $h=164\,\mathrm{pc}$ and $w\approx3.4\,\mathrm{kpc}$, while using the upper limit of $i=86\fdg 7$ implies $h=357\,\mathrm{pc} $ and $w\approx0.8\,\mathrm{kpc}$. Hence, in case the ring has a vertical thickness comparable to the thin disc scale height of the Milky Way ($\approx300\,\mathrm{pc}$), it has to be relatively narrow for the near and far side to be clearly resolved from each other; otherwise they must be notably flatter than the disc of our Galaxy.

Near the centre of the ring, we also observe emission that is much fainter and slightly more extended than the bright nuclear point source. This emission is most likely related to the central bar, kinematical proof of which was found by \citet{neininger96} and \citet{zschaechner12}, and which is oriented mainly along the line of sight \citep{kormendy10}. 
Since bars are usually connected to spiral arms or rings, the bar hence might be up to $\approx$11\,kpc long (the diameter of the ring), in which case it would constitute a substantial part of the disc.

\section{Thermal radio emission}
\label{thermalsep}

We produced maps of the thermal radio continuum emission using the H$\alpha$ and $24\mu$m-data mentioned in Sect.~\ref{ancillary}. For the dust ext inction correction of observed H$\alpha$ emission, \citet{calzetti07} and \citet{kennicutt09} established linear relations (each based on a large sample of nearby galaxies) of the form
\begin{equation}
\label{lhacorr}
 L_{\mathrm{H\alpha,corr}}=L_{\mathrm{H\alpha,obs}}+a\times\nu \, L_{\nu}(24\,\mu\mathrm{m})\,,
\end{equation}
where $L_{\mathrm{H\alpha,obs}}$ and $L_{\mathrm{H\alpha,corr}}$ are the bolometric lumiosities of the observed and extinction-corrected H$\alpha$ emission, respectively, and $\nu$ and $L_{\nu}$ are the frequency and monochromatic luminosity at 24\,$\mu\mathrm{m}$. 
Initially, we tried to correct for internal extinction 
adopting a 24\,$\mu\mathrm{m}$ scaling coefficient of $a=0.031\pm0.006$ where $\nu \, L_{\nu}(24\mu\mathrm{m})\ge3\times10^{38}\mathrm{erg/s}$ \citep[as found by][]{calzetti07}, and $a=0.02$ \citep{kennicutt09} otherwise.
We then followed \citet{tabatabaei07} to obtain the brightness temperature distribution $T_{b}$ of the thermal radio emission from the extinction-corrected H$\alpha$ maps (assuming an electron temperature of $10^4$\,K). Finally, we converted the brightness temperatures to intensities at $\nu=1.5\,\mathrm{GHz}$ and $\nu=6\,\mathrm{GHz}$ following \citep{pacholczyk70}.

Applying the above scaling coefficients for the extinction correction turned out to severely underpredict the thermal emission in NGC\,891, as the resulting non-thermal 
spectral indices along the disc plane (after creating non-thermal spectral index maps as described in Sect.~\ref{spectralindex}) were still considerably flatter at 6\,GHz 
($\approx\!-0.6$) than at 1.5\,GHz ($\approx\!-0.75$). 
To obtain a more reasonable estimate for the thermal contribution, 
we modified the scaling factor $a$ for $\nu \, L_{\nu}(24\mu\mathrm{m})\ge3\times10^{38}\mathrm{erg/s}$ such that the global thermal fractions agree with those found by \citet{dumke97}
and, for NGC\,4565, by \citet{niklas97}.
This was realised by the choice of $a=0.058$ for NGC\,891 and $a=0.068$ for NGC\,4565. We present the resulting thermal fraction maps at 1.5 and 6\,GHz in Fig.~\ref{fig:tp7}. In NGC\,891 we find maximum thermal fractions of 27\% at 1.5\,GHz and 49\% at 6\, GHz, while in NGC\,4565 the maximum is 28\% at 1.5\,GHz and 65\% at 6\,GHz.

\begin{table}[h]
\caption{Global thermal fractions and spectral indices.  
See text for details on the separation of the thermal and non-thermal emission.
}
{\small
\begin{center}
\begin{tabular}{lcc}
\toprule
\toprule[0.3pt]
 & NGC\,891 & NGC\,4565 \\
\hline
$f_{\mathrm{th,1.5GHz}}$ [\%] & $7.2\pm 3.1$ & $10.6\pm 4.9$ \\
$f_{\mathrm{th,6GHz}}$ [\%] & $18.4\pm 8.0$ & $32.9\pm 15.8$ \\
$\alpha_{\mathrm{tot,1.5-6GHz}}$ & $-0.80\pm 0.02$ & $-0.93\pm 0.02$ \\
$\alpha_{\mathrm{nth,1.5-6GHz}}$ & $-0.90 \pm 0.13$ & $-1.14\pm 0.25$ \\
\hline                                    
\end{tabular}
\label{tab:intfthalphas}
\end{center}
}
\end{table}

In Table~\ref{tab:intfthalphas} we present  global values of the thermal fractions and of the total and non-thermal spectral index between the two observing frequencies 
($\alpha_{\mathrm{tot,1.5-6GHz}}$ and $\alpha_{\mathrm{nth,1.5-6GHz}}$).  
The error we adopted in each case for the thermal fraction and non-thermal spectral index corresponds to the difference to the 
value obtained using $a=0.031$ for the disc. 

Meanwhile, \citet{vargas18} published a slightly different method to determine the thermal contribution, and derived a scaling coefficient of $a=0.042$ as a typical value for their  sub-sample of CHANG-ES galaxies (including NGC\,891). The thermal fractions they obtained for NGC\,891 agree with our results within the errors.

\begin{figure*}[h]
 \centering
\stackinset{l}{0pt}{t}{0pt}{\it a)}{\stackinset{l}{100pt}{t}{12pt}{\Large$\alpha_{\mathrm{tot}}$}{\includegraphics[scale=0.125,clip=true,trim=210pt 120pt 85pt 40pt]{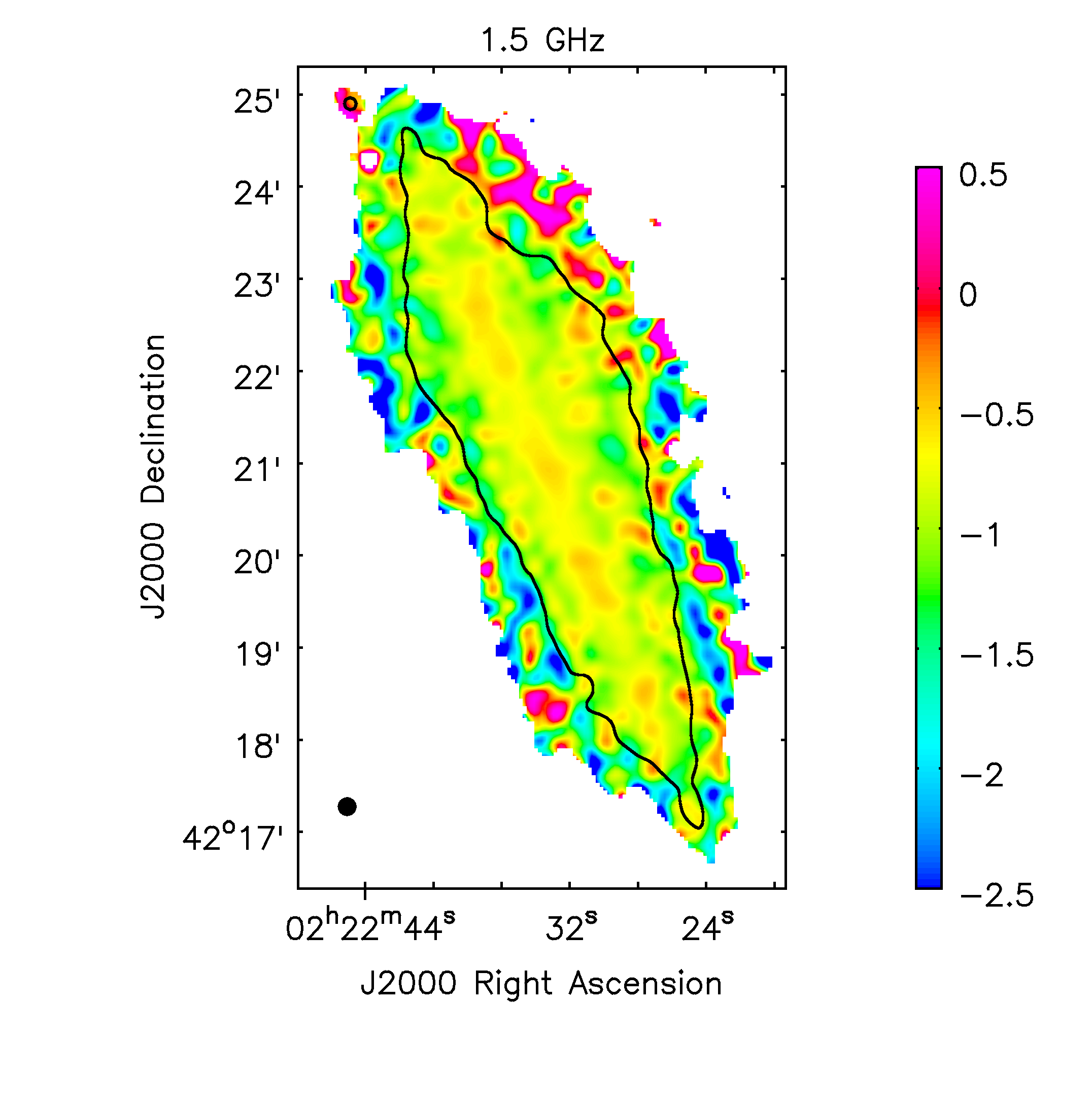}}}\stackinset{l}{0pt}{t}{0pt}{\it b)}{\stackinset{l}{92pt}{t}{12pt}{\Large$\alpha_{\mathrm{nth}}$}{\includegraphics[scale=0.125,clip=true,trim=270pt 120pt 85pt 40pt]{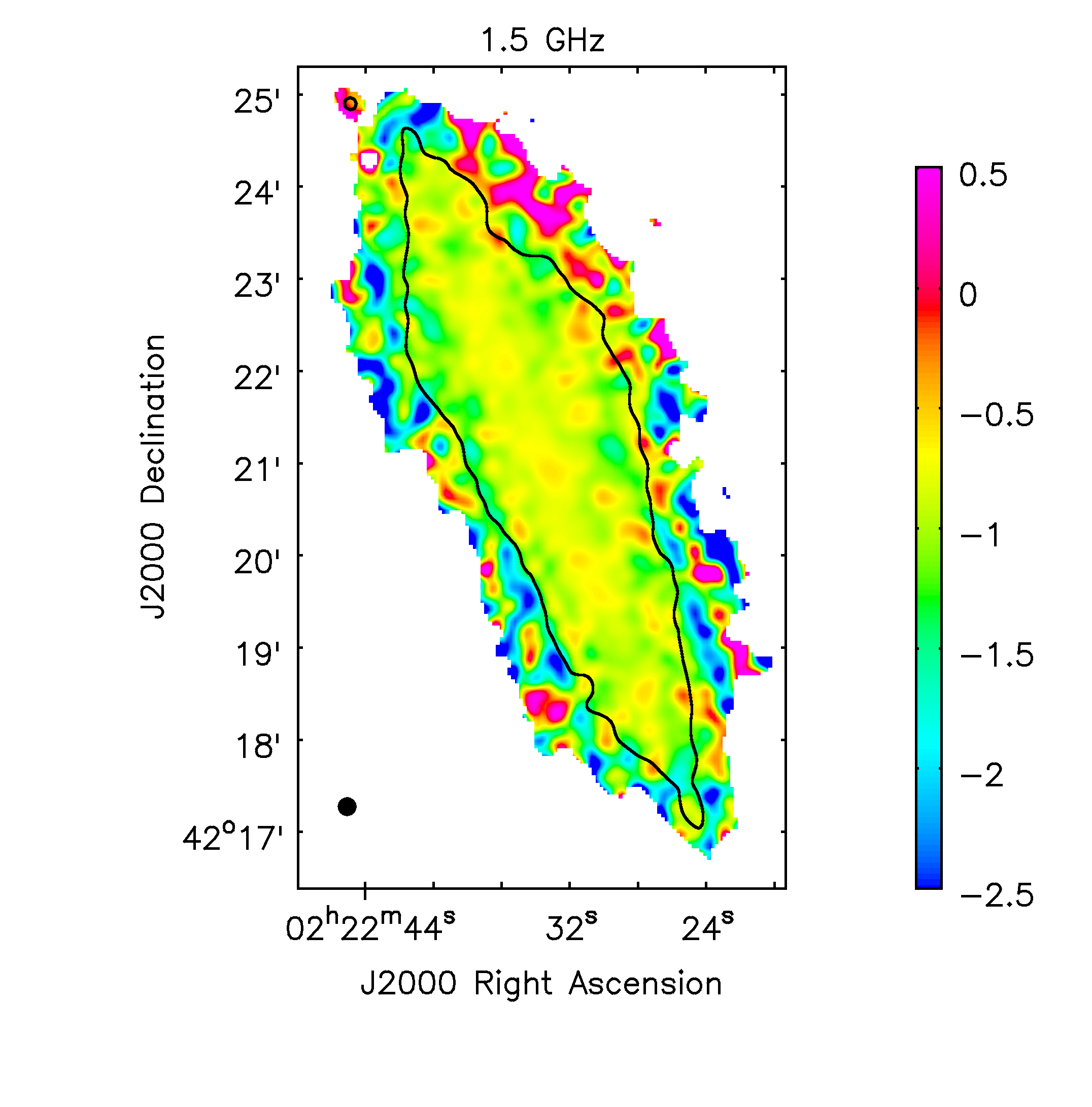}}}\stackinset{l}{0pt}{t}{0pt}{\it c)}{\stackinset{l}{93pt}{t}{10pt}{\Large$\Delta\alpha$}{\includegraphics[scale=0.125,clip=true,trim=250pt 120pt 85pt 40pt]{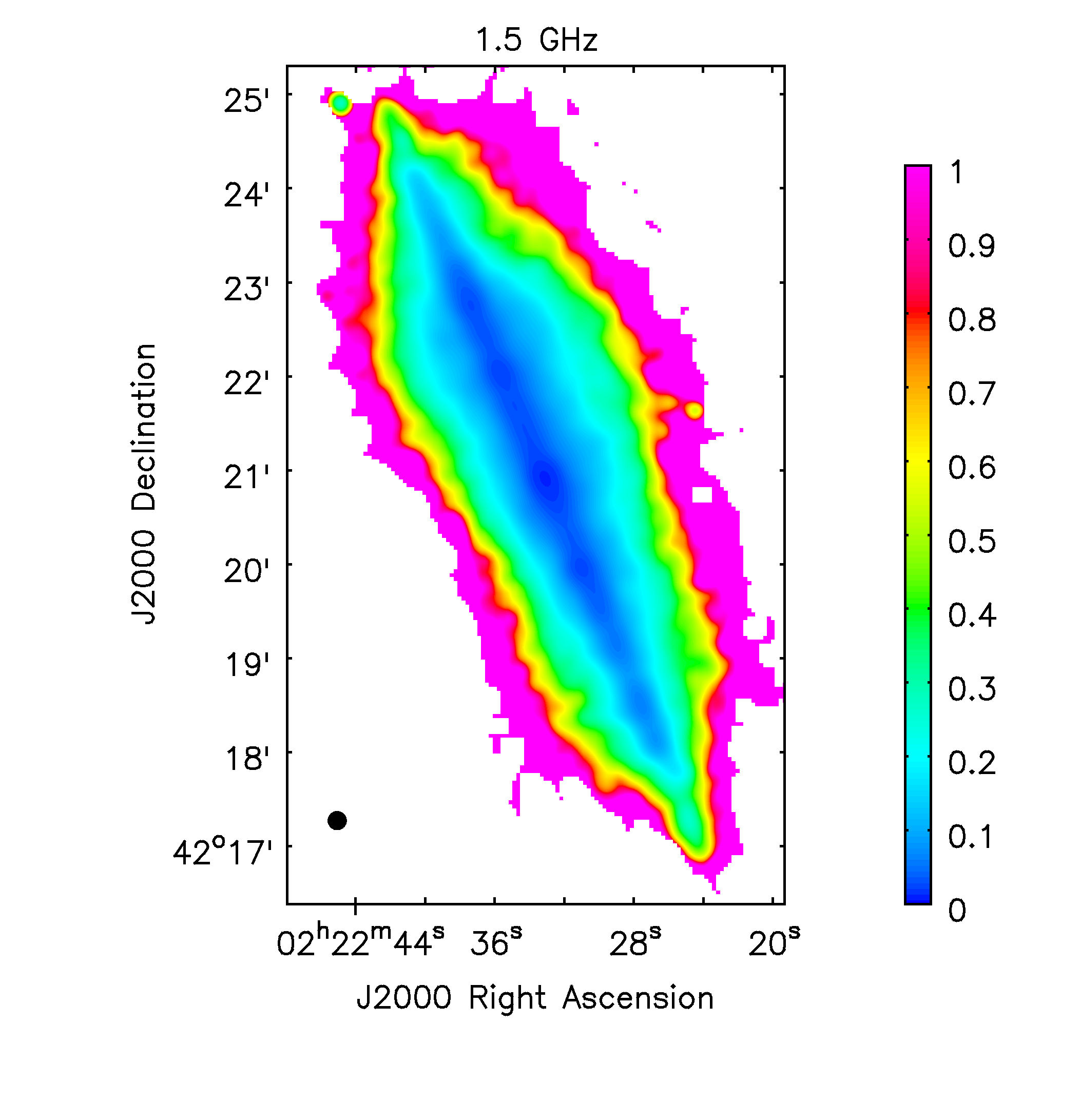}}}
\stackinset{l}{0pt}{t}{0pt}{\it d)}{\stackinset{l}{100pt}{t}{12pt}{\Large$\alpha_{\mathrm{tot}}$}{\includegraphics[scale=0.125,clip=true,trim=210pt 120pt 85pt 40pt]{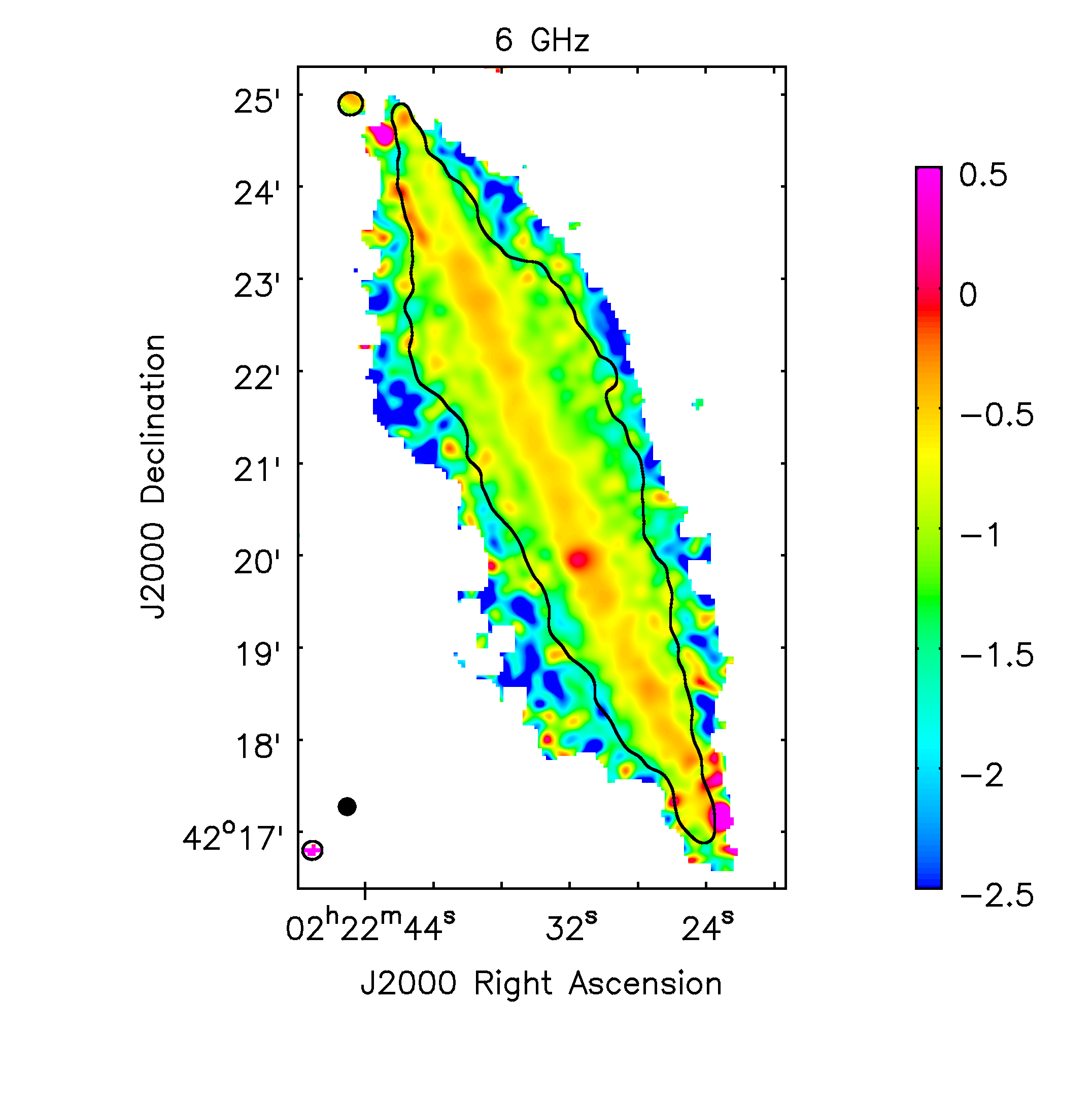}}}\stackinset{l}{0pt}{t}{0pt}{\it e)}{\stackinset{l}{92pt}{t}{12pt}{\Large$\alpha_{\mathrm{nth}}$}{\includegraphics[scale=0.125,clip=true,trim=270pt 120pt 85pt 40pt]{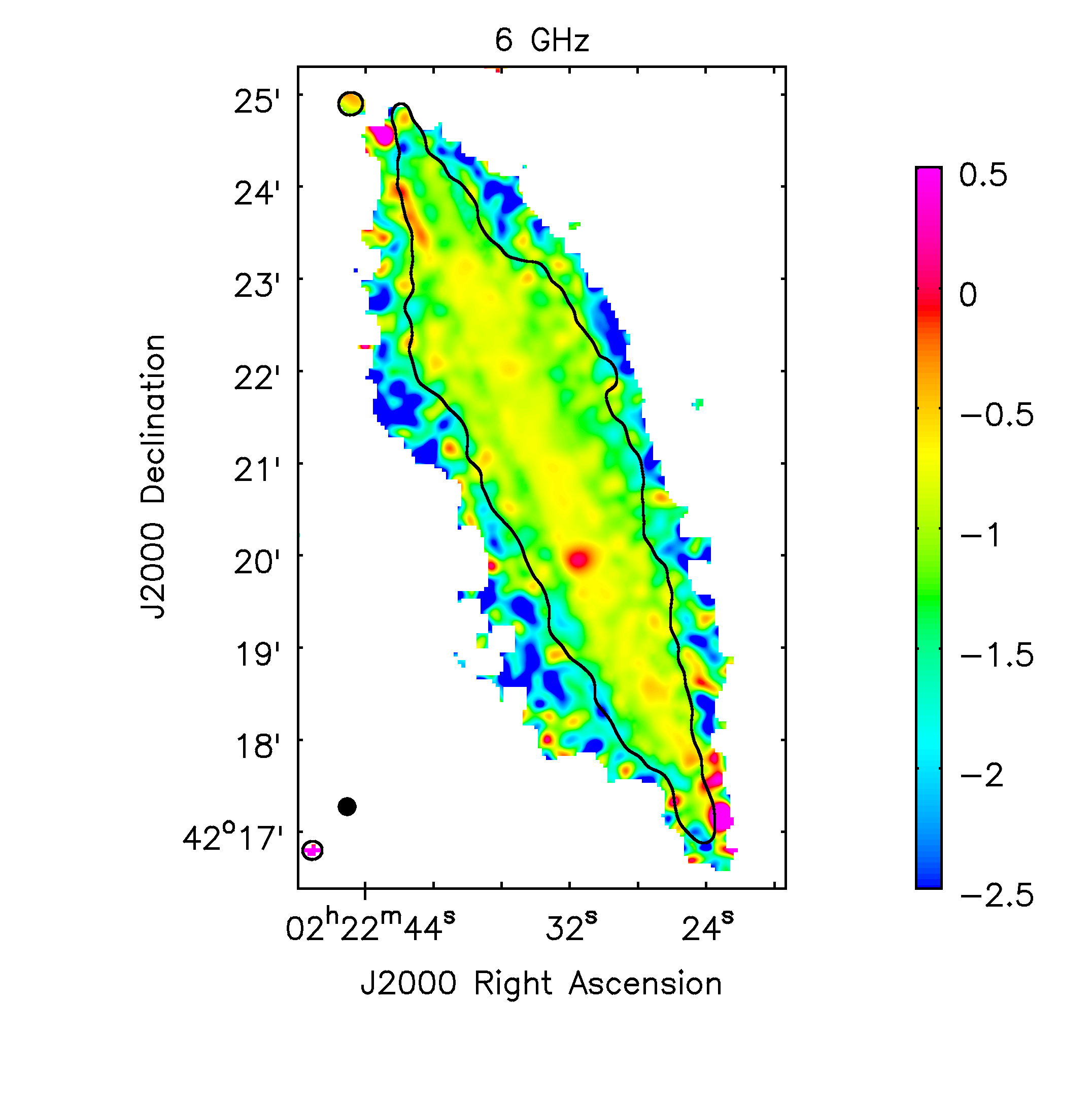}}}\stackinset{l}{0pt}{t}{0pt}{\it f)}{\stackinset{l}{93pt}{t}{10pt}{\Large$\Delta\alpha$}{\includegraphics[scale=0.125,clip=true,trim=250pt 120pt 85pt 40pt]{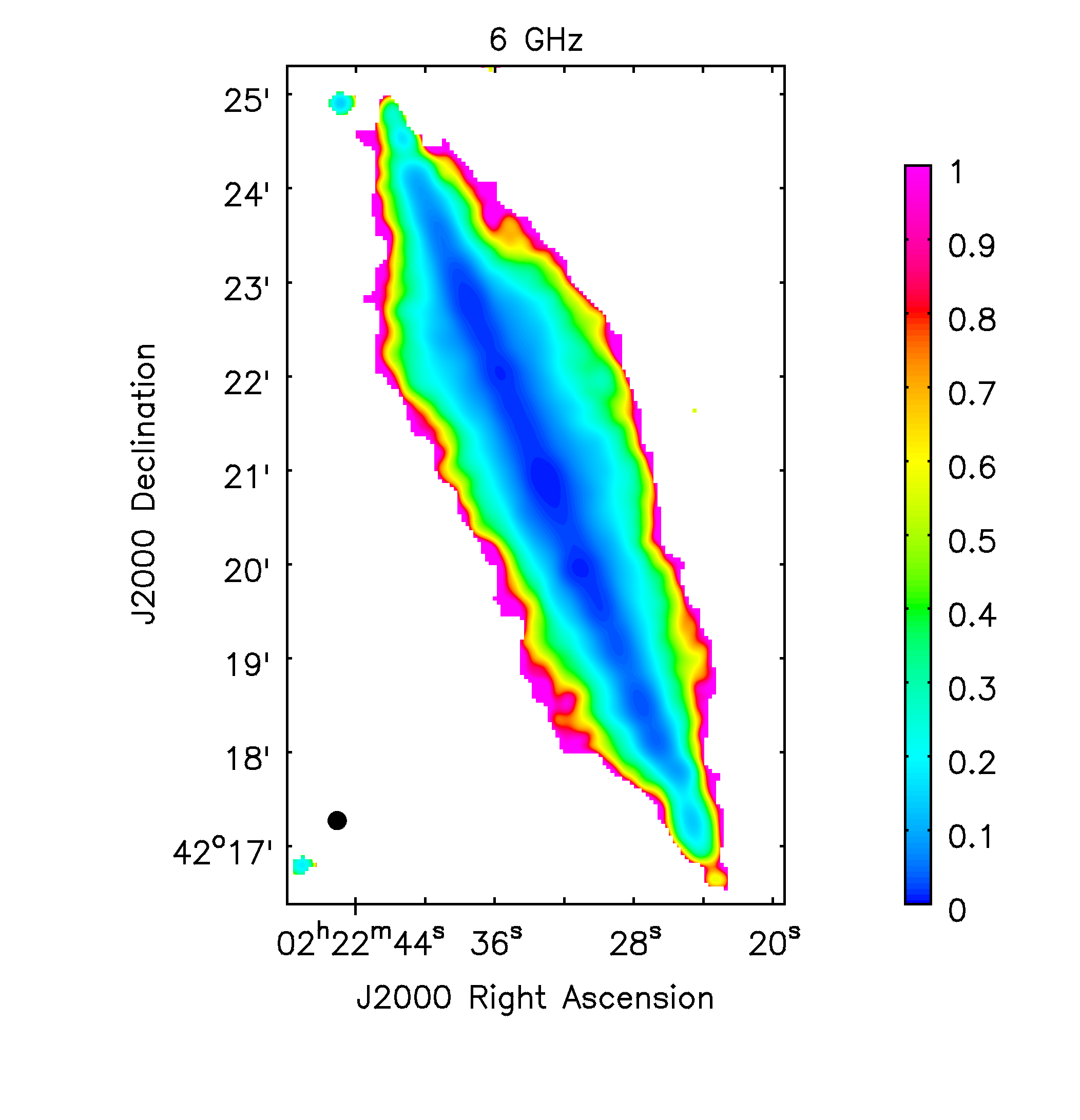}}}
\stackinset{l}{0pt}{t}{0pt}{\it g)}{\stackinset{l}{100pt}{t}{12pt}{\Large$\alpha_{\mathrm{tot}}$}{\includegraphics[scale=0.125,clip=true,trim=210pt 160pt 85pt 40pt]{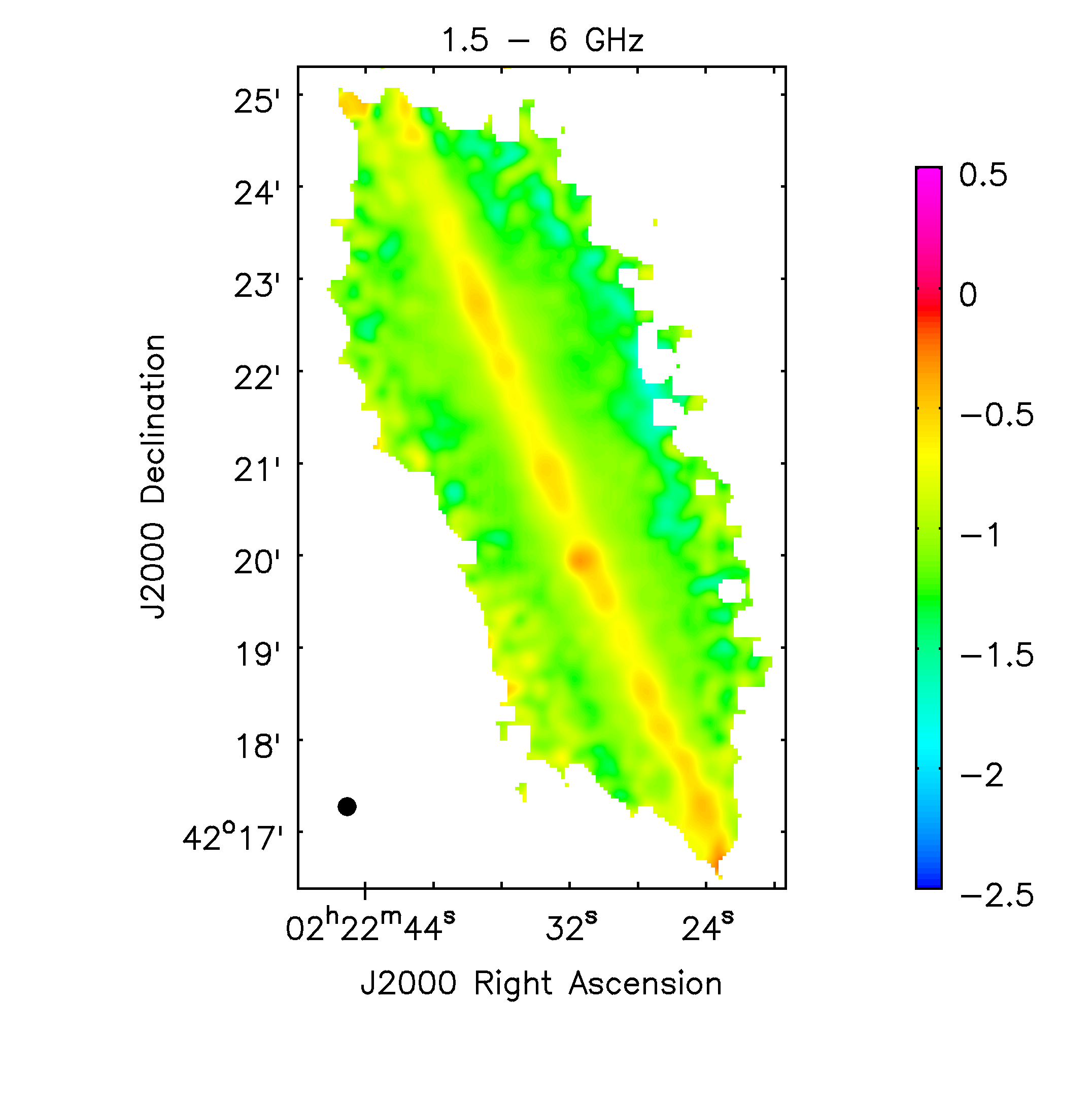}}}\stackinset{l}{0pt}{t}{0pt}{\it h)}{\stackinset{l}{92pt}{t}{12pt}{\Large$\alpha_{\mathrm{nth}}$}{\includegraphics[scale=0.125,clip=true,trim=270pt 160pt 85pt 40pt]{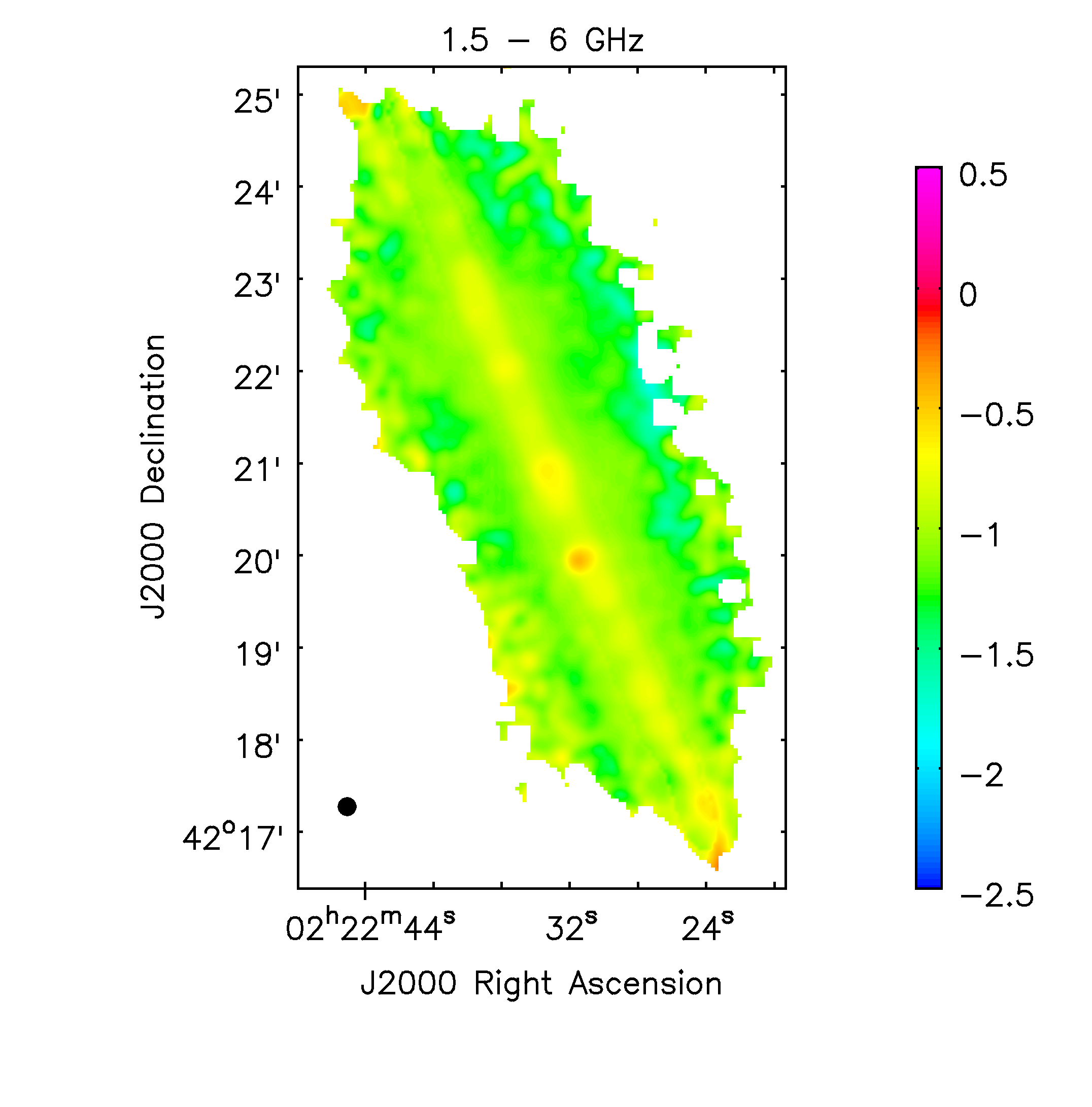}}}\stackinset{l}{0pt}{t}{0pt}{\it i)}{\stackinset{l}{93pt}{t}{10pt}{\Large$\Delta\alpha$}{\includegraphics[scale=0.125,clip=true,trim=270pt 160pt 85pt 40pt]{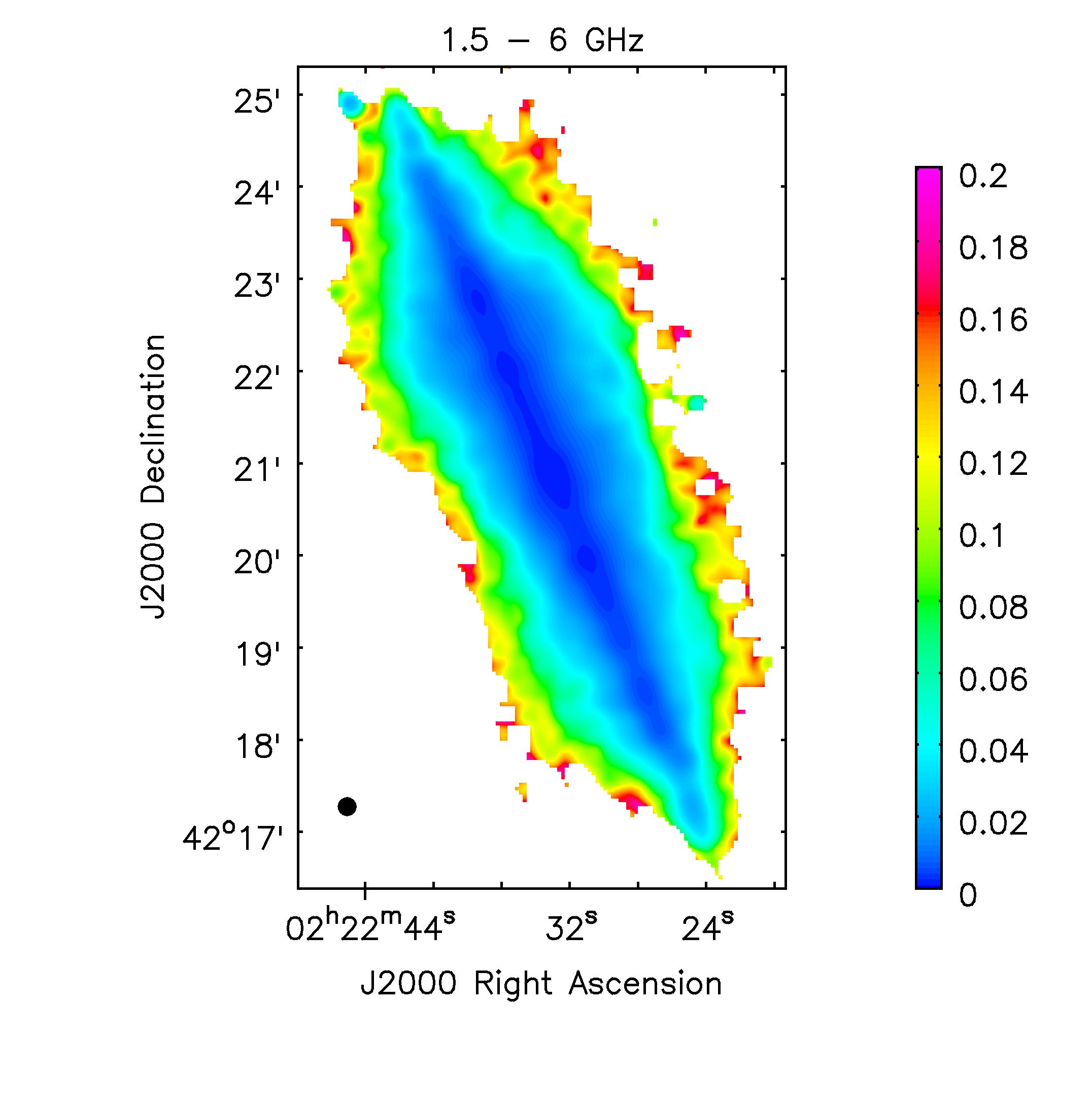}}}
 \caption{Short-spacing corrected spectral index maps of NGC\,891 and corresponding error maps.
 \it a\normalfont: 1.5\,GHz in-band total spectral index.
 \it b\normalfont: 1.5\,GHz in-band non-thermal spectral index.
 \it c\normalfont: 1.5\,GHz in-band spectral index error.
 \it d\normalfont: 6\,GHz in-band total spectral index.
 \it e\normalfont: 6\,GHz in-band non-thermal spectral index.
 \it f\normalfont: 6\,GHz in-band spectral index error.
 \it g\normalfont: Total spectral index between  1.5 and 6\,GHz.
 \it h\normalfont: Non-thermal spectral index between 1.5 and 6\,GHz.
 \it i\normalfont: Error of spectral index between 1.5 and 6\,GHz.
 The angular resolution in each panel is $12^{\prime\prime}$ FWHM
 (black circles). All maps extend to the $5\sigma$ level in the corresponding total intensity images. The black contours in the in-band maps are placed at the $30\sigma$ level, where the maps were cut off before creating the scatter plot in Fig.~\ref{fig:alpha3}~\it a\normalfont ; see text for details.}
 \label{fig:alpha12}
\end{figure*}

\begin{figure*}[h]
 \centering
 \stackinset{l}{0pt}{t}{0pt}{\it a)}{\stackinset{l}{26pt}{t}{12pt}{\Large$\alpha_{\mathrm{tot}}$}{\includegraphics[scale=0.13,clip=true,trim=200pt 200pt 80pt 135pt]{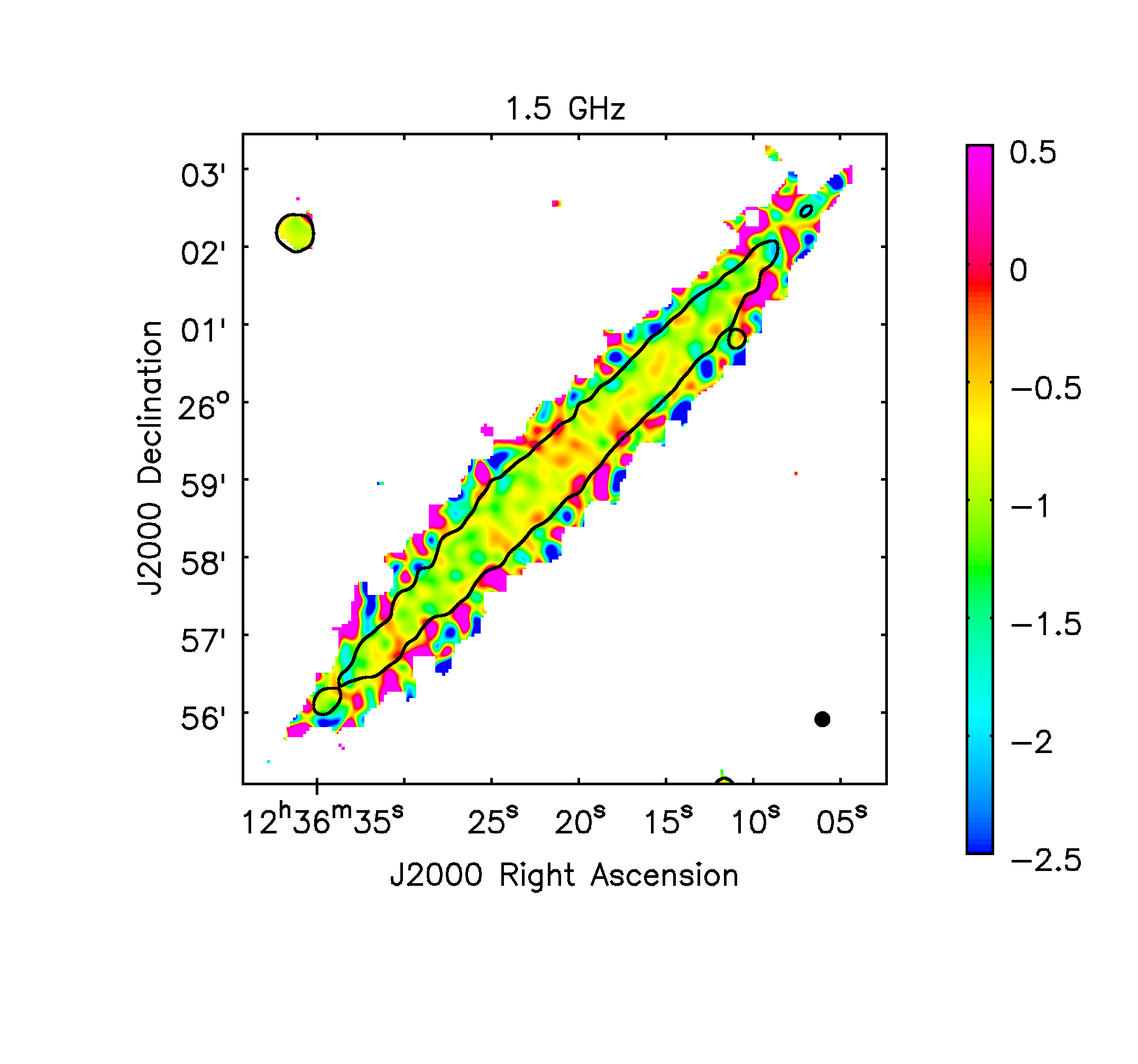}}}\stackinset{l}{0pt}{t}{0pt}{\it b)}{\stackinset{l}{22pt}{t}{12pt}{\Large$\alpha_{\mathrm{nth}}$}{\includegraphics[scale=0.13,clip=true,trim=237pt 200pt 80pt 135pt]{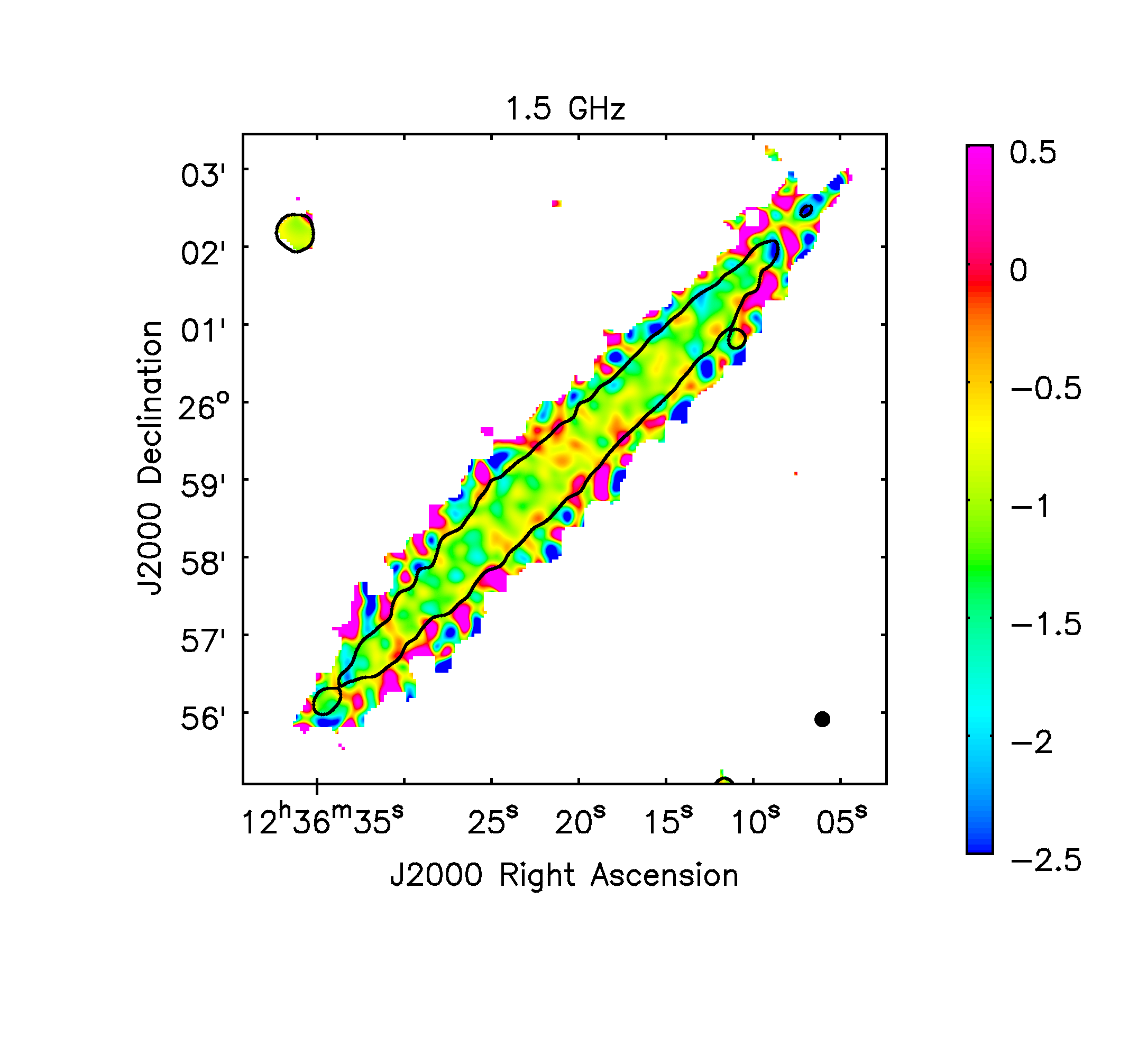}}}\stackinset{l}{0pt}{t}{0pt}{\it c)}{\stackinset{l}{22pt}{t}{10pt}{\Large$\Delta\alpha$}{\includegraphics[scale=0.13,clip=true,trim=237pt 200pt 80pt 135pt]{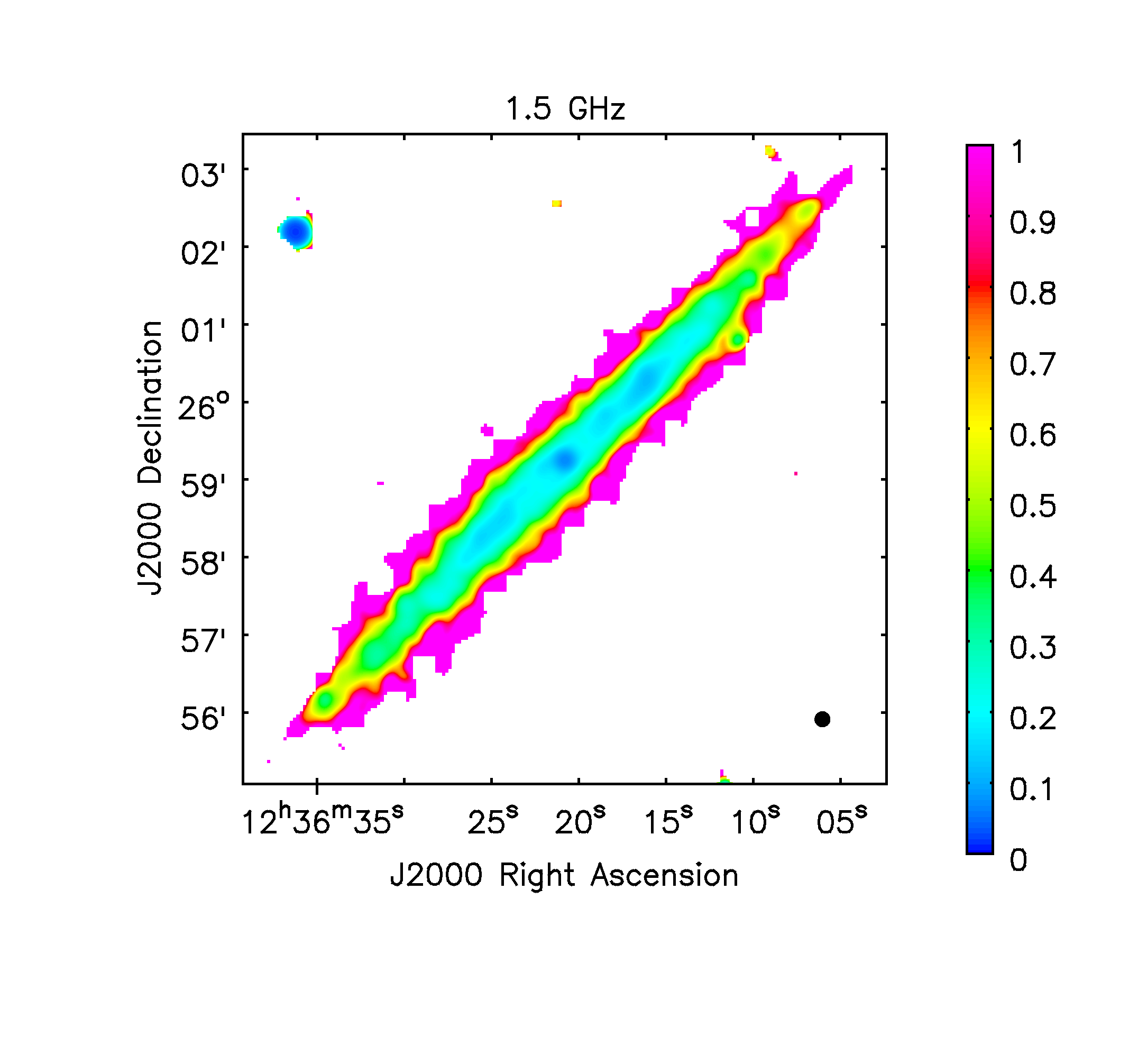}}}
 \stackinset{l}{0pt}{t}{0pt}{\it d)}{\stackinset{l}{26pt}{t}{12pt}{\Large$\alpha_{\mathrm{tot}}$}{\includegraphics[scale=0.13,clip=true,trim=200pt 200pt 80pt 135pt]{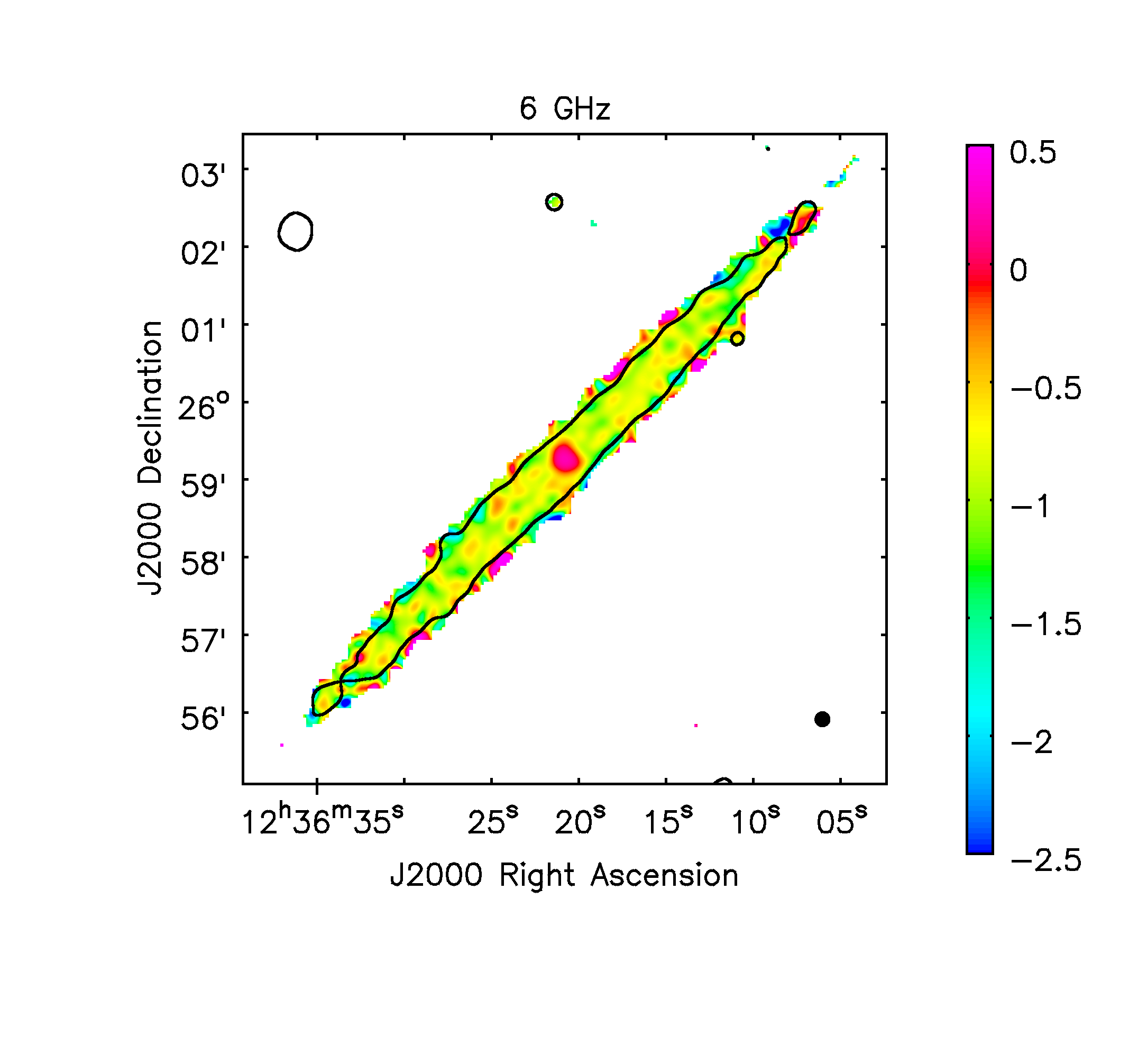}}}\stackinset{l}{0pt}{t}{0pt}{\it e)}{\stackinset{l}{22pt}{t}{12pt}{\Large$\alpha_{\mathrm{nth}}$}{\includegraphics[scale=0.13,clip=true,trim=237pt 200pt 80pt 135pt]{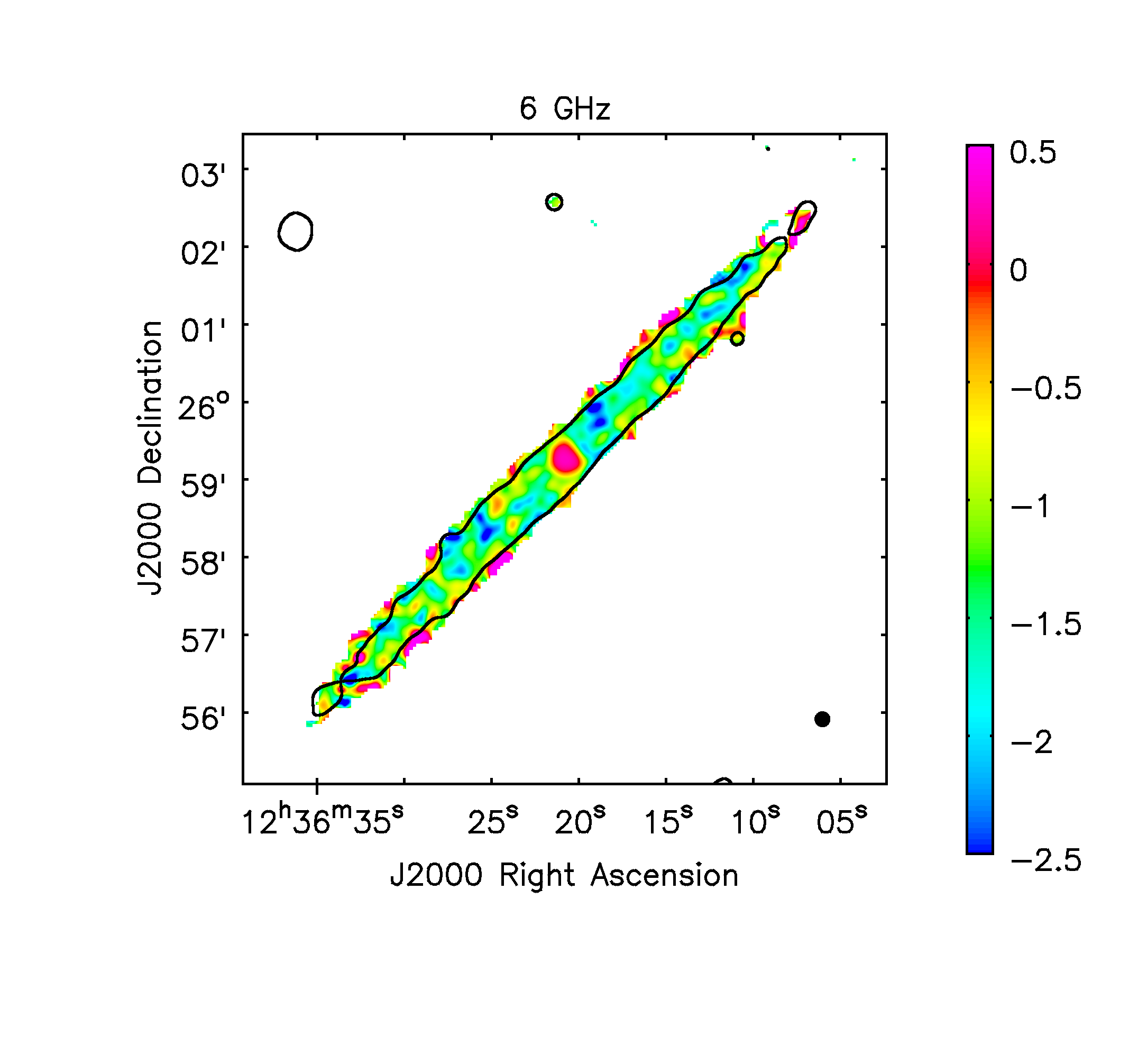}}}\stackinset{l}{0pt}{t}{0pt}{\it f)}{\stackinset{l}{22pt}{t}{10pt}{\Large$\Delta\alpha$}{\includegraphics[scale=0.13,clip=true,trim=237pt 200pt 80pt 135pt]{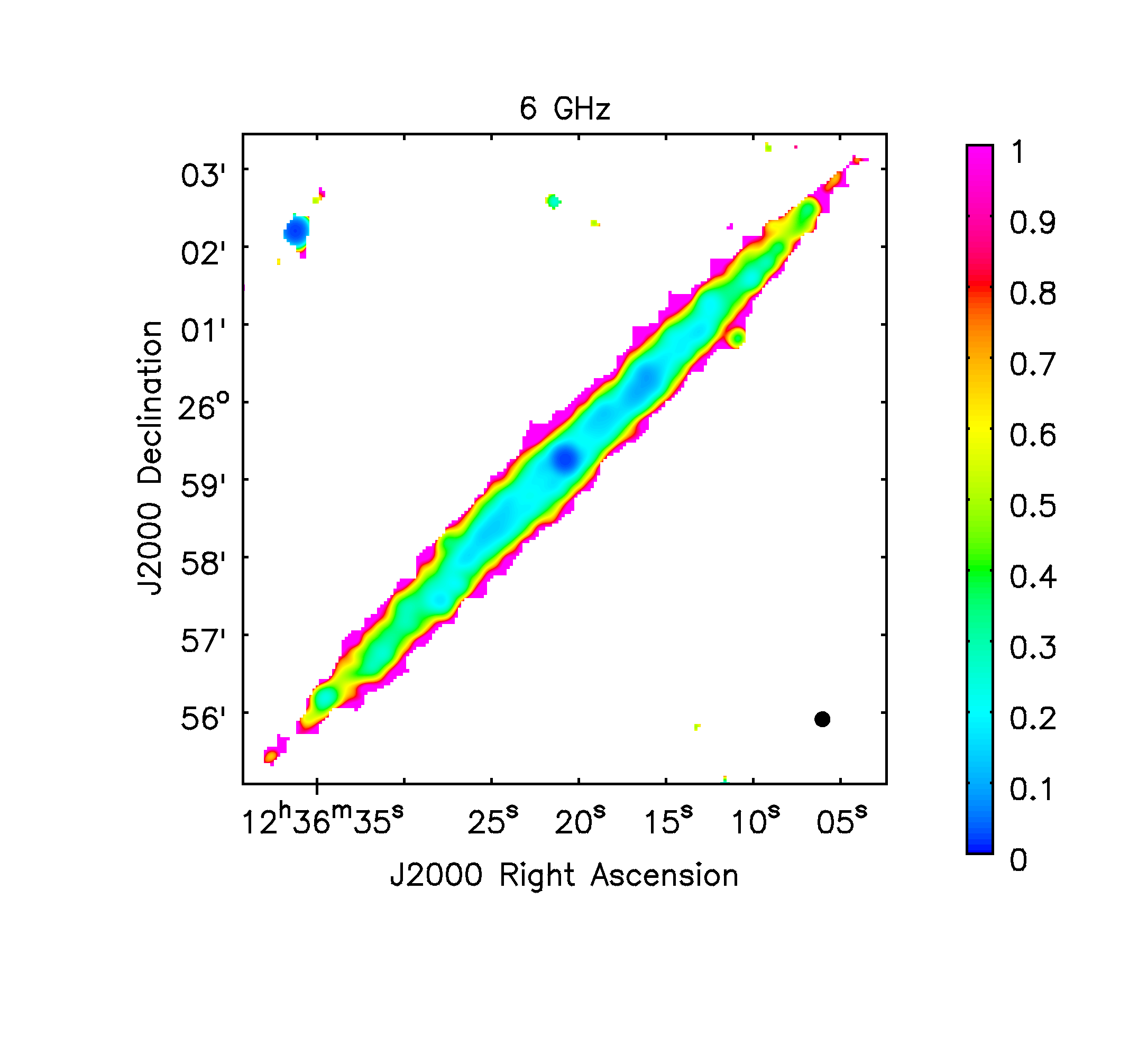}}}
 \stackinset{l}{0pt}{t}{0pt}{\it g)}{\stackinset{l}{26pt}{t}{12pt}{\Large$\alpha_{\mathrm{tot}}$}{\includegraphics[scale=0.13,clip=true,trim=200pt 200pt 80pt 135pt]{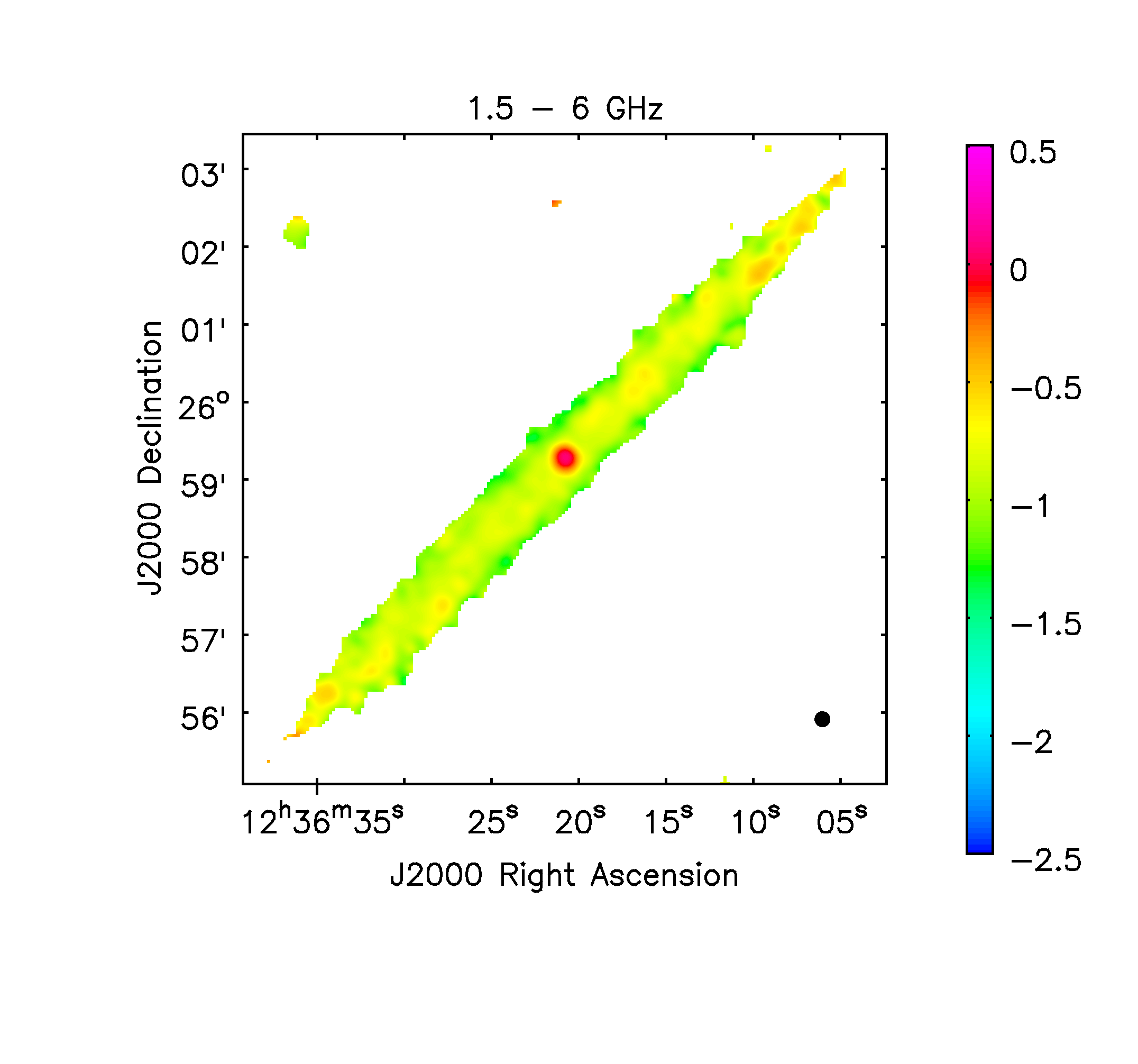}}}\stackinset{l}{0pt}{t}{0pt}{\it h)}{\stackinset{l}{22pt}{t}{12pt}{\Large$\alpha_{\mathrm{nth}}$}{\includegraphics[scale=0.13,clip=true,trim=237pt 200pt 80pt 135pt]{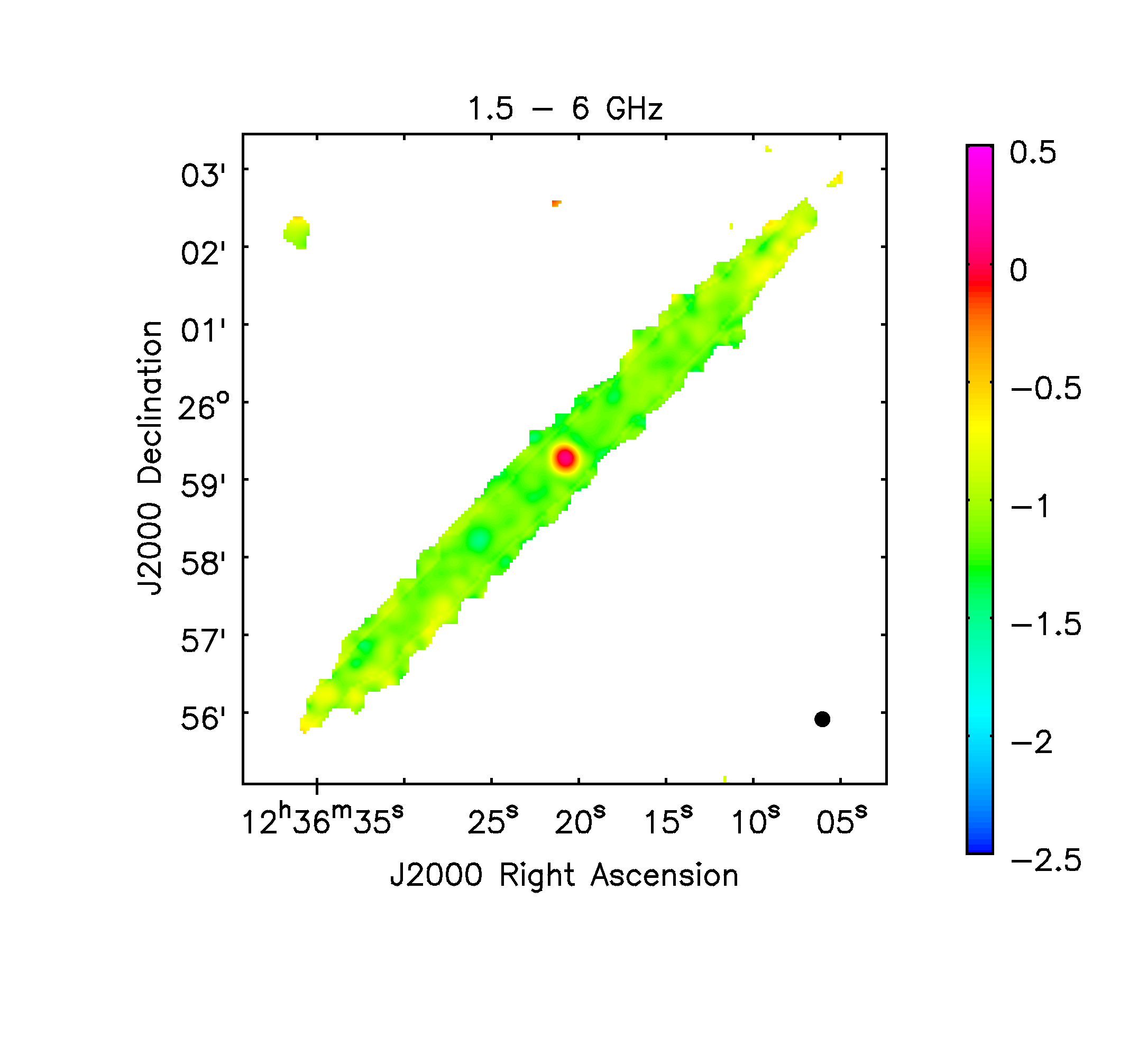}}}\stackinset{l}{0pt}{t}{0pt}{\it i)}{\stackinset{l}{22pt}{t}{10pt}{\Large$\Delta\alpha$}{\includegraphics[scale=0.13,clip=true,trim=237pt 200pt 80pt 135pt]{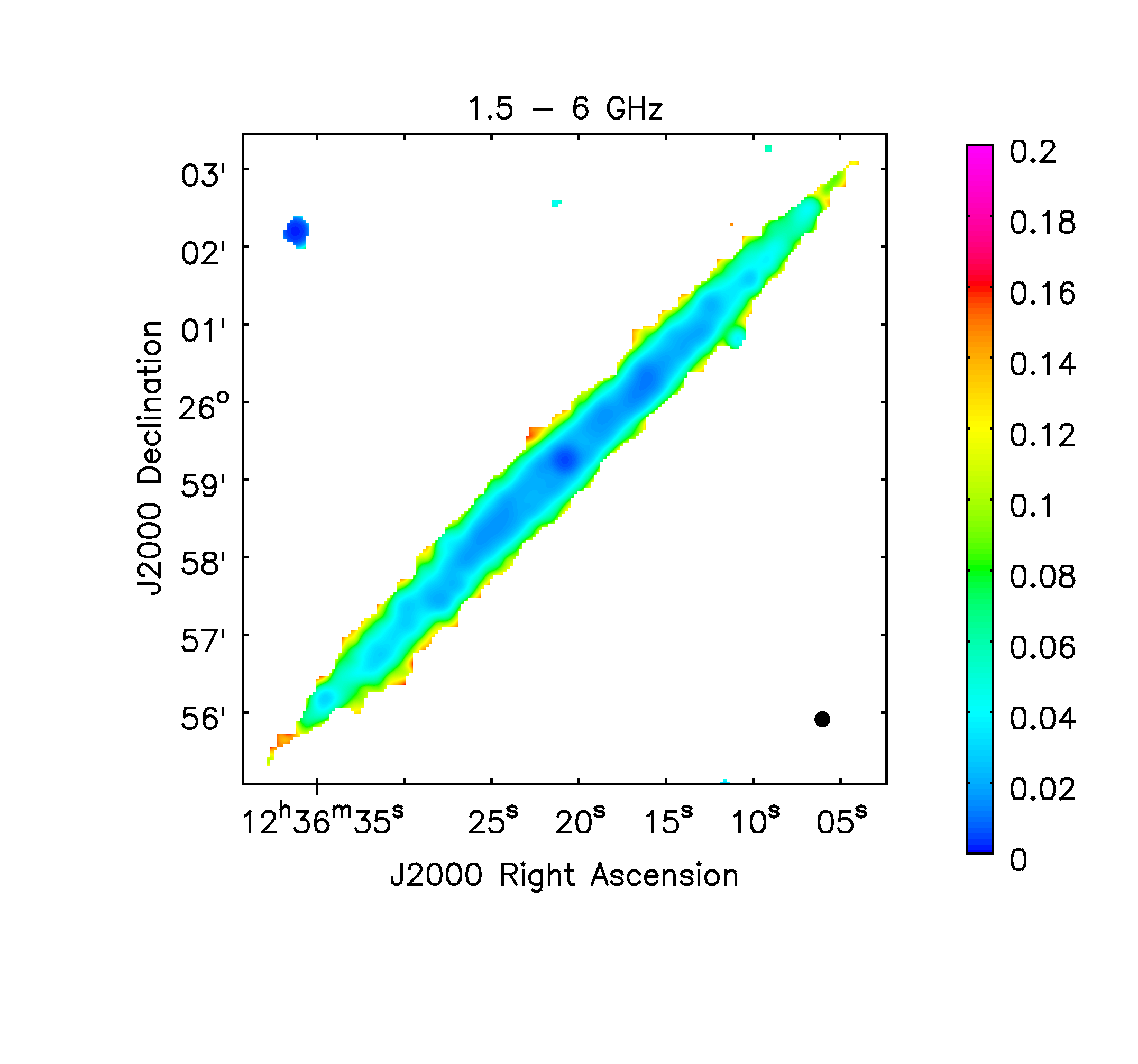}}}
 \caption{Short-spacing corrected spectral index maps of NGC\,4565 and corresponding error maps.
 \it a\normalfont: 1.5\,GHz in-band total spectral index.
 \it b\normalfont: 1.5\,GHz in-band non-thermal spectral index.
 \it c\normalfont: 1.5\,GHz in-band spectral index error.
 \it d\normalfont: 6\,GHz in-band total spectral index.
 \it e\normalfont: 6\,GHz in-band non-thermal spectral index.
 \it f\normalfont: 6\,GHz in-band spectral index error.
 \it g\normalfont: Total spectral index between 1.5 and 6\,GHz.
 \it h\normalfont: Non-thermal spectral index between 1.5 and 6\,GHz.
 \it i\normalfont: Error of spectral index between 1.5 and 6\,GHz.
 The angular resolution in each panel is $12^{\prime\prime}$ FWHM
 (black circles). All maps extend to the $5\sigma$ level in the corresponding total intensity images. The black contours in the in-band maps are placed at the $20\sigma$ level, where the maps were cut off before creating the scatter plot in Fig.~\ref{fig:alpha3}~\it b\normalfont. See text for more details.
 }
 \label{fig:alpha14}
\end{figure*}

\section{Spectral index distribution}
\label{spectralindex}

Before we can reliably interpret spectral index measurements from wide-band radio continuum data, a number of corrections and amendments during and after deconvolution need to be applied, which are not part of the standard imaging procedure described in Sect.~\ref{vlareduction}. In the first part of this section, we outline each of the necessary correction methods and their specific application to our data. We afterwards investigate our final results for the non-thermal spectral index distribution for consistency with the standard spectral ageing  models. In what follows, we restrict ourselves to the spectral index maps at intermediate resolution ($12\arcsec$), as the signal-to-noise ratio (S/N) of the high-resolution maps is insufficient to study extraplanar CRE propagation. Another reason is that the angular resolution of our far-infrared data is not matched to these high-resolution maps and we thus cannot correct for the thermal contribution.

\subsection{Post-imaging corrections}

\subsubsection{Inner $uv$-cutoff}
\label{uvcutoff}

For wide-band interferometric observations of extended sources, the lowest spatial frequencies (up to the lower bound $u_{\mathrm{min}}$ at the highest frequency $\nu_{\mathrm{max}}$) are only covered by a fraction of the frequency band, such that at large spatial scales there is progressively more flux missing as one moves to higher frequencies \citep[see][]{rau11}. This artificially steepens the spectrum measured at these scales. To avoid this bias towards steeper in-band spectral indices, we performed additional imaging runs excluding data at spatial frequencies below $u_{\mathrm{min}}$ at $\nu_{\mathrm{max}}$.\footnote{The fraction of visibility data lost in this way was less than $\approx3\%$ in each case.} Spatial frequency cutoffs at $0.25\,\mathrm{k}\lambda$ in L-band and at $0.8\,\mathrm{k}\lambda$ in C-band were thus applied for the in-band spectral index computation.

\subsubsection{Revised primary-beam correction}
\label{revpbcor}

In case of the C-band (D-array) observations, we found that the intensity and spectral index maps of the two separate pointings still showed major differences within their overlap region after primary-beam (PB) correction \citep[cf.][]{wiegert15}. We were able to improve the matching between the pointings by using the PB model at 6.25\,GHz (6.6\,GHz in case of the spectral index map of NGC\,891) instead of the centre frequency of 6.0\,GHz.

\subsubsection{Short-spacing corrections}
\label{alphashortspacings}

We computed `classical' (i.e. two-point) spectral index maps between 1.5 and 6\,GHz ($\alpha_{\mathrm{1.5-6GHz}}$) from the short-spacing corrected 1.5 and 6\,GHz total intensity maps, and accordingly between their non-thermal counterparts. These are shown in panels \it g \normalfont and \it h \normalfont of Figs.~\ref{fig:alpha12} and \ref{fig:alpha14}.

Also, we performed short-spacing corrections of the in-band $\alpha$ maps obtained by MFS imaging. In L-band, we performed the merging of B+C+D-array with D-array-only data (as described in Sect.~\ref{shortspacings}) for both Taylor-term images, to obtain short-spacing corrected maps of $\alpha_{\mathrm{tot,1.5GHz}}$ (Figs.~\ref{fig:alpha12}--\ref{fig:alpha14} \it a\normalfont). The corresponding $\alpha_{\mathrm{nth,1.5GHz}}$ maps (Figs.~\ref{fig:alpha12}--\ref{fig:alpha14} \it b\normalfont) were computed as
\begin{equation}
 \alpha_{\mathrm{nth,1.5GHz}}=\frac{I_{1}-\alpha_{\mathrm{th}} \, I_{0,\mathrm{th}}}{I_{0,\mathrm{nth}}}\,,
\end{equation}
where $I_{1}$ is the short-spacing corrected second Taylor-term map, $I_{0,\mathrm{th}}$ and $I_{0,\mathrm{nth}}$ are the thermal and short-spacing corrected non-thermal intensity maps, respectively, and $\alpha_{\mathrm{th}}=-0.1$.

As our Effelsberg data have a much narrower bandwidth than the VLA C-band observations, it is not possible to combine them during MFS imaging to produce short-spacing corrected in-band spectral index maps at 6\,GHz. Instead, we used MFS to image the C-band data at the beginning (5\,GHz) and end (7\,GHz) of the frequency band (applying the $0.8\,\mathrm{k}\lambda$ $uv$-cutoff), and merged the resulting total intensity maps\footnote{That is, after PB correction. Following the approach outlined in Sect.~\ref{revpbcor}, we used the PB models at 5.1\,GHz and 7.5\,GHz, which resulted in the best agreement between the two pointings in total intensity.} with the Effelsberg maps scaled to 5 and 7\,GHz, respectively\footnote{We note that this results in different sizes of the gap in the `$uv$-coverage' at the two frequencies, since the re-scaled Effelsberg maps have the same angular resolution and hence different effective dish diameters.}. 
From the short-spacing corrected 5 and 7\,GHz maps, we computed two-point $\alpha$ maps, which we adopt as the `in-band' $\alpha_{\mathrm{tot,6GHz}}$ distribution (Figs.~\ref{fig:alpha12}--\ref{fig:alpha14} \it d\normalfont). The same was done for the non-thermal emission maps, after computing $I_{0,\mathrm{th}}$ maps at 5 and 7\,GHz (Figs.~\ref{fig:alpha12}--\ref{fig:alpha14} \it e\normalfont).

\subsection{Spectral index maps}

\subsubsection{NGC\,891}
\label{finalalphaN891}

NGC\,891 shows the overall pattern typically observed in edge-on galaxies, that is rather flat mid-plane spectral indices and (on average)
steepening of the spectra with vertical height (Fig.~\ref{fig:alpha12}, left column). The angular resolution is sufficient to observe a relatively sharp transition between flat and steep spectral indices in the disc--halo interface region. An outstanding feature in the $\alpha_{\mathrm{6GHz}}$ and $\alpha_{\mathrm{1.5-6GHz}}$ maps is the radio supernova SN 1986J \citep{vangorkom86} at $\mathrm{RA}=02^{h}22^{m}31^{s}.32$, $\mathrm{Dec} = 42^{\circ}19^{\prime}57^{\prime\prime}.26$ \citep{bietenholz10}, with a spectral index at its centre of $\approx0.0$ at 6\,GHz and $\approx-0.5$ at 1.5\,GHz. In the mid-plane region of NGC\,891, $\alpha_{\mathrm{tot,6GHz}}$ is considerably flatter than $\alpha_{\mathrm{tot,1.5GHz}}$, due to the larger thermal contribution at 6\,GHz.

Around the edges, all in-band
$\alpha$ maps are dominated by either extremely steep, extremely flat, or even positive values, which are artefacts arising from instabilities of the spectral fit in regions of too low S/N. We were able to exclude these edge artefacts from our further analysis to a satisfactory extent by clipping the in-band $\alpha$ maps at the $30\sigma$ level.

After our thorough error consideration \citep[see also][]{schmidt16}, the corrections described in Sect.~\ref{uvcutoff} to Sect.~\ref{alphashortspacings}, and the high clipping level applied to our data we conclude that the fluctuations seen in the in-band spectral index maps (Figs.~\ref{fig:alpha12} a, b, d, and e) are real.
Around the mid-plane, $\alpha_{\mathrm{nth}}$ turns out to be in close agreement at both frequencies, which indicates that we found a realistic estimate of the thermal contribution in the disc.

\subsubsection{NGC\,4565}

Contrary to NGC\,891, where a relatively clear distinction between disc and halo is visible in the $\alpha$ distribution, only a very small portion of the halo is covered by our $\alpha$ maps of NGC\,4565 (Fig.~\ref{fig:alpha14}), because here low S/N values are reached already at much lower $z$ heights. 
While in the $\alpha_{\mathrm{1.5-6GHz}}$ the disc still shows a relatively homogeneous distribution of values around $\approx-0.7$ (total emission) and $\approx-1.1$ (non-thermal), the in-band $\alpha$ maps are characterised by stronger fluctuations on small scales, similar to the halo of NGC\,891.
Still, the ring-shaped structure seen in total intensity is also clearly identifiable in the $\alpha_{\mathrm{tot,6GHz}}$ map (and to a lesser degree in $\alpha_{\mathrm{nth,6GHz}}$), in the form of elongated flat-spectrum regions.

Overall, both total and non-thermal spectra in NGC\,4565 are found to be steeper than in the disc of NGC\,891. In particular,
$\alpha_{\mathrm{nth,6GHz}}$ in NGC\,4565 is considerably steeper than $\alpha_{\mathrm{nth,1.5GHz}}$ after subtracting the thermal contribution, whereas in the disc of NGC\,891 we measure similar non-thermal spectral indices at both frequencies. Even the $\alpha_{\mathrm{nth,1.5-6GHz}}$ map shows values as steep as in the lower halo ($\approx-1.2$) over a large portion of the disc. These remarkable differences between the non-thermal spectra of the two galaxies appear to be a direct consequence of their very different SFR surface densities, as we will discuss further below.

\begin{figure*}[h]
 \centering
 \normalsize NGC\,891 (1.5\,GHz) \\
 \includegraphics[scale=0.48,clip=true,trim=0pt 0pt 0pt 0pt]{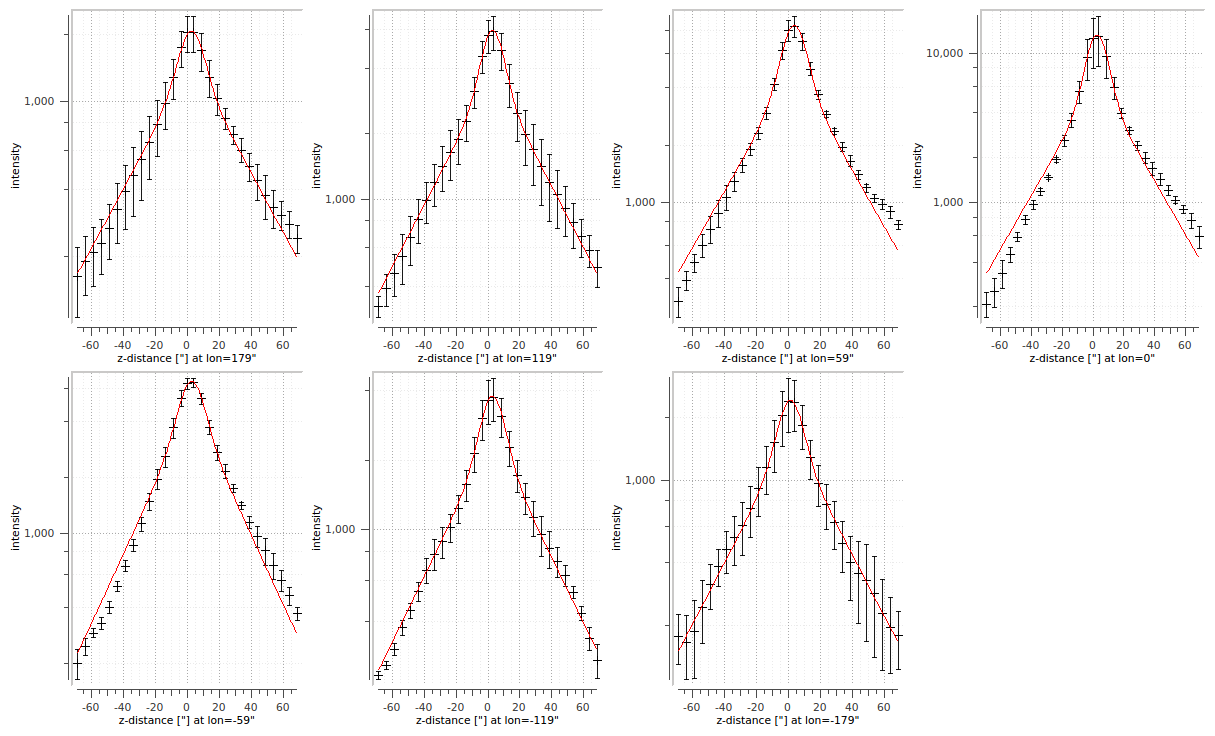} \\
 \vspace{3mm}
 \normalsize NGC\,891 (6\,GHz) \\
 \includegraphics[scale=0.48,clip=true,trim=0pt 0pt 0pt 0pt]{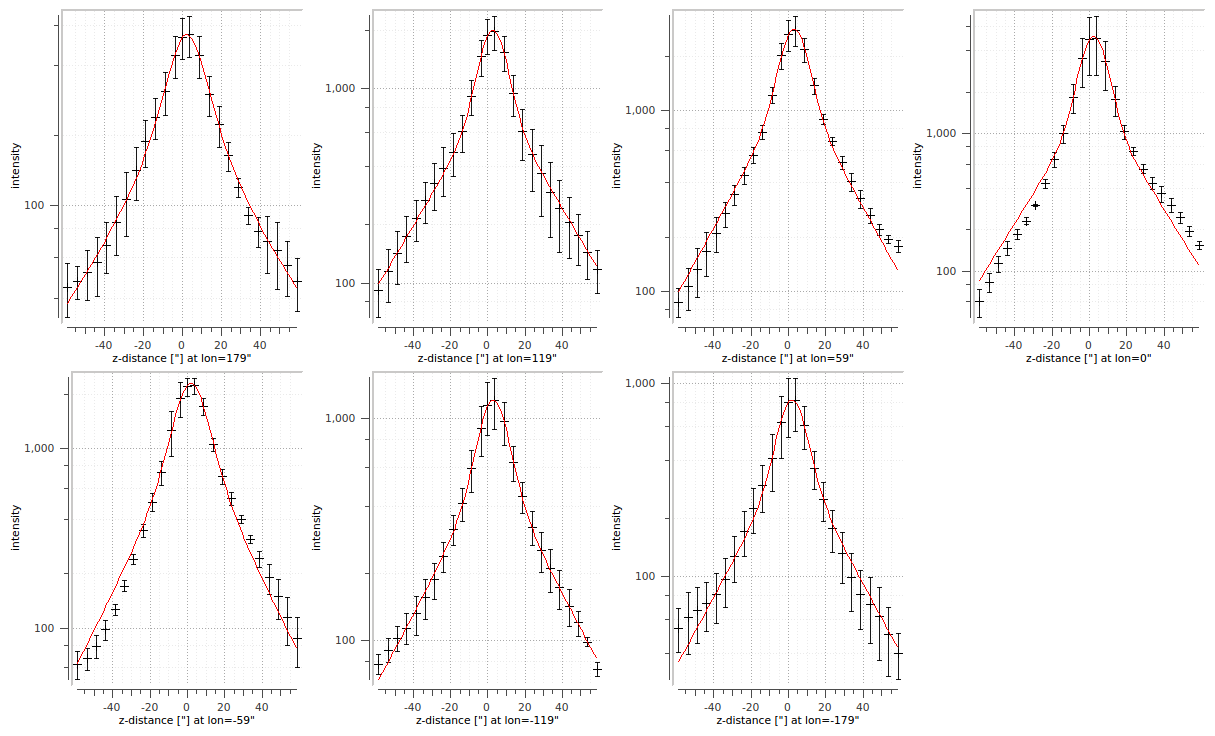}
 \caption{Vertical profiles of non-thermal intensity of NGC\,891 at 1.5 and 6\,GHz as function of distance $z$ to the galactic mid-plane, where positive $z$ values are on the north side and negative ones on the south side of the mid-plane. The galactic longitude at the centre of each vertical strip is indicated below each sub-panel ($1\arcsec\equiv44$\,pc). The data points correspond to the average intensity (in $\mu \mathrm{Jy\,beam^{-1}}$) within each rectangular box defined with \texttt{BoxModels}. The solid lines denote two-component exponential least-squares fits to the data (see Appendix~\ref{appendix_scaleh}).
 }
 \label{fig:cr2}
\end{figure*}

\begin{figure*}[h]
 \centering
 \normalsize NGC\,4565 (1.5\,GHz) \\
 \includegraphics[scale=0.501,clip=true,trim=0pt 0pt 0pt 0pt]{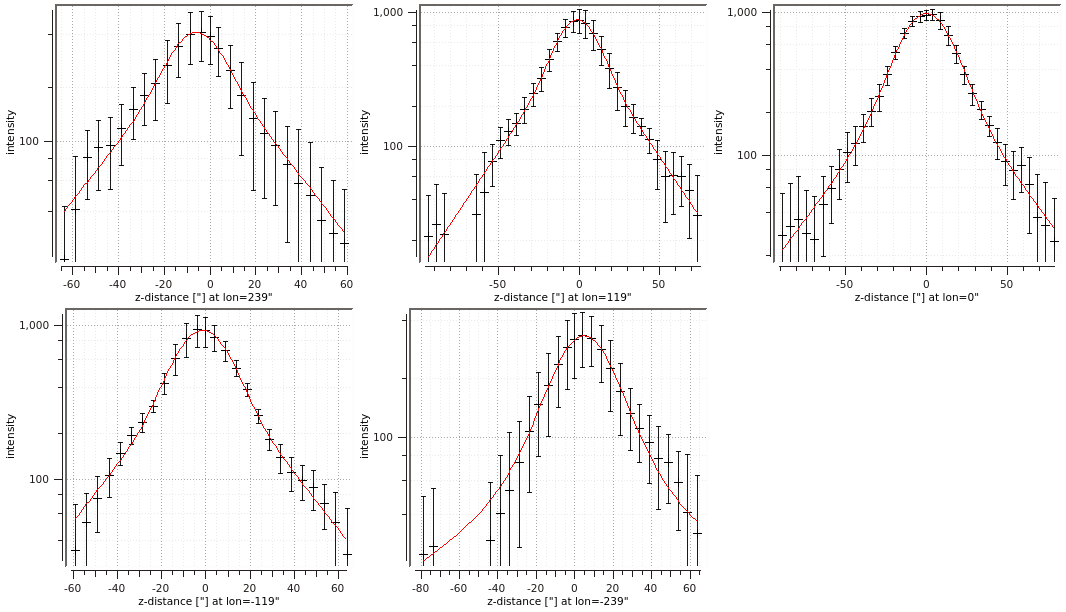} \\
 \vspace{3mm}
 \normalsize NGC\,4565 (6\,GHz) \\
 \includegraphics[scale=0.501,clip=true,trim=0pt 0pt 0pt 0pt]{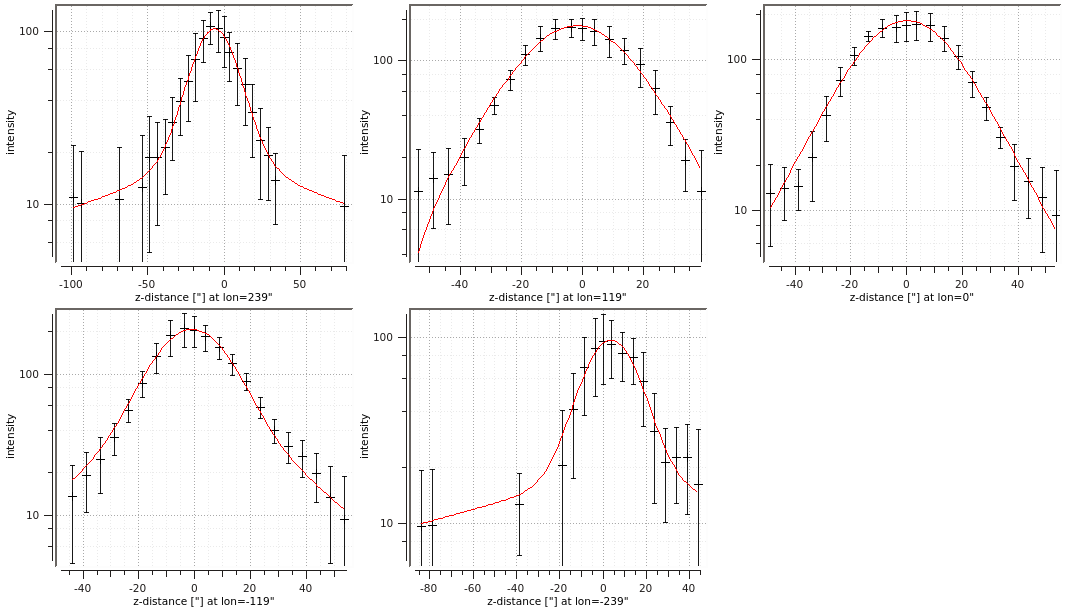}
 \caption{Vertical profiles of non-thermal intensity of NGC\,4565 at 1.5 and 6\,GHz as function of distance $z$ to the galactic mid-plane, where positive $z$ values are on the north side and negative ones on the south side of the mid-plane. The galactic longitude at the centre of each vertical strip is indicated below each sub-panel ($1\arcsec\equiv58$\,pc). The data points correspond to the average intensity (in $\mu \mathrm{Jy\,beam^{-1}}$) within each rectangular box defined with \texttt{BoxModels}. The solid lines denote two-component exponential least-squares fits to the data (see Appendix~\ref{appendix_scaleh}). In a few cases we did not use the fits to the entire vertical profile (as shown here), but instead fitted the parts at positive and negative $z$ separately and in each case adopted the mean value of the two resulting scale heights.
 }
 \label{fig:cr3}
\end{figure*}

\section{Synchrotron scale heights}
\label{scaleheights}

\subsection{Fitting the vertical intensity profiles}
\label{scalefitmethod}

We determined scale heights of the total and non-thermal (synchrotron) radio emission of NGC\,891 and 4565 by least-squares fitting of vertical ($z$-direction) intensity profiles \citep{dumke95,krause18}.
As input data we used the short-spacing corrected 1.5 and 6\,GHz images at $12\arcsec$ FWHM
resolution, from 
which we removed background sources close to the galaxies, as well as SN 1986J in NGC\,891 and the central source of NGC\,4565. The NOD3 task \texttt{BoxModels} 
\citep{mueller17} was used to fit $z$ profiles at different positions along the major axis of each galaxy. \texttt{BoxModels} computes the mean intensity within 
boxes of galactocentric radial width $\Delta r$ as projected onto the major-axis and vertical height interval $\Delta z$. To take the inclination of the galaxies into account and thus obtain deprojected 
scale heights, an effective beam size is automatically determined at each specified major-axis position.

In both galaxies, we sampled the vertical intensity distribution in steps of $\Delta z=5\arcsec$, corresponding to 0.22\,kpc at the distance of NGC\,891 and 0.29\,kpc in case 
of NGC\,4565.
To investigate the radial behaviour of scale heights for NGC\,891, we used seven strips of width $\Delta r=2.65$\,kpc, while for NGC\,4565, due to its much lower S/N, 
we chose five strips with $\Delta r=6.9$\,kpc. 
The emission of NGC\,891 appears to be superimposed on a plateau of extended low-level emission. We consider this plateau to be an artefact that may have been caused by inaccuracies 
in interpolating the Effelsberg map from 4.85 to 6\,GHz. We excluded it from our fitting because it would have otherwise systematically 
broadened our model profiles.

Describing the vertical emission distribution clearly requires two-component models (disc and halo) for NGC\,891, while for NGC\,4565 the situation is somewhat ambiguous, 
as we discuss below. As known from the studies of \citet{dahlem94} and \citet{dumke98}, we found the $z$ profiles of NGC\,891 to be better represented by 
two-component exponentials than by two-component Gaussians. 
For NGC\,4565, we found exponential and Gaussian models to provide similarly good fits. 
To enable the comparison of our scale heights to those of other edge-on galaxies, we adopt the results of the exponential fits for most of the subsequent analysis and 
discussion. However, we will also consider the Gaussian results for NGC\,4565 in the CRE transport modelling (Sect.~\ref{difflossmod}).

\begin{figure*}[h]
 \centering
 \topinset{\it a)}{\includegraphics[width=0.5\textwidth,clip=true,trim=15pt 0pt 0pt 5pt]{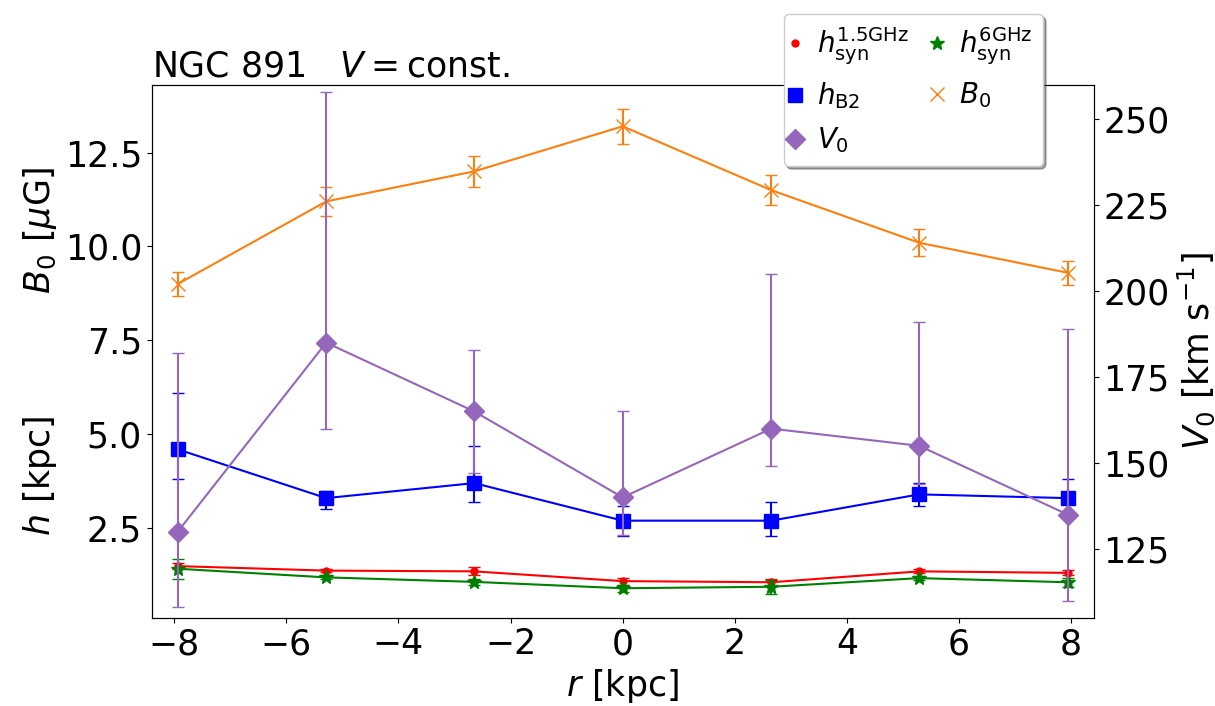}}{0pt}{0pt}\topinset{\it b)}{\includegraphics[width=0.5\textwidth,clip=true,trim=15pt 0pt 0pt 5pt]{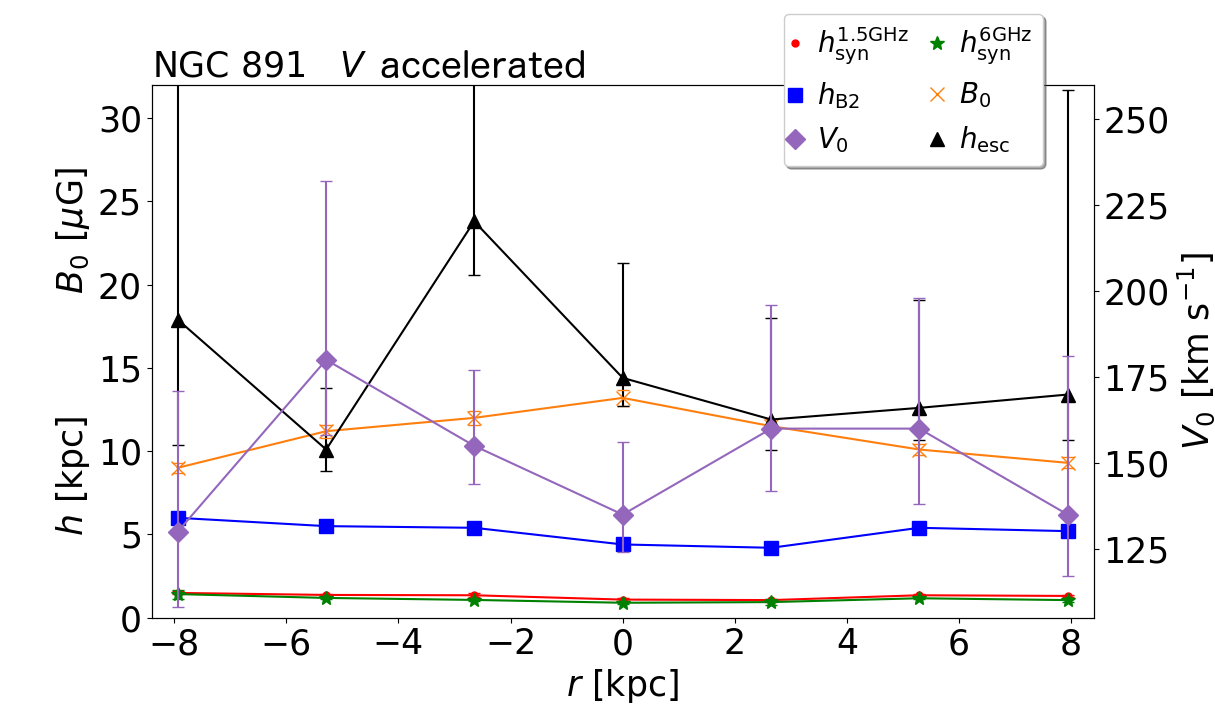}}{0pt}{0pt} \\
 \vspace{3mm}
 \topinset{\it c)}{\includegraphics[width=0.5\textwidth,clip=true,trim=15pt 0pt 0pt 5pt]{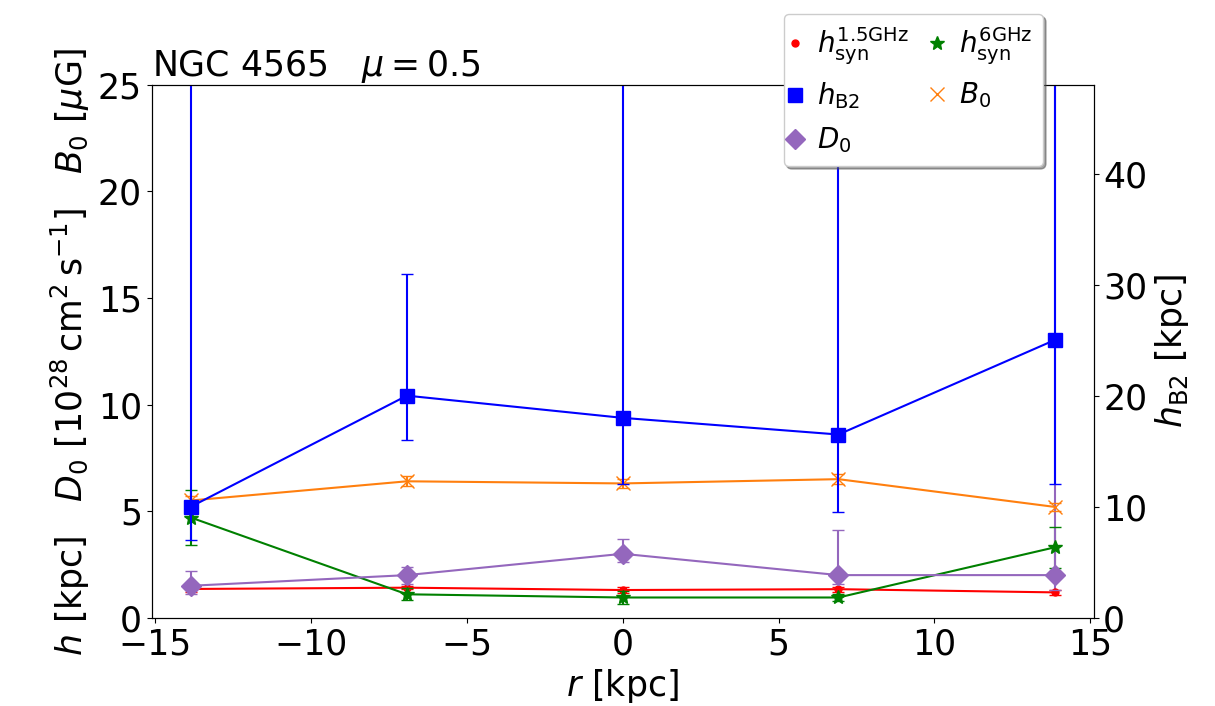}}{0pt}{0pt}\topinset{\it d)}{\includegraphics[width=0.5\textwidth,clip=true,trim=15pt 0pt 0pt 5pt]{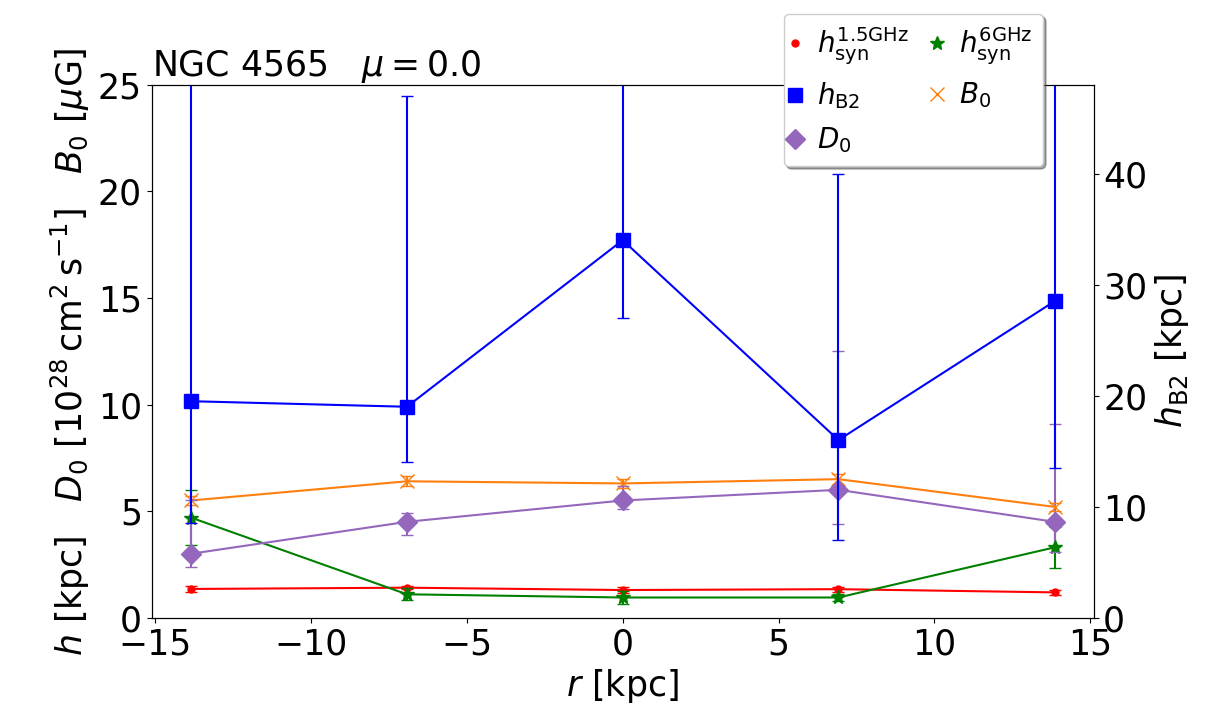}}{0pt}{0pt}
 \caption{Various quantities detemined in this work, shown as function of galactocentric radius $r$ as projected onto the major axis: synchrotron halo scale height $h_{\mathrm{syn}}$ at 1.5 and 6\,GHz, halo magnetic field scale height $h_\mathrm{B2}$, mid-plane total magnetic field strength $B_{0}$, advection speed $V_{0}$, CRE escape height $h_\mathrm{esc}$, and diffusion coefficient $D_{0}$. For NGC\,891, $h_\mathrm{B2}$ and $V_{0}$ are shown for constant (\it a\normalfont) and (including $h_\mathrm{esc}$) for accelerated CRE advection (\it b\normalfont). For NGC\,4565, $h_\mathrm{B2}$ and $D_{0}$ are shown for energy-dependent (\it c\normalfont) and energy-independent CRE diffusion (\it d\normalfont). In case of $V_{0}$, $D_{0}$, $h_\mathrm{B2}$, and $h_\mathrm{esc}$, the average between the northern and southern values is plotted. Negative values of $r$ are east and positive ones are west of the minor axis.
 }
 \label{fig:cr4}
\end{figure*}

\begin{figure*}[h]
 \centering
 \topinset{\it a)}{\includegraphics[scale=0.17,clip=true,trim=200pt 146pt 70pt 90pt]{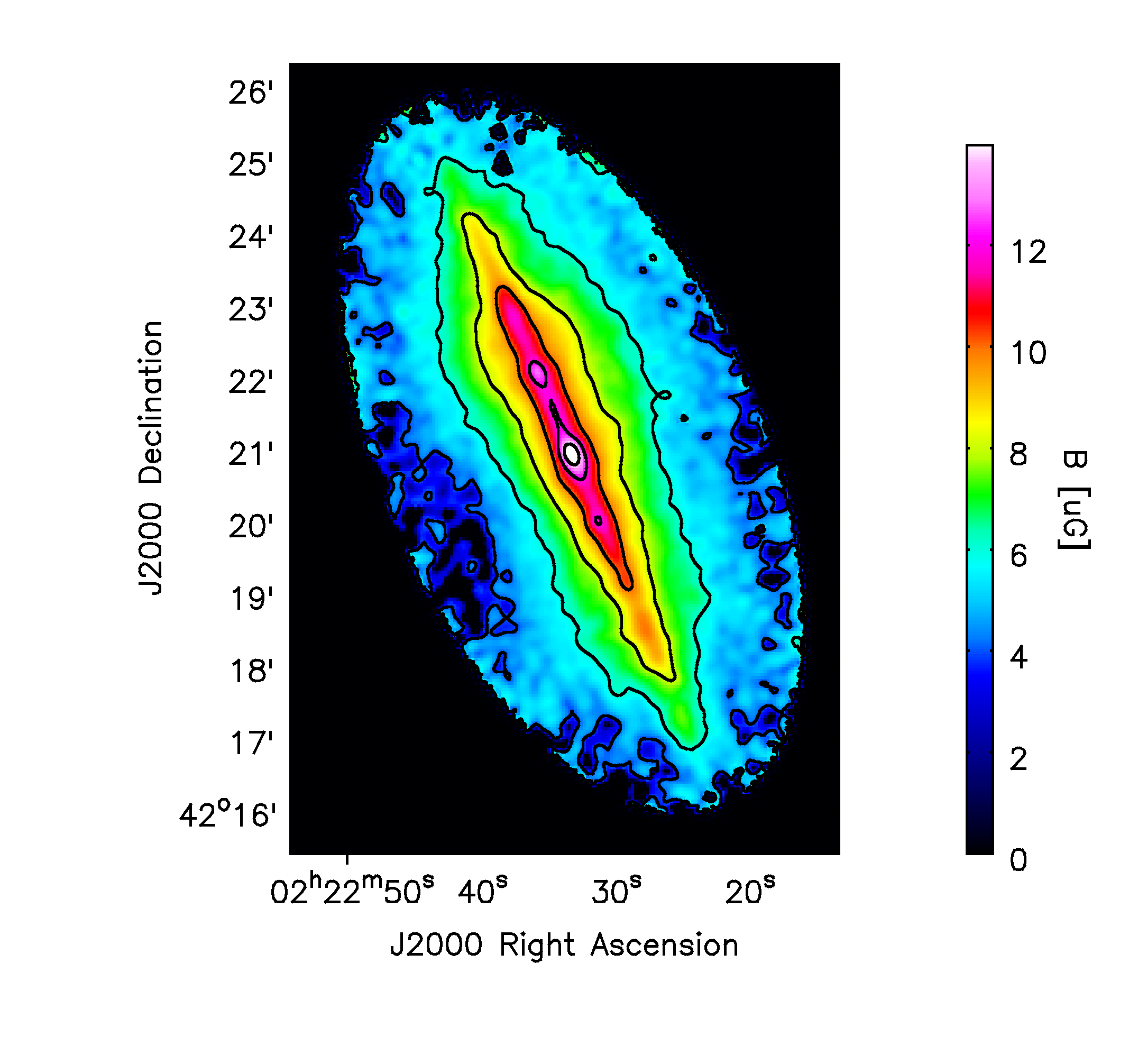}}{0pt}{0pt}\topinset{\it b)}{\includegraphics[scale=0.17,clip=true,trim=100pt 146pt 70pt 90pt]{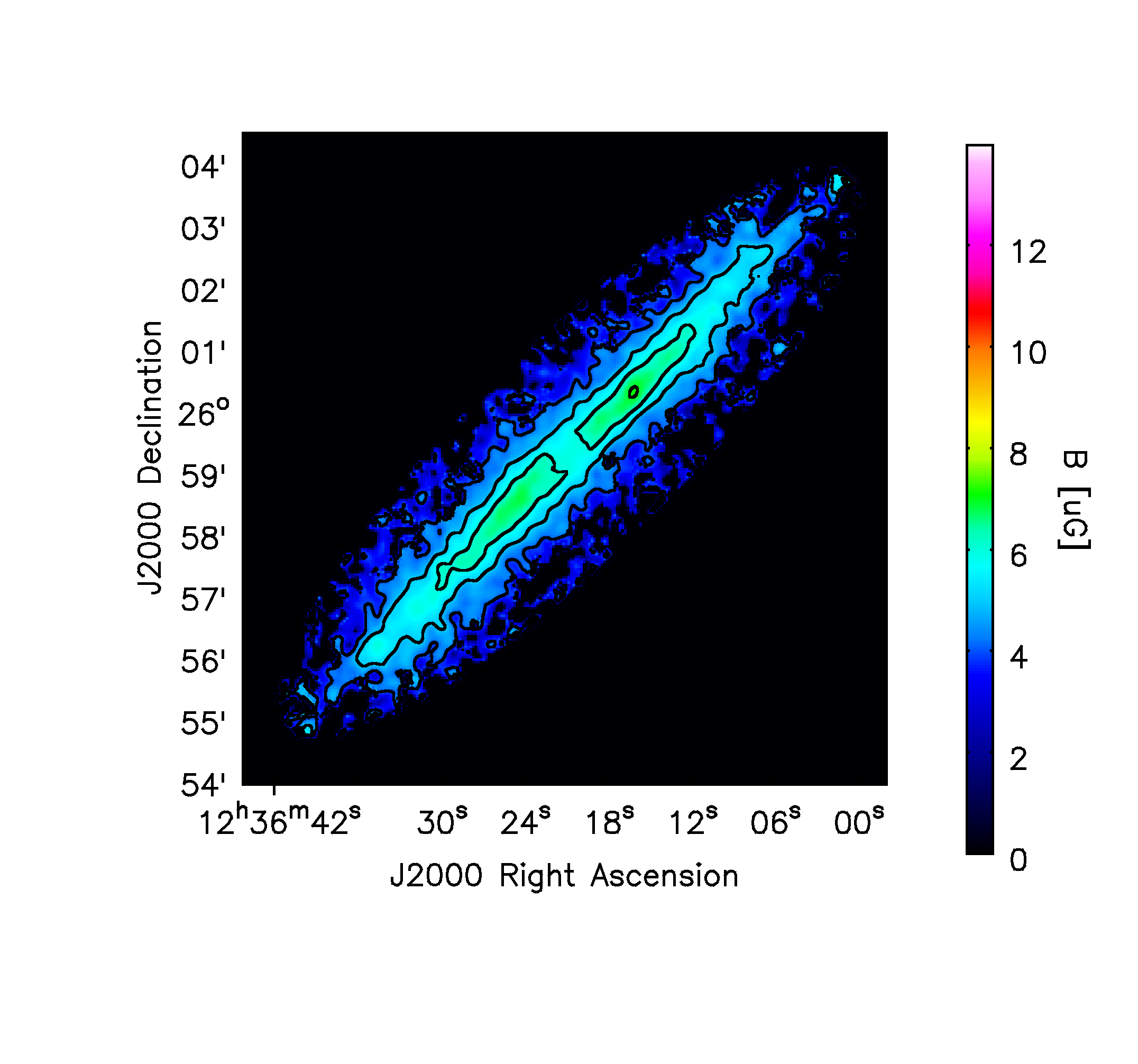}}{0pt}{15pt}
 \caption{Maps of total magnetic field strength in NGC\,891 (panel \it a\normalfont) and NGC\,4565 (panel \it b\normalfont), computed assuming energy equipartition and an oblate spheroid geometry (see text). The field strength was not calculated in regions where the spectral index between 1.5 and 6\,GHz is steeper than $-1.2$. The angular resolution in both panels is $12\arcsec$ FWHM.}
 \label{fig:cr5}
\end{figure*}

\subsection{Results}
\label{scalefitresults}

The measured intensity profiles (including the first ever results for the non-thermal scale heights of each galaxy\footnote{For brevity, we only present the non-thermal profiles here.}), along with the best-fitting models, are displayed in Figs.~\ref{fig:cr2} and \ref{fig:cr3}. The error of each data point is the average standard deviation within the single horizontal rows of pixels in the respective box. This is to avoid the effect of intensity gradients in $z$ on the errors. The resulting scale heights of the total and non-thermal emission in disc and halo, $h_{\mathrm{disc}}$ and $h_{\mathrm{halo}}$, 
are presented 
in Appendix~\ref{appendix_scaleh}. Corresponding plots of the non-thermal halo scale height versus galactocentric radius as projected onto the major-axis are shown in Fig.~\ref{fig:cr4}. 
For each strip, we attempted to fit a single profile to both sides above ($z>0$) and below ($z<0$) the mid-plane. For NGC\,4565 this type of fit was not always successful. In these cases, we performed separate fits for the northern and southern side and adopted the average scale height, replacing the formal fit errors by half the difference between the northern and southern value. The scale heights determined in this way are especially marked (see Appendix~\ref{appendix_scaleh} for the values). The large reduced $\chi^2$ values for NGC\,891 in the central strips are most probably resulting from the east--west asymmetry of its intensity distribution, that is from the different shapes of the $z$ profile on each side of the major axis. We found that separate fits for each side achieve much lower reduced $\chi^2$ values in the central strips for this galaxy, but yield average scale heights consistent with the single-fit results.

As evident from Figs.~\ref{fig:cr4} \it a \normalfont and \it b\normalfont, the dumbbell shape of the halo of NGC\,891 is reflected in the radial behaviour of its synchrotron scale heights. 
In line with the north--south asymmetry in total intensity, the scale heights are on average somewhat larger in the northern halo. The radial dependence of the scale heights in NGC\,891 is very similar at both frequencies. In NGC\,4565, on the other hand, there is a strong radial increase of the scale height at 6\,GHz (with large uncertainties, however) that is not observed at 1.5\,GHz.

The non-thermal halo scale heights $h_{\mathrm{syn}}$ of NGC\,891 increase from the central disc, where the total magnetic field strength in the mid-plane $B_{0}$ is largest, towards the outer disc with smaller values of $B_{0}$. The relation $h_{\mathrm{syn}} \propto B_{0}^{\xi}$ at 1.5 and 6\,GHz is shown in Fig.~\ref{fig:hvsB}. The fitted lines have slopes of $\xi=-1.39\pm0.64$ at 1.5\,GHz and $\xi=-1.32 \pm0.61$ at 6\,GHz. For NGC\,4565, no significant relation between $h_{\mathrm{syn}}$ and $B_{0}$ is found at either frequency.

The non-thermal halo scale heights at 1.5\,GHz are consistently larger than those at 6\,GHz in both galaxies. The average ratio between the halo scale heights of the non-thermal emission at the two frequencies is $1.17\pm0.07$ for the seven strips of NGC\,891 and $1.35\pm0.06$ for the three inner strips of NGC\,4565 that have acceptable uncertainties. These results indicate that energy losses of the CRE depend on magnetic field strength and on frequency, such as synchrotron losses (see Sect.~\ref{discussion:N891} for a discussion).

While in NGC\,891 the total and non-thermal scale heights are basically identical, the values for NGC\,4565 show substantial differences in this respect, which at least at 6\,GHz can be explained by its higher thermal fractions. 
Furthermore, in NGC\,4565 we measure extremely low values of $h_{\mathrm{disc}}$, which are hardly believable, since such small scales ($\approx$20\,pc) are far from being resolved by our observations. On the other hand, a reliable measurement of $h_{\mathrm{disc}}$ using our highest-resolution data is not possible due to the resolved ring structure of the inner disc.

In the south-eastern half of NGC\,4565, we obtain unusually large non-thermal disc scale heights at 6\,GHz. It is possible that these have been fitted erroneously, 
as here the 
non-thermal disc emission is not well aligned with the mid-plane from a radius of $\approx5$\,kpc onwards. In addition, the relatively large disc scale heights at 
the centre are probably induced by the partially resolved ring pattern.
The signature 
of the ring is visible in the central $z$ profiles ($\rm lon=0^{\prime\prime}$ in Fig.~\ref{fig:cr3}) as a flattening of the peak 
region, which makes it difficult to fit these profiles accurately. It should be noted that in these subplots the simultaneous fit of both sides of the major axis is displayed, 
even though in these cases we 
adopted as scale height the average result of the northern- and southern-side fits. Furthermore, we observe the position of the peak to shift from the mid-plane to 
negative galactic latitudes on the east side of the galaxy and to positive ones on the west side. This is likewise caused by the ring or spiral arm structure, as already 
noted by \citet{broeils85}.

The average halo scale height of the total emission of NGC\,4565 at 6\,GHz (1.6\,kpc) is in agreement with the global value found by \citet{dumke98}. For 
NGC\,891, however, we measure only 1.1\,kpc, whereas \citet{dumke98} measured 1.8\,kpc. This discrepancy certainly arises from the fact that these authors averaged 
over regions that extend out to a larger radius, while at the same time the central region was excluded. In addition, a slightly different inclination and distance 
were assumed in this study.

\section{Cosmic-ray transport in the halo}
\label{CRtransport}

\subsection{Total magnetic field strength distribution}
\label{bfeld}

A crucial input to CRE transport modelling is the total magnetic field strength in the disc. We assume equipartition between the energy densities of cosmic rays and magnetic fields. This assumption is valid in star-forming galaxies at spatial scales of more than about 1\,kpc \citep{seta19}. Using the revised equipartition relation of \citet{beck05}, we generated maps of the total field strength $B_{\mathrm{eq}}$ from the non-thermal 1.5\,GHz intensity maps and the non-thermal spectral index maps formed between 1.5 and 6\,GHz (all short-spacing corrected). 
We assumed a constant number-density ratio of CRE protons to CRE electrons of $K_{0}=100$ that is valid in the energy range relevant for the observed synchrotron emission. \footnote{This value is consistent with the expectation from CR acceleration models and also with direct observations in the local neighborhood \citep[see][for details]{beck05}. Even a large uncertainty of $K_{0}=100$ leads to a relatively small uncertainty of the magnetic field strength.}
It is reasonable to assume $K_{0}=100$ for the disc, but this value yields only a lower limit for $B_{\mathrm{eq}}$ in the halo. CR protons propagating into the halo are much less affected by energy losses than CR electrons, so that $K_{0}$ is expected to increase with $z$.
To estimate the pathlength through the galaxies, we assumed in both cases the geometry of an oblate spheroid with semi-major axis length $R$ and semi-minor axis length $Z$, and hence with a pathlength of $l=2R\sqrt{1-(r/R)^2-(z/Z)^2}$. 
This is strictly valid only for an inclination of $90\degr$, but still serves as a reasonable approximation for nearly edge-on galaxies such as NGC\,891 and 4565. Here, $R$ and $Z$ were determined for each galaxy from the $3\sigma$ levels in the L-band D-array map, with $Z$ being chosen as the average between the maximum and minimum $z$ extent above the central region. We obtained $R=14.1$\,kpc and $Z=7.1$\,kpc for NGC\,891, and  $R=22.9$\,kpc and $Z=5.8$\,kpc for NGC\,4565.

The resulting maps of the (line-of-sight averaged) total magnetic field strength are presented in Fig.~\ref{fig:cr5}. Except for a few edge artefacts in the map of NGC\,891 caused by low S/N and the region where we had subtracted the nucleus of NGC\,4565, $B_{\mathrm{eq}}$ decreases monotonically with $z$. The maximum magnetic field strength (averaged within the $12\arcsec$ beam) is found in the disc with $15\,\mu$G in NGC\,891 and $7\,\mu$G in NGC\,4565, while the averaged disc field strength is $10\,\mu$G in NGC\,891 and $6\,\mu$G in NGC\,4565. Averaged over the regions in which we determined the scale heights, 
we obtain field strengths of $8\,\mu$G and $5\,\mu$G for NGC\,891 and 4565, respectively. NGC\,891 thus features rather typical field strengths for normal galaxies, while those in NGC\,4565 are comparatively weak.

We used a Monte--Carlo method to estimate the magnetic field strength uncertainties, based on the rms noise in the input intensity maps \citep[see][]{basu13}. Uncertainties of the pathlength were not taken into account, since $B_{\mathrm{eq}}$ depends only weakly on the pathlength, and hence these errors ($\Delta l\lesssim3$\,kpc) do not make a significant contribution. The uncertainty in the proton-to-electron ratio contributes a systematic error. In the disc of NGC\,891, we find that the rms error of $B_{\mathrm{eq}}$ is typically 3.5\%, while at $z\approx 5$\,kpc it raises to 21\%. For NGC\,4565, typical rms errors are 3.6\% in the mid-plane and 28\% at $z\approx 3$\,kpc.

\subsection{Cosmic-ray transport modelling}
\label{difflossmod}

In what follows, we assume that the CREs are injected at $z=0$ and apply one-dimensional models describing either purely advective or purely diffusive vertical CRE transport. This is accomplished by numerically solving the 1D diffusion--loss equation using the SPectral INdex Numerical Analysis of K(c)osmic-ray Electron Radio-emission (\texttt{SPINNAKER})\footnote{https://www.github.com/vheesen/Spinnaker} software \citep{heesen16}. We follow the modelling procedure described in \citet{heesen18}, which we summarise here, indicating case-specific choices of model parameters where applicable.

\subsubsection{Parametrisation of advection and diffusion models}

In one dimension, the steady-state diffusion--loss equation \citep[e.g.][]{lerche80} for purely advective CRE transport at a constant speed $V$ is
\begin{equation}
\label{nezconv}
 \frac{\partial N(E,z)}{\partial z}=\frac{1}{V}\left(\frac{\partial}{\partial E}[b(E)N(E,z)]\right)\,.
\end{equation}
In case of no streaming, the advection speed $V$ is equivalent to the wind speed. Otherwise, adding the Alfv\'en velocity would not increase $V$ with respect to the general uncertainty of our model, although sometimes super-Alfv\'enic streaming is assumed as well \citep{farber+18}. For simplicity, we leave out this complication here and just study the advection speed. While cosmic ray-driven winds are expected to accelerate with height, they do so by less than an order of magnitude even within several $10$\,kpc \citep{everett08, breitschwerdt12}. Assuming a 
constant advection speed is hence worth considering as a first simple approximation. In addition to $V(z)=\mathrm{const.}$, we also model linearly accelerated 
velocity profiles of the form  
\begin{equation}
\label{vacc}
 V(z)=V_{0}\left(1+\frac{z}{h_{V}}\right)
\end{equation}
with launch velocity $V_{0}$ and velocity scale height $h_{V}$. Such a linearisation is a useful approximation for the expected velocity profiles of pressure-driven wind models that go through a critical point.

For pure diffusion, the transport equation is given by
\begin{equation}
\label{nezdiff}
 \frac{\partial^{2}N(E,z)}{\partial z^{2}}=\frac{1}{D(E)}\left(\frac{\partial}{\partial E}[b(E)N(E,z)]\right)\,,
\end{equation}
where the diffusion coefficient $D$ depends on the energy as $D(E)=D_{0}\,(E/\mathrm{GeV})^{\,\mu}$.
While \citet{strong07} found  $0.3\lesssim\mu\lesssim0.6$ from modelling CRE propagation in the Milky Way, and \citet{murphy12} found (for a star-forming region in the Large Magellanic Cloud) a strong energy dependence of $D$ for CRs between $\approx$3 and 70\,GeV, there are recent indications that the energy dependence may not be significant for CRE energies $\lesssim10$\,GeV \citep{recchia16b,mulcahy16}. We therefore fitted for $D_{0}$ using different choices of $\mu$, as stated below.

Here, $b(E)$ is the combined rate of synchrotron, IC, and (in case of advection) adiabatic losses\footnote{The contribution of ionisation and bremsstrahlung losses can be neglected, as we show in Appendix~\ref{ionbrems}.} of a single CRE:
\begin{equation}
\label{be}
 b(E)=-\left(\frac{\mathrm{d}E}{\mathrm{d}t}\right)=\frac{4}{3}\sigma_{\mathrm{T}}c\left(\frac{E}{m_{\mathrm{e}}c^{2}}\right)^{2}(U_{\mathrm{mag}}+U_{\mathrm{rad}})+\frac{1}{3}\left(\frac{\mathrm{d}V}{\mathrm{d}z}\right)\,,
\end{equation}
where $\sigma_{\mathrm{T}}=6.6\times10^{-25}\,\mathrm{cm}^{2}$ is the Thomson cross section and $U_{\mathrm{rad}}$ and $U_{\mathrm{mag}}$ are the radiation field and magnetic energy density, respectively, which we calculated as described in Appendix~\ref{uradumag}. For the numerical integration  of Eq.~(\ref{nezconv}) or (\ref{nezdiff}), we use the boundary condition at $z=0$ of $N(E,0)=N_{0}E^{\,\gamma_{\,\mathrm{inj}}}$, where $\gamma_{\mathrm{inj}}=2\alpha_{\mathrm{inj}}-1$ is the injection value of the CRE energy spectral index.

Motivated by the shape of the vertical intensity profiles observed for most galaxies \citep[e.g.][]{krause17,heesen18}, a two-component exponential distribution of the total magnetic field strength is assumed:
\begin{equation}
 B(z)=B_{1}\times\exp{(-|z|/h_{\mathrm{B1}})} +(B_{0}-B_{1})\times\exp{(-|z|/h_{\mathrm{B2}})}\,,
\end{equation}
where $h_{\mathrm{B1}}$ and $h_{\mathrm{B2}}$ are the magnetic field scale heights of the disc and halo component, $B_{0}$ is the total magnetic field strength in the mid-pane (at $z=0$), and $B_{1}$ is the difference between $B_{0}$ and the amplitude of the halo field component.

CRE transport by advection and diffusion may occur simultaneously in the halo. Some models predict that advection dominates over diffusion soon after the CREs break out of the disc \citep{ptuskin97,recchia16a}.
Here we only consider either pure advection or pure diffusion models.

\begin{figure*}[h]
 \centering
 \includegraphics[clip=true,trim=0pt 0pt 0pt 0pt,scale=0.29]{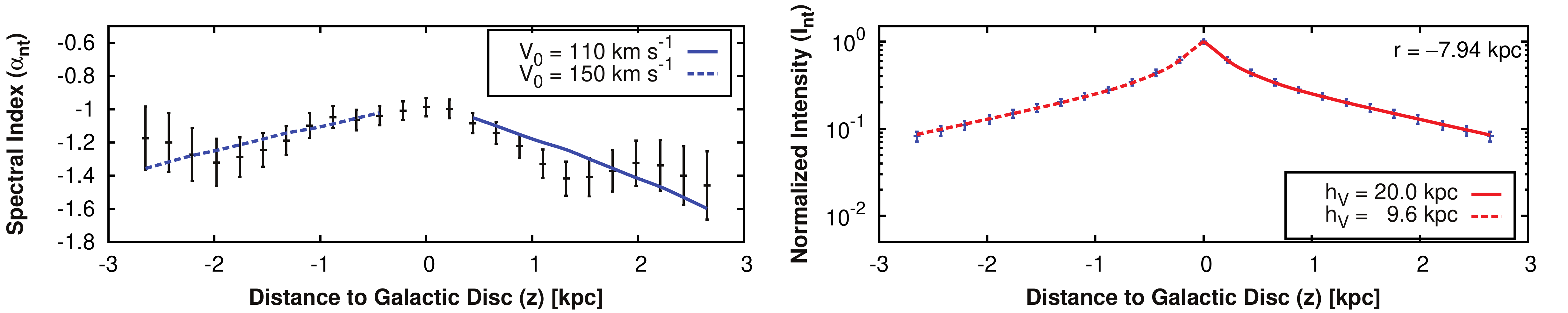}
 \includegraphics[clip=true,trim=0pt 0pt 0pt 0pt,scale=0.29]{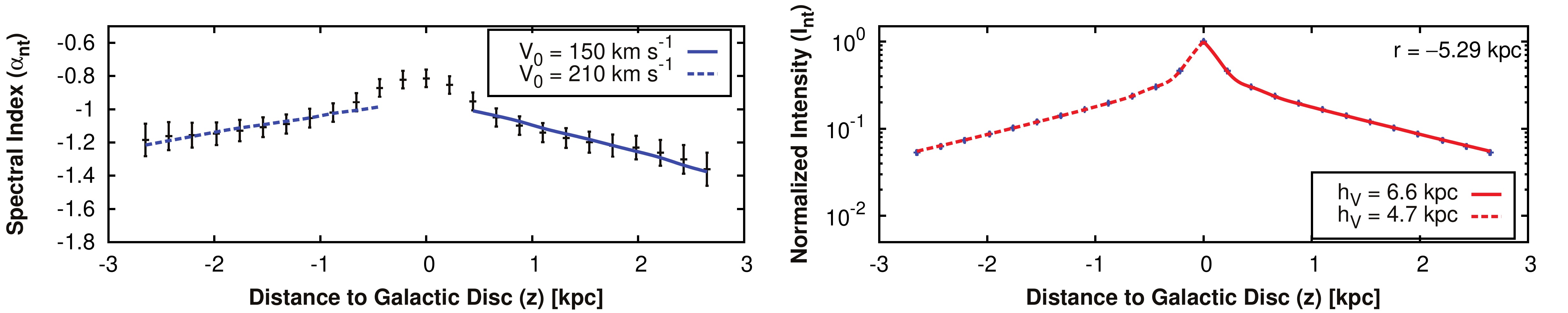}
 \includegraphics[clip=true,trim=0pt 0pt 0pt 0pt,scale=0.29]{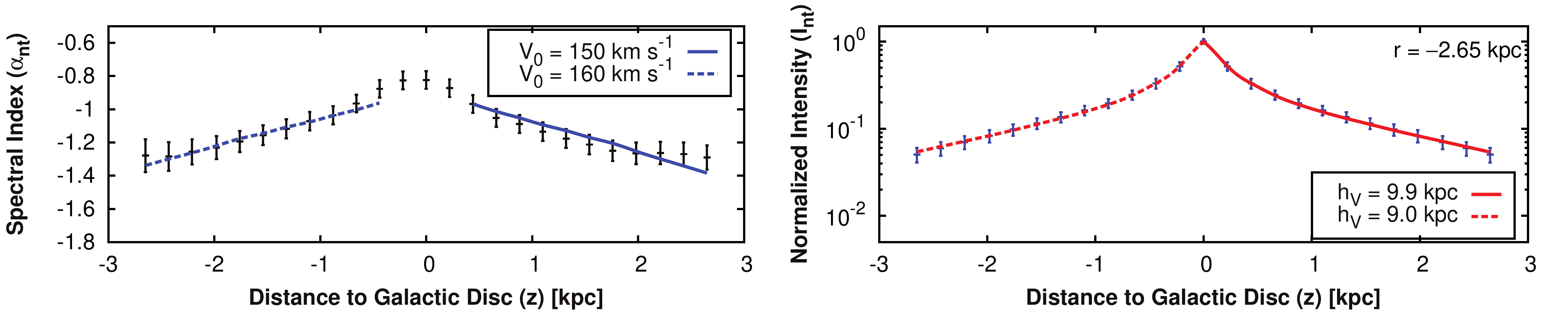}
 \includegraphics[clip=true,trim=0pt 0pt 0pt 0pt,scale=0.29]{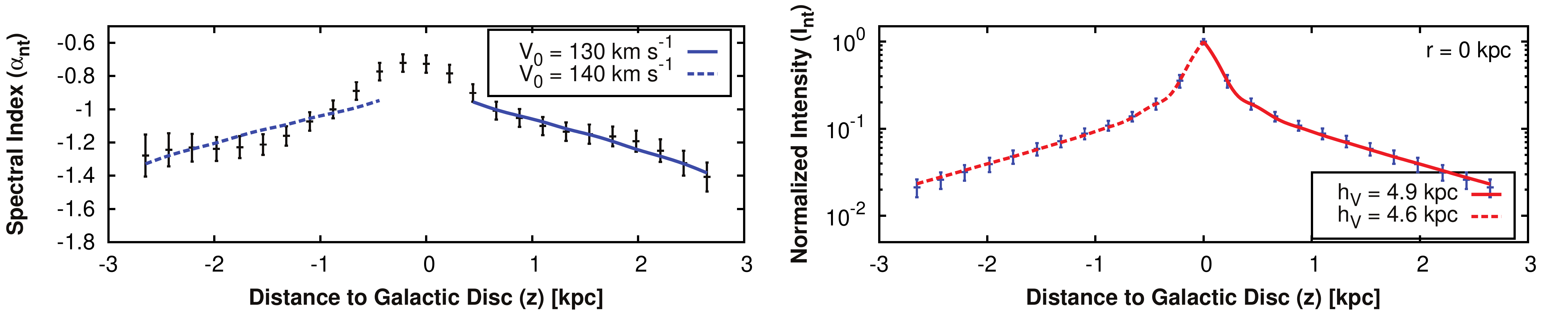}
 \includegraphics[clip=true,trim=0pt 0pt 0pt 0pt,scale=0.29]{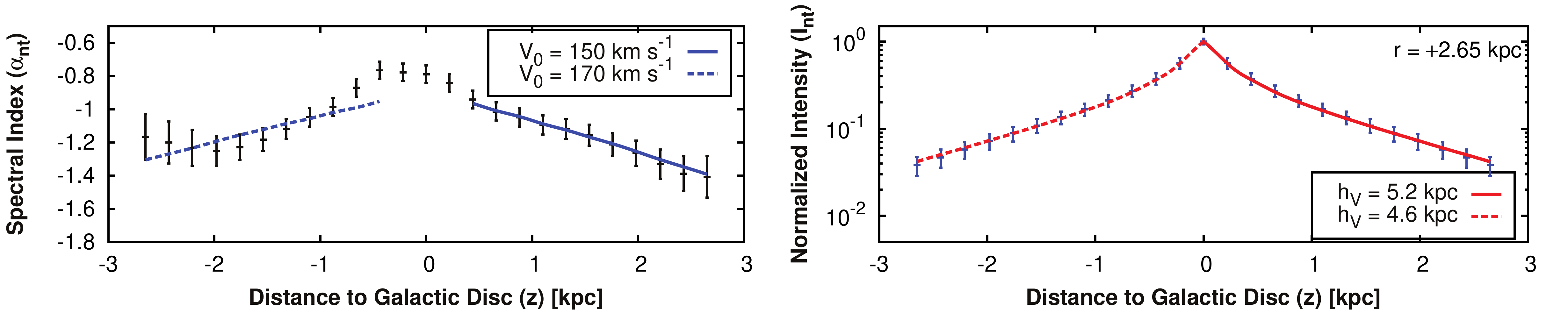}
 \includegraphics[clip=true,trim=0pt 0pt 0pt 0pt,scale=0.29]{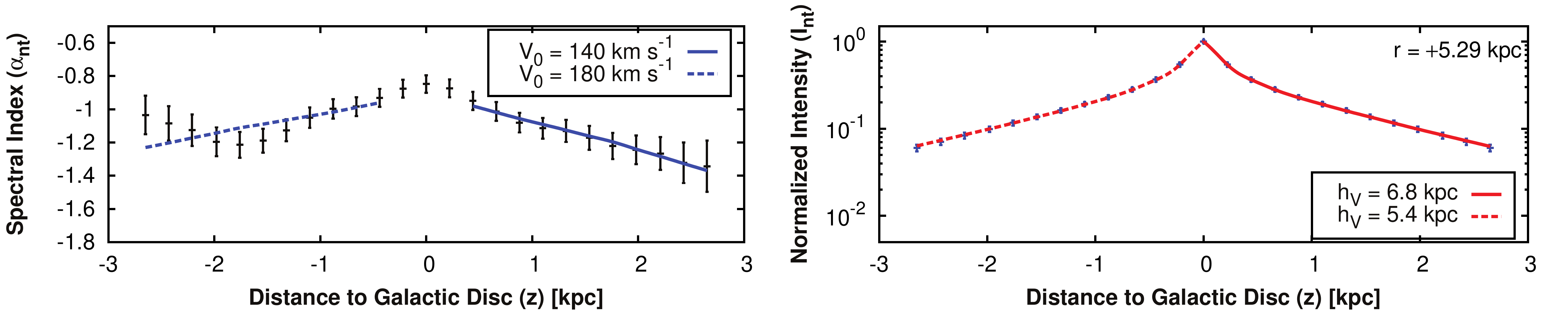}
 \includegraphics[clip=true,trim=0pt 0pt 0pt 0pt,scale=0.29]{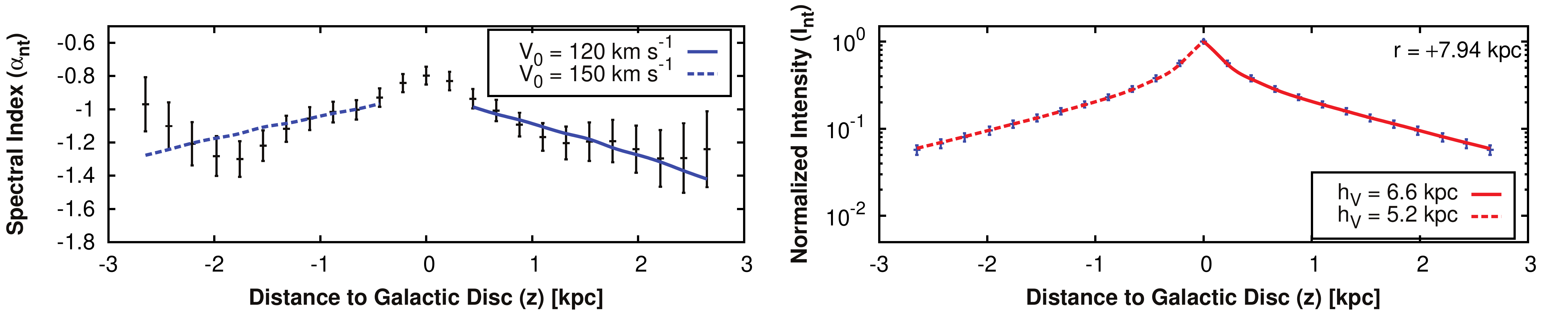}
\caption{Accelerated advection models for NGC\,891. Data points denote the vertical profile of the non-thermal spectral index between 1.5 and 6\,GHz (left panels) and the exponential model of the non-thermal intensity profile at 1.5\,GHz (right panels). The radial position of each profile is given in the right-hand-side plot; $r<0$ is east of the minor axis and $r>0$ is west of the minor axis. Positive $z$ values are on the north side and negative ones on the south side of the mid-plane. Solid lines show the best-fitting advection models.}
 \label{fig:N891adv2_CR}
\end{figure*}

\begin{figure*}[h]
 \centering
 \includegraphics[clip=true,trim=0pt 0pt 0pt 0pt,scale=0.29]{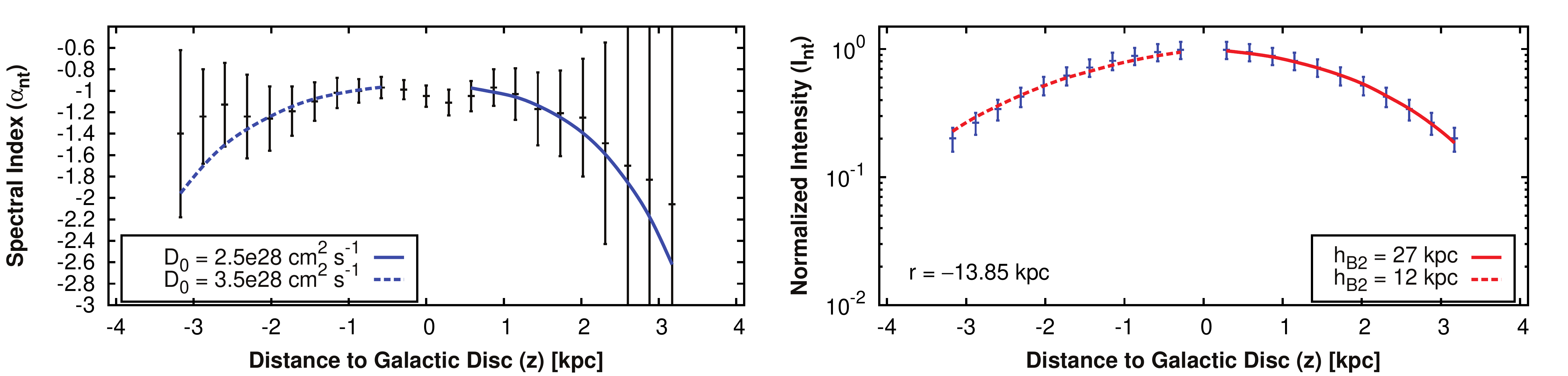}
 \includegraphics[clip=true,trim=0pt 0pt 0pt 0pt,scale=0.29]{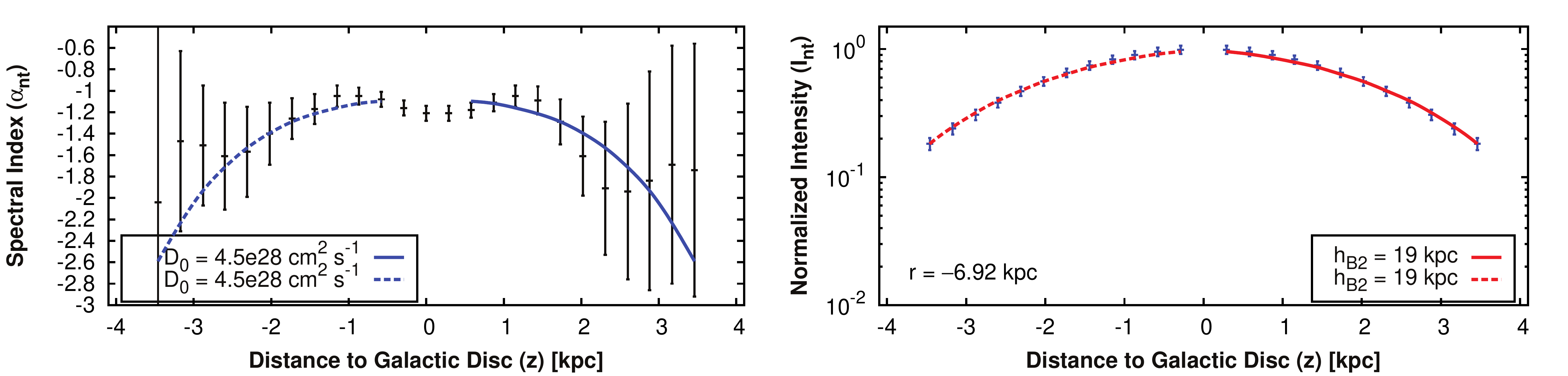}
 \includegraphics[clip=true,trim=0pt 0pt 0pt 0pt,scale=0.29]{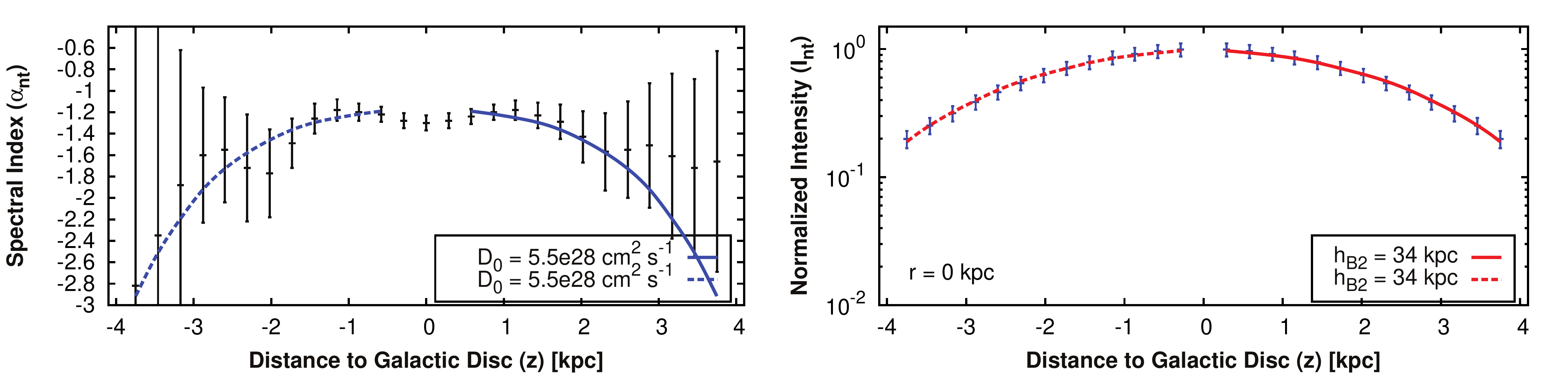}
 \includegraphics[clip=true,trim=0pt 0pt 0pt 0pt,scale=0.29]{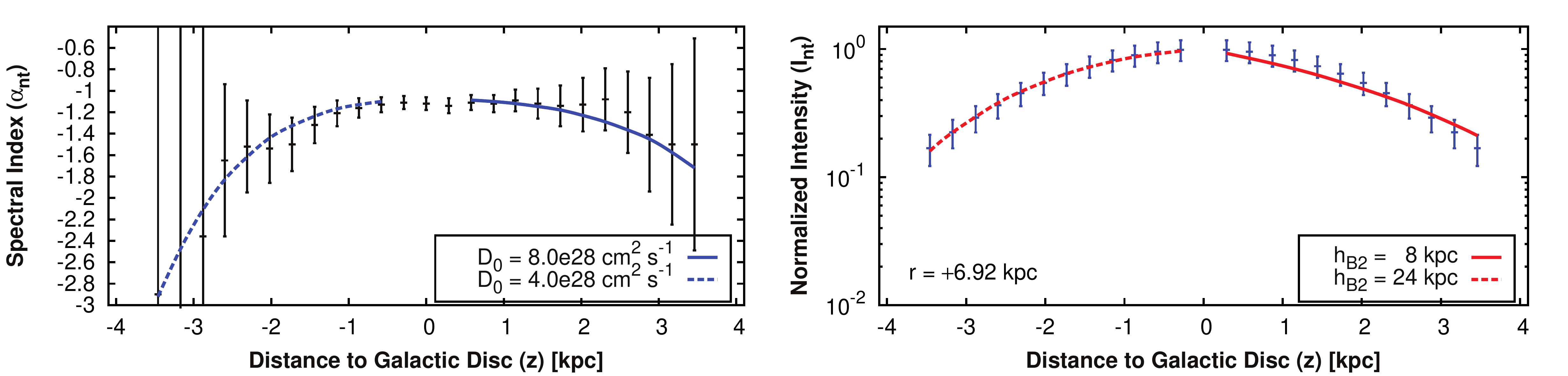}
 \includegraphics[clip=true,trim=0pt 0pt 0pt 0pt,scale=0.29]{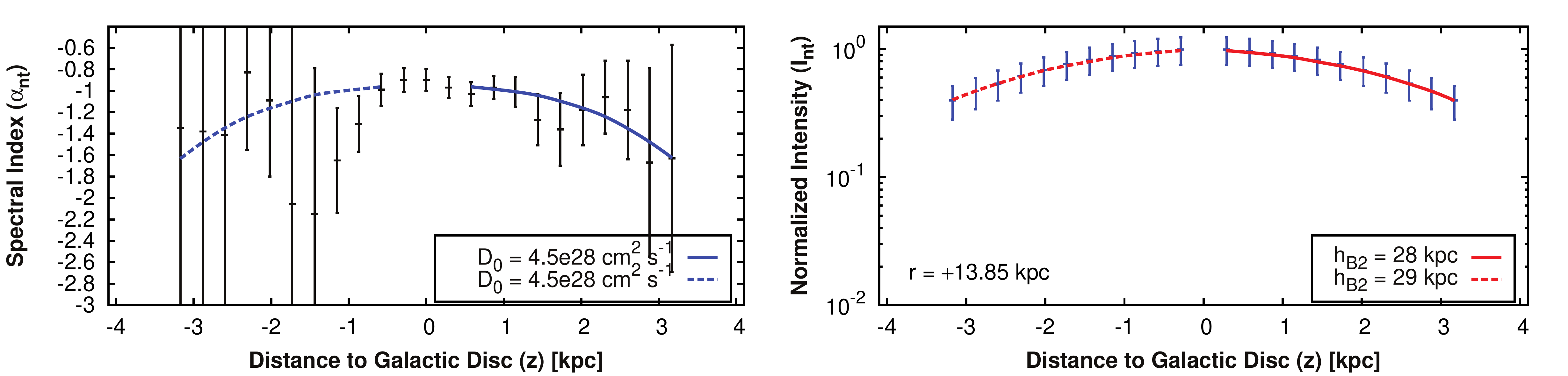}
\caption{Diffusion models for NGC\,4565 (assuming no energy dependence for the diffusion coefficient). Data points denote the vertical profile of the non-thermal spectral index between 1.5 and 6\,GHz (left panels) and the Gaussian model of the non-thermal intensity profile at 1.5\,GHz (right panels). The radial position of each profile along the major axis is given in the right-hand-side plot; $r<0$ is east of the minor axis and $r>0$ is west of the minor axis. Positive $z$ values are on the north side and negative ones on the south side of the mid-plane. Solid lines show the best-fitting diffusion models.}
 \label{fig:N4565diff00_CR}
\end{figure*}

\subsubsection{Fitting procedure}

For each galaxy, we attempted to fit both advection and diffusion models. To study the radial behaviour of advection speed, diffusion coefficient, and magnetic field 
scale height, we modelled the CRE transport at the same major-axis positions for which the synchrotron scale heights were determined.

As input data for each vertical strip, we used the (two-point) spectral index profile derived from the non-thermal intensity profiles at both frequencies, as well 
as the intensity \emph{model} profiles at 1.5\,GHz as computed in Sect.~\ref{scaleheights}. We prefer using the model profiles over fitting directly to the data as 
the former are deconvolved from the effective beam and thus have a resolved disc component. The error bars of these intensity values were derived from the amplitude 
and scale height uncertainties. In all cases, we use the 1.5\,GHz intensity profiles, as their amplitudes and scale heights (hereafter $h_{\mathrm{syn}}$) are 
constrained better than those at 6\,GHz due to the higher S/N in the 1.5\,GHz maps.  

Since the observed spectral index distributions are not symmetric with respect to the mid-plane, we produced separate models for the northern and southern side of 
each galaxy. Moreover, we restricted the fitting of the spectral index profiles to the halo regime ($z\gtrsim600$\,pc for NGC\,891 and $z\gtrsim800$\,pc for 
NGC\,4565), as the effective beam cannot be deconvolved from these profiles. 

As the above-mentioned plateau of faint emission in the 6\,GHz data of NGC\,891 would artificially flatten the outer parts of the spectral index profiles, we 
performed the overall modelling only out to $z\approx3$\,kpc for this galaxy. For NGC\,4565, we omitted the data point at $z=0$ in the intensity profiles, as 
the scale heights of the (mostly) extremely thin disc component modelled in Sect.~\ref{scaleheights} are smaller than the FWHM of the synthesised beam by more 
than an order of magnitude. We thus only fitted a single intensity component for NGC\,4565, which, for consistency, we keep referring to as the halo component.

To obtain the mid-plane total magnetic field strength $B_0$ at each radial position, we ran \texttt{BoxModels} on the $B_{\mathrm{eq}}$ map and adopted the average 
value within the central box of each respective strip. We then performed a simultaneous fit of the spectral index profile and the intensity profile, varying the 
parameters $V$ (or $D_{0}$), $B_{1}$, $h_{\mathrm{B1}}$, $h_{\mathrm{B2}}$ (or $h_{V}$, s.b.), and $\gamma_{\mathrm{inj}}$ until we found a reasonable initial 
guess model. As the fit of the intensity profile depends mainly on the magnetic field profile, and conversely the fit of the spectral index profile depends almost 
exclusively on the advection speed or diffusion coefficient, the parameters could be constrained very well by the simultaneous fitting in most cases. For each model, 
$B_{1}$ and $h_{\mathrm{B1}}$ were kept fixed at the initial guess value (for NGC\,4565 we set both to zero for the reasons given above). We then probed a range of 
values for $V$ (or $D_{0}$) and $h_{\mathrm{B2}}$ to find the minimum $\chi^2$ for the intensity and spectral index fits. For the accelerated advection 
models (Eq.~\ref{vacc}), we kept $h_{\mathrm{B2}}$ fixed at the expected equipartition value $h_{\mathrm{B,eq}}=h_{\mathrm{syn}}(3-\alpha_{\mathrm{nth}})$ (assuming 
an average $\alpha_{\mathrm{nth}}$ of $-1$ in the halo of NGC\,891 and $-1.15$ for NGC\,4565) and varied $V_{0}$ and $h_{V}$.\footnote{Similar to the behaviour of 
$V$ and $h_{\mathrm{B2}}$ in the constant-velocity models, the choice of $V_{0}$ mostly affects the spectral index profile, while $h_{V}$ (at a 
fixed $h_{\mathrm{B2}}$) mostly affects the intensity profile.} In each case, the set of parameters which minimises the sum of both reduced $\chi^2$
values ($\chi^{2}_{I}+\chi^{2}_{\alpha}$) was adopted as the best-fitting solution.

For $\gamma_{\mathrm{inj}}$, we enforced an upper limit in that we did not allow values steeper than the lower error margin of the innermost fitted spectral index data point in each case. However, we did allow values that may be steeper than the true injection spectral index (i.e. $-0.7\lesssim\alpha_{\mathrm{inj}}\lesssim-0.5$), as the disc contains CREs of various spectral ages, with older populations occurring in inter-arm regions \citep[e.g.][]{tabatabaei13} and possibly in superbubbles \citep{heesen15}.

While for the diffusion models of NGC\,891 the best fits were achieved using the commonly assumed energy exponent of $\mu=0.5$, NGC\,4565 could also be fitted with lower values of $\mu$. Hence, for comparison we additionally modelled this galaxy with $\mu=0.3$ and $\mu=0$. Diffusion produces approximately Gaussian intensity profiles whereas advection leads to exponential ones. Therefore, since the synchrotron emission of NGC\,4565 is described equally well by Gaussian and exponential profiles, we used the former for modelling diffusion and the latter for modelling advection.
For each model, we adopted the range of $V$ (or $D_{0}$), and $h_{\mathrm{B2}}$ (or $h_{V}$) values for which both $\chi^{2}_{I}\leq(\chi^{2}_{\mathrm{I,min}}+1)$ and $\chi^{2}_{\alpha}\leq(\chi^{2}_{\mathrm{\alpha,min}}+1)$ as their upper and lower error margins.

\subsubsection{Results}

We present the vertical profiles of the non-thermal spectral index $\alpha_{\mathrm{nth,1.5-6GHz}}$ and the non-thermal intensity $I_{\mathrm{nth,1.5GHz}}$ at each major-axis position, along with our best-fitting advection and diffusion models, in Figs.~\ref{fig:N891adv2_CR} and \ref{fig:N4565diff00_CR} for our preferred solutions. The remaining models are presented in Appendix~\ref{appendix_CRprop}, where we also list the key parameters for all models. In order to illustrate the dependence with galactocentric radius, we present radial profiles of these parameters for a subset of models in Fig.~\ref{fig:cr4}.

For NGC\,891, advection is clearly favoured over diffusion, mainly because the data are more consistent with the linear spectral index profiles produced by advection.
We find advection speeds consistently around $\approx$150\,$\mathrm{km\,s^{-1}}$, with a tendency towards high upper error limits at large radii. 
For both advection and diffusion, the halo magnetic field scale height $h_{\mathrm{B2}}$ shows a similar behaviour to the synchrotron scale heights, however 
with a stronger increase at the eastern edge ($r=-7.9$\,kpc), where it reaches about twice the value at the centre.
The diffusion models for NGC\,891 fit the data considerably worse than the advection models (as evident in particular on the south side of the central strip, where 
the best achievable $\chi^{2}_{\alpha}$ is 3.6), as they predict a shape of the $\alpha_{\mathrm{nth,1.5-6GHz}}$ profile opposite to what is observed: our data show 
a rather steep decline of $\alpha_{\mathrm{nth,1.5-6GHz}}$ in the lower halo, followed by a more shallow profile at high $z$ (with the tendency to flatten out in most cases), whereas diffusion results in a shallow 
decline at low $z$, followed by a steep one towards the outer halo. Apart from this, we do obtain reasonable diffusion coefficients of 
$1.5\times10^{28}\,\mathrm{cm^2\,s^{-1}}<D_{0}<3.5\times10^{28}\,\mathrm{cm^2\,s^{-1}}$.

In the case of NGC\,4565, we find that all of our models yield reasonable fits in terms of constantly low $\chi^{2}$. 
However, for diffusion with $\mu=0.5$, this is mainly due to the large spectral index uncertainties for $z\gtrsim2$\,kpc. From $z\gtrsim1.5$\,kpc onwards, the $\alpha_{\mathrm{nth,1.5-6GHz}}$ profiles steepen more rapidly than predicted by these models, so that for diffusion the best-fitting solutions are rather found with $\mu=0$ (and in some cases $\mu=0.3$). The lower the assumed energy dependence of $D$, the higher $D_{0}$ needs to be to fit the $\alpha_{\mathrm{nth,1.5-6GHz}}$ profiles. For $\mu=0$, $D_{0}$ still turns out to be in a realistic range (up to $5.5\times\mathrm{10^{28}\,cm^{2}\,s^{-1}}$; except for the northern halo at $r=+6.9$\,kpc, where we find $D_{0}=8\times\mathrm{10^{28}\,cm^{2}\,s^{-1}}$). 
For advection, $V$ turns out to be rather constant with radius, as for NGC\,891, however with values of only around $\approx$90\,$\mathrm{km\,s^{-1}}$ (again with the exception of the northern side at $r=+6.9$\,kpc, where the spectral index remains comparatively flat for some reason). The corresponding $h_{\mathrm{B2}}$ values are radially constant as well (around 3.5\,kpc), but would have increased at the disc edges if we had fitted to the $I_{\mathrm{nth,6GHz}}$ profiles instead (cf. Fig.~\ref{fig:cr4}). Still, our upper error limits for $h_{\mathrm{B2}}$ are highest at the edges.
For diffusion, the $h_{\mathrm{B2}}$ 
values are much higher than for advection, since here we used the Gaussian $I_{\mathrm{nth,1.5GHz}}$ models. They are, however, badly constrained in this case, as above a certain value, increasing $h_{\mathrm{B2}}$ has a negligible effect on fitting the intensity profiles. In several cases, the intensity error bars are in fact large enough to allow for the magnetic field strength to be practically constant in $z$. 

In case of accelerated advection, the initial velocities $V_{0}$ for both galaxies are either equal to or marginally lower than the velocities in the 
respective constant-advection models. On the other hand, various amounts of acceleration (controlled by $h_{V}$) are required to fit the intensity profiles.

\section{Discussion}
\label{discussion}

\begin{figure*}[h]
 \centering
 \topinset{\it a) \qquad\qquad\qquad\qquad\qquad \normalfont\large NGC\,891}{\includegraphics[scale=0.395,clip=true,trim=15pt 24pt 10pt 40pt]{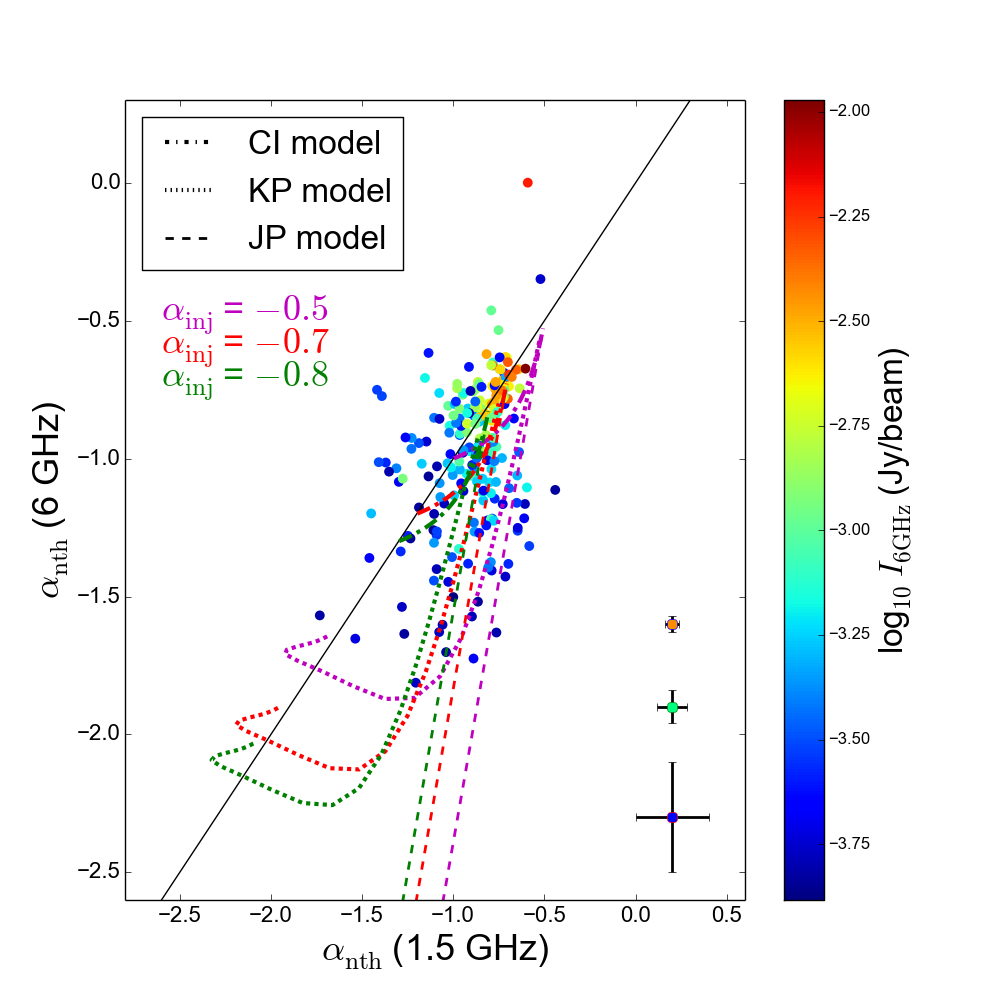}}{0pt}{0pt}\topinset{\it b) \qquad\qquad\qquad\qquad\qquad \normalfont\large NGC\,4565}{\includegraphics[scale=0.395,clip=true,trim=15pt 24pt 10pt 40pt]{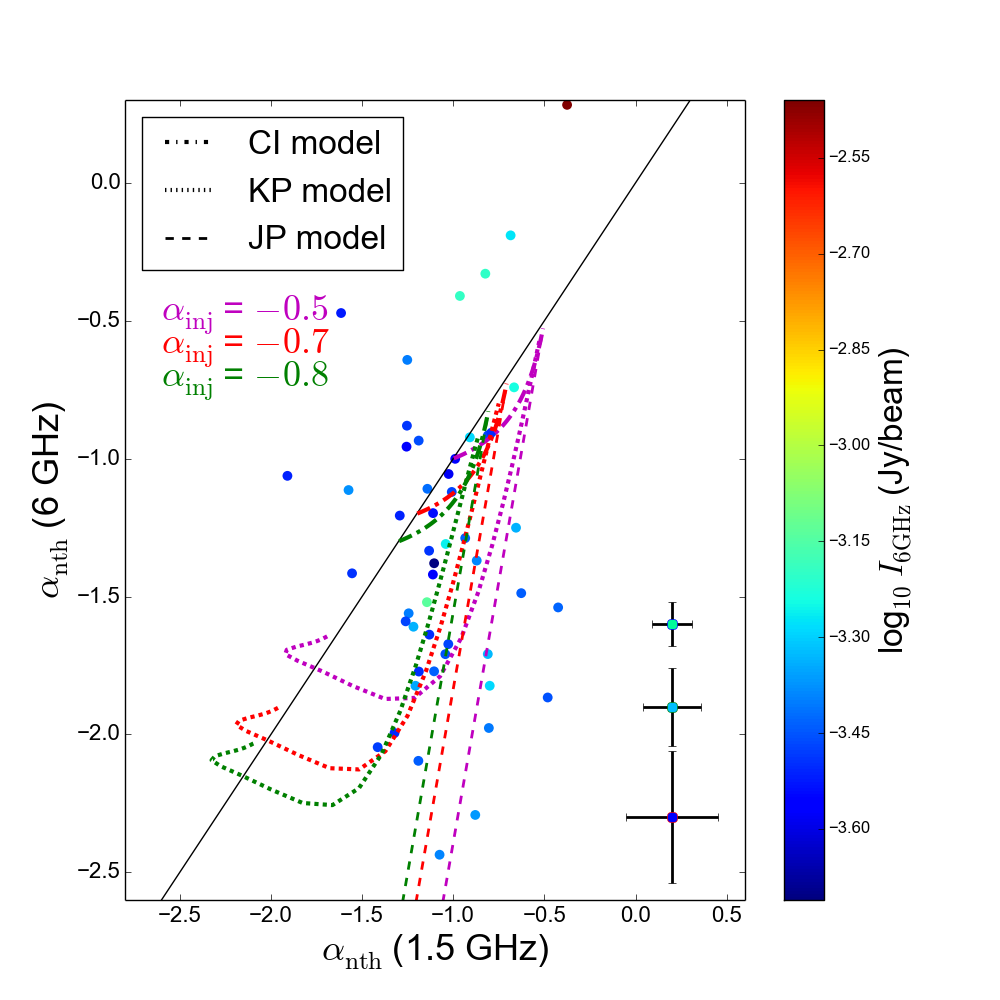}}{0pt}{0pt}
 \caption{Scatter plots of the non-thermal in-band spectral index distributions (6\,GHz vs. 1.5\,GHz; both are equal along the solid black line), in comparison to different spectral ageing models.
 \it a\normalfont: NGC\,891 (Fig.~\ref{fig:alpha12}~\it b \normalfont vs. Fig.~\ref{fig:alpha12}~\it e\normalfont).
 \it b\normalfont: NGC\,4565 (Fig.~\ref{fig:alpha14}~\it b \normalfont vs. Fig.~\ref{fig:alpha14}~\it e\normalfont).
 All $\alpha_{\mathrm{nth}}$ maps were cut off at $30\sigma$ (NGC\,891) or $20\sigma$ (NGC\,4565) of the respective total intensity maps. Each data point represents the 1.5 and 6\,GHz spectral index averaged within one synthesised beam and is colour-coded based on the 6\,GHz total intensity map (Figs.~\ref{fig:tp2}~\it b \normalfont and \ref{fig:tp5}~\it b\normalfont). The dash-dotted, dotted, and dashed lines show the expected positions in the plot for the CI, KP, and JP models, respectively, for different break frequencies $\nu_{\mathrm{br}}$ and three different injection spectral indices $\alpha_{\mathrm{inj}}$.  
 The symbols with error bars represent the $\Delta\alpha$ maps (panels \it c \normalfont and \it f \normalfont of Figs.~\ref{fig:alpha12}-\ref{fig:alpha14}) and are shown for three different intensity values.
 }
 \label{fig:alpha3}
\end{figure*}

\subsection{CRE injection and spectral ageing}
\label{alphatoKPJPCI}

Cosmic rays are injected into the ISM with a power-law spectrum in energy that leads to a power-law spectrum of synchrotron intensity $I_{\nu}$ with radio frequency $\nu$ ($I_{\nu}\propto\nu^{{\alpha}_{\mathrm{inj}}}$). For diffusive shock acceleration \citep[see\,e.g.][]{caprioli15}, an initial radio spectral index $\alpha_{\mathrm{inj}}$ 
between $-0.7$ and $-0.5$ is expected. At radio frequencies of typically $\gtrsim1$\,GHz, synchrotron and inverse-Compton (IC) losses are expected to induce a break in the local CRE emission spectrum. The shape of this spectral break (and its evolution with time) depends on the CRE injection process (i.e. $\alpha_{\mathrm{inj}}$ and injection time-scale) and their energy loss rate, which in turn depends on the magnetic field strength and on the distribution of their pitch angles with respect to the magnetic field lines. Here, the standard theory distinguishes between two models for a discrete epoch of particle injection: according to the Kardashev--Pacholczyk (KP) model \citep{kardashev62,pacholczyk70}, the individual electrons maintain their original pitch angles and the spectrum declines as a power law with spectral index $4/3\alpha_{\mathrm{inj}}-1$, whereas in the Jaffe--Perola (JP) model \citep{jaffe73} an isotropic distribution of pitch angles is generated, causing the spectrum to cut off exponentially. For continuous 
injection (CI) of CREs (at a constant injection rate), the spectrum above the break frequency is a power law with spectral index $\alpha_{\mathrm{inj}}-0.5$ (and thus less steep than for the discrete-epoch injection models).\footnote{The JP model is consistent with the \texttt{SPINNAKER} advection model, where the CREs are injected in the disc which is equivalent to a discrete-epoch injection model. Further, an isotropic pitch angle distribution of the CREs is assumed.}

We note that the above loss mechanisms, especially described by JP and KP models, are valid for the scenario when cosmic rays are injected on time-scales significantly larger than the typical CRE energy loss time-scales. Moreover, when large volumes of galaxies are averaged by the telescope beam, a combination of loss mechanisms and spatially varying CRE lifetimes are encompassed, resulting in strong spectral fall-offs to become smoother. The injection time-scale of cosmic rays is similar to the rate of supernovae of massive stars ($M>8\,\rm M_\odot$). Assuming a Kroupa-type initial mass function \citep{kroupa01} and a typical SFR surface density of up to $10^{-1}\rm \, M_{\odot} \,yr^{-1} \,kpc^{-2}$ for edge-on galaxies, the injection times-scale of CREs is $\gtrsim 10^8\,\rm yr \, kpc^{-2}$. Therefore, at the $12\arcsec$ angular resolution of our investigation, equating to spatial resolutions of 530 and 700\,pc in NGC\,891 and 4565, respectively, we expect that JP- or KP-type models for CRE energy loss are applicable. However, recent relativistic electromagnetic particle simulation by \citet{holcomb18} demonstrates that interaction of cosmic rays with self-generated Alfv\'en waves results in pitch angle scattering, suggesting the KP model to be a less likely model for CRE energy loss.

Having determined the spatially resolved non-thermal spectral index distribution at two well-separated frequencies, it is possible to put constraints on the shape of the spectrum for a given line of sight. In Fig.~\ref{fig:alpha3} we show scatter plots of $\alpha_{\mathrm{nth,6GHz}}$ vs. $\alpha_{\mathrm{nth,1.5GHz}}$ for both galaxies. The maps were first clipped below $30\sigma$ (NGC\,891) and $20\sigma$ (NGC\,4565) in total intensity, to reduce the effect of noise-based edge artefacts. Each point represents a beam-averaged value and is colour-coded based on $\log_{10}(I_{\mathrm{nth,6GHz}})$, so that red (NGC\,891) or light-blue (NGC\,4565) data points are associated with the disc, while dark-blue points represent the halo.

To help interpreting these plots, we constructed synthetic KP, JP, and CI spectra for different choices of break frequency $\nu_{\mathrm{br}}$ and injection spectral index $\alpha_{\mathrm{inj}}$ \citep[also see][]{basu15}. 
In both panels of Fig.~\ref{fig:alpha3} we show the trajectories in the $\alpha_{\mathrm{nth,1.5GHz}}$-$\alpha_{\mathrm{nth,6GHz}}$ plane that the different models predict for an injection spectral index of $-0.5$, $-0.7$, and $-0.8$. Moving from flatter to steeper values of $\alpha_{\mathrm{nth,1.5GHz}}$ and $\alpha_{\mathrm{nth,6GHz}}$ along these trajectories corresponds to a decrease in $\nu_{\mathrm{br}}$ and hence a forward movement in time; in turn an expected increase in $z$ for edge-on galaxies.

As for the CI model a maximum difference between $\alpha_{\mathrm{nth,1.5GHz}}$ and $\alpha_{\mathrm{nth,6GHz}}$ is reached at $\nu_{\mathrm{br}}=3.75\,\mathrm{GHz}$, we find for both galaxies that the majority of data points below the 1:1 line  
cannot be explained by a continuous injection process, but are reproduced well by the KP and JP models. For NGC\,891, the maximum steepening among the plotted data points corresponds to $\nu_{\mathrm{br}}\approx5\,\mathrm{GHz}$ for the JP model and $\nu_{\mathrm{br}}\approx1.5\,\mathrm{GHz}$ for the KP model. In NGC\,4565, of the three models only the JP model matches the points of maximum steepening, which are found at $\nu_{\mathrm{br}}\approx3\,\mathrm{GHz}$.

In NGC\,891, spectral indices steeper than $\approx-1.3$ above the $30\sigma$ level seem hardly plausible (at least at 1.5\,GHz), as we pointed out above on the basis of the $\alpha_{\mathrm{nth,1.5-6GHz}}$ map. If we ignore all values steeper than $-1.3$ in Fig.~\ref{fig:alpha3}~\it a \normalfont, the remaining data points are within their errors roughly consistent with the CI model, with $\alpha_{\mathrm{inj}}$ lying in the typically expected range between $-0.8$ and $-0.5$. When ignoring only $\alpha_{\mathrm{nth,1.5GHz}}<-1.3$, most of the remaining points at steeper $\alpha_{\mathrm{nth,6GHz}}$ are still found in that same range of $\alpha_{\mathrm{inj}}$ if a KP or JP model is considered. We note that the error bars plotted in Fig.~\ref{fig:alpha3} are based on the $\Delta\alpha$ maps, which represent only statistical errors and thus do not account for any systematic errors related to 
short-spacing corrections or uncertainties in estimating the thermal contribution.

Even though for NGC\,4565 the noise-induced spectral index errors are quite large, a clear trend towards stronger steepening between the two frequencies than in NGC\,891 is seen, especially considering that all data points shown in Fig.~\ref{fig:alpha3}~\it b \normalfont are still located in the disc of the galaxy. As previously indicated, this trend is already evident by comparing the $\alpha_{\mathrm{nth,1.5-6GHz}}$ maps of the two galaxies, which are affected much less by uncertainties than the corresponding in-band maps. As a consequence of the strong spectral steepening, the data points in the scatter plot of NGC\,4565 predominantly conform to the JP and KP models. 
We will briefly get back to discussing spectral ageing effects in Sect.~\ref{discussion:N4565}.

\begin{table*}
\caption{Radio diameters $d_{\mathrm{r}}$, 
average exponential (total and non-thermal) scale heights in the disc, $\overline{h}_{\mathrm{tot,disc}}$ and  $\overline{h}_{\mathrm{syn,disc}}$, and halo, $\overline{h}_{\mathrm{tot,halo}}$ and $\overline{h}_{\mathrm{syn,halo}}$, respectively. Further, we present the normalised halo scale heights $\tilde{h}$ for the total emission (see text for details).}
{\small
\begin{center}
\begin{tabular}{lcccccc}
\toprule
\toprule[0.3pt]
Galaxy  & $d_{\mathrm{r}}$ [kpc] & $\overline{h}_{\mathrm{tot,disc}}$ [kpc] & $\overline{h}_{\mathrm{tot,halo}}$ [kpc] & $\overline{h}_{\mathrm{syn,disc}}$ [kpc] & $\overline{h}_{\mathrm{syn,halo}}$ [kpc] & $\tilde{h}$ [kpc] \\
\hline
  NGC\,891 (1.5 GHz)  & $27.6\pm 2.8$ & 0.17 $\pm$ 0.02 & 1.28 $\pm$ 0.14 & 0.19 $\pm$ 0.03 & 1.29 $\pm$ 0.15 & 4.64 $\pm\,$ 0.69  \\
  NGC\,891 (6 GHz)    & $26.6\pm 2.7$ & 0.17 $\pm$ 0.04 & 1.09 $\pm$ 0.10 & 0.20 $\pm$ 0.06 & 1.11 $\pm$ 0.16 & 4.10 $\pm\,$ 0.56 \\
  NGC\,4565 (1.5 GHz) & $47.1\pm 4.7$ & 0.02 $\pm$ 0.01 & 1.60 $\pm$ 0.17 & 0.05 $\pm$ 0.07 & 1.35 $\pm$ 0.06 & 3.40 $\pm\,$ 0.50 \\
  NGC\,4565 (6 GHz)   & $43.8\pm 4.4$ & 0.08 $\pm$ 0.13 & 1.36 $\pm$ 0.39 & 0.41 $\pm$ 0.20 & 1.00 $\pm$ 0.09 & 3.11 $\pm\,$ 0.94 \\
\hline                                    
\end{tabular}
\label{tab:diams}
\end{center}
} 
{\footnotesize 
}
\end{table*}

\subsection{Normalised scale heights}
\label{sect:hnorm}

\citet{krause18} found that the halo scale height of the total emission from a galaxy primarily depends on its (radio) disc diameter. In order to 
eliminate this diameter dependence of halo scale heights, they defined a normalised scale height as $\tilde{h}=100\times h_{\mathrm{halo}}/d_{\mathrm{r}}$, where 
$d_{\mathrm{r}}$ is the radio diameter at a given frequency. We determined the radio diameters of both galaxies at each frequency from the $5\sigma$ contour in the 
respective total emission maps that were used for the scale height measurements. To calculate $\tilde{h}$, we used the average total emission scale height $\overline{h}_{\mathrm{tot,disc}}$, where for 
NGC\,4565 we averaged only over the three inner strips at both frequencies. The resulting diameters and normalised scale heights are listed in Table~\ref{tab:diams}. 
\begin{figure}[h]
 \centering
 \includegraphics[width=0.5\textwidth,clip=true,trim=0pt 0pt 0pt 0pt]{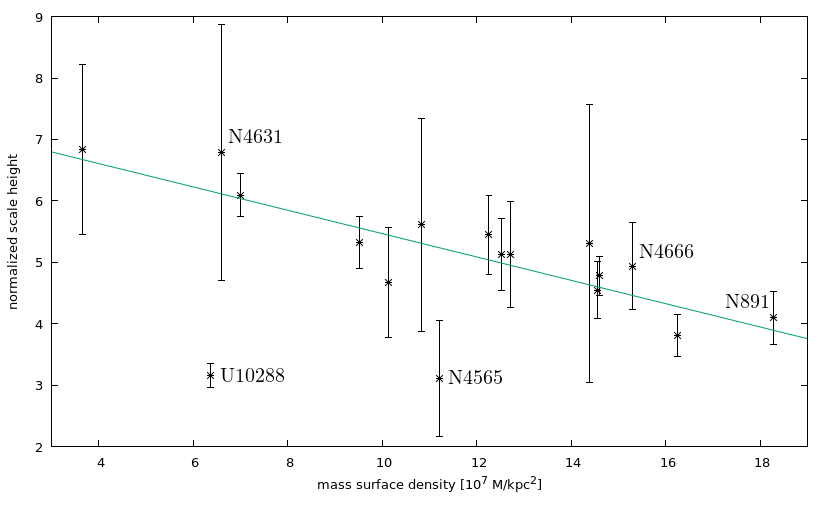}
 \caption{Normalised scale heights $\tilde{h}$ at 6\,GHz vs. mass surface densities for the CHANG-ES galaxy sample of \citet{krause18}, as well as the CHANG-ES galaxies NGC\,891, 4565 (both this work), 4631 (Mora et al., submitted), and 4666 \citep{stein19}. The weighted linear fit to the data (excluding NGC\,4565 and UGC\,10288) has a slope of $0.18\pm0.02$ (reduced $\chi^2 = 0.38$).
 }
 \label{fig:msd}
\end{figure}

Moreover, \citet{krause18} observed a tight anti-correlation between $\tilde{h}$ and the total mass surface density $MSD$, which indicates that a gravitational 
deceleration of the vertical CRE transport occurs. To check how NGC\,891 and 4565 relate to the tight anti-correlation we computed the latter as 
$MSD=M/(\pi \, r_{25}^2$), 
where $M$ is the total mass \citep[taken from][]{irwin12a} and $r_{25}$ is the blue-light radius (calculated from the blue diameter $d_{25}$ as given in Table~\ref{tab:basicparms}). We obtained 
$MSD=18.2 \times 10^7 \,$M$_{\odot} \, $kpc$^{-2}$ for NGC\,891 and $MSD=11.3 \times 10^7 \,$M$_{\odot} \, $kpc$^{-2}$ for NGC\,4565. Figure~\ref{fig:msd} shows that on the one hand 
NGC\,891 fulfills the anti-correlation, which strengthens the evidence for a gravitational deceleration, since with this galaxy we have extended the sample of \citet{krause18} to higher mass surface densities.

On the other hand, with NGC\,4565 we have determined a second outlier besides UGC\,10288, both having rather low normalised scale heights but no exceptionally high mass durface densities. Notably, both NGC\,4565 and UGC\,10288 have a significantly lower average total magnetic field strength ($\approx$6\,$\mu$G) than the rest of the sample. This suggests that either the $\tilde{h}$--$MSD$ anti-correlation breaks down for galaxies with weak magnetic fields, or that there is a similar anti-correlation for this type of galaxy, with an offset regarding the normalised scale heights.

\begin{figure*}[h]
 \centering
 \includegraphics[clip=true,trim=50pt 0pt 50pt 0pt,width=0.5\textwidth]{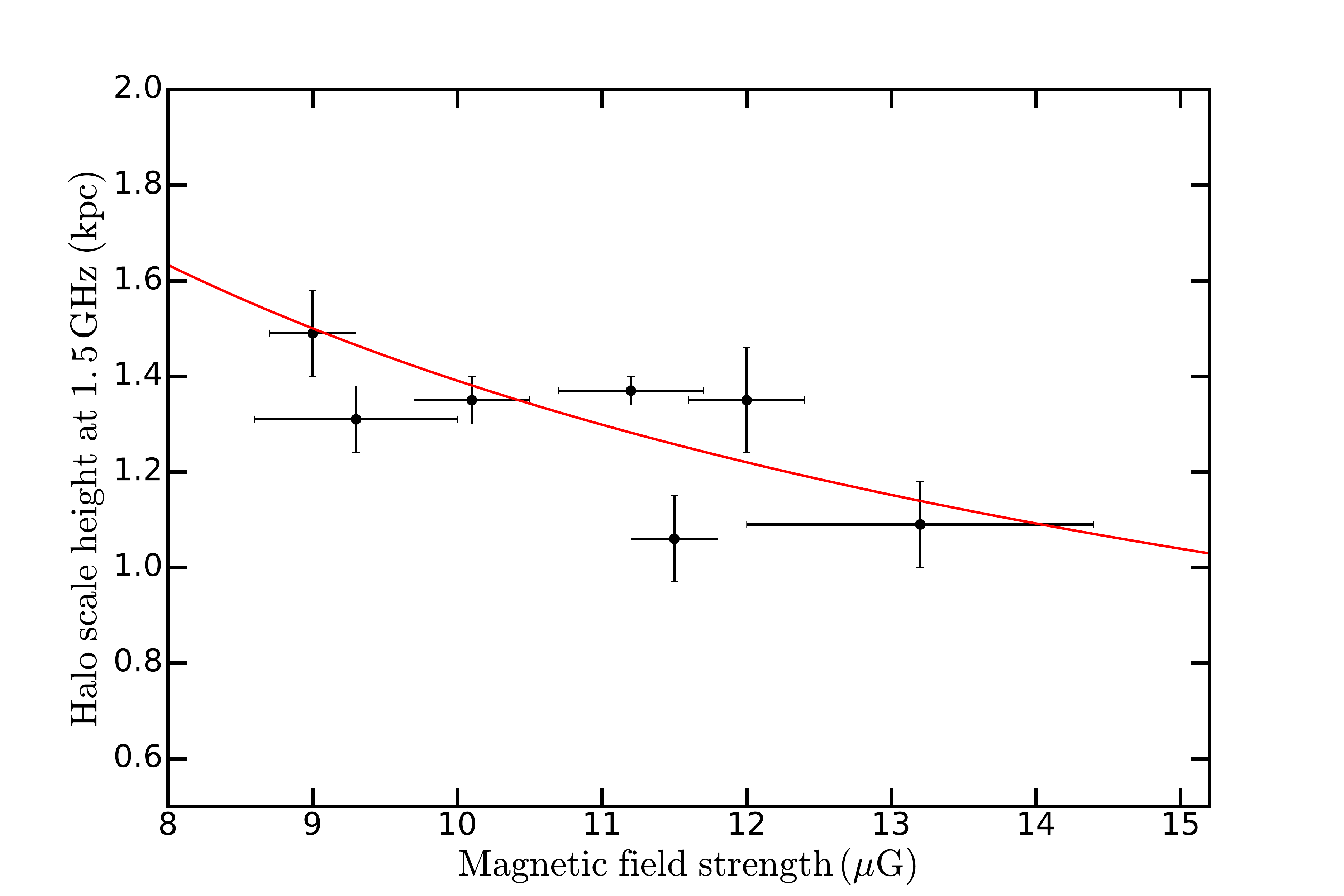}\includegraphics[clip=true,trim=50pt 0pt 50pt 0pt,width=0.5\textwidth]{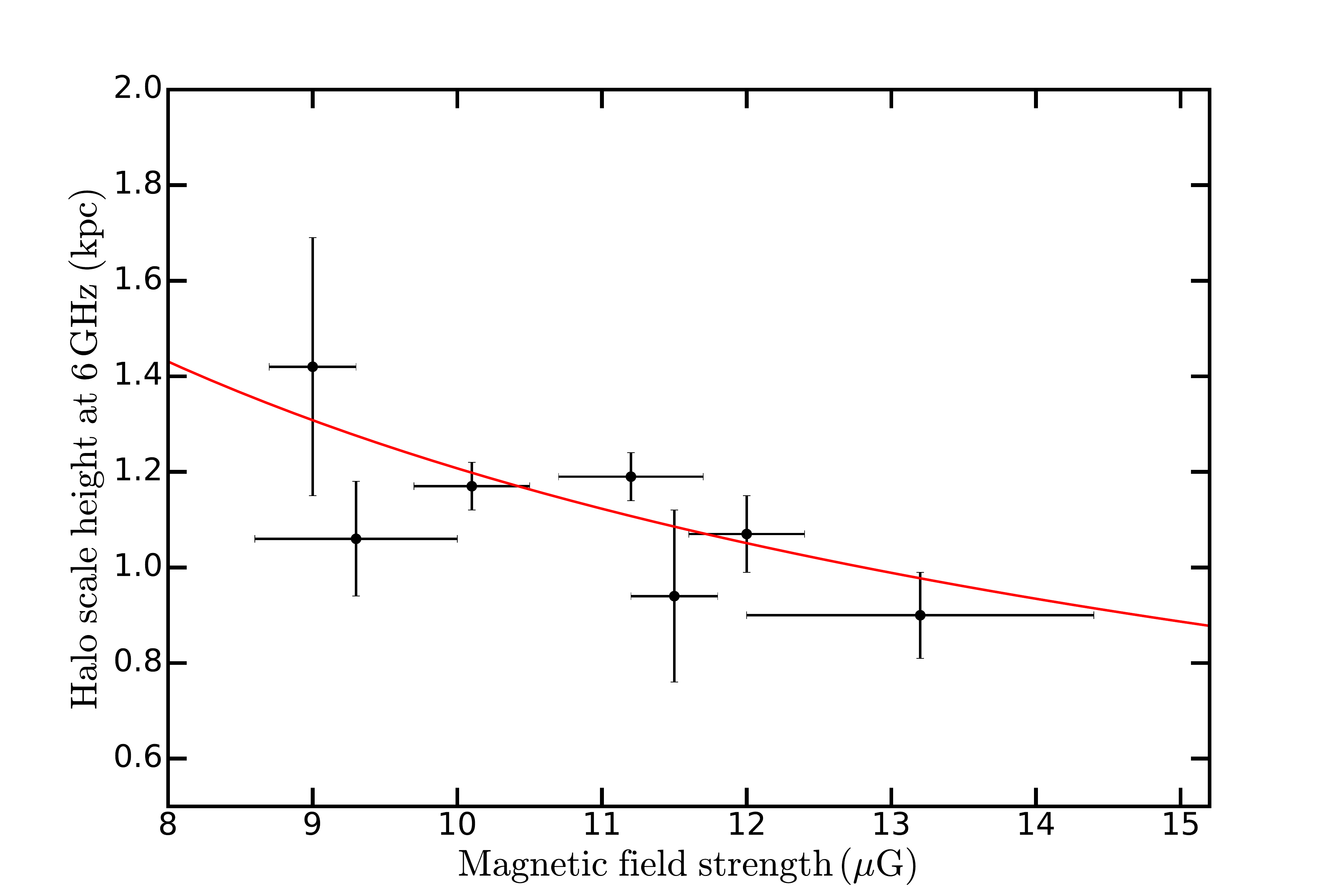}
\caption{Non-thermal halo scale height of NGC\,891 at 1.5\,GHz (left) and 6\,GHz (right) as a function of the total magnetic field strength in the mid-plane. Each data point shows the measurement in the 7 strips along the major axis. Red lines shows the best-fitting power-law function, with an exponent of $-1.39\pm0.64$ at 1.5\,GHz and $-1.32\pm0.61$ at 6\,GHz.}
 \label{fig:hvsB}
\end{figure*}

\subsection{NGC\,891: Accelerated advection flow}
\label{discussion:N891}

The analysis in Sect.~\ref{CRtransport} has shown that the CRE transport in the halo of NGC\,891 is advection-dominated. However, models with constant advection speeds find $V\approx 150\,\mathrm{km\,s^{-1}}$, which is below the escape velocity near the mid-plane
$V_{\mathrm{esc}}=\sqrt{2[1+\ln(r_{\star}/r)]}\times V_{\mathrm{rot}}\geq429\,\mathrm{km\,s^{-1}}$ \citep[where $r_{\star}\approx 18$\,kpc is the radial extent of the flat part of the rotation curve;][]{oosterloo07}. This would imply that the galaxy does not host a galactic wind. \citet{heesen18} 
found a much higher value of $600\,\mathrm{km\,s^{-1}}$, mainly because their larger beam size restricted the fit of the spectral index profile to 
$z\gtrsim1.5$\,kpc.  
Another finding of our models that deserves closer attention is that with a constant advection speed the halo magnetic field scale heights $h_{\mathrm{B2}}$ are systematically 
lower (mostly by $\approx$35\%) than what is expected for energy equipartition; 
in contrast, the diffusion models are mostly consistent with $h_{\mathrm{B2}}=h_{\mathrm{B2,eq}}$.

In case the magnetic field and CREs are coupled to the ionised gas,  
the scale height of the X-ray emission relates to the CRE scale height $h_{\mathrm{e}}$ as $h_{\mathrm{e}}=2h_{\mathrm{X-ray}}$.  
For equipartition, we expect $h_{\mathrm{e}}\approx2h_{\mathrm{syn}}$ 
and hence $h_{\mathrm{X-ray}}\approx h_{\mathrm{syn}}$ and $h_{\mathrm{B,eq}}\approx4h_{\mathrm{X-ray}}$. For NGC\,891, \citet{li13a} found globally
$h_{\mathrm{X-ray}}\approx1.3$\,kpc, which is consistent with our average values of $h_{\mathrm{syn}}$. 
The result of \citet{li13a} is possibly biased by the inclusion of the bulge component and may therefore even underestimate the X-ray scale height of the halo 
\citep{hodges-kluck13}. This suggests that we underestimated $h_{\mathrm{B2}}$ by enforcing $V(z)=\mathrm{const}$ and hence equipartition in the halo cannot 
be ruled out.

As mentioned above, models with fixed equipartition scale heights for the halo can indeed be fitted by assuming an accelerated advection flow. Then the 
$h_{\mathrm{B2}}$ values for $V(z)=\mathrm{const}$ (i.e. $h_{V}=\infty$) can be considered lower limits, as the CRE gas is progressively diluted due to the 
acceleration resulting in adiabatic energy loss of the CREs, which lowers the synchrotron intensity at high $z$. To compensate for this effect, a larger magnetic field 
scale height is required to fit the intensity profile.

For the accelerated advection model, the height at which the advection speed exceeds the escape velocity, $h_{\mathrm{esc}}$, is plotted in Fig.~\ref{fig:cr4} as function of galactocentric radius as projected onto the
major axis (see Appendix~\ref{appendix_CRprop} for tabulated values). We find that this occurs in the upper halo, mostly at heights of
$9\,\mathrm{kpc}\lesssim h_{\mathrm{esc}}\lesssim17\,\mathrm{kpc}$. 

The observed increase of the halo scale height with decreasing (disc) field strength (see Fig.~\ref{fig:hvsB}) indicates that synchrotron losses indeed play an important role in NGC\,891. A likely scenario is that the CREs suffer from synchrotron losses in the inner halo, causing the steepening of the spectral index and the dependence of scale height on field strength, whereas the accelerating wind causes the observed flattening of the spectral index profile towards the outer halo (where adiabatic expansion losses possibly dominate over synchrotron losses).

Based on an observation with the Low Frequency Array (LOFAR) at 146\,MHz and our CHANG-ES data at 1.5\,GHz, \citet{mulcahy18} found an average ratio of the scale heights $h_{\rm syn}$ of the non-thermal halo emission of $1.7\pm0.3$, which gives a scaling with frequency as $h_{\rm syn}\propto\nu^\zeta$ with $\zeta=-0.23\pm0.08$. This points towards dominating synchrotron losses together with diffusive CR propagation, for which $\zeta=-0.25$ is expected if the diffusion coefficient is not energy dependent. On the other hand, the vertical spectral index profile they measured is almost linear (like in our case) and therefore indicates advective propagation.

As far as scale heights are concerned, a more suitable indicator of CR transport mode is the ratio of the normalised scale heights, since these have been corrected for their diameter dependence. A reliable correction (normalisation) for the diameter dependence of the scale height is difficult at 146\,MHz because the LOFAR map of \citet{mulcahy18} is less sensitive (lower S/N) than the VLA maps. Between 1.5 and 6\,GHz, we found a ratio of scale heights $h_{\rm syn}$ of the non-thermal emission of $1.17\pm0.07$ (see Sect.~\ref{scalefitresults}) and a ratio of normalised scale heights $\tilde{h}$ of $1.12\pm0.07$, which means that $h_{\rm syn}$ and $\tilde{h}$ scale with frequency as $h_{\rm syn}\propto\nu^\zeta$ with $\zeta=-0.11\pm0.04$ and $\tilde{h}\propto\nu^\zeta$ with $\zeta=-0.08\pm0.04$, respectively. These values suggest synchrotron losses with diffusive propagation with an energy-dependent diffusion coefficient ($\zeta=-0.125$ expected for $\mu=0.5$). Alternatively, the data may be explained by a mixture of escape losses ($\zeta=0$) and synchrotron losses with advective propagation ($\zeta=-0.5$).

The observed relation between the non-thermal halo scale heights $h_{\mathrm{syn}}$ and total magnetic field strengths $B_{0}$ with exponents of $\xi = -1.39\pm0.64$ at 1.5\,GHz, $\xi = -1.32\pm0.61$ at 6\,GHz (Fig.~\ref{fig:hvsB}), and $\xi = -1.2\pm0.6$ at 146\,MHz \citep{mulcahy18}, is consistent with either (1) synchrotron losses with advective propagation ($\xi=-1.5$), or (2) synchrotron losses with diffusive propagation with or without energy-dependent diffusion ($\xi=-0.875$ expected for $\mu=0.5$ and $\xi=-0.75$ for $\mu=0$).
While the significant dependence of halo scale height on magnetic field strength supports that synchrotron losses of the CRE are dominating, the uncertainties do not allow us to further constrain the type of CRE propagation. Most notably, the relatively weak frequency dependence of the halo scale heights appears to be in conflict with our above results from modelling that advection dominates the CR transport and that an accelerating wind is present. However, our modelling allows that a mixture of synchrotron and escape losses is occurring and diffusion may still make a significant contribution. What we possibly observe is a mixture of regions with substantially different magnetic field strengths and advection speeds within a given resolution element and line-of-sight. In this scenario -- which is also suggested by the observed radial variations in $V_{0}$ and $h_{V}$ -- both synchrotron-loss- and escape-loss-dominated regions may be present in NGC\,891, and likewise the predominance of advective or diffusive CRE transport may regionally vary throughout the galaxy.

\subsection{NGC\,4565: Advection or diffusion?}
\label{discussion:N4565}

The model results for NGC~4565 are somewhat ambigious as to whether diffusion or advection is the dominant CR transport process.
This allows to consider the possibility that NGC\,4565 is actually an intermediate case, where both mechanisms make a significant contribution in the halo. During its lifetime $t_{\rm CRE}$ a CRE can travel to a distance $z$ either by diffusion, $z \propto \sqrt{t_{\rm CRE}}$, or by advection, $z \propto t_{\rm CRE}$. Hence, advection will take over from diffusion at a certain height as the dominating transport process for a non-zero advection speed. The 
critical height $z_{\star}$ above which advection takes over diffusion can be estimated by equating the diffusive and advective time-scales \citep[e.g.][]{recchia16a}:
\begin{equation}
 \frac{z_{\star}^{2}}{D}\approx\frac{z_{\star}}{V}\Rightarrow z_{\star}=0.3\times\frac{D_{28}}{V_{100}}\,\mathrm{kpc}\,,
\end{equation}
with $D_{28}$ being the diffusion coefficient in units of $10^{28}\,\mathrm{cm^2\,s^{-1}}$ and $V_{100}$ the advection speed in units of 100\,$\mathrm{km\,s^{-1}}$. 
Inserting our average fit values of $D_{0}=3.8\times10^{28}\,\mathrm{cm^2\,s^{-1}}$ (for $\mu\leq0.3$) and $V=90\,\mathrm{km\,s^{-1}}$ results in 
$z_{\star}\approx1.3$\,kpc, which is equal to the average $h_{\mathrm{syn}}$ at 1.5\,GHz. The velocity scale heights in our accelerated advection models 
($4\,\mathrm{kpc}\lesssim h_{V}\lesssim14\,\mathrm{kpc}$) are quite large by comparison, therefore $z_{\star}$ would not change significantly in this case. 
The rather constant $\alpha_{\mathrm{nth,1.5-6GHz}}$ profiles up to $z\approx1$\,kpc suggest that diffusion indeed dominates at least up to this height. Therefore, 
the advection speeds we obtained have to be considered as upper limits. However, given the large spectral index uncertainties in the halo, we are unable to formally distinguish between advective and diffusive CRE transport at $z\gtrsim1$\,kpc.

The ratio of scale heights $h_{\rm syn}$ of the non-thermal emission between 1.5 and 6\,GHz is $1.35\pm0.06$ (Sect.~\ref{scalefitresults}) and the ratio of normalised scale heights $\tilde{h}$ is $1.26\pm0.06$. Hence, $h_{\rm syn}$ and $\tilde{h}$ scale with frequency with exponents $\zeta=-0.22\pm0.04$ and $\zeta=-0.17\pm0.04$, respectively, suggesting synchrotron losses with diffusive propagation, with or without an energy-dependent diffusion coefficient ($\zeta$ between $-0.25$ and $-0.125$ for $\mu$ between 0 and 0.5).

As in the case of NGC\,891, $h_{\mathrm{B2}}$ is $\approx$30\% lower than $h_{\mathrm{B2,eq}}$ for advection at a constant speed. 
Contrary for diffusion,
$h_{\mathrm{B2}}$ is even higher (on average by a factor of $\approx$3.5) than the equipartition values derived from the Gaussian synchrotron scale heights; this suggests that 
the galaxy might be advection-dominated from a certain height onwards. The X-ray scale heights measured by \citet{li13a} would predict $h_{\mathrm{B2}}=9.3$\,kpc 
(north), $h_{\mathrm{B2}}=4$\,kpc (south), which is loosely compatible with the expected $h_{\mathrm{B2,eq}}$ of 5--6\,kpc. Hence, we again assume equipartition in the 
following considerations.

To study whether a galactic wind may be present in case of advection, we again determined the heights $h_{\mathrm{esc}}$ where the escape velocity of 
$V_{\mathrm{esc}}\geq437\,\mathrm{km\,s^{-1}}$ is reached (see Appendix~\ref{appendix_CRprop}). 
Except on the north side at $r=+6.9$\,kpc, $h_{\mathrm{esc}}$ is an order of magnitude larger than the detected vertical halo extent, therefore the presence of a wind can be safely ruled out for this galaxy.

We recall that NGC\,4565 is the only galaxy besides UGC\,10288 known to fall outside the $\tilde{h}$--$MSD$ anti-correlation found by \citet{krause18}, and that both galaxies have in common a low total magnetic field strength. Notably, all galaxies from the \citet{krause18} sample except UGC\,10288 seem to be in agreement with an advection-dominated halo, which suggests that a certain minimum magnetic field strength may be required for a strong advection flow and hence for a galactic wind.

The absence of a wind as well as the absence of a bright and extended halo also appears to be plausible in view of our findings regarding the CRE 
injection in the disc (Sect.~\ref{spectralindex}).
Considering the low $\Sigma_{\mathrm{SFR}}$ of $0.73\times10^{-3} \, \mathrm{M_{\odot}yr^{-1}kpc^{-2}}$, and unless we severely overestimated the thermal emission 
component, it is very likely that in this galaxy the CRE injection rates are lower than the synchrotron loss rates, which would explain why the spectra are best 
described by models for discrete-epoch injection (JP or KP). These low injection rates are apparently not sufficient to produce a strong advection flow.

A scenario that might explain the strong spectral steepening we observe in NGC\,4565 is that it is currently in a quiescent phase following a former period of 
higher star-forming activity. 
To investigate whether SN rates  are indeed low enough to cause significant spectral ageing, they 
need to be compared to the CRE energy loss rates. However, determining how SN rates \textit{locally} influence the observed CRE spectra in an edge-on galaxy is not 
straightforward due to the long lines of sight through the disc.

\section{Summary and conclusions}
\label{sumcon}

In this paper, we used wide-band VLA radio continuum observations at 1.5 and 6\,GHz to study the CRE propagation in the edge-on galaxies NGC\,891 and 4565. We corrected the total emission for the thermal contribution estimated from H$\alpha$ and IR data and applied short-spacing corrections at 6\,GHz using Effelsberg 
observations. Then we analyzed the spectral index distribution between the two observing frequencies as well as within each frequency band and compared these data to 
standard models of CRE spectral ageing. Further, we determined radio scale heights at various galactocentric radii as projected onto the major axis -- for the first time also for the 
non-thermal radio continuum emission --, and generated maps of the total magnetic field strength assuming energy equipartition between the CREs and the magnetic field. Finally, 
we modelled advective and diffusive CRE transport 
by solving the 1D diffusion--loss equation. This was again done at different major-axis positions, to study the radial dependence of advection speeds, diffusion 
coefficients, and synchrotron and magnetic field scale heights. Our main findings are the following:
\begin{itemize}
 \item The non-thermal spectral index in NGC\,4565 is consistently steeper than in NGC\,891. Between $1.5$ and 6\,GHz, the disc-averaged non-thermal spectral index is 
 $-0.85\pm0.09$ in NGC\,891 and $-1.10\pm0.14$ in NGC\,4565.
 \item The overall non-thermal spectral index distribution of NGC\,891 is mostly consistent with continuous CRE injection. In NGC\,4565, most of the local 
 synchrotron spectra (also in the disc) feature a break between 1.5 and 6\,GHz and are thus more in line with discrete-epoch injection (JP or KP models). 
 This implies low CRE injection rates (as expected due to the low SFR and $\Sigma_{\mathrm{SFR}}$), which are possibly lower than the synchrotron loss rates.
 \item We detect a thin radio ring associated with the previously known ring of molecular gas and dust. The ring has a vertical thickness of $260\pm100$\,pc and a width in the galactic plane of $2.2\pm1.3$\,kpc.
 \item The synchrotron halo scale heights of NGC\,891 increase with radius, which is in line with the dumbbell shape of its radio continuum halo and indicates significant synchrotron losses of the CREs. Still, there may be regions where CRE escape losses dominate.
 \item Assuming energy equipartition, the average (maximum) total magnetic field strength in the disc of NGC\,891 is 10\,$\mu$G (15\,$\mu$G). In the disc of 
 NGC\,4565 we measure 6\,$\mu$G (7\,$\mu$G).
 \item The CRE transport in the halo of NGC\,891 is probably dominated by advection. This advection flow is an accelerated galactic wind of velocity 
 $\approx150\,\mathrm{km\,s^{-1}}$ near the mid-plane, and reaches the escape velocity at a height of 9 to 17\,kpc, depending on major-axis position. Moreover, this wind is likely to coexist with diffusion-dominated regions. Vertical emission gradients are dominated by the dilution of the CRs resulting from the adiabatic expansion, while synchrotron losses cause a steepening of the spectral index with height. 
 \item NGC\,4565 is diffusion-dominated at least to heights of $z\mkern-5mu=\mkern-5mu1$\,kpc (and probably throughout the halo to even larger heights), with a diffusion coefficient $\geq2\times10^{28}\,\mathrm{cm^2\,s^{-1}}$. 
 \item The results for NGC\,4565 indicate that at CRE energies of a few GeV  
 the diffusion coefficient is only weakly energy-dependent ($D(E)=D_{0}E^{\,\mu}$ with $0\leq \mu\leq 0.3$).
 \item Our results for NGC\,891 are compatible with energy equipartition if the advection flow is accelerated. For NGC\,4565 this is the case as well 
 if both advection and diffusion make a significant contribution to the CRE transport in the halo. 
\end{itemize}

Our analysis demonstrates that the data quality currently provided by the VLA is sufficient for studying the spatial and spectral behaviour of various key 
quantities related to CRE propagation in galactic discs and haloes. 
Future work could examine how common dumbbell-shaped  haloes are among galaxies, and how the presence of a dumbbell shape correlates with synchrotron-loss- 
and escape-dominated  haloes and with diffusive and advective CRE transport. Related to this, it is still unclear if and how scale heights, advection speeds, and 
diffusion coefficients depend on the local $\Sigma_{\rm SFR}$ and $MSD$ within a galaxy.

A step forward in answering such open questions would be to apply the methods presented here to the remaining galaxies in the CHANG-ES sample. This should also 
include the spectral ageing analysis for each galaxy, to test whether the expected correlation of spectral steepening with SFR, advection speed, and halo extent holds 
true. We plan to extend our study to a higher number of observing frequencies (with low frequency data being provided by for instance LOFAR) and thus properly sample the 
spatially resolved CRE spectral energy distribution. This would allow us to test more sophisticated spectral evolution models than those considered in this work. Making 
use of multi-frequency data may also help to distinguish better between advective and diffusive CRE transport in cases such as NGC\,4565, and to put better 
constraints on magnetic field scale heights. The latter is feasible in particular if the full polarisation information of the CHANG-ES observations is utilised. 
An upcoming comprehensive study of the extraplanar rotation measures (RMs) in all CHANG-ES galaxies is expected to provide valuable information on the 3D magnetic field structure in galactic haloes.

\begin{acknowledgements}
We thank Hans-Rainer Kl\"ockner for his valuable comments on the manuscript.
We thank Fatemeh Tabatabaei for the reduction of the 8.35\,GHz Effelsberg data. 
The KPNO H$\alpha$ maps were provided to us with the kind permission of Maria Patterson.\\
This  research  has
used  the  Karl  G.  Jansky  Very  Large  Array  operated  by  the
National  Radio  Astronomy  Observatory  (NRAO).  NRAO  is
a  facility  of  the  National  Science  Foundation  operated  under
co-operative  agreement  by  Associated  Universities,  Inc. 
Moreover, this research is partly based on observations with the 100-m telescope of the MPIfR (Max-Planck-Institut f\"{u}r Radioastronomie) at Effelsberg. We thank the operators for their help during the observations.
\end{acknowledgements}

\bibliographystyle{aa} 
\bibliography{literature} 


\appendix

\section{Energy densities}
\label{uradumag}

To quantify synchrotron and IC losses in our CRE transport models (see Eq.~\ref{be}), we need an estimate of the following two quantities:
While synchrotron losses depend on the magnetic energy density $U_{\mathrm{mag}}=B^2/(8\pi)$, $U_{\mathrm{rad}}=U_{\mathrm{IRF}}+U_{\mathrm{CMB}}$ is the energy density of the radiation that is relevant for inverse-Compton scattering. The energy density of the interstellar radiation field (IRF), in turn, can be expressed as the sum of the stellar and dust radiation fields, $U_{\mathrm{IRF}}=U_{\mathrm{star}}+U_{\mathrm{dust}}$.

$U_{\mathrm{dust}}=L_{\mathrm{FIR}}/(2\pi R^{2}c)$ can be determined from the total far-infrared luminosity $L_{\mathrm{FIR}}$ and the radial extent $R$ of the galaxy, which we determined in the previous section. We used the $L_{\mathrm{FIR}}$ values from \citet{sanders03} after scaling them for each galaxy by the squared ratio between our assumed distance and the distance these authors used. 
Moreover, we assumed $U_{\mathrm{\mathrm{star}}}=1.73U_{\mathrm{\mathrm{dust}}}$, as found for the solar neighbourhood \citep{draine11}, and thus $U_{\mathrm{IRF}}=2.73U_{\mathrm{dust}}$. With $U_{\mathrm{CMB}}=4.2\times10^{-13}\,\mathrm{erg\,cm^{-3}}$ (at redshift 0), we obtain global radiation energy densities of $U_{\mathrm{rad}}=1.04\times10^{-12}\,\mathrm{erg\,cm^{-3}}$ for NGC\,891 and $U_{\mathrm{rad}}=4.98\times10^{-13}\,\mathrm{erg\,cm^{-3}}$ for NGC\,4565. Using the average equipartition field strengths specified in Sect.\ref{bfeld} yields $U_{\mathrm{mag}}=2.55\times10^{-12}\,\mathrm{erg\,cm^{-3}}$ and $U_{\mathrm{mag}}=1.08\times10^{-12}\,\mathrm{erg\,cm^{-3}}$, respectively. Hence, for NGC\,891 $U_{\mathrm{rad}}/U_{\mathrm{mag}}=0.41$, and for NGC\,4565 $U_{\mathrm{rad}}/U_{\mathrm{mag}}=0.46$, which means that on average synchrotron losses are clearly dominating over IC losses in both galaxies.

\section{CRE energy losses}
\label{ionbrems}

We calculate the combined synchrotron and IC loss rate as \citep[e.g.][]{heesen16}
\begin{equation}
\label{synICtime}
 t_{\mathrm{syn+IC}}=1.0815\times10^{9}\left(\frac{\nu}{\mathrm{GHz}}\right)^{-0.5}\left(\frac{B}{\mathrm{\mu G}}\right)^{-1.5}\left(1+\frac{U_{\mathrm{rad}}}{U_{\mathrm{mag}}}\right)^{-1}\,\mathrm{yr}\,.
\end{equation}
In addition, ionisation losses occur on a time-scale of \citep{murphy09}
\begin{equation}
 t_{\mathrm{ion}}=4.1\times10^{9}\left(\frac{\langle n\rangle}{\mathrm{cm}^{-3}}\right)^{-1}\left(\frac{E}{\mathrm{GeV}}\right) \, \left[3 \, \mathrm{ln}\left(\frac{E}{\mathrm{GeV}}\right)+42.5\right]^{-1}\,\mathrm{yr}\,,
\end{equation}
as well as bremsstrahlung losses, the time-scale of which is given by
\begin{equation}
 t_{\mathrm{brems}}=3.96\times10^{7}\left(\frac{\langle n\rangle}{\mathrm{cm}^{-3}}\right)^{-1}\,\mathrm{yr}\,,
\end{equation}
where $\langle n\rangle$ is the average particle number density of neutral gas in the ISM, and the CRE energy can be written as $E(\mathrm{GeV})=(\nu/16.1\,\mathrm{MHz})^{1/2}\,(B/\mathrm{\mu G})^{-1/2}$. 
We estimate $\langle n\rangle$ in both galaxies from the total atomic gas masses given in Table~\ref{tab:basicparms}, assuming that this amount of gas is distributed in an oblate spheroid of radius $R$ as given in Sect.~\ref{bfeld} and semi-minor axis length corresponding to the H\,{\sc i} disc scale height of $1.25$\,kpc for NGC\,891 \citep{oosterloo07} and $0.3$\,kpc for NGC\,4565 \citep{zschaechner12}. 
Assuming for simplicity that all atoms are hydrogen nuclei ($m_{\mathrm{p}}=1.673\times10^{-24}$\,g), we find $\langle n\rangle=0.22\,\mathrm{cm^{-3}}$ for NCG\,891 and $\langle n\rangle=0.65\,\mathrm{cm^{-3}}$ for NGC\,4565. Further, we use the maximum total magnetic field strengths of $13\,\mu$G and $6.5\,\mu$G, respectively (as determined in Sect.~\ref{bfeld}). 
In the case of NGC\,891, we thus have $t_{\mathrm{syn+IC}}^{\mathrm{1.5GHz}}=1.3\times10^{7}$\,yr, $t_{\mathrm{syn+IC}}^{\mathrm{6GHz}}=6.7\times10^{6}$\,yr, $t_{\mathrm{ion}}^{\mathrm{1.5GHz}}=1.0\times10^{9}$\,yr, $t_{\mathrm{ion}}^{\mathrm{6GHz}}=2.0\times10^{9}$\,yr, and $t_{\mathrm{brems}}=1.8\times10^{8}$\,yr. For NGC\,4565 we obtain $t_{\mathrm{syn+IC}}^{\mathrm{1.5GHz}}=3.6\times10^{7}$\,yr, $t_{\mathrm{syn+IC}}^{\mathrm{6GHz}}=1.8\times10^{7}$\,yr, $t_{\mathrm{ion}}^{\mathrm{1.5GHz}}=4.9\times10^{8}$\,yr, $t_{\mathrm{ion}}^{\mathrm{6GHz}}=9.5\times10^{8}$\,yr, and $t_{\mathrm{brems}}=6.1\times10^{7}$\,yr. 
Thus, in both galaxies, ionisation loss rates are at least one order of magnitude lower than those of synchrotron and IC losses, and can hence be neglected. The same is true for bremsstrahlung losses in NGC\,891, while in NGC\,4565 bremsstrahlung losses may have a minor effect on the spectrum at 1.5\,GHz.

\FloatBarrier

\section{Radio continuum scale heights}
\label{appendix_scaleh}

In this appendix, we present the scale heights in the individual strips in both NGC\,891 and 4565. In each strip, we have measured scale heights in the thin disc and thick disc (halo), fitting two-component exponential functions to the vertical intensity profiles. Scale heights are presented for both $1.5$ and 6~GHz, separately for the total and non-thermal radio continuum emission. Tables~\ref{tab:expN891} and \ref{tab:expN4565} contain the exponential scale heights of NGC\,891 and 4565, respectively.  Table~\ref{tab:gaussN4565} contains the Gaussian scale heights of NGC\,4565. In each strip, we have measured Gaussian scale heights in the thin disc and thick disc (halo), fitting two-component Gaussian functions to the vertical intensity profiles.

\begin{table*}[h!]
\caption{Exponential scale heights of NGC 891}
{\small
\begin{center}
\begin{tabular}{c|lll|llll}
\toprule
\toprule[0.3pt]
\multicolumn{1}{l}{} & \multicolumn{2}{l}{\underline{1.5 GHz, total}}   & \multicolumn{1}{l}{} &  & \multicolumn{2}{l}{\underline{1.5 GHz, non-thermal}} & \multicolumn{1}{l}{} \\ 
$r$ [kpc] & $h_{\mathrm{disc}}$ [kpc] & $h_{\mathrm{halo}}$ [kpc] & $\chi^{2}_{\mathrm{red}}$ &  & $h_{\mathrm{disc}}$ [kpc] & $h_{\mathrm{halo}}$ [kpc] & $\chi^{2}_{\mathrm{red}}$ \\ 
\hline
$-7.94$ & 0.20 $\pm$ 0.03 & 1.47 $\pm$ 0.08 & 0.28 &  & 0.23 $\pm$ 0.04 & 1.49 $\pm$ 0.09 & 0.24 \\ 
$-5.29$ & 0.14 $\pm$ 0.01 & 1.36 $\pm$ 0.03 & 0.13 &  & 0.15 $\pm$ 0.01 & 1.37 $\pm$ 0.03 & 0.11 \\ 
$-2.65$ & 0.18 $\pm$ 0.03 & 1.34 $\pm$ 0.12 & 4.58 &  & 0.20 $\pm$ 0.04 & 1.35 $\pm$ 0.11 & 3.95 \\ 
0 & 0.14 $\pm$ 0.02 & 1.08 $\pm$ 0.09 & 11.06 &  & 0.14 $\pm$ 0.03 & 1.09 $\pm$ 0.09 & 10.07 \\ 
2.65 & 0.19 $\pm$ 0.04 & 1.06 $\pm$ 0.09 & 6.45 &  & 0.21 $\pm$ 0.06 & 1.06 $\pm$ 0.09 & 5.58 \\ 
5.29 & 0.17 $\pm$ 0.02 & 1.35 $\pm$ 0.06 & 1.16 &  & 0.19 $\pm$ 0.02 & 1.35 $\pm$ 0.05 & 0.86 \\ 
7.94 & 0.17 $\pm$ 0.02 & 1.30 $\pm$ 0.14 & 0.08 &  & 0.21 $\pm$ 0.03 & 1.31 $\pm$ 0.07 & 0.07 \\ 
\hline
average & \multicolumn{1}{l}{0.17 $\pm$ 0.02} & \multicolumn{1}{l}{1.28 $\pm$ 0.14} & &  & \multicolumn{1}{l}{0.19 $\pm$ 0.03} & \multicolumn{1}{l}{1.29 $\pm$ 0.15} & \multicolumn{1}{l}{} \\ 
\hline
\multicolumn{1}{l}{} &  &  & \multicolumn{1}{l}{} &  &  &  & \multicolumn{1}{l}{} \\ 
\multicolumn{1}{l}{} &  &  & \multicolumn{1}{l}{} &  &  &  & \multicolumn{1}{l}{} \\ 
\toprule
\toprule[0.3pt]
\multicolumn{1}{l}{} & \multicolumn{2}{l}{\underline{6 GHz, total}}    & \multicolumn{1}{l}{} &  & \multicolumn{2}{l}{\underline{6 GHz, non-thermal}}  & \multicolumn{1}{l}{} \\ 
$r$ [kpc] & $h_{\mathrm{disc}}$ [kpc] & $h_{\mathrm{halo}}$ [kpc] & $\chi^{2}_{\mathrm{red}}$ &  & $h_{\mathrm{disc}}$ [kpc] & $h_{\mathrm{halo}}$ [kpc] & $\chi^{2}_{\mathrm{red}}$ \\ 
\hline
$-7.94$ & 0.20 $\pm$ 0.01 & 1.21 $\pm$ 0.09 & 0.25 &  & 0.33 $\pm$ 0.06 & 1.42 $\pm$ 0.27 & 0.25 \\ 
$-5.29$ & 0.14 $\pm$ 0.01 & 1.16 $\pm$ 0.03 & 0.01 &  & 0.16 $\pm$ 0.01 & 1.19 $\pm$ 0.05 & 0.02 \\ 
$-2.65$ & 0.16 $\pm$ 0.01 & 1.05 $\pm$ 0.09 & 1.94 &  & 0.18 $\pm$ 0.02 & 1.07 $\pm$ 0.08 & 1.54 \\ 
0 & 0.13 $\pm$ 0.01 & 0.88 $\pm$ 0.09 & 11.69 &  & 0.14 $\pm$ 0.02 & 0.90 $\pm$ 0.09 & 10.39 \\ 
2.65 & 0.25 $\pm$ 0.04 & 1.08 $\pm$ 0.29 & 4.77 &  & 0.25 $\pm$ 0.05 & 0.94 $\pm$ 0.18 & 4.34 \\ 
5.29 & 0.17 $\pm$ 0.01 & 1.17 $\pm$ 0.05 & 0.28 &  & 0.20 $\pm$ 0.01 & 1.17 $\pm$ 0.05 & 0.33 \\ 
7.94 & 0.15 $\pm$ 0.01 & 1.05 $\pm$ 0.09 & 0.15 &  & 0.17 $\pm$ 0.04 & 1.06 $\pm$ 0.12 & 0.17 \\ 
\hline
average & \multicolumn{1}{l}{0.17 $\pm$ 0.04} & \multicolumn{1}{l}{1.09 $\pm$ 0.10} &  &  & \multicolumn{1}{l}{0.20 $\pm$ 0.06} & \multicolumn{1}{l}{1.11 $\pm$ 0.16} & \multicolumn{1}{l}{} \\ 
\hline
\end{tabular}
\label{tab:expN891}
\end{center}
}
\end{table*}

\begin{table*}
\caption{Exponential scale heights of NGC 4565}
{\small 
\begin{center}
\begin{tabular}{c|lll|llll}
\toprule
\toprule[0.3pt]
\multicolumn{1}{l}{} & \multicolumn{2}{l}{\underline{1.5 GHz, total}}   & \multicolumn{1}{l}{} &  & \multicolumn{2}{l}{\underline{1.5 GHz, non-thermal}} & \multicolumn{1}{l}{} \\ 
$r$ [kpc] & $h_{\mathrm{disc}}$ [kpc] & $h_{\mathrm{halo}}$ [kpc] & $\chi^{2}_{\mathrm{red}}$ &  & $h_{\mathrm{disc}}$ [kpc] & $h_{\mathrm{halo}}$ [kpc] & $\chi^{2}_{\mathrm{red}}$ \\ 
\hline
$-13.85$ & 0.02 $\pm$ 0.01 & \textit{1.50 $\pm$ 0.16} & 0.13 &  & 0.02 $\pm$ 0.11 & \textit{1.35 $\pm$ 0.12} & 0.09 \\ 
$-6.92$ & 0.02 $\pm$ 0.01 & 1.70 $\pm$ 0.21 & 0.29 &  & 0.02 $\pm$ 0.09 & 1.41 $\pm$ 0.08 & 0.21 \\ 
0 & 0.03 $\pm$ 0.13$^{\star}$ & 1.41 $\pm$ 0.13$^{\star}$ & 0.18 &  & 0.18 $\pm$ 0.11$^{\star}$ & 1.30 $\pm$ 0.15$^{\star}$ & 0.16 \\ 
6.92 & 0.02 $\pm$ 0.01 & 1.70 $\pm$ 0.22 & 0.52 &  & 0.02 $\pm$ 0.09 & 1.34 $\pm$ 0.11 & 0.19 \\ 
13.85 & 0.02 $\pm$ 0.01 & \textit{1.36 $\pm$ 0.21} & 0.16 &  & 0.02 $\pm$ 0.01 & \textit{1.19 $\pm$ 0.11} & 0.12 \\ 
\hline
average & 0.02 $\pm$ 0.01 & 1.60 $\pm$ 0.17 &  &  & 0.05 $\pm$ 0.07 & 1.35 $\pm$ 0.06 &  \\ 
\hline
\multicolumn{1}{l}{} &  &  & \multicolumn{1}{l}{} &  &  &  & \multicolumn{1}{l}{} \\ 
\multicolumn{1}{l}{} &  &  & \multicolumn{1}{l}{} &  &  &  & \multicolumn{1}{l}{} \\ 
\toprule
\toprule[0.3pt]
\multicolumn{1}{l}{} & \multicolumn{2}{l}{\underline{6 GHz, total}}    & \multicolumn{1}{l}{} &  & \multicolumn{2}{l}{\underline{6 GHz, non-thermal}}  & \multicolumn{1}{l}{} \\ 
$r$ [kpc] & $h_{\mathrm{disc}}$ [kpc] & $h_{\mathrm{halo}}$ [kpc] & $\chi^{2}_{\mathrm{red}}$ &  & $h_{\mathrm{disc}}$ [kpc] & $h_{\mathrm{halo}}$ [kpc] & $\chi^{2}_{\mathrm{red}}$ \\ 
\hline
$-13.85$ & 0.02 $\pm$ 0.010 & \textit{2.13 $\pm$ 0.40} & 0.21 &  & 0.53 $\pm$ 0.09 & \textit{4.70 $\pm$ 1.30} & 0.13 \\ 
$-6.92$ & 0.04 $\pm$ 0.20 & 0.92 $\pm$ 0.16 & 0.24 &  & 0.57 $\pm$ 0.16 & 1.10 $\pm$ 0.28 & 0.22 \\ 
0 & 0.32 $\pm$ 0.10$^{\star}$ & 1.64 $\pm$ 0.41$^{\star}$ & 0.23 &  & 0.52 $\pm$ 0.06$^{\star}$ & 0.95 $\pm$ 0.30$^{\star}$ & 0.26 \\ 
6.92 & 0.02 $\pm$ 0.04$^{\star}$ & 1.53 $\pm$ 0.29$^{\star}$ & 0.51 &  & 0.02 $\pm$ 0.32 & 0.95 $\pm$ 0.18 & 0.25 \\ 
13.85 & 0.02 $\pm$ 0.08 & \textit{2.47 $\pm$ 1.07} & 0.23 &  & 0.40 $\pm$ 0.12 & \textit{3.30 $\pm$ 0.96} & 0.22 \\ 
\hline
average & 0.08 $\pm$ 0.13 & 1.36 $\pm$ 0.39 &  &  & \multicolumn{1}{l}{0.41 $\pm$ 0.20} & \multicolumn{1}{l}{1.00 $\pm$ 0.09} &  \\ 
\hline
\end{tabular}
\label{tab:expN4565}
\end{center}
}
{\footnotesize \textit{Notes.} Only the inner three strips were used to compute the scale heights of the halo emission.\\
$^{\star}$Average values of the separately determined scale heights on the northern and southern side of the mid-plane.}
\end{table*}

\begin{table*}
\caption{Gaussian scale heights of NGC 4565}
{\small
\begin{center}
\begin{tabular}{c|lll|llll}
\toprule
\toprule[0.3pt]
\multicolumn{1}{l}{} & \multicolumn{2}{l}{\underline{1.5 GHz, total}}   & \multicolumn{1}{l}{} &  & \multicolumn{2}{l}{\underline{1.5 GHz, non-thermal}} & \multicolumn{1}{l}{} \\ 
$r$ [kpc] & $h_{\mathrm{disc}}$ [kpc] & $h_{\mathrm{halo}}$ [kpc] & $\chi^{2}_{\mathrm{red}}$ &  & $h_{\mathrm{disc}}$ [kpc] & $h_{\mathrm{halo}}$ [kpc] & $\chi^{2}_{\mathrm{red}}$ \\ 
\hline
$-13.85$ & 0.01 $\pm$ 0.01 & 2.76 $\pm$ 0.42 & 0.13 &  & 0.02 $\pm$ 0.01 & 2.50 $\pm$ 0.09 & 0.08 \\ 
$-6.92$ & 0.02 $\pm$ 0.01 & 3.05 $\pm$ 0.21 & 0.18 &  & 0.02 $\pm$ 0.16 & 2.65 $\pm$ 0.12 & 0.13 \\ 
0 & 0.29 $\pm$ 0.10 & 2.87 $\pm$ 0.22 & 0.13 &  & 0.58 $\pm$ 0.06 & 2.95 $\pm$ 0.16 & 0.10 \\ 
6.92 & 0.01 $\pm$ 0.02 & 3.34 $\pm$ 0.80 & 0.59 &  & 0.02 $\pm$ 0.02 & 2.59 $\pm$ 0.29 & 0.26 \\ 
13.85 & 0.03 $\pm$ 0.01 & 2.84 $\pm$ 0.40 & 0.16 &  & 0.67 $\pm$ 0.13 & 3.30 $\pm$ 0.59 & 0.12 \\ 
\hline
average & 0.07 $\pm$ 0.11 & 2.97 $\pm$ 0.21 &  &  & 0.26 $\pm$ 0.30 & 2.80 $\pm$ 0.29 &  \\ 
\hline
\multicolumn{1}{l}{} &  &  & \multicolumn{1}{l}{} &  &  &  & \multicolumn{1}{l}{} \\ 
\multicolumn{1}{l}{} &  &  & \multicolumn{1}{l}{} &  &  &  & \multicolumn{1}{l}{} \\ 
\toprule
\toprule[0.3pt]
\multicolumn{1}{l}{} & \multicolumn{2}{l}{\underline{6 GHz, total}} & \multicolumn{1}{l}{} &  & \multicolumn{2}{l}{\underline{6 GHz, non-thermal}}  & \multicolumn{1}{l}{} \\ 
$r$ [kpc] & $h_{\mathrm{disc}}$ [kpc] & $h_{\mathrm{halo}}$ [kpc] & $\chi^{2}_{\mathrm{red}}$ &  & $h_{\mathrm{disc}}$ [kpc] & $h_{\mathrm{halo}}$ [kpc] & $\chi^{2}_{\mathrm{red}}$ \\ 
\hline
$-13.85$ & 0.02 $\pm$ 0.27 & 4.56 $\pm$ 0.64 & 0.16 &  & 0.70 $\pm$ 0.07 & 6.16 $\pm$ 1.24 & 0.06 \\ 
$-6.92$ & 0.13 $\pm$ 0.10 & 1.71 $\pm$ 0.38 & 0.12 &  & 1.03 $\pm$ 0.10 & 5.50 $\pm$ 1.13 & 0.16 \\ 
0 & 0.59 $\pm$ 0.07 & 3.35 $\pm$ 1.06 & 0.24 &  & 1.06 $\pm$ 0.15 & 1.13 $\pm$ 0.21 & 0.23 \\ 
6.92 & 0.20 $\pm$ 0.12$^{\star}$ & 2.73 $\pm$ 0.19$^{\star}$ & 0.09 &  & 0.28 $\pm$ 0.23$^{\star}$ & 2.13 $\pm$ 0.36$^{\star}$ & 0.03 \\ 
13.85 & 0.02 $\pm$ 0.15 & 7.39 $\pm$ 2.82 & 0.17 &  & 0.57 $\pm$ 0.14 & 9.61 $\pm$ 1.54 & 0.16 \\ 
\hline
average & 0.19 $\pm$ 0.21 & 3.95 $\pm$ 1.95 &  &  & \multicolumn{1}{l}{0.73 $\pm$ 0.29} & \multicolumn{1}{l}{4.92 $\pm$ 3.03} &  \\ 
\hline
\end{tabular}
\label{tab:gaussN4565}
\end{center}
}
{\footnotesize \textit{Notes.}\\
$^{\star}$ Average values of the separately determined scale heights on the northern and southern side of the mid-plane. 
}
\end{table*}


\section{CR transport models}
\label{appendix_CRprop}

\subsection{NGC\,891}
\begin{figure*}
 \centering
 \includegraphics[clip=true,trim=0pt 0pt 0pt 0pt,scale=0.29]{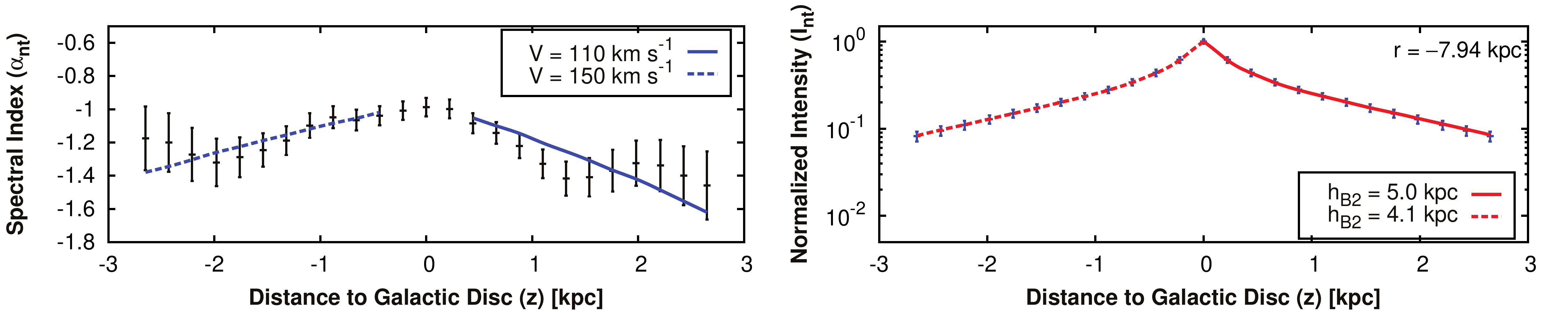}
 \includegraphics[clip=true,trim=0pt 0pt 0pt 0pt,scale=0.29]{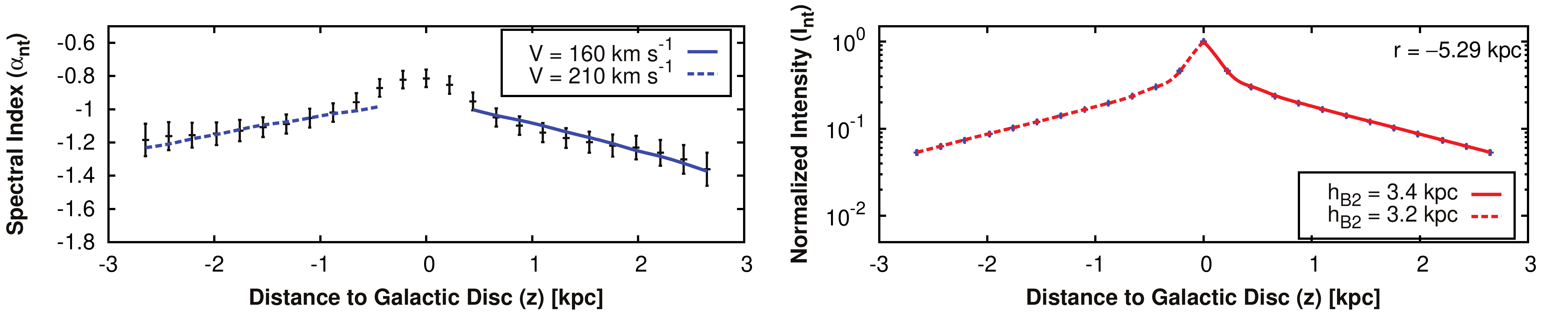}
 \includegraphics[clip=true,trim=0pt 0pt 0pt 0pt,scale=0.29]{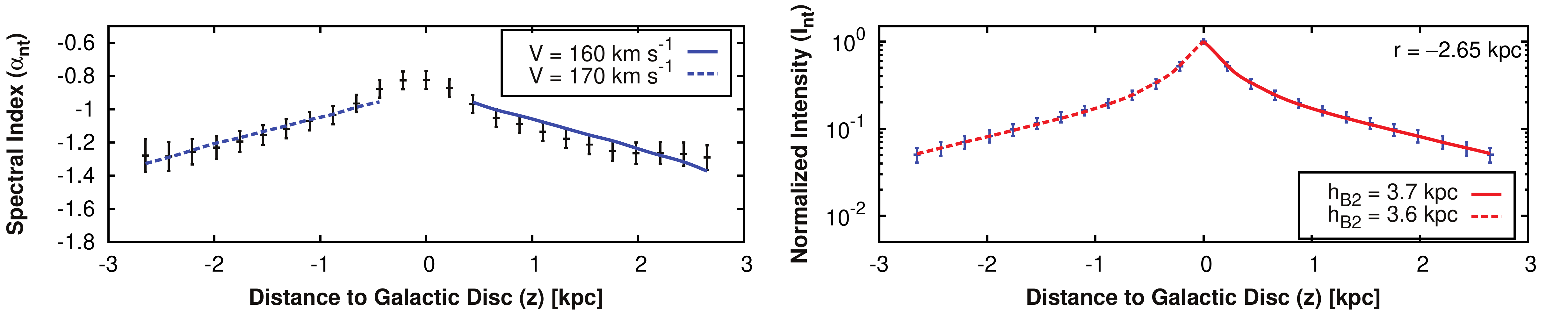}
 \includegraphics[clip=true,trim=0pt 0pt 0pt 0pt,scale=0.29]{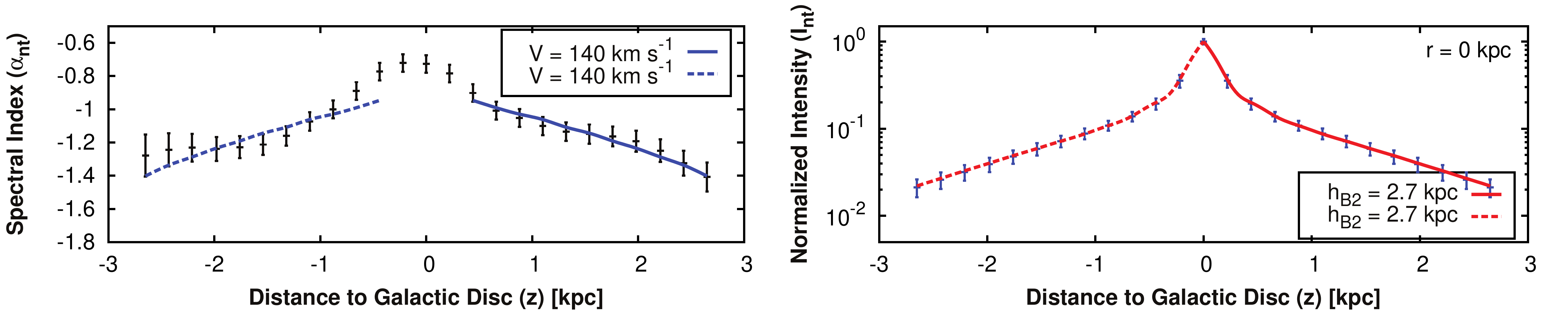}
 \includegraphics[clip=true,trim=0pt 0pt 0pt 0pt,scale=0.29]{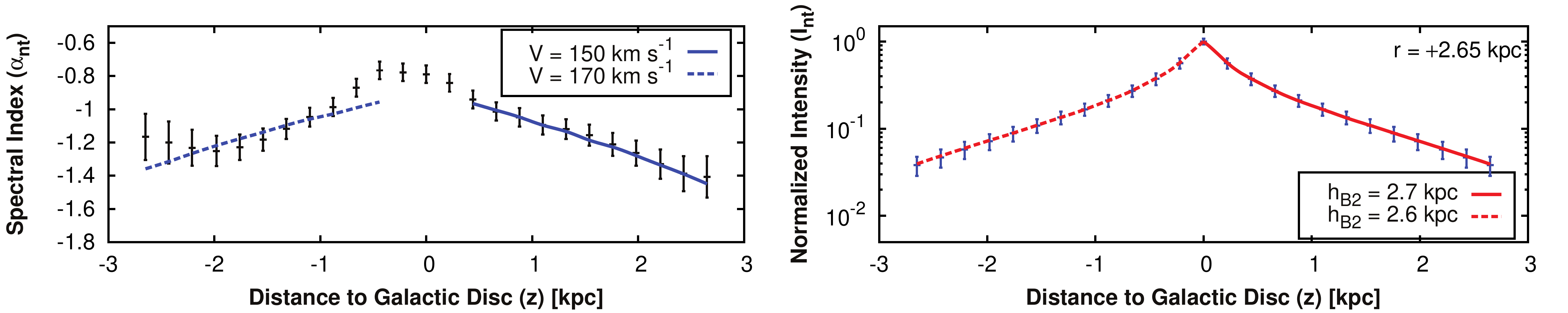}
 \includegraphics[clip=true,trim=0pt 0pt 0pt 0pt,scale=0.29]{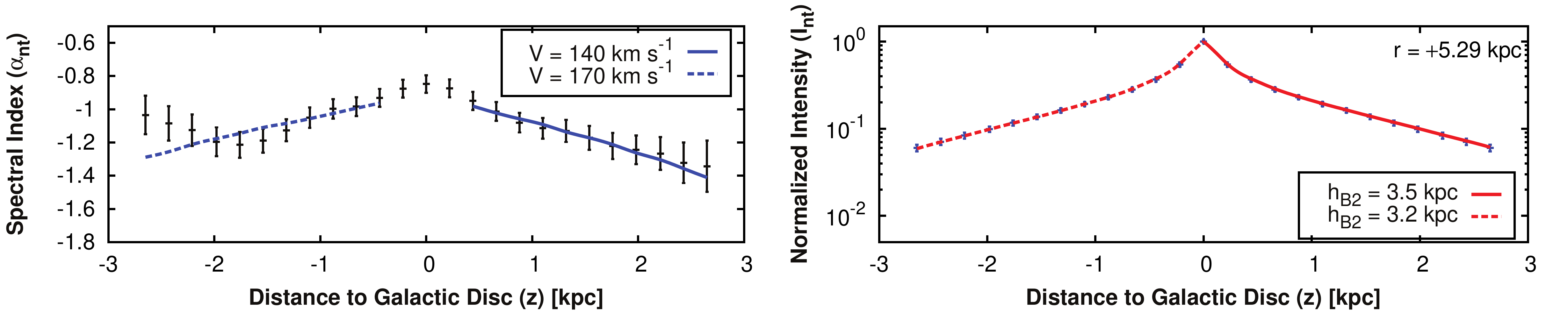}
 \includegraphics[clip=true,trim=0pt 0pt 0pt 0pt,scale=0.29]{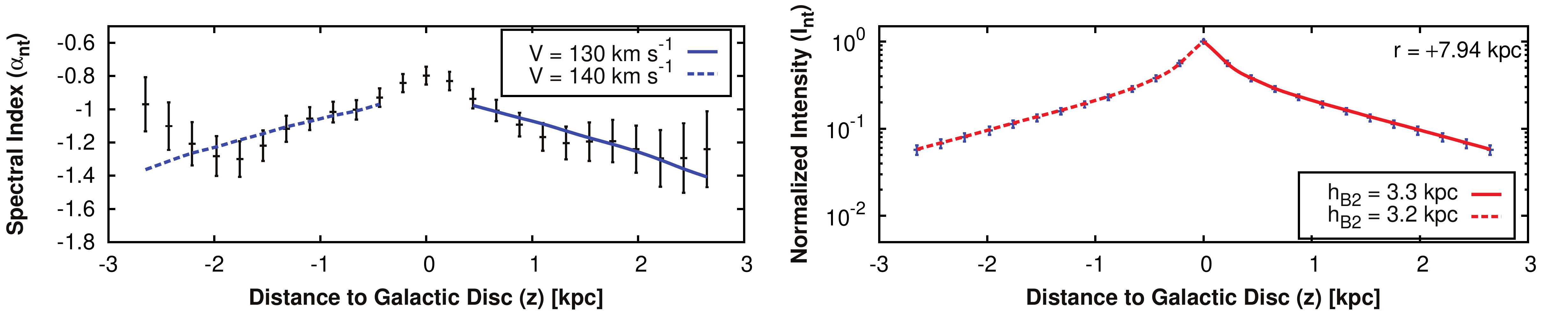}
\caption{Advection models with a constant advection speed for NGC\,891. Data points denote the vertical profile of the non-thermal spectral index between 1.5 and 6\,GHz (left panels) and the exponential model of the non-thermal intensity profile at 1.5\,GHz (right panels). The radial position of each profile is given in the right-hand-side plot; $r<0$ is east of the minor axis and $r>0$ is west of the minor axis. Positive $z$ values are on the north side and negative ones on the south side of the mid-plane. The solid lines show the best-fitting advection models.}
 \label{fig:N891adv1}
\end{figure*}

\begin{figure*}
 \centering
 \includegraphics[clip=true,trim=0pt 0pt 0pt 0pt,scale=0.29]{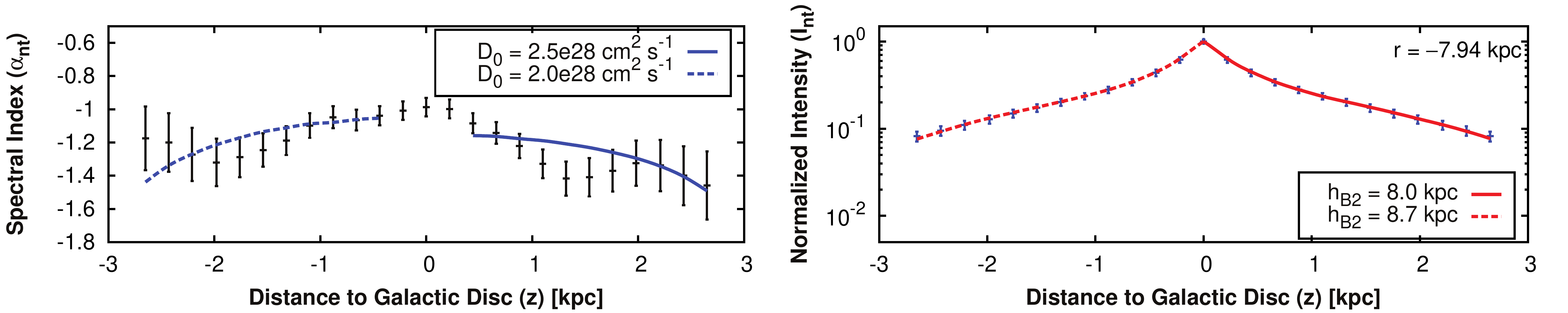}
 \includegraphics[clip=true,trim=0pt 0pt 0pt 0pt,scale=0.29]{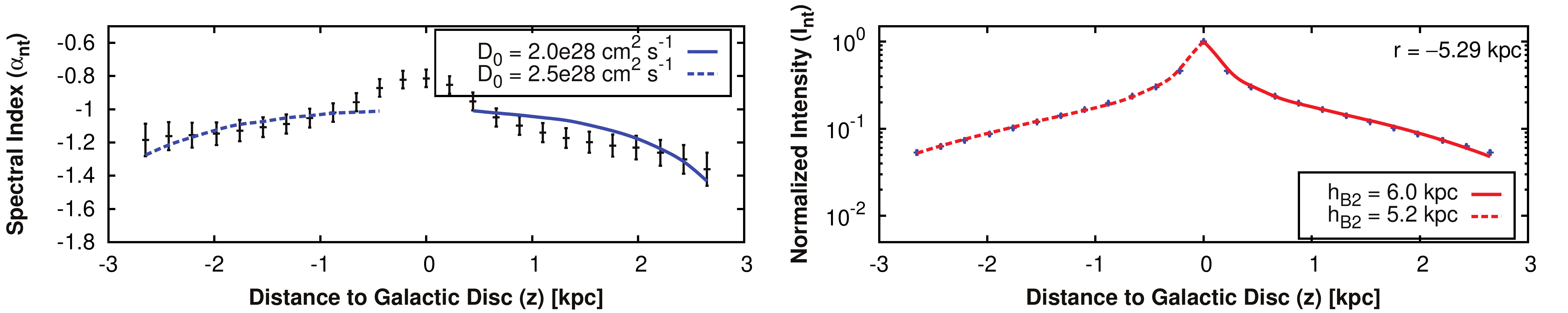}
 \includegraphics[clip=true,trim=0pt 0pt 0pt 0pt,scale=0.29]{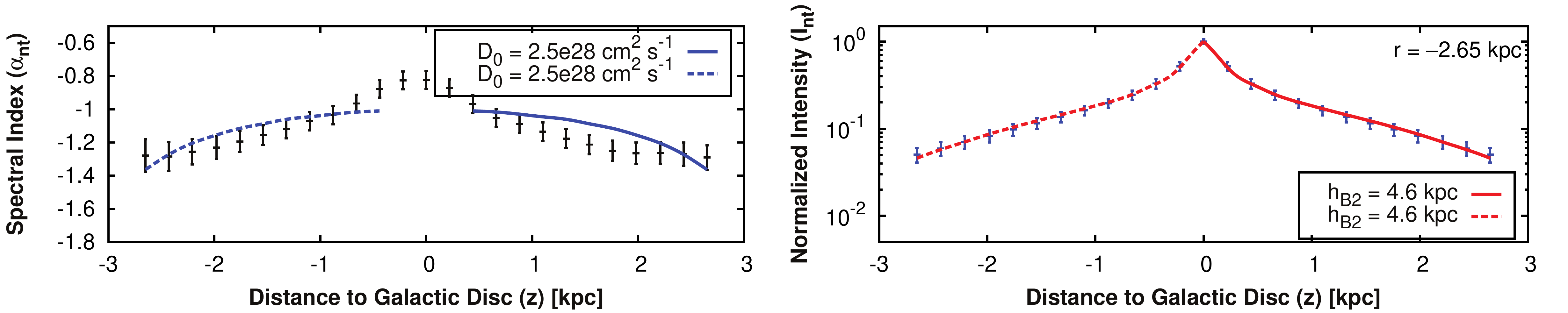}
 \includegraphics[clip=true,trim=0pt 0pt 0pt 0pt,scale=0.29]{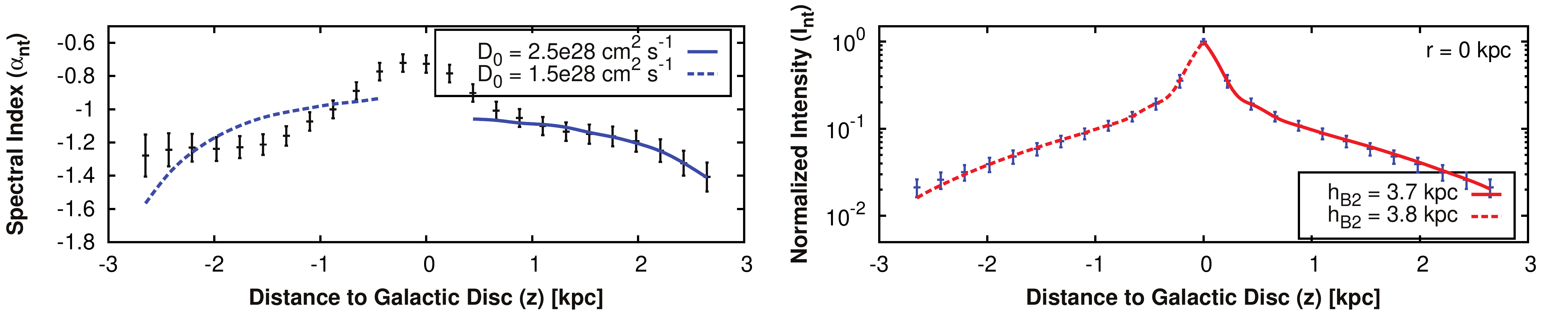}
 \includegraphics[clip=true,trim=0pt 0pt 0pt 0pt,scale=0.29]{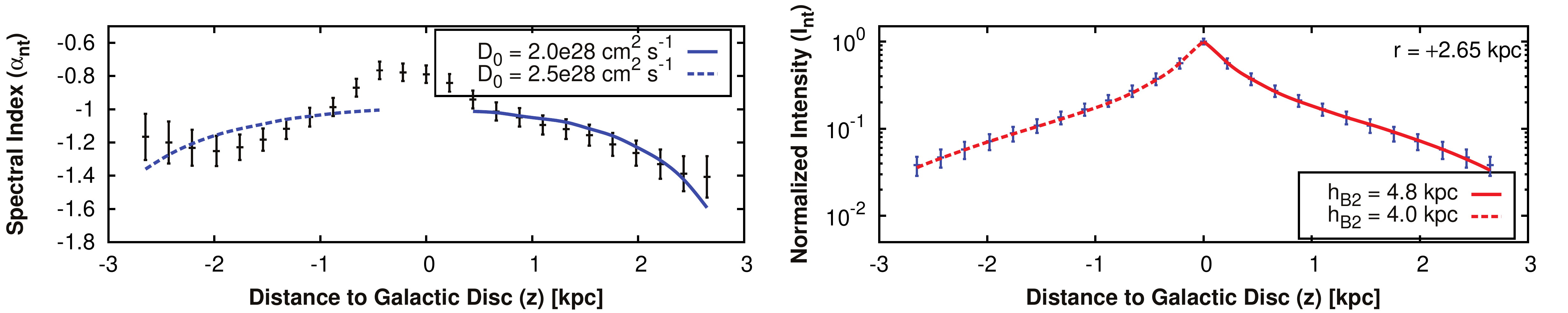}
 \includegraphics[clip=true,trim=0pt 0pt 0pt 0pt,scale=0.29]{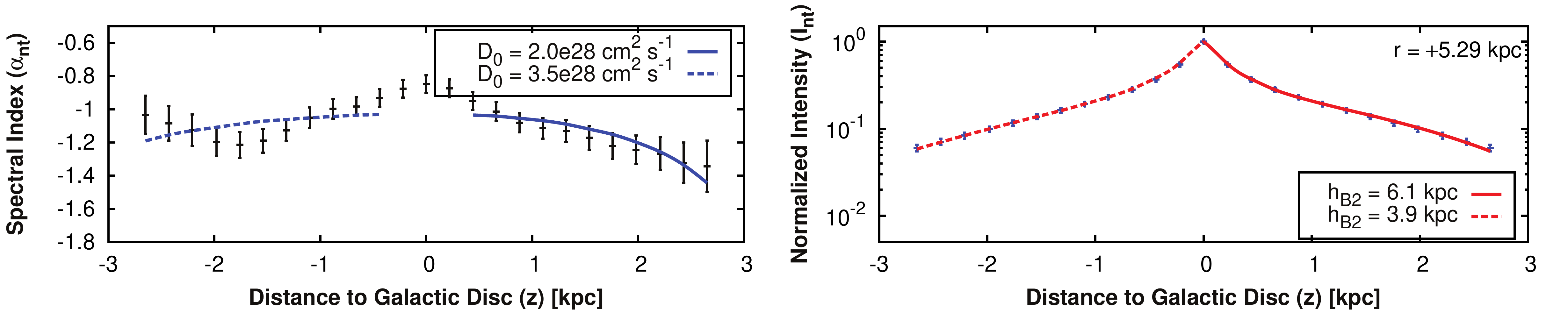}
 \includegraphics[clip=true,trim=0pt 0pt 0pt 0pt,scale=0.29]{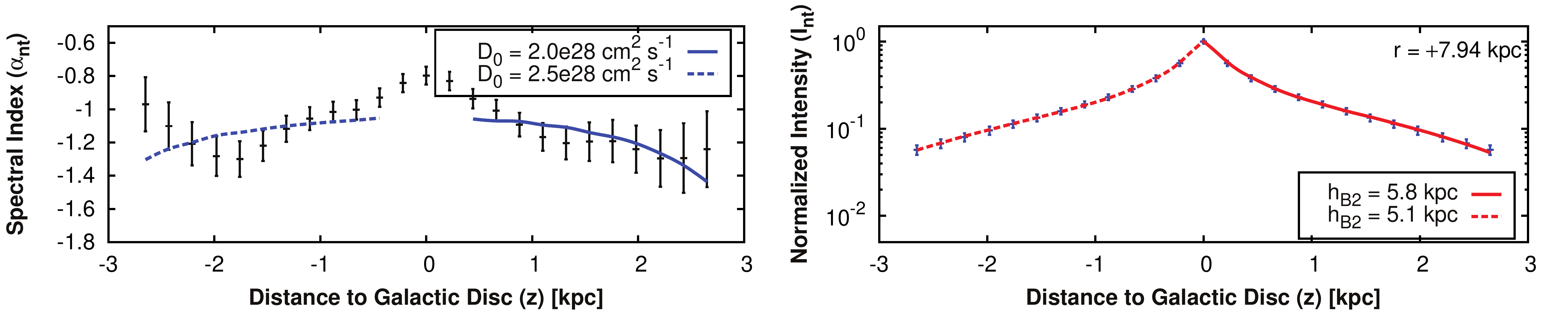}
\caption{Diffusion models with an energy-dependent diffusion coefficient [$D(E)=D_0(E/{1~\rm GeV})^{\mu}$] for NGC\,891 (assuming $\mu=0.5$). Data points denote the vertical profile of the non-thermal spectral index between 1.5 and 6\,GHz (left panels) and the exponential model of the non-thermal intensity profile at 1.5\,GHz (right panels). The radial position of each profile is given in the right-hand-side plot; $r<0$ is east of the minor axis and $r>0$ is west of the minor axis. Positive $z$ values are on the north side and negative ones on the south side of the mid-plane. Solid lines show the best-fitting diffusion models.}
 \label{fig:N891diff}
\end{figure*}

In this appendix, we present the best-fitting advection and diffusion cosmic-ray transport models in the individual strips in NGC\,891. In Table~\ref{tab:N891adv1}, the best-fitting parameters are presented for a constant advection speed; the corresponding best-fitting intensity and spectral index profiles are presented in Fig.~\ref{fig:N891adv1}. In Table~\ref{tab:N891adv2} we present the best-fitting parameters for advection with a linearly accelerating advection velocity; the corresponding best-fitting intensity and spectral index profiles are presented in Fig.~\ref{fig:N891adv2_CR}. In Table~\ref{tab:N891diff} we present the best-fitting parameters for diffusion with an energy dependent diffusion coefficient assuming $\mu = 0.5$; the corresponding best-fitting intensity and spectral index profiles are presented in Fig.~\ref{fig:N891diff}.

\subsection{NGC\,4565}
\begin{figure*}
 \centering
 \includegraphics[clip=true,trim=0pt 0pt 0pt 0pt,scale=0.29]{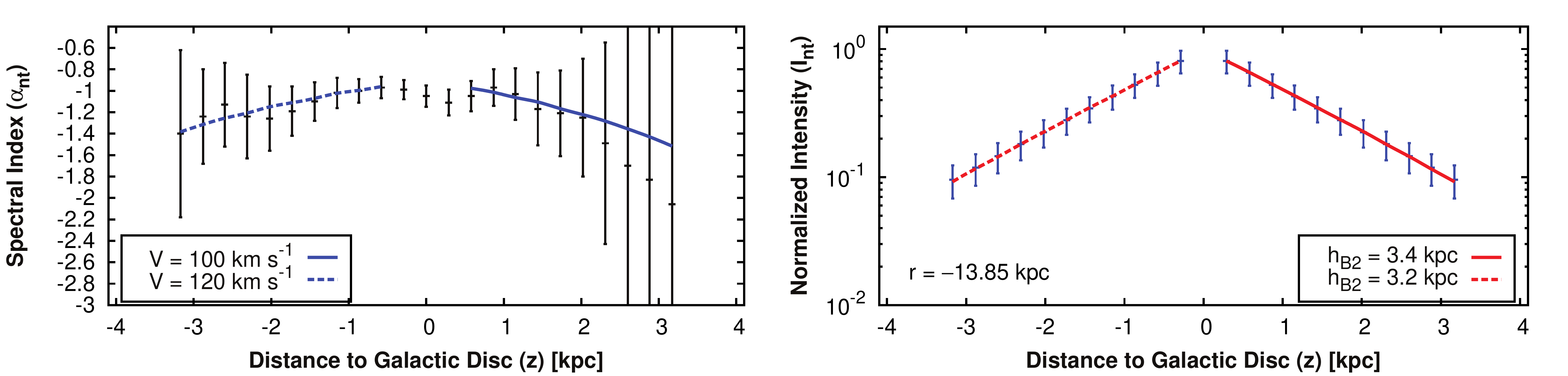}
 \includegraphics[clip=true,trim=0pt 0pt 0pt 0pt,scale=0.29]{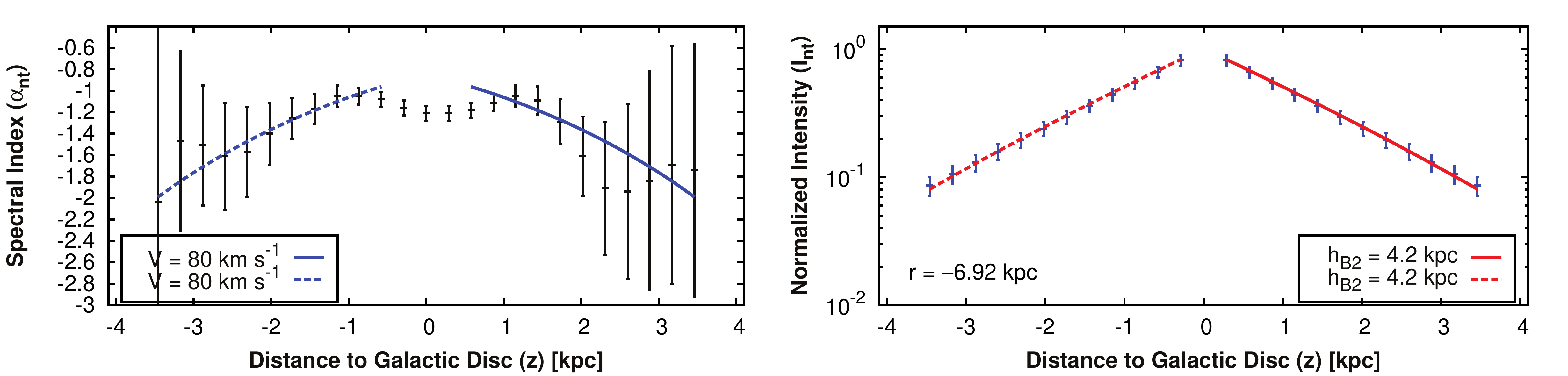}
 \includegraphics[clip=true,trim=0pt 0pt 0pt 0pt,scale=0.29]{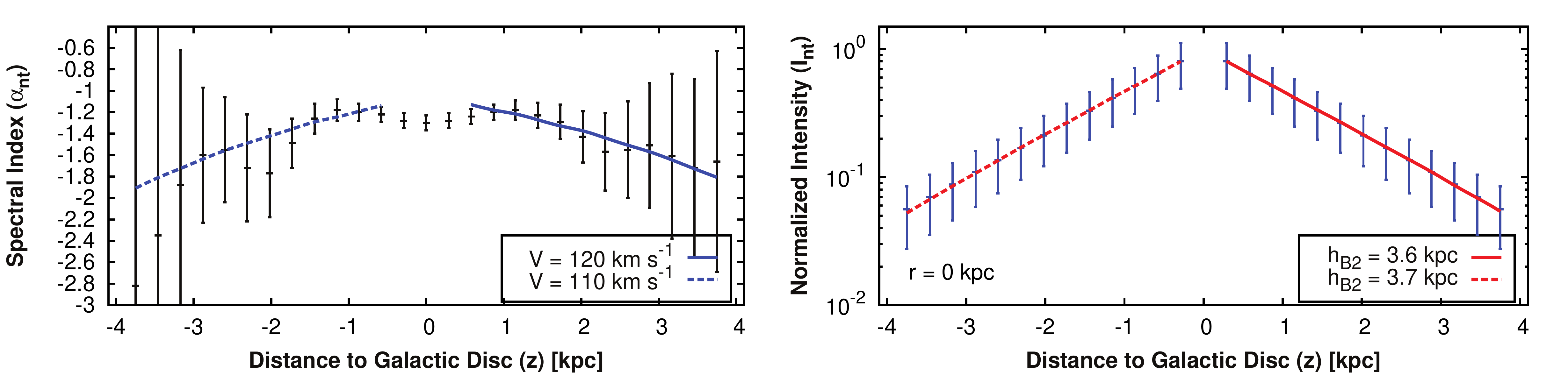}
 \includegraphics[clip=true,trim=0pt 0pt 0pt 0pt,scale=0.29]{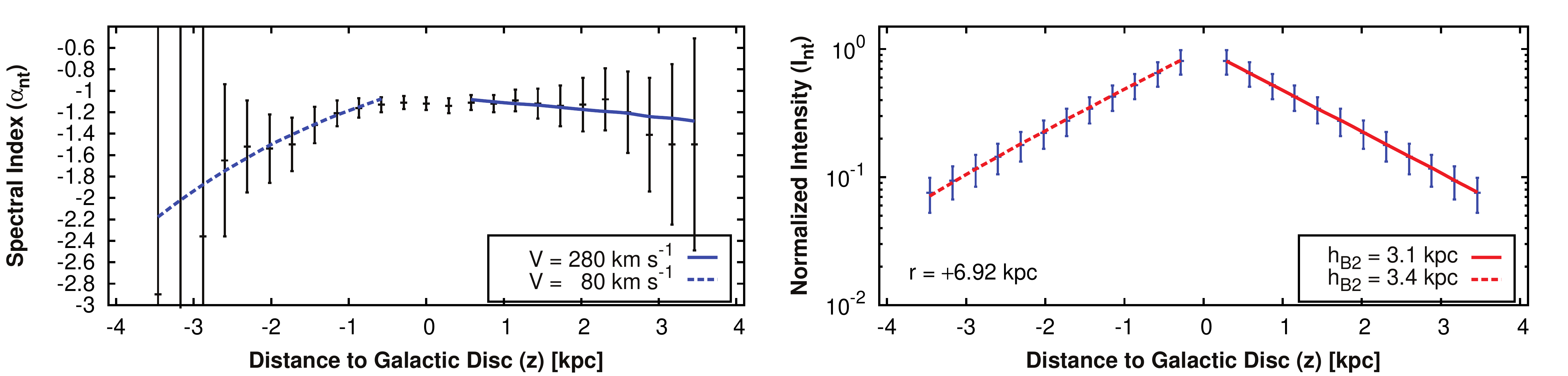}
 \includegraphics[clip=true,trim=0pt 0pt 0pt 0pt,scale=0.29]{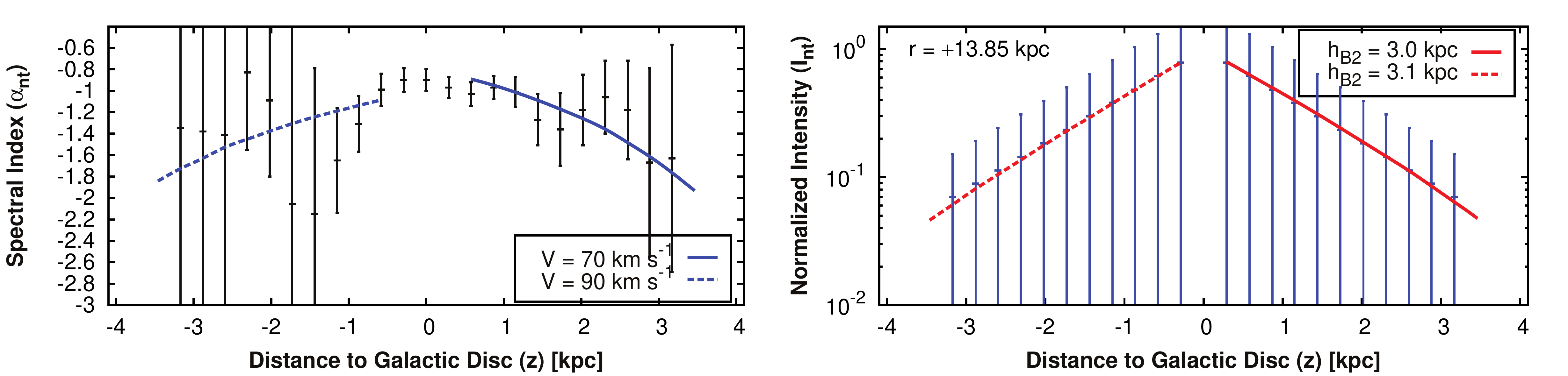} \\
\caption{Advection models with a constant advection speed for NGC\,4565. Data points denote the vertical profile of the non-thermal spectral index between 1.5 and 6\,GHz (left panels) and the exponential model of the non-thermal intensity profile at 1.5\,GHz (right panels). The radial position of each profile is given in the right-hand-side plot; $r<0$ is east of the minor axis and $r>0$ is west of the minor axis. Positive $z$ values are on the north side and negative ones on the south side of the mid-plane. Solid lines show the best-fitting advection models.}
 \label{fig:N4565adv1}
\end{figure*}

\begin{figure*}
 \centering
 \includegraphics[clip=true,trim=0pt 0pt 0pt 0pt,scale=0.29]{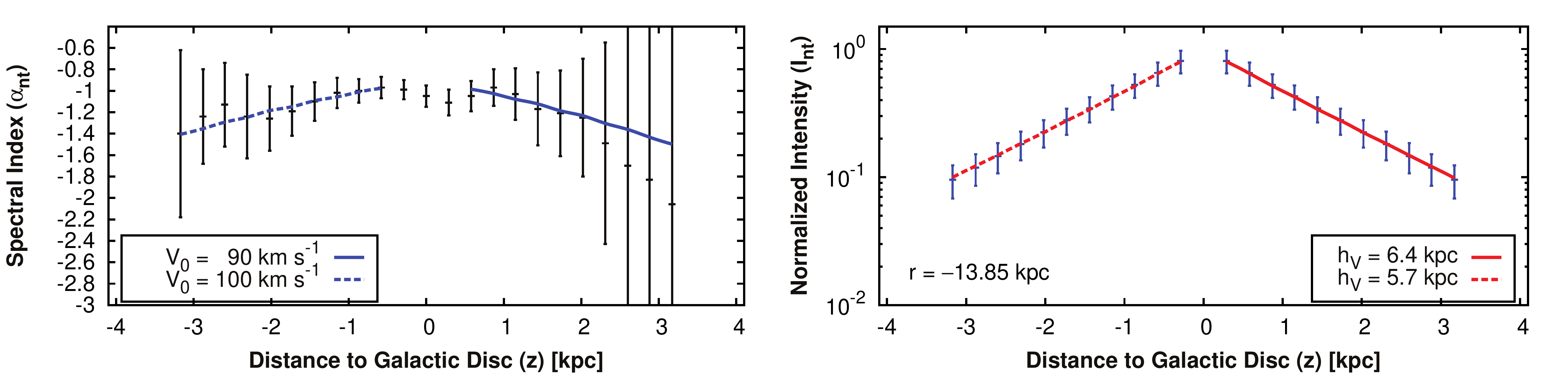}
 \includegraphics[clip=true,trim=0pt 0pt 0pt 0pt,scale=0.29]{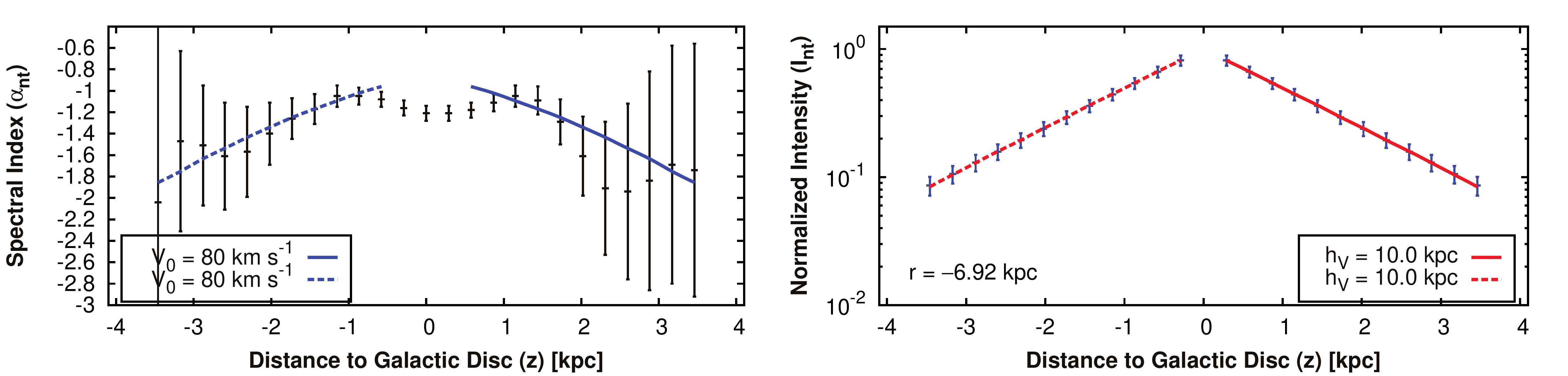}
 \includegraphics[clip=true,trim=0pt 0pt 0pt 0pt,scale=0.29]{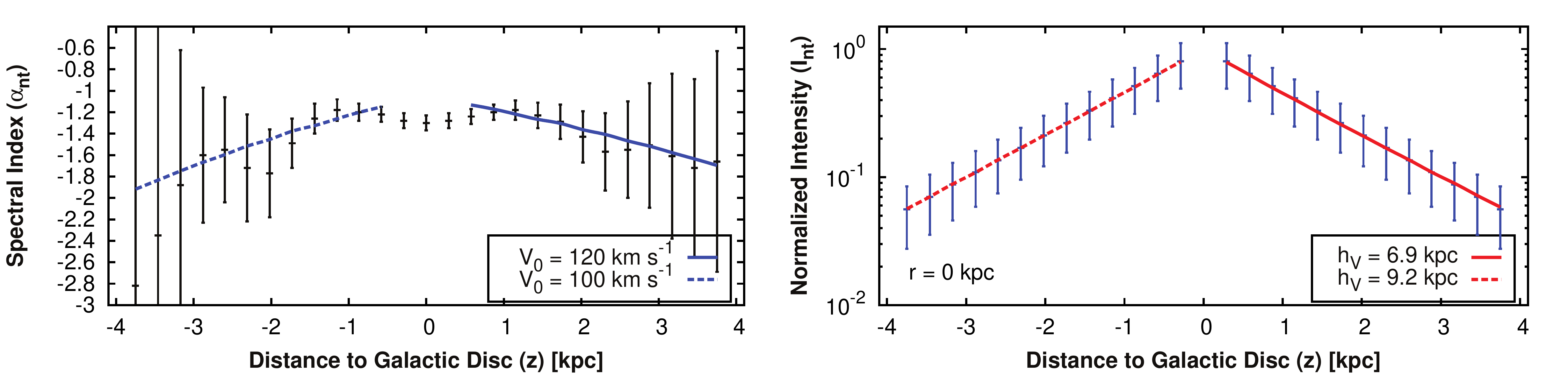}
 \includegraphics[clip=true,trim=0pt 0pt 0pt 0pt,scale=0.29]{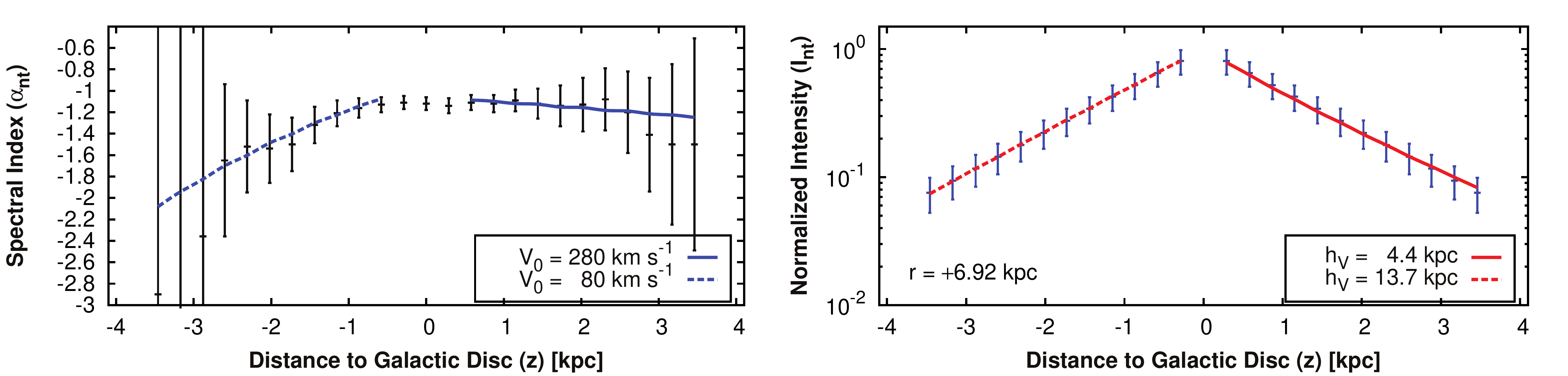}
 \includegraphics[clip=true,trim=0pt 0pt 0pt 0pt,scale=0.29]{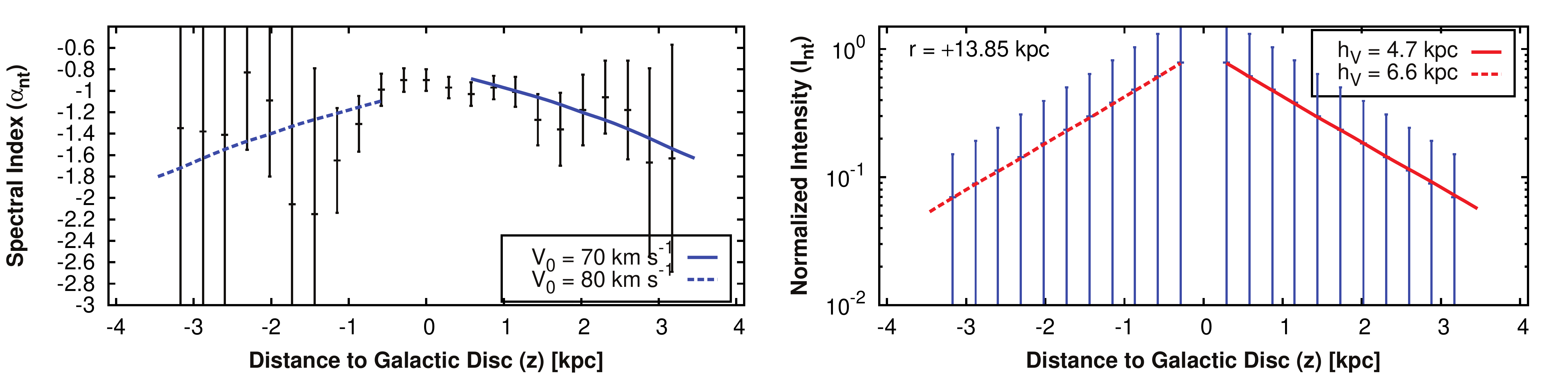} \\
\caption{Advection models with an accelerating advection speed for NGC\,4565. Data points denote the vertical profile of the non-thermal spectral index between 1.5 and 6\,GHz (left panels) and the exponential model of the non-thermal intensity profile at 1.5\,GHz (right panels). The radial position of each profile is given in the right-hand-side plot; $r<0$ is east of the minor axis and $r>0$ is west of the minor axis. Positive $z$ values are on the north side and negative ones on the south side of the mid-plane. Solid lines show the best-fitting advection models.}
 \label{fig:N4565adv2}
\end{figure*}

\begin{figure*}
 \centering
 \includegraphics[clip=true,trim=0pt 0pt 0pt 0pt,scale=0.29]{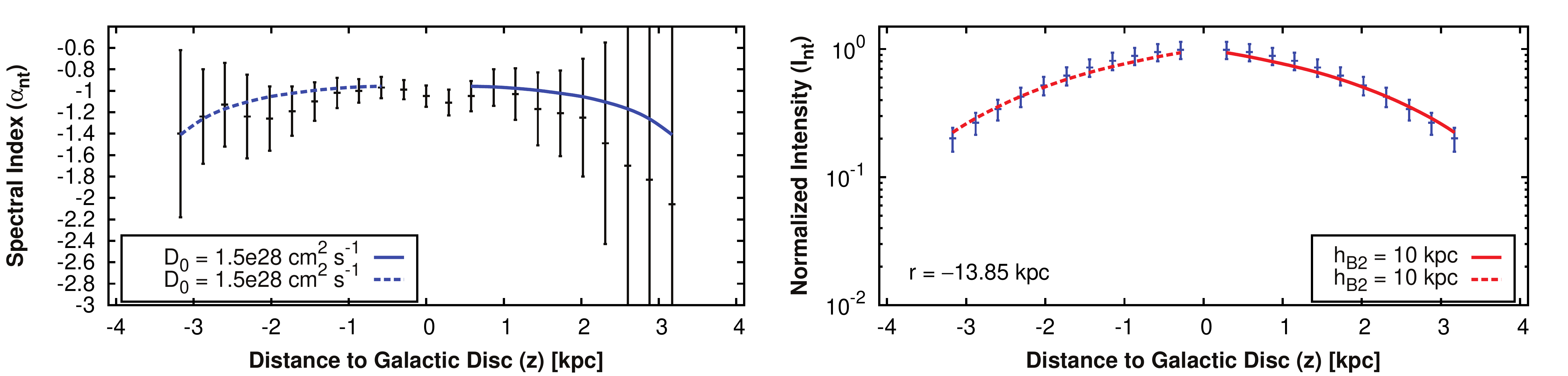}
 \includegraphics[clip=true,trim=0pt 0pt 0pt 0pt,scale=0.29]{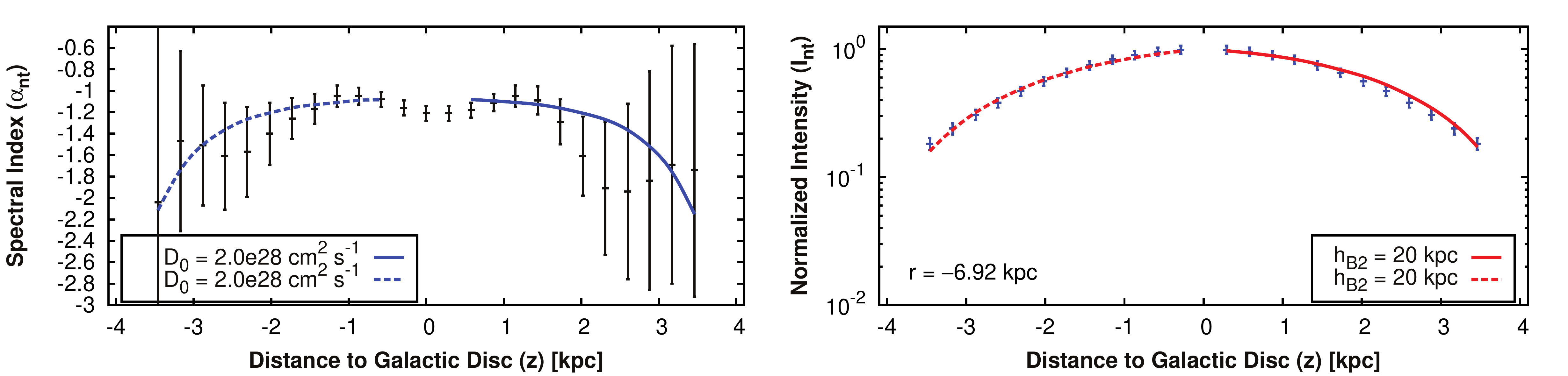}
 \includegraphics[clip=true,trim=0pt 0pt 0pt 0pt,scale=0.29]{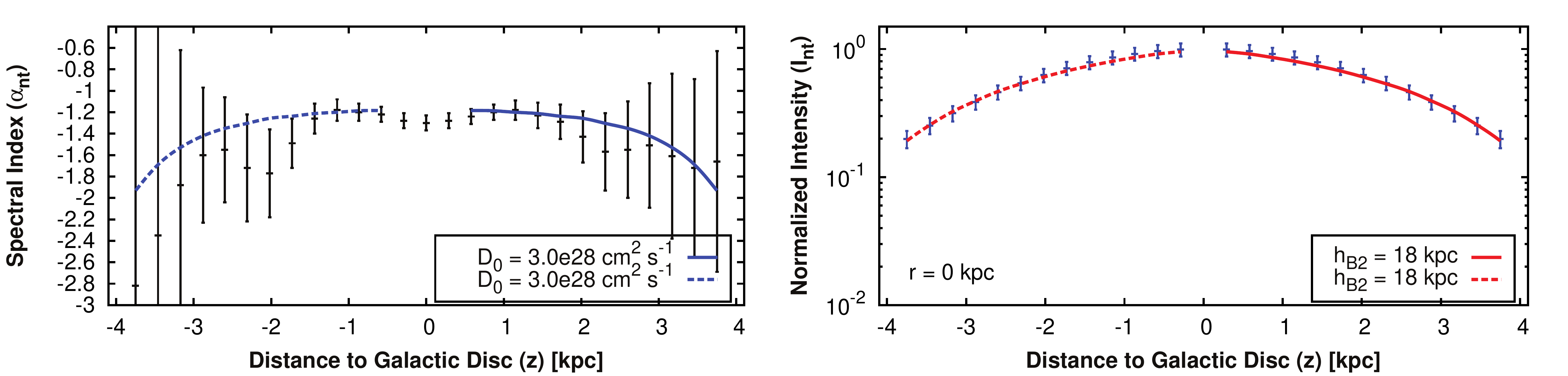}
 \includegraphics[clip=true,trim=0pt 0pt 0pt 0pt,scale=0.29]{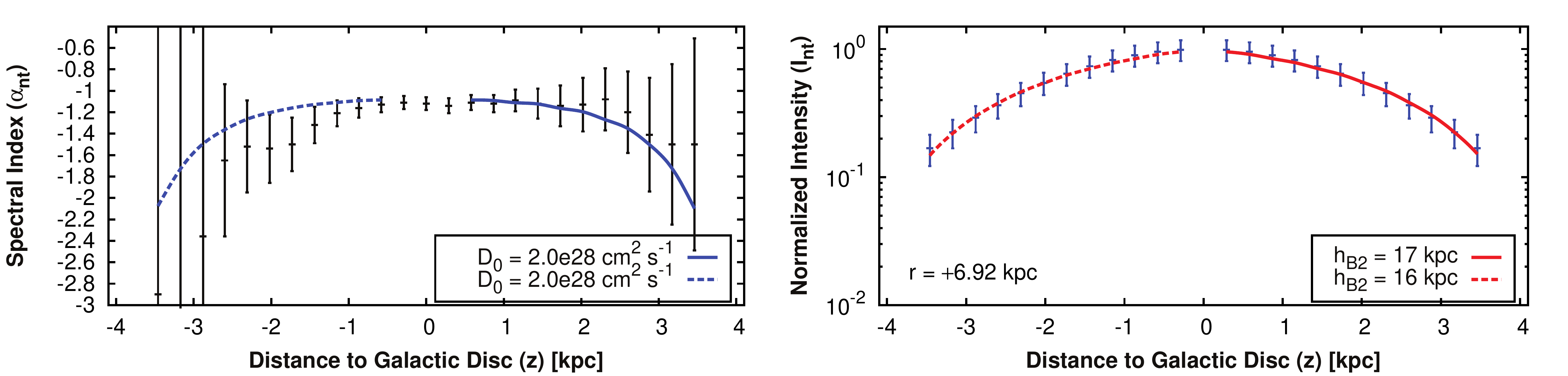}
 \includegraphics[clip=true,trim=0pt 0pt 0pt 0pt,scale=0.29]{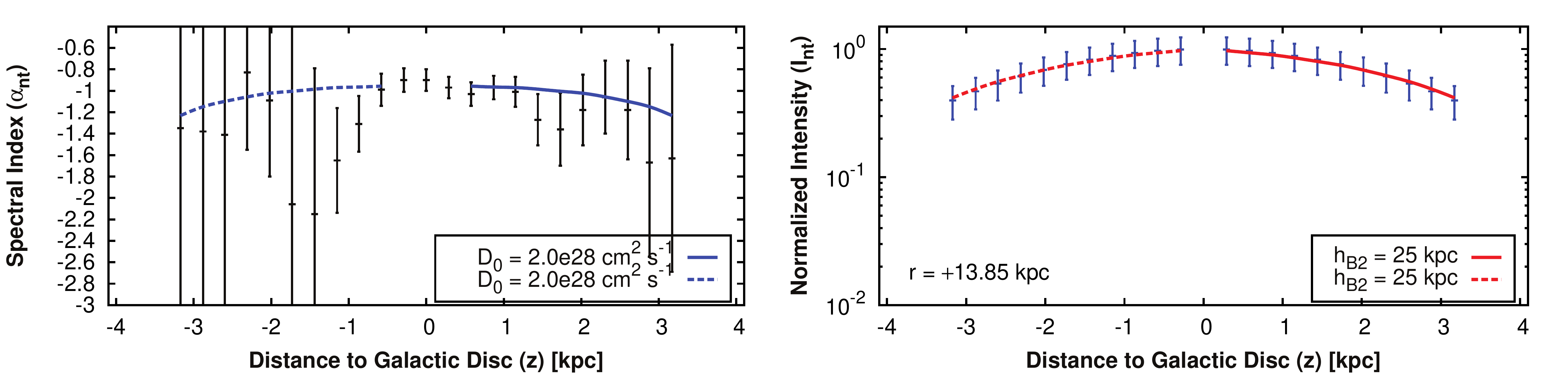}
\caption{Diffusion models with an energy-dependent diffusion coefficient [$D(E)=D_0(E/{1~\rm GeV})^{\mu}$] for NGC\,4565 (assuming $\mu=0.5$). Data points denote the vertical profile of the non-thermal spectral index between 1.5 and 6\,GHz (left panels) and the Gaussian model of the non-thermal intensity profile at 1.5\,GHz (right panels). The radial position of each profile is given in the right-hand-side plot; $r<0$ is east of the minor axis and $r>0$ is west of the minor axis. Positive $z$ values are on the north side and negative ones on the south side of the mid-plane. Solid lines show the best-fitting diffusion models.}
 \label{fig:N4565diff05}
\end{figure*}

\begin{figure*}
 \centering
 \includegraphics[clip=true,trim=0pt 0pt 0pt 0pt,scale=0.29]{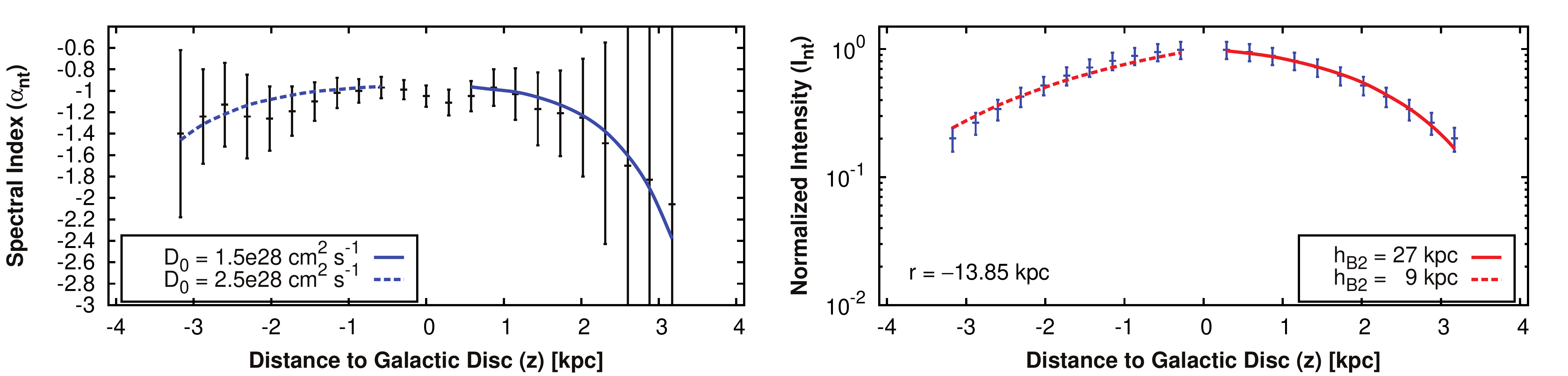}
 \includegraphics[clip=true,trim=0pt 0pt 0pt 0pt,scale=0.29]{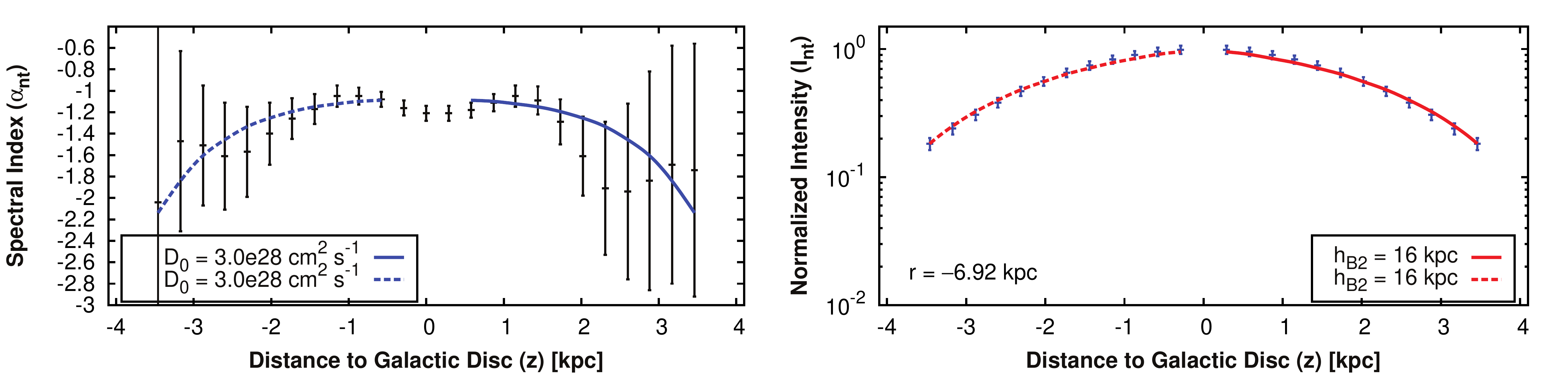}
 \includegraphics[clip=true,trim=0pt 0pt 0pt 0pt,scale=0.29]{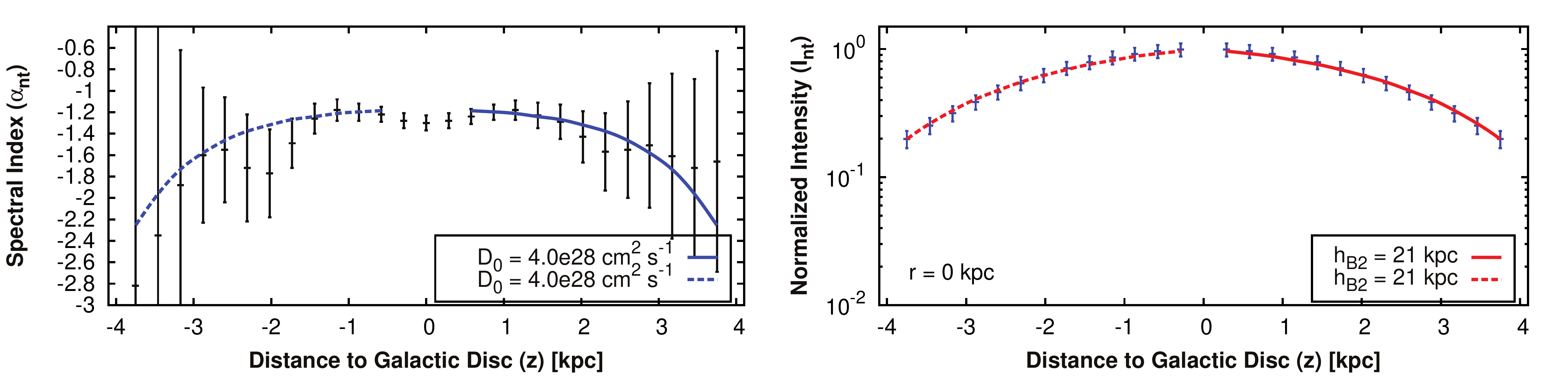}
 \includegraphics[clip=true,trim=0pt 0pt 0pt 0pt,scale=0.29]{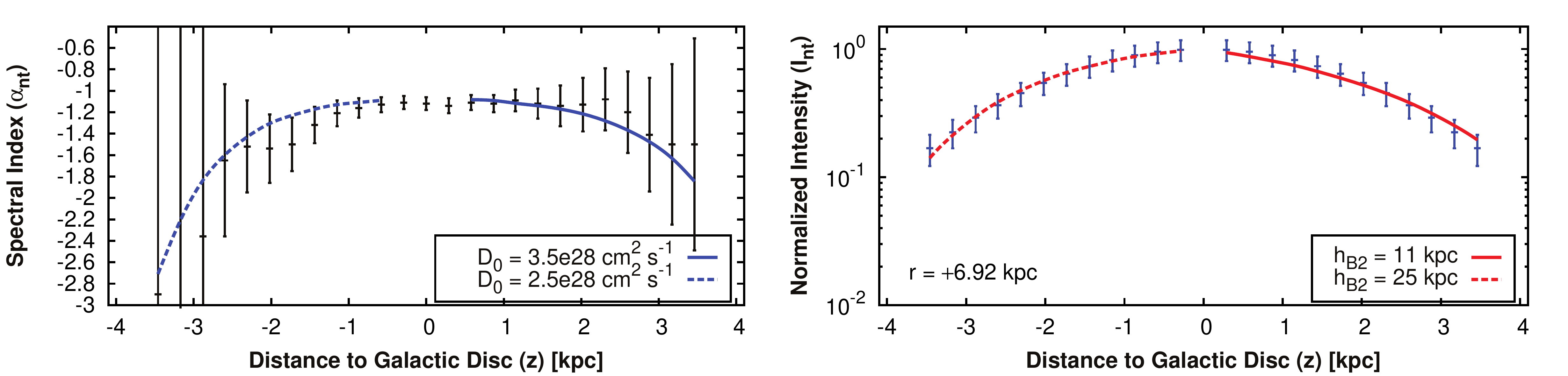}
 \includegraphics[clip=true,trim=0pt 0pt 0pt 0pt,scale=0.29]{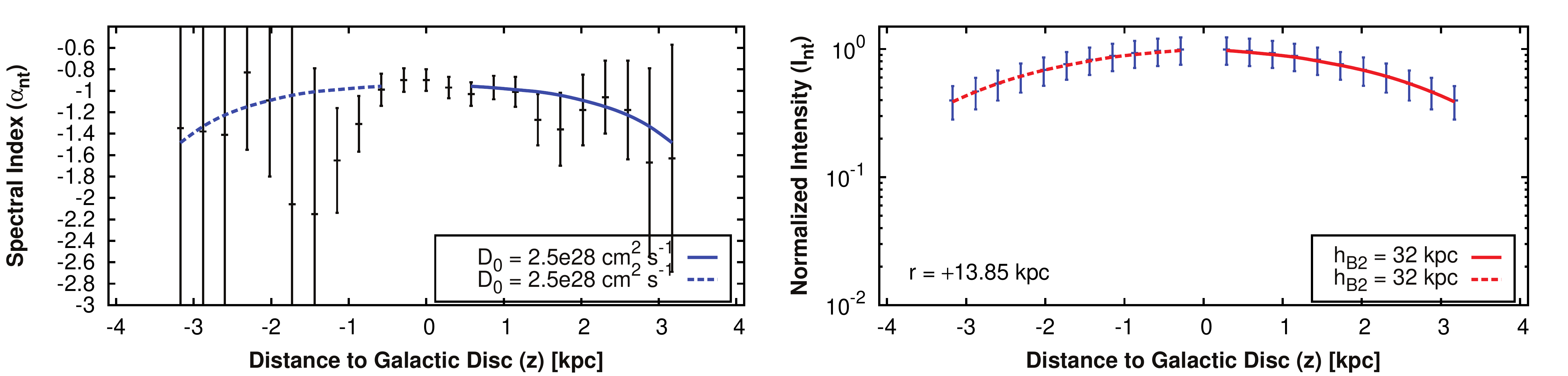}
\caption{Diffusion models with an energy-dependent diffusion coefficient [$D(E)=D_0(E/{1~\rm GeV})^{\mu}$] for NGC\,4565 (assuming $\mu=0.3$). Data points denote the vertical profile of the non-thermal spectral index between 1.5 and 6\,GHz (left panels) and the Gaussian model of the non-thermal intensity profile at 1.5\,GHz (right panels). The radial position of each profile is given in the right-hand-side plot; $r<0$ is east of the minor axis and $r>0$ is west of the minor axis. Positive $z$ values are on the north side and negative ones on the south side of the mid-plane. Solid lines show the best-fitting diffusion models.}
 \label{fig:N4565diff03}
\end{figure*}

In this appendix, we present the best-fitting advection and diffusion cosmic-ray transport models in the individual strips in NGC\,4565. In Table~\ref{tab:N4565adv1}, the best-fitting parameters are presented for a constant advection speed; the corresponding best-fitting intensity and spectral index profiles are presented in Fig.~\ref{fig:N4565adv1}. In Table~\ref{tab:N4565adv2} we present the best-fitting parameters for advection with a linearly accelerating advection velocity; the corresponding best-fitting intensity and spectral index profiles are presented in Fig.~\ref{fig:N4565adv2}. In Table~\ref{tab:N4565diff05} we present the best-fitting parameters for diffusion with an energy dependent diffusion coefficient assuming $\mu = 0.5$; the corresponding best-fitting intensity and spectral index profiles are presented in Fig.~\ref{fig:N4565diff05}. In Table~\ref{tab:N4565diff03} we present the best-fitting parameters for diffusion with an energy dependent diffusion coefficient assuming $\mu = 0.3$; the corresponding best-fitting intensity and spectral index profiles are presented in Fig.~\ref{fig:N4565diff03}. In Table~\ref{tab:N4565diff00} we present the best-fitting parameters for diffusion with an energy independent diffusion coefficient; the corresponding best-fitting intensity and spectral index profiles are presented in Fig.~\ref{fig:N4565diff00_CR}. 

\begin{table*}
\caption{Advection models with a constant advection speed for NGC\,891.}
{\small
\begin{center}
\begin{tabular}{c|lccclcc}
\toprule
\toprule[0.3pt]
$r$ [kpc] & $V$ [$\mathrm{km\,s^{-1}}$] & $B_{0}$ [$\mu$G] & $h_{\mathrm{B1}}$ [kpc] & $h_{\mathrm{B1}}/h_{\mathrm{B1,eq}}$ & $h_{\mathrm{B2}}$ [kpc] & $h_{\mathrm{B2}}/h_{\mathrm{B2,eq}}$ & $\chi^{2}_{I}+\chi^{2}_{\alpha}$ \\ 
\hline
$-7.94$ N & 110 $^{+30}_{-20}$  & 9.0 &   0.30 & 0.34 &  5.0 $^{+2.5}_{-1.3}$  & 0.83 & 1.22 \\ 
$-7.94$ S & 150 $^{+100}_{-40}$  & 9.0 &   0.30 & 0.34 &  4.1 $^{+1.8}_{-1.0}$  & 0.68 & 0.41 \\ 
$-5.29$ N & 160 $^{+40}_{-30}$  & 11.2 &   0.20 & 0.35 &  3.4 $^{+0.4}_{-0.4}$  & 0.62 & 0.30 \\ 
$-5.29$ S & 210 $^{+110}_{-40}$  & 11.2 &   0.20 & 0.35 &  3.2 $^{+0.2}_{-0.3}$  & 0.58 & 0.26 \\ 
$-2.65$ N & 160 $^{+20}_{-20}$  & 12.0 &   0.30 & 0.39 &  3.7 $^{+1.3}_{-0.7}$  & 0.69 & 1.00 \\ 
$-2.65$ S & 170 $^{+30}_{-30}$  & 12.0 &   0.30 & 0.39 &  3.6 $^{+1.4}_{-0.7}$  & 0.67 & 0.16 \\ 
    0 N & 140 $^{+30}_{-10}$  & 13.2 &   0.20 & 0.38 &  2.7 $^{+0.5}_{-0.5}$  & 0.61 & 0.29 \\ 
    0 S & 140 $^{+40}_{-20}$  & 13.2 &   0.20 & 0.38 &  2.7 $^{+0.6}_{-0.6}$  & 0.61 & 1.14 \\ 
 2.65 N & 150 $^{+40}_{-10}$  & 11.5 &   0.30 & 0.37 &  2.7 $^{+0.7}_{-0.6}$  & 0.64 & 0.07 \\ 
 2.65 S & 170 $^{+80}_{-20}$  & 11.5 &   0.30 & 0.37 &  2.6 $^{+0.6}_{-0.6}$  & 0.62 & 1.30 \\ 
 5.29 N & 140 $^{+40}_{-20}$  & 10.1 &   0.25 & 0.35 &  3.5 $^{+0.5}_{-0.5}$  & 0.65 & 0.18 \\ 
 5.29 S & 170 $^{+60}_{-10}$  & 10.1 &   0.25 & 0.35 &  3.2 $^{+0.3}_{-0.4}$  & 0.59 & 1.36 \\ 
 7.94 N & 130 $^{+60}_{-30}$  & 9.3 &  0.25  & 0.31 &  3.3 $^{+0.9}_{-0.6}$  & 0.63 & 0.36 \\ 
 7.94 S & 140 $^{+90}_{-20}$  & 9.3 &  0.25  & 0.31 &  3.2 $^{+0.5}_{-0.7}$  & 0.62 & 1.36 \\ 
\hline
\end{tabular}
\label{tab:N891adv1}
\end{center}
}
{\footnotesize 
\textit{Notes.} "N" denotes north of the mid-plane, "S" denotes south of the mid-plane; $V$: advection speed; $B_{0}$: mid-plane magnetic field strength; $h_{\mathrm{B1}}$: scale height of the disc magnetic field; $h_{\mathrm{B2}}$: scale height of the halo magnetic field; $h_{\mathrm{B1,eq}},h_{\mathrm{B2,eq}}$: scale height of the disc or halo magnetic field if energy equipartition is assumed; $\chi^{2}_{I}$: reduced $\chi^{2}$ of the fit to the intensity profile; $\chi^{2}_{\alpha}$: reduced $\chi^{2}$ of the fit to the spectral index profile. 
}
\end{table*}

\begin{table*}
\caption{Advection models with an accelerating advection speed for NGC\,891.}
{\small
\begin{center}
\begin{tabular}{c|lccccllc}
\toprule
\toprule[0.3pt]
$r$ [kpc] & $V_{0}$ [$\mathrm{km\,s^{-1}}$] & $B_{0}$ [$\mu$G] & $h_{\mathrm{B1}}$ [kpc] & $h_{\mathrm{B1}}/h_{\mathrm{B1,eq}}$ & $h_{\mathrm{B2}}$ ($\overset{!}{=}h_{\mathrm{B2,eq}}$) [kpc] & $h_{V}$ [kpc] & $h_{\mathrm{esc}}$ [kpc] & $\chi^{2}_{I}+\chi^{2}_{\alpha}$ \\ 
\hline
$-7.94$ N & 110 $^{+20}_{-20}$  & 9.0 &   0.30 & 0.34 & 6.0  & 20.0 $^{+\infty}_{-16.0}$ & 58.0 $^{+\infty}_{-32.1}$ &  1.20 \\ 
$-7.94$ S & 150 $^{+80}_{-40}$  & 9.0 &   0.30 & 0.34 & 6.0 & 9.6 $^{+56.9}_{-4.6}$ & 17.9 $^{+64.7}_{-7.5}$ &  0.44 \\ 
$-5.29$ N & 150 $^{+30}_{-20}$  & 11.2 &   0.20 & 0.35 & 5.5  & 6.6 $^{+5.2}_{-1.0}$ & 14.3 $^{+6.5}_{-2.2}$ &  0.28 \\ 
$-5.29$ S & 210 $^{+100}_{-40}$  & 11.2 &   0.20 & 0.35 & 5.5  & 4.7 $^{+2.2}_{-0.2}$ & 5.9 $^{+3.6}_{-1.4}$ &  0.28 \\ 
$-2.65$ N & 150 $^{+20}_{-10}$  & 12.0 &   0.30 & 0.39 & 5.4  & 9.9 $^{+\infty}_{-4.0}$ & 25.9 $^{+\infty}_{-4.7}$  & 0.84 \\ 
$-2.65$ S & 160 $^{+40}_{-20}$  & 12.0 &   0.30 & 0.39 & 5.4  & 9.0 $^{+213}_{-3.6}$ & 21.6 $^{+211}_{-4.2}$  & 0.19 \\ 
    0 N & 130 $^{+30}_{-10}$  & 13.2 &   0.20 & 0.38 & 4.4  & 4.9 $^{+6.8}_{-1.6}$ & 15.6 $^{+10.2}_{-2.5}$ &  0.29 \\ 
    0 S & 140 $^{+30}_{-20}$  & 13.2 &   0.20 & 0.38 & 4.4  & 4.6 $^{+7.1}_{-1.5}$ & 13.2 $^{+9.3}_{-2.4}$ & 1.21 \\ 
 2.65 N & 150 $^{+30}_{-20}$  & 11.5 &   0.30 & 0.37 & 4.2  & 5.2 $^{+8.8}_{-2.1}$ & 13.6 $^{+10.1}_{-2.8}$ & 0.09 \\ 
 2.65 S & 170 $^{+60}_{-30}$  & 11.5 &   0.30 & 0.37 & 4.2  & 4.6 $^{+7.1}_{-1.9}$ & 10.1 $^{+6.9}_{-2.2}$ & 1.26 \\ 
 5.29 N & 140 $^{+30}_{-20}$  & 10.1 &   0.25 & 0.35 & 5.4  & 6.8 $^{+9.0}_{-1.6}$ & 16.3 $^{+11.9}_{-3.0}$ &  0.12 \\ 
 5.29 S & 180 $^{+70}_{-40}$  & 10.1 &   0.25 & 0.35 & 5.4  & 5.4 $^{+4.3}_{-1.1}$ & 8.9 $^{+5.0}_{-2.3}$ &  1.20 \\ 
 7.94 N & 120 $^{+60}_{-20}$  & 9.3 &  0.25  & 0.31 & 5.2  & 6.6 $^{+20.6}_{-2.3}$ & 17.0 $^{+35.1}_{-4.8}$ &  0.30 \\ 
 7.94 S & 150 $^{+70}_{-30}$  & 9.3 &  0.25  & 0.31 & 5.2  & 5.2 $^{+8.3}_{-1.3}$ & 9.7 $^{+10.6}_{-2.6}$ &  1.20 \\ 
\hline
\end{tabular}
\label{tab:N891adv2}
\end{center}
}
{\footnotesize 
\textit{Notes.} "N" denotes north of the mid-plane, "S" denotes south of the mid-plane; $V_{0}$: initial advection speed in the mid-plane; $B_{0}$: mid-plane magnetic field strength; $h_{\mathrm{B1}}$: scale height of the disc magnetic field; $h_{\mathrm{B2}}$: scale height of the halo magnetic field; $h_{\mathrm{B1,eq}},h_{\mathrm{B2,eq}}$: scale height of the disc or halo magnetic field if energy equipartition is assumed; $h_{V}$: height where $V=2V_{0}$; $h_{\mathrm{esc}}$: height where $V=V_{\mathrm{esc}}$; $\chi^{2}_{I}$: reduced $\chi^{2}$ of the fit to the intensity profile; $\chi^{2}_{\alpha}$: reduced $\chi^{2}$ of the fit to the spectral index profile. 
}
\end{table*}

\begin{table*}
\caption{Diffusion models with an energy-dependent diffusion coefficient ($\mu=0.5$) for NGC\,891.}
{\small
\begin{center}
\begin{tabular}{c|cccclcc}
\toprule
\toprule[0.3pt]
$r$ [kpc] & $D_{0}$ [$\mathrm{10^{28}\,cm^{2}\,s^{-1}}$] & $B_{0}$ [$\mu$G] & $h_{\mathrm{B1}}$ [kpc] & $h_{\mathrm{B1}}/h_{\mathrm{B1,eq}}$ & $h_{\mathrm{B2}}$ [kpc] & $h_{\mathrm{B2}}/h_{\mathrm{B2,eq}}$ & $\chi^{2}_{I}+\chi^{2}_{\alpha}$ \\ 
\hline
$-7.94$ N &  2.5 $^{+2.5}_{-0.8}$  & 9.0 &  0.40  & 0.43  &  8.0 $^{+10.5}_{-3.2}$  & 1.33 & 1.35 \\ 
$-7.94$ S &  2.0 $^{+7.5}_{-0.4}$  & 9.0 &  0.40  & 0.43  &  8.7 $^{+9.7}_{-4.3}$  & 1.45 & 0.84 \\ 
$-5.29$ N &  2.0 $^{+0.5}_{-0.2}$  & 11.2 &  0.25  & 0.42  &  6.0 $^{+0.9}_{-1.2}$  & 1.09 & 2.49 \\ 
$-5.29$ S &  2.5 $^{+1.0}_{-0.5}$  & 11.2 &  0.25  & 0.42  &  5.2 $^{+1.2}_{-0.7}$  & 0.95 & 1.03 \\ 
$-2.65$ N &  2.5 $^{+0.5}_{-0.5}$  & 12.0 &  0.25  & 0.31  &  4.6 $^{+2.1}_{-1.2}$  & 0.85 & 2.86 \\ 
$-2.65$ S &  2.5 $^{+0.5}_{-0.5}$  & 12.0 &  0.25  & 0.31  &  4.6 $^{+1.3}_{-1.2}$  & 0.85 & 1.06 \\ 
    0 N &  2.5 $^{+1.0}_{-0.4}$  & 13.2 &  0.20  & 0.36  &  3.7 $^{+1.2}_{-1.0}$  & 0.84 & 0.31 \\ 
    0 S &  1.5 $^{+0.5}_{-0.1}$  & 13.2 &  0.20  & 0.36  &  3.8 $^{+1.5}_{-1.1}$  & 0.86 & 3.63 \\ 
 2.65 N &  2.0 $^{+0.5}_{-0.2}$  & 11.5 &  0.35  & 0.42  &  4.8 $^{+3.0}_{-1.6}$  & 1.14 & 0.59 \\ 
 2.65 S &  2.5 $^{+2.0}_{-0.5}$  & 11.5 &  0.35  & 0.42  &  4.0 $^{+2.9}_{-1.2}$  & 0.95 & 2.15 \\ 
 5.29 N &  2.0 $^{+1.0}_{-0.3}$  & 10.1 &  0.30  & 0.39  &  6.1 $^{+1.8}_{-1.7}$  & 1.13 & 0.63 \\ 
 5.29 S &  3.5 $^{+54.0}_{-1.0}$  & 10.1 &  0.30  & 0.39  &  3.9 $^{+1.0}_{-0.9}$  & 0.72 & 1.43 \\ 
 7.94 N &  2.0 $^{+11.0}_{-0.5}$  & 9.3 &  0.35  & 0.42  &  5.8 $^{+3.4}_{-2.3}$  & 1.12 & 0.55 \\ 
 7.94 S &  2.5 $^{+\infty}_{-0.5}$  & 9.3 &  0.35  & 0.42  &  5.1 $^{+1.4}_{-1.9}$  & 0.98 & 1.54 \\ 
\hline
\end{tabular}
\label{tab:N891diff}
\end{center}
}
{\footnotesize 
\textit{Notes.} "N" denotes north of the mid-plane, "S" denotes south of the mid-plane; $D_{0}$: diffusion coefficient [$D(E)=D_{0}(E/1~{\rm GeV})^{\,\mu}$]; $B_{0}$: mid-plane magnetic field strength; $h_{\mathrm{B1}}$: scale height of the disc magnetic field; $h_{\mathrm{B2}}$: scale height of the halo magnetic field; $h_{\mathrm{B1,eq}},h_{\mathrm{B2,eq}}$: scale height of the disc or halo magnetic field if energy equipartition is assumed; $\chi^{2}_{I}$: reduced $\chi^{2}$ of the fit to the intensity profile; $\chi^{2}_{\alpha}$: reduced $\chi^{2}$ of the fit to the spectral index profile. 
}
\end{table*}


\begin{table*}
\caption{Advection models with a constant advection speed for NGC\,4565.}
{\small
\begin{center}
\begin{tabular}{c|cclcc}
\toprule
\toprule[0.3pt]
$r$ [kpc] & $V$ [$\mathrm{km\,s^{-1}}$] & $B_{0}$ [$\mu$G] & $h_{\mathrm{B2}}$ [kpc] & $h_{\mathrm{B2}}/h_{\mathrm{B2,eq}}$ & $\chi^{2}_{I}+\chi^{2}_{\alpha}$ \\ 
\hline
$-13.85$ N &  100 $^{+\infty}_{-50}$  & 5.5 &  3.4 $^{+6.2}_{-1.2}$  & 0.61 & 0.05 \\ 
$-13.85$ S &  120 $^{+3700}_{-50}$  & 5.5 &  3.2 $^{+2.1}_{-1.0}$  & 0.57 & 0.06 \\ 
 $-6.92$ N &  80 $^{+50}_{-20}$  & 6.4 &  4.2 $^{+2.9}_{-1.2}$  & 0.72 & 0.46 \\ 
 $-6.92$ S &  80 $^{+60}_{-10}$  & 6.4 &  4.2 $^{+1.3}_{-1.3}$  & 0.72 & 0.17 \\ 
     0 N &  120 $^{+140}_{-50}$  & 6.3 &  3.6 $^{+3.5}_{-1.5}$  & 0.67 & 0.07 \\ 
     0 S &  110 $^{+250}_{-20}$  & 6.3 &  3.7 $^{+2.4}_{-1.6}$  & 0.69 & 0.20 \\ 
  6.92 N &  280 $^{+\infty}_{-160}$  & 6.5 &  3.1 $^{+1.5}_{-0.8}$  & 0.55 & 0.08 \\ 
  6.92 S &  80 $^{+70}_{-20}$  & 6.5 &  4.4 $^{+6.0}_{-1.7}$  & 0.79 & 0.05 \\ 
 13.85 N &  70 $^{+610}_{-20}$  & 5.2 &  3.0 $^{+5.9}_{-2.4}$  & 0.61 & 0.31 \\ 
 13.85 S &  90 $^{+\infty}_{-50}$  & 5.2 &  3.1 $^{+116}_{-2.6}$  & 0.63 & 0.41 \\ 
\hline
\end{tabular}
\label{tab:N4565adv1}
\end{center}
}
{\footnotesize 
\textit{Notes.} "N" denotes north of the mid-plane, "S" denotes south of the mid-plane; $V$: advection speed; $B_{0}$: mid-plane magnetic field strength; $h_{\mathrm{B1}}$: scale height of the disc magnetic field; $h_{\mathrm{B2}}$: scale height 
of the halo magnetic field; $h_{\mathrm{B1,eq}},h_{\mathrm{B2,eq}}$: scale height of the disc or halo magnetic field if energy equipartition is assumed; 
$\chi^{2}_{I}$: reduced $\chi^{2}$ of the fit to the intensity profile; $\chi^{2}_{\alpha}$: reduced $\chi^{2}$ of the fit to the spectral index profile. 
}
\end{table*}

\begin{table*}
\caption{Advection models with an accelerating advection speed for NGC\,4565.}
{\small
\begin{center}
\begin{tabular}{c|cccllc}
\toprule
\toprule[0.3pt]
$r$ [kpc] & $V_{0}$ [$\mathrm{km\,s^{-1}}$] & $B_{0}$ [$\mu$G] & $h_{\mathrm{B2}}$ ($\overset{!}{=}h_{\mathrm{B2,eq}}$) [kpc] & $h_{V}$ [kpc] & $h_{\mathrm{esc}}$ [kpc] & $\chi^{2}_{I}+\chi^{2}_{\alpha}$ \\ 
\hline
$-13.85$ N &  90 $^{+\infty}_{-40}$  & 5.5 &  5.6  & 6.4 $^{+\infty}_{-4.3}$ & 24.7 $^{+\infty}_{-16.4}$ & 0.05 \\ 
$-13.85$ S & 100 $^{+2200}_{-30}$  & 5.5 &  5.6  & 5.7 $^{+74.6}_{-3.5}$ & 19.2 $^{+472}_{-10.5}$ & 0.07 \\ 
 $-6.92$ N &  80 $^{+50}_{-20}$  & 6.4 &  5.8  & 10.0 $^{+\infty}_{-6.0}$ & 55.5 $^{+\infty}_{-22.7}$ & 0.42 \\ 
 $-6.92$ S &  80 $^{+50}_{-10}$  & 6.4 &  5.8  & 10.0 $^{+62.0}_{-6.2}$ & 55.5 $^{+208}_{-21.4}$ & 0.09 \\ 
     0 N & 120 $^{+220}_{-50}$  & 6.3 &  5.4  & 6.9 $^{+\infty}_{-5.0}$ & 23.2 $^{+\infty}_{-12.6}$ & 0.09 \\ 
     0 S & 100 $^{+230}_{-30}$  & 6.3 &  5.4  & 9.2 $^{+\infty}_{-7.3}$ & 39.0 $^{+\infty}_{-20.4}$ & 0.22 \\ 
  6.92 N & 280 $^{+\infty}_{-160}$  & 6.5 &  5.6  & 4.4 $^{+16.2}_{-2.1}$ & 3.8 $^{+\infty}_{-1.0}$ & 0.11 \\ 
  6.92 S &  80 $^{+60}_{-20}$  & 6.5 &  5.6  & 13.7 $^{+\infty}_{-10.6}$ & 76.0 $^{+\infty}_{-38.3}$ & 0.05 \\ 
 13.85 N &  70 $^{+130}_{-50}$  & 5.2 &  4.9  & 4.7 $^{+\infty}_{-4.6}$ & 24.7 $^{+\infty}_{-18.6}$ & 0.27 \\ 
 13.85 S &  80 $^{+\infty}_{-60}$  & 5.2 &  4.9  & 6.6 $^{+\infty}_{-6.5}$ & 29.5 $^{+\infty}_{-22.0}$ & 0.40 \\ 
\hline
\end{tabular}
\label{tab:N4565adv2}
\end{center}
}
{\footnotesize 
\textit{Notes.} "N" denotes north of the mid-plane, "S" denotes south of the mid-plane; $V_{0}$: initial advection speed in the mid-plane; $B_{0}$: mid-plane magnetic field strength; $h_{\mathrm{B1}}$: scale height of the disc magnetic field; $h_{\mathrm{B2}}$: 
scale height of the halo magnetic field; $h_{\mathrm{B1,eq}},h_{\mathrm{B2,eq}}$: scale height of the disc or halo magnetic field if energy equipartition is 
assumed; $h_{V}$: height where $V=2V_{0}$; $h_{\mathrm{esc}}$: height where $V=V_{\mathrm{esc}}$; $\chi^{2}_{I}$: reduced $\chi^{2}$ 
of the fit to the intensity profile; $\chi^{2}_{\alpha}$: reduced $\chi^{2}$ of the fit to the spectral index profile. 
}
\end{table*}

\begin{table*}
\caption{Diffusion models with an energy-dependent diffusion coefficient for NGC\,4565 ($\mu=0.5$).}
{\small
\begin{center}
\begin{tabular}{c|cclcc}
\toprule
\toprule[0.3pt]
$r$ [kpc] & $D_{0}$ [$\mathrm{10^{28}\,cm^{2}\,s^{-1}}$] & $B_{0}$ [$\mu$G] & $h_{\mathrm{B2}}$ [kpc] & $h_{\mathrm{B2}}/h_{\mathrm{B2,eq}}$ & $\chi^{2}_{I}+\chi^{2}_{\alpha}$ \\ 
\hline
$-13.85$ N &  1.5 $^{+1.0}_{-0.5}$  & 5.5 &  10 $^{+\infty}_{-4}$  & 1.96 & 0.40 \\ 
$-13.85$ S &  1.5 $^{+1.0}_{-0.5}$  & 5.5 &  10 $^{+\infty}_{-4}$  & 1.96 & 0.48 \\ 
 $-6.92$ N &  2.0 $^{+0.5}_{-0.5}$  & 6.4 &  20 $^{+15}_{-6}$  & 3.70 & 1.43 \\ 
 $-6.92$ S &  2.0 $^{+0.5}_{-0.5}$  & 6.4 &  20 $^{+15}_{-6}$  & 3.70 & 1.25 \\ 
     0 N &  3.0 $^{+1.0}_{-0.5}$  & 6.3 &  18 $^{+\infty}_{-8}$  & 3.00 & 0.32 \\ 
     0 S &  3.0 $^{+1.0}_{-0.5}$  & 6.3 &  18 $^{+\infty}_{-8}$  & 3.00 & 0.59 \\ 
  6.92 N &  2.0 $^{+3.0}_{-0.5}$  & 6.5 &  17 $^{+83}_{-11}$  & 3.21 & 0.22 \\ 
  6.92 S &  2.0 $^{+3.0}_{-0.5}$  & 6.5 &  16 $^{+84}_{-10}$  & 3.02 & 0.83 \\ 
 13.85 N &  2.0 $^{+\infty}_{-1.0}$  & 5.2 &  25 $^{+\infty}_{-19}$  & 3.73 & 0.49 \\ 
 13.85 S &  2.0 $^{+\infty}_{-1.0}$  & 5.2 &  25 $^{+\infty}_{-19}$  & 3.73 & 0.72 \\ 
\hline
\end{tabular}
\label{tab:N4565diff05}
\end{center}
}
{\footnotesize 
\textit{Notes.} "N" denotes north of the mid-plane, "S" denotes south of the mid-plane; $D_{0}$: diffusion coefficient [$D(E)=D_{0}(E/{\rm 1~GeV})^{\,\mu}$]; $B_{0}$: mid-plane magnetic field strength; $h_{\mathrm{B1}}$: scale height of the disc magnetic field;
$h_{\mathrm{B2}}$: scale height of the halo magnetic field; $h_{\mathrm{B1,eq}},h_{\mathrm{B2,eq}}$: scale height of the disc or halo magnetic field if energy 
equipartition is assumed; $\chi^{2}_{I}$: reduced $\chi^{2}$ of the fit to the intensity profile; $\chi^{2}_{\alpha}$: reduced $\chi^{2}$ of the fit to the spectral index profile.}
\end{table*}

\begin{table*}
\caption{Diffusion models with an energy-dependent diffusion coefficient for NGC\,4565 ($\mu=0.3$).}
{\small
\begin{center}
\begin{tabular}{c|cclcc}
\toprule
\toprule[0.3pt]
$r$ [kpc] & $D_{0}$ [$\mathrm{10^{28}\,cm^{2}\,s^{-1}}$] & $B_{0}$ [$\mu$G] & $h_{\mathrm{B2}}$ [kpc] & $h_{\mathrm{B2}}/h_{\mathrm{B2,eq}}$ & $\chi^{2}_{I}+\chi^{2}_{\alpha}$ \\ 
\hline
$-13.85$ N &  1.5 $^{+3.0}_{-0.5}$  & 5.5 &  27 $^{+\infty}_{-19}$  & 5.29 & 0.13 \\ 
$-13.85$ S &  2.5 $^{+3.0}_{-1.5}$  & 5.5 &  9 $^{+\infty}_{-2}$  & 1.76 & 0.60 \\ 
 $-6.92$ N &  3.0 $^{+0.5}_{-0.5}$  & 6.4 &  16 $^{+54}_{-6}$  & 2.96 & 0.74 \\ 
 $-6.92$ S &  3.0 $^{+0.5}_{-0.5}$  & 6.4 &  16 $^{+54}_{-6}$  & 2.96 & 0.57 \\ 
     0 N &  4.0 $^{+2.0}_{-0.5}$  & 6.3 &  21 $^{+\infty}_{-11}$  & 3.50 & 0.23 \\ 
     0 S &  4.0 $^{+2.0}_{-0.5}$  & 6.3 &  21 $^{+\infty}_{-11}$  & 3.50 & 0.41 \\ 
  6.92 N &  3.5 $^{+3.5}_{-1.0}$  & 6.5 &  11 $^{+\infty}_{-6}$  & 2.08 & 0.35 \\ 
  6.92 S &  2.5 $^{+4.5}_{-0.5}$  & 6.5 &  25 $^{+\infty}_{-19}$  & 4.72 & 0.48 \\ 
 13.85 N &  2.5 $^{+\infty}_{-1.0}$  & 5.2 &  32 $^{+\infty}_{-25}$  & 4.78 & 0.35 \\ 
 13.85 S &  2.5 $^{+\infty}_{-1.0}$  & 5.2 &  32 $^{+\infty}_{-25}$  & 4.78 & 0.69 \\ 
\hline
\end{tabular}
\label{tab:N4565diff03}
\end{center}
}
{\footnotesize 
\textit{Notes.} "N" denotes north of the mid-plane, "S" denotes south of the mid-plane; $D_{0}$: diffusion coefficient [$D(E)=D_{0}(E/{\rm 1~GeV})^{\,\mu}$]; $B_{0}$: mid-plane magnetic field strength; $h_{\mathrm{B1}}$: scale height of the disc magnetic field;
$h_{\mathrm{B2}}$: scale height of the halo magnetic field; $h_{\mathrm{B1,eq}},h_{\mathrm{B2,eq}}$: scale height of the disc or halo magnetic field if energy 
equipartition is assumed; $\chi^{2}_{I}$: reduced $\chi^{2}$ of the fit to the intensity profile; $\chi^{2}_{\alpha}$: reduced $\chi^{2}$ of the fit to the spectral index profile.}
\end{table*}

\begin{table*}
\caption{Diffusion models with no energy dependency of the diffusion coefficient for NGC\,4565 ($\mu=0$).}
{\small
\begin{center}
\begin{tabular}{c|cclcc}
\toprule
\toprule[0.3pt]
$r$ [kpc] & $D_{0}$ [$\mathrm{10^{28}\,cm^{2}\,s^{-1}}$] & $B_{0}$ [$\mu$G] & $h_{\mathrm{B2}}$ [kpc] & $h_{\mathrm{B2}}/h_{\mathrm{B2,eq}}$ & $\chi^{2}_{I}+\chi^{2}_{\alpha}$ \\ 
\hline
$-13.85$ N &  2.5 $^{+4.0}_{-0.5}$  & 5.5 &  27 $^{+\infty}_{-21}$  & 5.29 & 0.06 \\ 
$-13.85$ S &  3.5 $^{+3.0}_{-1.0}$  & 5.5 &  12 $^{+32}_{-5}$  & 2.35 & 0.58 \\ 
 $-6.92$ N &  4.5 $^{+0.5}_{-1.0}$  & 6.4 &  19 $^{+48}_{-7}$  & 3.52 & 0.67 \\ 
 $-6.92$ S &  4.5 $^{+0.5}_{-0.5}$  & 6.4 &  19 $^{+48}_{-7}$  & 3.52 & 0.65 \\ 
     0 N &  5.5 $^{+1.0}_{-0.5}$  & 6.3 &  34 $^{+100}_{-10}$  & 5.67 & 0.59 \\ 
     0 S &  5.5 $^{+1.0}_{-0.5}$  & 6.3 &  34 $^{+100}_{-10}$  & 5.67 & 0.28 \\ 
  6.92 N &  8.0 $^{+7.0}_{-3.0}$  & 6.5 &  8 $^{+24}_{-2}$  & 1.51 & 0.58 \\ 
  6.92 S &  4.0 $^{+11.0}_{-1.0}$  & 6.5 &  24 $^{+\infty}_{-18}$  & 4.53 & 0.17 \\ 
 13.85 N &  4.5 $^{+6.5}_{-2.0}$  & 5.2 &  28 $^{+\infty}_{-20}$  & 4.18 & 0.29 \\ 
 13.85 S &  4.5 $^{+6.5}_{-2.0}$  & 5.2 &  29 $^{+\infty}_{-21}$  & 4.33 & 0.67 \\ 
\hline
\end{tabular}
\label{tab:N4565diff00}
\end{center}
}
{\textit{Notes.} "N" denotes north of the mid-plane, "S" denotes south of the mid-plane; $D_{0}$: diffusion coefficient [$D(E)=D_{0}(E/{\rm 1~GeV})^{\,\mu}$]; $B_{0}$: mid-plane magnetic field strength; $h_{\mathrm{B1}}$: scale height of the disc magnetic field;
$h_{\mathrm{B2}}$: scale height of the halo magnetic field; $h_{\mathrm{B1,eq}},h_{\mathrm{B2,eq}}$: scale height of the disc or halo magnetic field if energy 
equipartition is assumed; $\chi^{2}_{I}$: reduced $\chi^{2}$ of the fit to the intensity profile; $\chi^{2}_{\alpha}$: reduced $\chi^{2}$ of the fit to the spectral index profile.}
\end{table*}

\FloatBarrier

\clearpage

\end{document}